\documentclass[aps,twocolumn,pra,superscriptaddress]{revtex4-2}
\newcounter{defcounter}
\usepackage{graphicx}
\usepackage{amsmath}

\usepackage{multirow}
\usepackage{xcolor}
\usepackage[normalem]{ulem}
\usepackage{color}

\usepackage{relsize}
\usepackage{amssymb}
\usepackage{ccaption}

\usepackage{algpseudocode}

\newif\ifcomment
\commentfalse
\ifcomment
\newcommand{\add}[1]{\textcolor{blue}{#1}}
\newcommand{\delete}[1]{\textcolor{red}{\sout{#1}}}
\newcommand{\edit}[2]{\textcolor{red}{\sout{#1}} \textcolor{blue}{#2}}
\newcommand{\mnote}[1]{\marginpar{\textcolor{green}{\textbf{#1}}}}
\else
\newcommand{\add}[1]{#1}
\newcommand{\delete}[1]{}
\newcommand{\edit}[2]{#2}
\newcommand{\mnote}[1]{}
\fi
\let\oldaddcontentsline\addcontentsline
\renewcommand{\addcontentsline}[3]{}
\makeatletter
\def\maketitle{
	\@author@finish
	\title@column\titleblock@produce
	\suppressfloats[t]}
\makeatother

\newcommand{\be}{\begin{equation}}

\newcommand{\ee}{\end{equation}}
\newcommand{\bme}{\begin{myequation}}
	
	\newcommand{\eme}{\end{myequation}}

\newcommand{\bea}{\begin{eqnarray}}
\newcommand{\eea}{\end{eqnarray}}
\newcommand{\Eq}[1]{Eq.(\ref{#1})}
\newcommand{\Fig}[1]{Fig.\,\ref{#1}}
\newcommand{\Sec}[1]{Sec.\,\ref{#1}}
\newcommand{\Tab}[1]{Table \,\ref{#1}}
\newcommand{\Onlinecite}[1]{Ref.\,\cite{#1}} 

\newcommand{\bD}{\ensuremath{\mathbf{D}}}

\newcommand{\bDw}{\ensuremath{\mathbf{D}^{\rm w}}}
\newcommand{\Dw}{\ensuremath{D^{\rm w}}}

\newcommand{\bT}{\ensuremath{\mathbf{T}}}
\newcommand{\bbT}{\ensuremath{\mathbf{T}^{\rm f}}}
\newcommand{\bTw}{\ensuremath{\mathbf{T}^{\rm w}}}
\newcommand{\bTd}{\ensuremath{\mathbf{T}^{\rm d}}}
\newcommand{\bTa}{\ensuremath{\mathbf{T}^{\rm a}}}

\newcommand{\bTAF}{\ensuremath{\mathbf{T}^{\rm u}}}

\newcommand{\bS}{\ensuremath{\mathbf{S}}}
\newcommand{\bSw}{\ensuremath{\mathbf{S}^{\rm w}}}

\newcommand{\bTp}[1]{\ensuremath{\mathbf{T}^{\rm #1}}}
\newcommand{\bSp}[1]{\ensuremath{\mathbf{S}^{\rm #1}}}
\newcommand{\hT}[1]{\ensuremath{\hat{T}^{\rm #1}}}
\newcommand{\bhT}[1]{\ensuremath{\hat{\mathbf{T}}^{\rm #1}}}
\newcommand{\tT}[1]{\ensuremath{\tilde{T}^{\rm #1}}}
\newcommand{\btT}[1]{\ensuremath{\tilde{\mathbf{T}}^{\rm #1}}}
\newcommand{\T}[1]{\ensuremath{T^{\rm #1}}}
\newcommand{\R}[1]{\ensuremath{R^{\rm #1}}}

\newcommand{\Sw}{\ensuremath{S^{\rm w}}}
\newcommand{\Tw}{\ensuremath{T^{\rm w}}}
\newcommand{\Sd}{\ensuremath{\mathbf{S}^{\rm d}}}
\newcommand{\Sa}{\ensuremath{\mathbf{S}^{\rm a}}}
\newcommand{\Sf}{\ensuremath{\mathbf{S}^{\rm f}}}
\newcommand{\St}{\ensuremath{\mathbf{S}^{\rm f}}}
\newcommand{\SAF}{\ensuremath{\mathbf{S}^{\rm u}}}

\newcommand{\Se}{\ensuremath{\mathbf{S}^{\rm e}}}

\newcommand{\Sav}{\ensuremath{\bar{\mathbf{S}}}}
\newcommand{\Tav}{\ensuremath{\bar{\mathbf{T}}}}

\newcommand{\tb}{\ensuremath{t_{\rm b}}}
\newcommand{\rb}{\ensuremath{r_{\rm b}}}

\newcommand{\Id}{\ensuremath{I^{\rm d}}}
\newcommand{\Ia}{\ensuremath{I^{\rm a}}}
\newcommand{\Idap}{\ensuremath{I^{\rm f}}}
\newcommand{\Ib}{\ensuremath{I^{\rm b}}}
\newcommand{\IAF}{\ensuremath{I^{\rm u}}}

\newcommand{\gb}{\ensuremath{\bar{\gamma}}}
\newcommand{\gbf}{\ensuremath{\bar{\gamma}_{\rm r}}}
\newcommand{\gbs}{\ensuremath{\bar{\gamma}_{\rm s}}}
\newcommand{\qf}{\ensuremath{q_{\rm r}}}
\newcommand{\qs}{\ensuremath{q_{\rm s}}}
\newcommand{\gD}{\ensuremath{\gamma_{\rm D}}}
\newcommand{\gA}{\ensuremath{\gamma_{\rm A}}}

\newcommand{\sg}{\ensuremath{\sigma}}
\newcommand{\sgf}{\ensuremath{\sigma_{\rm r}}}
\newcommand{\sgs}{\ensuremath{\sigma_{\rm s}}}

\newcommand{\Nt}{\ensuremath{N_{\rm t}}}
\newcommand{\Ns}{\ensuremath{N_{\rm s}}}
\newcommand{\Nc}{\ensuremath{N_{\rm c}}}

\newcommand{\Ert}{\ensuremath{\langle\epsilon\rangle}}
\newcommand{\Erd}{\ensuremath{\langle\epsilon_{\rm d}\rangle}}
\newcommand{\Era}{\ensuremath{\langle\epsilon_{\rm a}\rangle}}
\newcommand{\Erf}{\ensuremath{\langle\epsilon_{\rm f}\rangle}}

\newcommand{\pf}{\ensuremath{p^{\rm f}}}
\newcommand{\taud}{\ensuremath{\tau_{\rm d}}}
\newcommand{\tauf}{\ensuremath{\tau_{\rm f}}}
\newcommand{\ff}{\ensuremath{f^{\rm f}}}
\newcommand{\fd}{\ensuremath{f^{\rm d}}}

\newcommand{\ld}{\ensuremath{\lambda_{\rm d}}}

\newcommand{\bF}[1]{\ensuremath{\mathbf{F}_{#1}}}
\newcommand{\lex}[1]{\ensuremath{\lambda^{\rm ex}_{#1}}}

\newcommand{\Itot}{\ensuremath{I^{\rm t}}}

\newenvironment{myequation}{%
	\addtocounter{equation}{-1}
	\refstepcounter{defcounter}
	\renewcommand\theequation{m\thedefcounter}
	\begin{equation}}
{\end{equation}}

\begin{document}


\title{uFLIM -- Unsupervised analysis of FLIM-FRET microscopy data}



\author{Francesco Masia}\email{masiaf@cf.ac.uk}
\affiliation{School of Physics and Astronomy, Cardiff University, The Parade, Cardiff CF24 3AA, UK}
\affiliation{School of Biosciences, Cardiff University, Museum Avenue, Cardiff CF10 3AX, UK}
\author{Walter Dewitte}
\affiliation{School of Biosciences, Cardiff University, Museum Avenue, Cardiff CF10 3AX, UK}
\author{Paola Borri}
\affiliation{School of Biosciences, Cardiff University, Museum Avenue, Cardiff CF10 3AX, UK}
\author{Wolfgang Langbein}
\affiliation{School of Physics and Astronomy, Cardiff University, The Parade, Cardiff CF24 3AA, UK}

\begin{abstract}
	Despite their widespread use in cell biology, fluorescence lifetime imaging microscopy (FLIM) data-sets are challenging to analyse, because each spatial position can contain a superposition of multiple fluorescent components.  \edit{Here, we present a data analysis method making the most out of the information in the available photon budget, while being fast and unbiased.}{Here, we present a data analysis method employing all information in the available photon budget, as well as being fast.} The method, called uFLIM, determines spatial distributions and temporal dynamics of multiple fluorescent components with no prior knowledge. It goes significantly beyond current approaches which either assume the functional dependence of the dynamics, e.g. an exponential decay, or require dynamics to be known\add{, or calibrated}. Its efficient non-negative matrix factorization algorithm allows for real-time data processing. We \edit{show}{validate {\it in silico}} that uFLIM is capable to disentangle the spatial distribution and spectral properties of \edit{5}{five} fluorescing probes, from only two excitation and detection channels and a photon budget of 100 detected photons per pixel. By adapting the method to data exhibiting F\"orster resonant energy transfer (FRET), we retrieve the spatial and transfer rate distribution of the bound species, without constrains on donor and acceptor dynamics. \delete{This powerful unsupervised method allows to extract information previously inaccessible from FLIM-FRET data.} 	
\end{abstract}

\maketitle{}

t\section{Introduction}
Fluorescence microscopy is a widely used tool to study the distribution of biomolecules in living cells and tissues, with high contrast, specificity, and spatial resolution. The decay dynamics of the fluorescence intensity following pulsed excitation can reveal information on the local environment of the emitting fluorophore. This concept is used in fluorescence lifetime imaging microscopy (FLIM), where spatially-resolved emission dynamics are recorded \cite{BerezinCR10}. Typically, the emission intensity is measured as \add{a} function of the delay after an excitation pulse, but there are also frequency-domain implementations \cite{RaspeNMe16}. Spatially-resolved fluorescence dynamics have been used to sense local variation of temperature\,\cite{OkabeNC12}, pH\,\cite{OrteACSN13,SchmittBBA14}, and \edit{ionic}{ion} concentration\,\cite{AgronskaiaJBO04}. FLIM can also be used to distinguish multiple spectrally overlapping fluorophores via their different decay dynamics\,\cite{NiehorsterNM16}.

Among the various processes which alter the lifetime of an emitter, F\"orster resonant energy transfer (FRET) offers the possibility of studying protein-protein interaction\,\cite{SunNP11,MargineanuSR16}. Here, two proteins of interest are tagged with different fluorophores, called donor and acceptor. The emission spectrum of the donor spectrally overlaps with the absorption spectrum of the acceptor, and the \delete{light} excitation is \edit{resonant to}{spectrally overlapping with} the absorption of the donor. If the distance of the two fluorophores is small enough, typically in the nanometre range, significant non-radiative transfer of the excitation occurs from the donor to the acceptor. Such energy transfer can be detected by a quenching of the donor emission and a corresponding enhancement of the acceptor emission. For high accuracy and sensitivity, a method to detect the transfer not relying on absolute intensities is preferable, and this can be achieved by measuring the change of the fluorescence dynamics in FLIM. The energy transfer provides an additional \edit{relaxation}{loss} channel for the donor, increasing its decay rate, and a corresponding delayed excitation of the acceptor. Notably, FLIM-FRET is not affected by absolute intensity changes, typically present due to photobleaching, illumination inhomogeneity and/or concentration distributions.

To \edit{analyze}{analyse} FLIM, a common approach is to fit the signal decay assuming a mono- or bi-exponential decay behaviour.  Recently, global analysis methods offering faster algorithms compared to pixel by pixel fitting have been reported\,\cite{DattaJBO20}, and a clustering step can be introduced to further speed up the analysis\,\cite{BrodwolfT20,LiBOE21}.  Most of these methods assume exponential decay dynamics, and the instrument response function in time-domain needs to be known to extract the exponential time constants. FRET is observed as an additional decay rate and can be extracted from the fit parameters\,\cite{LaptenokJSS07,WarrenPO13}. However, while being a convenient mathematical function to use, and the simplest solution of rate equation models, an exponential decay is only approximately representing the physical behaviour of a fluorophore embedded in a heterogeneous environment.   

Phasor analysis is an alternative simple and widely used approach \,\cite{ClaytonJM04,DigmanBJ08,StringariPNAS11}. In this method, each FLIM pixel is represented by two quantities, namely the real and imaginary part of the Fourier coefficient of the first harmonic (typically referring to the excitation repetition rate) \edit{normalized}{normalised} to the amplitude of the zeroth harmonic. These values are then interpreted as coordinates in the resulting "phasor plot" in the complex plane. Pure exponential decay dynamics of varying decay times are forming a semi-circle in this plot. Due to the linearity of the transform, mixed exponential decay dynamics are resulting in  averages of pure component phasors, and thus have amplitudes inside the circle. The phasor analysis provides a useful tool when applied to FLIM-FRET data. The occurrence of energy transfer can be identified as a deviation of the phasors from the values obtained in regions of the sample occupied only by unbound donor molecules.\edit{More quantitative information on the FRET process can be obtained if some assumptions}{For a quantitative analysis of FRET some assumptions} are imposed, e.g. \edit{a mono-exponential decay of the unbound donor fluorescence}{the FRET efficiency trajectory is obtained by approximating the unbound donor fluorescence as mono-exponential}.
 
FLIM data can also be \edit{analyzed}{analysed} by linear unmixing of the intensity decay on the basis of selected reference patterns\,\cite{GregorBook15} measured {\it a priori} in samples with similar properties as the sample under investigation. In this method, each fluorescence decay is approximated as a linear combination of reference decay curves by minimizing the Kullback-Leibler discrepancy (KLD)\,\cite{LeeBook01}, which \edit{maximizes}{maximises} the likelihood of the model for data showing Poisson noise, \edit{i.e. the noise from the photon emission}{which is the expected photon detection} statistics. Typically, a gradient descent method using multiplicative update rules is used to find the non-negative fractional concentrations of the reference patterns\,\cite{LeeBook01}. The reference patterns are either extracted from singly labelled control samples or by selecting regions of the image which are assumed to show the individual components. This approach has been applied to the analysis of multispectral time-domain FLIM, and up to nine different fluorophores could be visualised\,\cite{NiehorsterNM16}. The supervised determination of the components and their dynamics \edit{is generally complicating}{complicates} the analysis, and introduces a bias. Specifically in the analysis of FRET-FLIM data, where the different components interact, \edit{reliably determining}{a reliable determination of} the individual component dynamics is challenging.

Recently, a deep learning method to analyse FLIM and FLIM-FRET datasets was developed \cite{SmithPNAS19}. \edit{However, the benefits of the free-fit approach are accompanied by the typical shortcoming of deep learning, i.e. the long computational time required to generate the training set and to train the neural network,}{However, the benefits of the fit-free approach are accompanied by the typical shortcoming of deep learning, i.e. the need to generate the training set and to train the neural network,} which is a topic of further investigation\,\cite{DongJSTQE21}. \delete{Notably, generating the training dataset requires prior knowledge, constraining the expected responses - in }\Onlinecite{SmithPNAS19} \delete{a single or bi-exponential fluorescence decay and the instrument response were used.}\add{Notably, generating the training dataset requires prior knowledge. For example, in }\Onlinecite{SmithPNAS19} \add{ a single or bi-exponential fluorescence decay and the instrument response were used to create the training set, which is defining the expected responses, and restricting the retrieved parameters to two time-constants and two amplitudes}

In this work, we \edit{demonstrate}{propose} an unsupervised FLIM analysis (uFLIM) method, using a fast non-negative matrix factorization (NMF) algorithm\,\cite{KimProcICDM09} and random initial guesses for both the spatial distribution and decay traces of the factorization components. Similar to the pattern unmixing, the NMF method decomposes the data into a linear combination of few components, but differently from  pattern unmixing\add{,} it does not require prior knowledge of the component patterns, which instead can be deduced as part of the factorization. The method thus offers the advantages of pattern unmixing, {\it i.e.} the absence of assumptions on fluorescence dynamics and \add{of} prior knowledge of the instrument response function, while operating at higher speed and additionally dropping the prior knowledge of reference patterns. We demonstrate the performance of uFLIM in distinguishing multiple spectrally overlapping fluorescing proteins, showing that the method can retrieve the spatial distribution and dynamics of \edit{5}{five} fluorescent protein probes using data from a simple FLIM set-up with only two excitation lasers and two detection channels.

Building on this method, we introduce a FRET analysis, which uses the donor and acceptor dynamics determined by uFLIM from samples or sample regions not showing FRET. Energy transfer is quantitatively characterised by the quantum efficiencies of donor and acceptor emission into the detection channels, as well as the mean and variance of the transfer rate distribution, here assumed to be log-normal\,\cite{Balakrishnan_Book94}. The analysis determines the values of these quantities, and thus the \edit{FRET}{donor-acceptor} pair (DAP) dynamics, together with the spatial distribution of the \edit{FRET pairs}{DAPs}, by minimizing the NMF factorization error. Other components, such as autofluorescence, can be retrieved at the same time without prior knowledge. Notably, uFLIM-FRET does not assume a functional dependence of the fluorescence dynamics, but calculates the non-exponential donor and acceptor dynamics in the \edit{FRET pair}{DAP} from their unbound dynamics using the distribution of FRET rates.

\section{Method}\label{sec:methods}

\subsection{uFLIM}\label{sec:uFLIM_method}
Measured FLIM data are reshaped as \edit{a}{an} ($\Ns\times \Nt$) matrix \bD\, where \Ns\ and \Nt\ indicate the number of spatial and temporal points, respectively. Then, a number of components \Nc\ much smaller \add{than} \Nt\ is chosen to represent the data, and NMF is used to determine the spatial distribution matrix \bS\  of $\Ns \times \Nc$ elements, and the dynamics matrix \bT\ of $\Nc \times \Nt$ elements. If present, multiple spectral channels are stacked in the \Nt\ dimension, and multiple data are stacked in the \Ns\ dimension, keeping track of the ordering for later decomposition.  

We assume in the following that \bD\ is given as \add{the} number of detected photons, which has Poissonian noise with a standard \edit{devation}{deviation} given by $\sqrt{D_{ij}}$. We utilise a fast NMF algorithm that minimizes the residual $||\bD-\bS\bT||_2$, where $||.||_2$ indicates the Frobenius norm\,\cite{KimProcICDM09}. This method provides the decomposition of maximum likelihood in \add{the} case of Gaussian white noise in the data, i.e. a noise independent of the data value. To be able to use this algorithm, which is 2-3 orders of magnitude faster than gradient descent methods accounting for non-white noise, we partially whiten the data before factorization by applying a scaling as follows. 
We generate the  time-averaged image \Sav\  and the spatially averaged dynamics \Tav\ by averaging \bD\ along the temporal and spatial points, respectively, 
\be \bar{S}_i=\frac{1}{\Nt}\sum_{j} D_{ij}, \quad \bar{T}_j=\frac{1}{\Ns}\sum_{i} D_{ij}, \ee

For average counts below unity, the photon counting statistics deviates significantly from \edit{gaussian}{Gaussian} noise, and the above whitening is not representing the required whitening well. We therefore limit $\bar{S}_i$ and $\bar{T}_j$ to a minimum of $\xi$ in the whitening. The background-subtracted, partially whitened data $\bDw$ are then defined as 
\be\Dw_{ij} = \frac{D_{ij}-b}{\sqrt{\bar{S}_{i}}\sqrt{\bar{T}_{j}}}\,,\label{eq:pw}\ee
with the average dark counts $b$, which can be measured independently. We assume here that $b$ is \edit{independent of  position, time, and spectral channel}{equal across the position, time, and spectral channels}, as it is typically the case for scanning time-correlated single photon counting, but also note that inhomogeneous dark counts can be subtracted in the same fashion. In \bDw\add{,} the data has been divided by the expected standard deviation of the data when factorized into the average spatial and temporal dependence. This method whitens spatially dependent time-integrated intensities, as well as spatially integrated time-dependent intensities. $\bDw$ is then factorized by NMF, minimizing $E=||\bDw-\bSw\bTw||_2$, and the resulting  decomposition $\bSw$ and $\bTw$ is de-whitened to recover the factorization of the original data
\be S_{ij} = \Sw_{ij}\sqrt{\bar{S}_i}\,,\quad T_{ij} = \Tw_{ij}\sqrt{\bar{T}_j}\,,\ee
so that $\bD\approx\bS\bT+b$.
We will see that this treatment of noise is providing equivalent results to minimizing the KLD for the data considered. When showing \bS\ in this work, it refers to a normalized \bT, such that \bS\ represents the number of photons detected at each spatial point. 
	
\subsection{uFLIM-FRET}\label{sec:uFLIMFRET_method}
Beyond the unsupervised analysis of FLIM data, we have extended the algorithm to retrieve the spatial distribution of FRET pairs. In the literature, FRET efficiencies are often derived from fitting the measured dynamics by exponential decays and comparing the resulting decay times with the decay constant measured in samples where only the donor is present. These methods are limited by the assumption of exponential decay dynamics, and require the knowledge of the instrument response function. 

In uFLIM, the temporal dynamics \edit{is}{are} retrieved without prior knowledge or assumption of an exponential decay. Therefore uFLIM can be applied to data showing pure donor and acceptor dynamics as components, providing the normalized pure donor and acceptor dynamics, which we call \bTp{d}\ and \bTp{a}, with $\{\bTd\}=\{\bTa\}=1$ where $\{ . \}$ indicates the 1-norm. We note that the emission of a molecule is proportional to the probability to be in its excited state. FRET occurs when donor and acceptor are in close proximity, forming a \edit{donor-acceptor pair (DAP)}{DAP}. The FRET process introduces a non-radiative \edit{decay}{excitation transfer} channel from the donor to the acceptor, \edit{with}{characterised by a} \delete{decay} rate $\gamma$ (a sketch of the energy diagram is shown in \Fig{FigSM_FRETLevelScheme}). \edit{Therefore, we can describe the dynamics of the probability of the donor  to be in its excited state in the presence of FRET by subtracting the FRET transfer at points in the past, propagated to the present using \bTp{d}. Similarly, we add the transfer to the probability of the acceptor to be in its excited state, and propagate it to the present using \bTp{a}.}{Therefore, the fluorescence intensity of the donor in the DAP at a given time point can be calculated  by subtracting from \bTp{d} the FRET to the acceptor up to that time point. Equivalently, the intensity of the acceptor in the DAP can be calculated from \bTp{a} by adding the FRET from the donor.}
The modified donor dynamics $\btT{d}$ in the DAP is accordingly calculated using
\be \tT{d}_i(\gamma)=\T{d}_i-\sum_{j=1}^{i-1}f_j(\gamma)\hT{d}_{i-j+1} \label{eq:Td} \ee
iterating along the temporal channel $i=1,2,..,l$. This expression contains the dynamics
\be  \hT{d}_k= \begin{cases}
	\T{d}_{m+k-1}/\T{d}_m & \text{for $k\leq l-m+1$} \\
	\hT{d}_{l-m}\T{d}_l/\T{d}_{2l-m-k} & \text{for $k>l-m+1$}
\end{cases} \label{eq:Thatd} \ee
where $m$ is the temporal channel at which $\bTd$ is maximum. In \Eq{eq:Thatd}, we have extrapolated the donor excitation decay beyond the last measured point $l$ using the decay observed over the extrapolation time interval prior to $l$. The FRET transfer $f_j(\gamma)$ at time $j$ is given by the modified occupation of the donor excited state at that time, the time-step $\Delta$, and the transfer rate $\gamma$, 
\be f_j(\gamma)=\tT{d}_j \gamma\Delta\,.\ee
These equations determine the effect of FRET on the donor excitation, by subtracting the FRET transfer at points in the past, propagated to the present using the  response function $\bhT{d}$. We approximate $\bhT{d}$ by the measured donor emission dynamics, normalized to its maximum and starting from its maximum as time zero of the response. This is adequate for FRET rates smaller than the inverse time resolution of the measurements\delete{,} and is consistent with the finite resolution of the data for which it is used. To support this statement\add{,} we have compared the resulting dynamics with the analytical solution of the donor excitation modified by a single FRET process in the simple condition of a mono-exponential decay for the pure donor and a Gaussian instrument response function (IRF), as shown in the supplementary information (SI) \Sec{sec:analytic}. The time-resolution limitation can be controlled by refining the system dynamics, for example\add{,} by deconvolution of a response function before analysis. Note\add{,} however\add{,} that \add{the} deconvolution is modifying the noise of the data from the simple Poisson distribution of photon counts.

\begin{figure}
	\includegraphics*[width=\columnwidth]{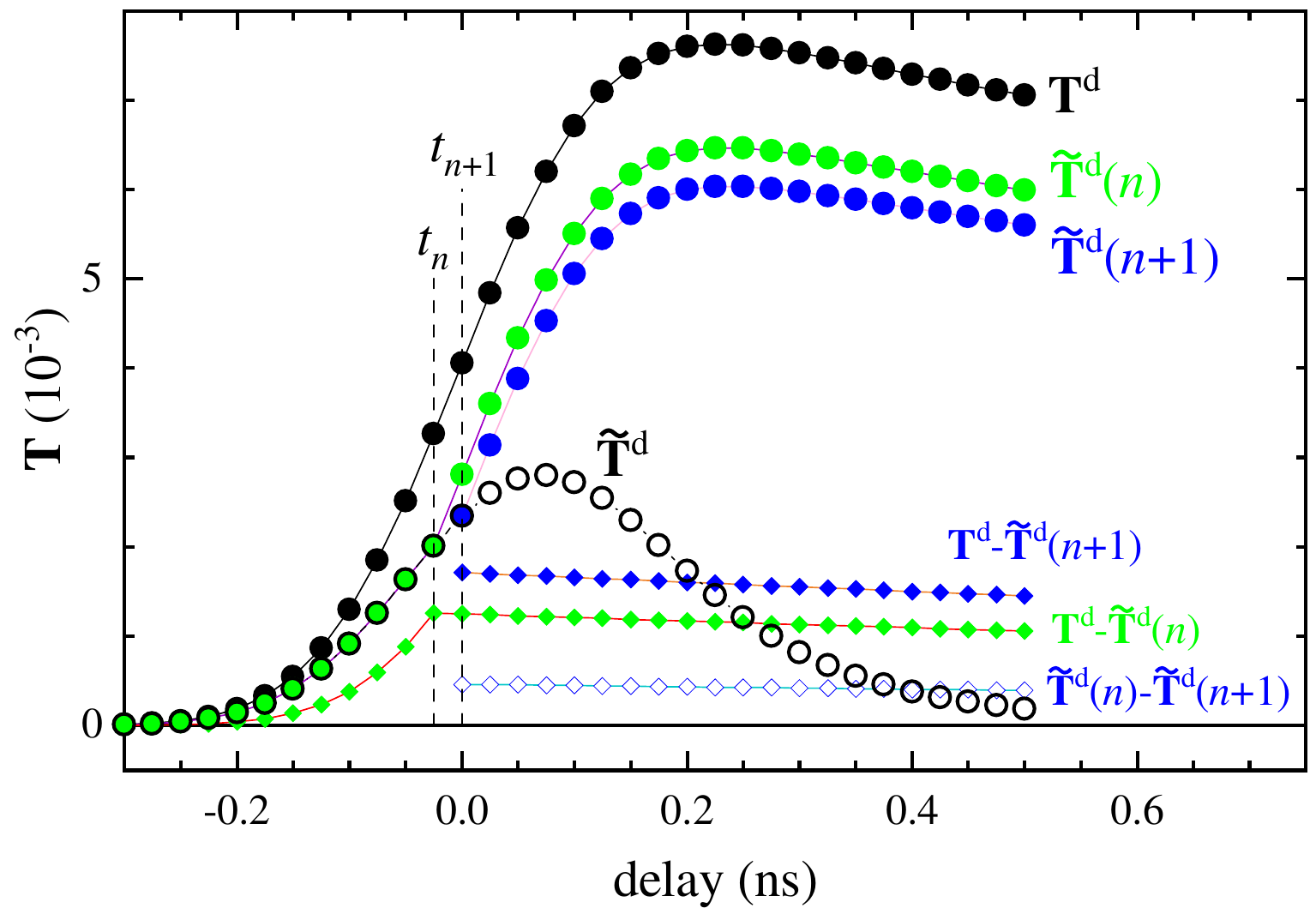}
	\caption{Illustration of \Eq{eq:Td} calculating the modified donor dynamics in the DAP undergoing FRET, \btT{d}, from the free donor dynamics \bTp{d}. Intermediate results subtracting only the transfer occurring before time step $n$ ($\btT{d}(n)$) and $n+1$ ($\btT{d}(n+1)$) are shown, together with the corresponding subtracted transfer $\bTp{d}-\btT{d}(n)$ and $\bTp{d}-\btT{d}(n+1)$, and the additional transfer occurring between $t_n$ and $t_{n+1}$, given by $\btT{d}(n)-\btT{d}(n+1)$.}
	\label{FigSM_FRETModelSketch}
\end{figure}

\Fig{FigSM_FRETModelSketch} illustrates the iterative calculation of \btT{d} from \bTp{d} by \Eq{eq:Td}. The modified dynamics $\btT{d}(n)$ including only the contributions of previous temporal points up to $n-1$ \edit{is}{are} shown \edit{(}{ in} green filled circles\edit{),}{ and are} given by\add{:}
\be \tT{d}_i(n)=\T{d}_i-\sum_{j=1}^{{\rm min}(n,i)-1}f_j\hT{d}_{i-j+1}. \ee
Including the additional temporal point $n$ in $\btT{d}(n+1)$ (blue filled circles), the dynamics for $i>n$ \edit{is}{are} decreased by the contribution of the excitation transferred between the time point $n$ and $n+1$ (\edit{blue empty}{empty blue} diamonds), given by\add{:}
\be \tT{d}_i(n)-\tT{d}_i(n+1)=  \begin{cases}
	f_n\hT{d}_{i-n+1}, & \text{for } i \geq n \\
	0 & \text{otherwise } \\
\end{cases}\,. \ee
Including all previous temporal points, we recover the correct modified dynamics $\btT{d}$.

The modified acceptor excitation dynamics \edit{is}{are} calculated using the same approach, resulting in\add{:}
\be	\tT{a}_i(\gamma)=\kappa\T{a}_{i}+\sum_{j=1}^{i-1}f_j(\gamma)\hT{a}_{i-j+1},\label{eq:Ta} \ee
with the normalized and zero-centered acceptor dynamics\add{:}
\be \hT{a}_k=
\begin{cases}
	\T{a}_{m+k-1}/\T{a}_m & \text{for $k\leq l-m+1$} \\
	\hT{a}_{l-m} \T{a}_l/\T{a}_{2l-m-k} & \text{for $k>l-m+1$}
\end{cases} \,. \ee
The prior normalisation of \bTp{d}\ and \bTp{a} ensures the conservation of the number of excitations by the transfer from donor to acceptor in \Eq{eq:Ta}, which also contains the direct excitation of the acceptor by the laser (see \Fig{FigSM_FRETLevelScheme}) quantified by $\kappa$. While $\kappa$ can be included in the parameters to be retrieved by the method, we assume in the following that $\kappa$ is known {\it a priori}, noting that it is given by the relative absorption crossection of acceptor and donor at the excitation wavelength\delete{,} and can be determined independently. We assume to have two spectral channels\delete{,} and that the donor and acceptor emission is detected dominantly by the respective channels, given by the fraction of donor \R{d} (acceptor \R{a}) emission detected by the donor (acceptor) channel, respectively. The dynamics \btT{D} (\btT{A}) detected in the donor (acceptor) channel for a donor-acceptor pair undergoing FRET with rate $\gamma$ is then given by\add{:}
\be
\begin{split}
	\btT{D}(\gamma,q)=\R{d}\btT{d}(\gamma)+q\left(1-\R{a}\right)\btT{a}(\gamma)\\
	\btT{A}(\gamma,q)=\left(1-\R{d}\right)\btT{d}(\gamma)+q\R{a}\btT{a}(\gamma),
\end{split}\label{eq:fret}
\ee
respectively. Here, we have introduced the ratio $q$ between acceptor and donor, of the detection probability (summed over both channels) of an excitation decay, to take into account the different quantum efficiency of the acceptor and donor, and the different probability of detecting an emitted photon in the two channels, including detector efficiency and filter performance. The values of $\R{d}$ and $\R{a}$ can be simply measured \delete{using samples of only donor or acceptor} from the ratio of the number of detected photons in donor versus acceptor channel\add{ using samples of only donor or acceptor}.
Determining $q$ instead requires to additionally determine the relative excitation rates of the donor versus acceptor molecules in the two samples, which in turn requires knowledge of relative molar concentration and relative absorption $\kappa$. We have considered here the situation where $\R{d}$ and $\R{a}$ have been measured, while $q$ is retrieved as part of the retrieval process.

Typically, the acceptor has a small absorption at the excitation wavelength, so $\kappa \ll 1$, and is hardly detected by the donor channel, so $1-\R{a} \ll 1$. Later in the manuscript\add{,} we show the more challenging condition of $\kappa=1$, where three components (donor, acceptor, and DAP) need to be included, while the simpler case $\kappa=0$, showing an improved retrieval for a given photon budget, is given in the SI. 

In the NMF, the temporal points of \add{the} donor and acceptor channel are concatenated into the temporal dimension of $\bD$. Three NMF components in $\bT{}$ are used, given by the donor, $\left[\R{d}\bTd,\left(1-\R{d}\right)\bTd\right]$, the acceptor $\kappa q \left[\left(1-\R{a}\right)\bTa,\R{a}\bTa\right]$  and the DAP $\btT{f}(\gamma,q)=\left[\btT{D}(\gamma,q),\btT{A}(\gamma,q)\right]$. The latter is a function of the FRET rate $\gamma$ and  the ratio $q$. The spatial distributions \Sd, \Sa, and \Sf\ of these components are determined from the data by NMF. Additional components can be added to the NMF analysis, for example\add{,} to take into account autofluorescence, determining their temporal dynamics and spatial distribution without prior knowledge, as shown\add{,} for example\add{,} in \Sec{sec:resultsFRETAF}.

The most likely values of $\gamma$ and $q$\add{,} given the data\add{,} are the ones minimizing the residuals of the NMF. The model can be expanded to several FRET components with different rates, which is a typical situation in FRET due to the variation in \add{the} distance and \add{the} relative orientation of the donor and acceptor transition dipoles \cite{GopichPNAS12}. Such a variation can be efficiently rationalized using a distribution of rates $P(\gamma;\gb,\sg)$ of mean value \gb\ and relative standard deviation \sg, resulting in the FRET dynamics\add{:} 
\be\label{eq:fretint}\bbT{}(\gb,\sg,q)=\int P(\gamma;\gb,\sg) \btT{f}(\gamma,q)d\gamma\,.\ee
Again, the most likely values of the parameters \gb, \sg, and $q$ \edit{are minimizing}{minimise} the residual of the NMF, which are found using a computationally efficient method detailed in the SI \Sec{sec:FRETparmin}.

In the following, we consider a log-normal distribution\cite{Balakrishnan_Book94} of rates with mean $\gb$ and standard deviation $\gb\sigma$, which can be written as
\be 
P_{\rm ln}(\gamma) = \frac{1}{\gamma\zeta\sqrt{2\pi}}\exp\left(-\frac{1}{8}\left(\frac{2}{\zeta}\ln\left( \frac{\gamma}{\gb}\right)+\zeta\right)^2\right),\label{eq:lognorm}
\ee
where $\zeta=\sqrt{\ln\left(\sg^2+1\right)}$. Interestingly, this distribution can also determine the mean and standard deviation of the donor-acceptor distance. In the dipole approximation, and for a given relative donor and acceptor orientation or fast orientational averaging, the FRET rate is simply expressed as $\gamma=\gD(R_0/R)^6$, with the F\"{o}rster radius $R_0$, the free donor decay rate \gD, and the donor-acceptor distance $R$. Using this expression, the extracted log-normal distribution in the FRET rate $\gamma$ of mean \gb\ and standard deviation $\sg \gb$ can be analytically expressed by a log-normal distribution in distance given by  
\begin{align} 
P(R)=&\frac{6}{R\zeta\sqrt{2\pi}} \times\\ \nonumber
&\exp\left(-\frac{1}{8}\left(\frac{2}{\zeta}\left(\ln\left( \frac{\gD}{\gb}\right)+6\ln\left(\frac{R}{R_0}\right)\right)+\zeta\right)^2\right)\,.
\end{align}
The first and second moments of this distribution can be calculated as
\be \bar{R}=\int_0^{\infty} R\,P(R)dR=R_0\sqrt[6]{\frac{\gD}{\gb}}\left(\sigma^2+1\right)^{-\frac{5}{72}}\ee
and
\be \overline{R^2}=\int_0^{\infty} R^2\,P(R)dR=R_0^2\sqrt[3]{\frac{\gD}{\gb}}\left(\sigma^2+1\right)^{-\frac{1}{9}}.\ee
so that the standard deviation $\sigma_R$ in distance can be determined using $\sigma_R^2=\overline{R^2}-\bar{R}^2$ as 
\be \sigma_R = R_0\sqrt[6]{\frac{\gD}{\gb}}\sqrt{\left(\sigma^2+1\right)^{-\frac{1}{9}}-\left(\sigma^2+1\right)^{-\frac{5}{36}}}.\ee
The mean $\bar{R}$ and the standard deviation $\sigma_R$ of the donor-acceptor distance is therefore obtained analytically by the parameters of the log-normal rate distribution determined by uFLIM-FRET.
\section{Results and Discussion}
\label{sec:results}

\subsection{uFLIM application I: Single spectral channel datasets}
\label{sec:resultsDye}
\begin{figure}	
	\includegraphics[width=\columnwidth]{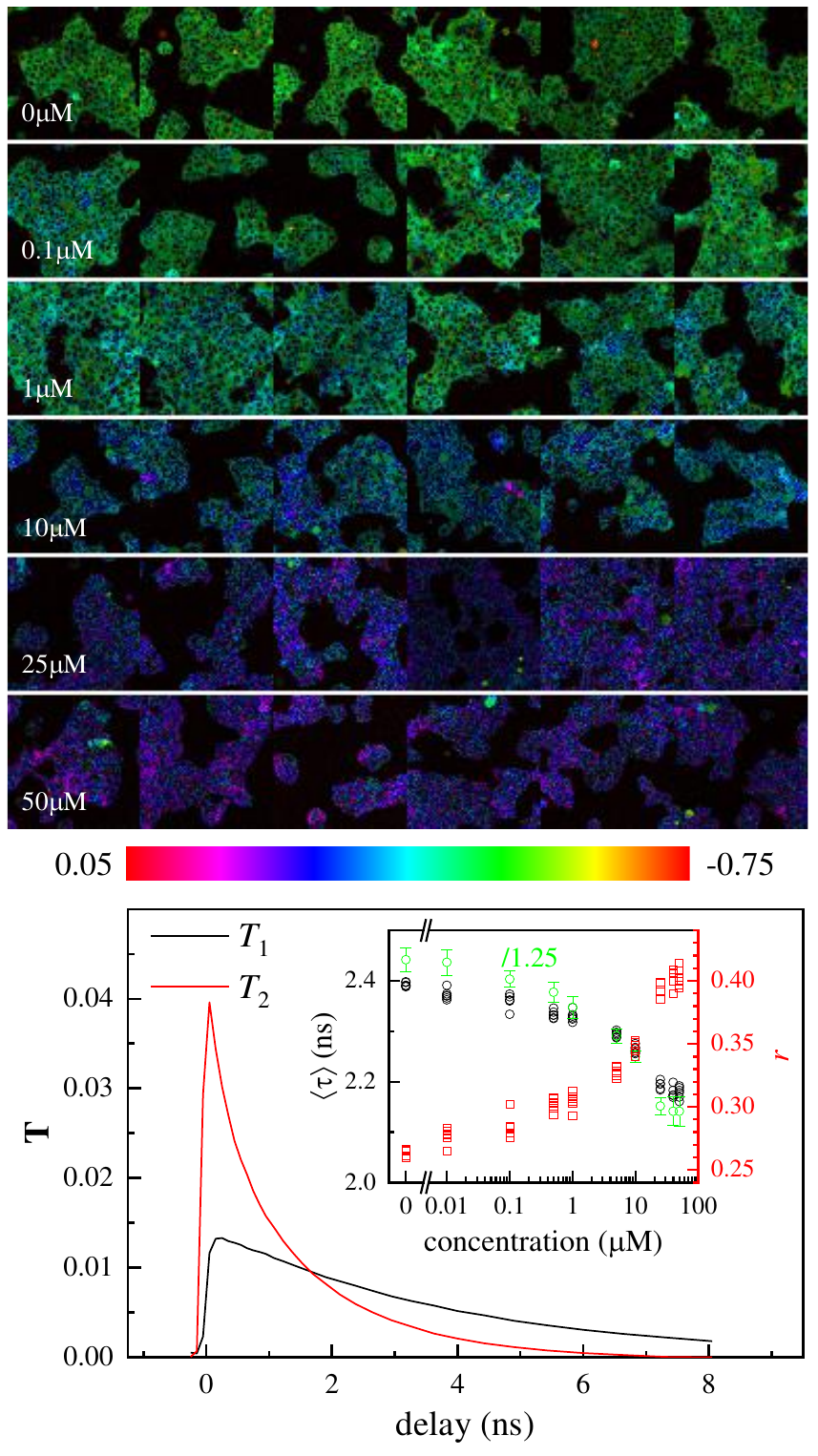}
	\caption{Results of the uFLIM algorithm applied on a dataset from \Onlinecite{ChennelSen16} of TCSPC FLIM on HepG2 expressing the T2-AMPKAR compound as a function of the concentration of the 991 activator. In these measurements, a single channel detects the dynamics of the T2-AMPKAR compound. The data have been factorised into two components which show different dynamics. Top: Concentrations $\bS_1$ and $\bS_2$ displayed with \edit{a}{an} HSV mapping as discussed in the text for different concentrations of 991 as indicated. The contrast is encoded as \add{the} hue of the \edit{color}{colour}\delete{, as shown}. Bottom: Temporal dynamics \bT\ of the \edit{three}{two} retrieved components. Inset: Weighted average lifetime $\langle\tau\rangle$ (black symbols) and fraction of $\bS_2$ in each image $r=\{\bS_{2}\}/(\{\bS_{1}\}+\{\bS_{2}\})$ (red symbols) for the different fields of view versus \add{the} activator concentration. The green symbols show the lifetime estimated in \Onlinecite{ChennelSen16}\add{, obtained by global least square fitting,} divided by 1.25.}
	\label{Fig_T2AMPKAR}
\end{figure}

Here\add{,} we demonstrate the uFLIM analysis of experimental data reported in \Onlinecite{ChennelSen16}, in which the fluorescence lifetime of a dye changes due to variations in the environmental conditions, specifically the T2-AMPKAR construct in \add{the} presence of the 991 activator resulting in FRET. In these measurements, a single channel detects the dynamics of the T2-AMPKAR compound using time-correlated single photon counting (TCSPC) with $50$\,ps time bins. We have analysed the data in Fig.\,4 of \Onlinecite{ChennelSen16}, with $4\times 4$ spatial binning and a temporal binning as discussed in \ref{sec:binning}\add{,} using \tb=100\,ps and $\rb=0.1$. Data have been factorised by uFLIM into two components using a whitening threshold $\xi=1$, as shown in  \Fig{Fig_T2AMPKAR} for a selection of activator concentrations (complete results are shown in \Fig{FigSM_T2AMPKAR}). We have measured a computational time of about 0.6\,\textmu s/pixel for a single uFLIM step on an Intel i7-8700 CPU. We found that convergence (error change below 1\textperthousand\ per iteration) was reached within about 10 iterations. Further computational times reported below refer to the same CPU.

The dynamics $\bT_{1,2}$ of the components (see \Fig{Fig_T2AMPKAR} bottom) suggest\delete{s} that the first component, showing a slower decay, \edit{is representing}{represents} the emission of T2-AMPKAR without 991, while the second \edit{is representing}{represents} the T2-AMPKAR - 991 pair. For visualization, the spatial distributions $\bS_{1,2}$ are encoded using a hue-saturation-value (HSV) colour mapping at maximum saturation. The value (V), which is the brightness, is taken as \add{the} square root of $\bS_1+\bS_2$, normalised for each image. The hue (H) is given by the point-wise contrast $(\bS_2-\bS_1)/(\bS_1+\bS_2)$, offset and scaled as indicated. We observe a change of colour of the HSV maps from green to violet with \add{an} increasing concentration of 991, showing \add{an} increasing fraction of T2-AMPKAR with 991 attached. To compare with the global fitting exponential decay analysis in \Onlinecite{ChennelSen16}, the average lifetime $\langle\tau\rangle = ( \{\bS_{1}\}\tau_1+\{\bS_{2}\}\tau_2 ) / ( \{\bS_{1}\}+\{\bS_{2}\} )$ is given in the inset of the bottom panel in \Fig{Fig_T2AMPKAR}, where $\{ . \}$ indicates the 1-norm, and $\tau_2$ and $\tau_3$ are the lifetimes of the individual components given by the first moments of their dynamics. The resulting $\langle\tau\rangle$ exhibits a dependence on the activator concentration consistent with\,\Onlinecite{ChennelSen16}.
The applied spatial and temporal binning \edit{is increasing}{increases} the average number of photons per point well above one, from 0.19 in the original data to 17 in the binned data, improving the outcome of the factorisation, as we detail in the supplementary information \Sec{sec:AMPCAR}. 

This example shows that uFLIM is able to analyze FLIM experiments with \add{the} resulting weighted average lifetime showing
a similar dependence as the value obtained by \add{the} global exponential fitting, yet providing the dynamics of the components not constrained to an exponential decay. As \add{an} additional example, we show in the SI \Sec{sec:resultsFLIM} the uFLIM analysis of time-gated FLIM images of mixtures of two different dyes, and its ability to recover their dynamics and distribution.

\subsection{uFLIM application II: Multiple spectral channels and unmixing of many fluorescent proteins}\label{sec:sFLIM}

Imaging living cells which are expressing multiple fluorescent proteins (FPs) is crucial when disentangling the protein interaction network.  Here, we explore the capability of uFLIM to extract the spatial distribution of a large number of FPs, by unmixing their spectral and temporal profiles. A similar question was asked in \Onlinecite{NiehorsterNM16} using the pattern-matching algorithm on spectrally-resolved fluorescence lifetime imaging microscopy (sFLIM) data. sFLIM was employed with sequential excitation at three wavelengths and detection over 32 spectral channels. Up to nine fluorescent probes could be separated, for data having a photon budget of around 1000 photons per pixel in the bright regions. However, this result required prior knowledge of fluorescence decay and spectral signature patterns, a constrain that can be lifted with uFLIM. 

To test the performance of uFLIM on sFLIM datasets, we generated synthetic data combining several FPs. Since the large number of excitation and detection channels used in \Onlinecite{NiehorsterNM16} are not available in most FLIM experimental set-ups, we simulate here a much simpler system with only two excitation lasers (at wavelengths of 460\,nm and 490\,nm) and two detection channels (over wavelength ranges of 500--550\,nm and 550--700\,nm). We use an excitation repetition rate of $r=40$\,MHz, a detection range from -1\,ns to 24\,ns with $l=1000$ temporal channels, and a Gaussian instrument response function $\exp(-t^2/w^2)$ with $w=141.4$\,ps. 

We first consider \edit{8}{eight} known FPs numbered by the index $f$ (see SI \Tab{tab:8FPs}), with spatial distributions given by selected paintings \cite{Wikipedia}, which were cropped and resized to $256\times256$ pixels, converted to greyscale using a gamma of 1.5\delete{,} and normalized to have unity mean, yielding the distribution matrix \bF{f}. For each combination of excitation wavelength (index $e$) and detection channel (index $d$), we define a scaled spatial distribution $\bF{def}=c_{def}\bF{f}$, where $c_{def}$ accounts for the quantum efficiency and the extinction coefficient of the FP, and the fraction of photon emission by FP $f$ detected by channel $d$, see SI. We also define the fraction of photons detected in a given channel as $\hat{c}_{def}=c_{def}/\sum_{d,e}c_{def}$, and the fraction of detected photons contributed by a given FP as $\hat{c}_{f}=\sum_{d,e}c_{def}/\sum_{d,e,f}c_{def}$. The measured FP dynamics over the time $t$, represented by the matrix $\bT_f$, are calculated as the convolution between the Gaussian IRF and a mono-exponential decay with a decay rate $\gamma_f$ given by the inverse lifetime $\tau$, see SI \Sec{subsec:gensyn}.   
The noiseless sFLIM synthetic data are then obtained by multiplying the paintings with the FP dynamics, and summing the resulting FP emission, assuming equal spatially-integrated numbers of each FP, yielding\add{:}
\be \bD^{\rm s}_{ed}=A\sum_f\bF{def}\bT_f\,, \ee
where the normalization $A$ is ensuring $\{\bD^{\rm s}\}=\Ns\Itot$, and the average number of photons \Itot\ per spatial point was chosen to be 100 or $10^4$ in the results shown. To simulate photon counting detection and corresponding noise, the integer values of a random variable following Poisson statistics with a mean value given by the noiseless sFLIM data are taken as sFLIM data. Computational time was reduced by partially binning the 1000 time channels in $\bT_f$ according to the method described in the SI \Sec{sec:binning}, using \rb=0.05 and \tb=25\,ps. 

\begin{figure}
	\includegraphics[width=\columnwidth]{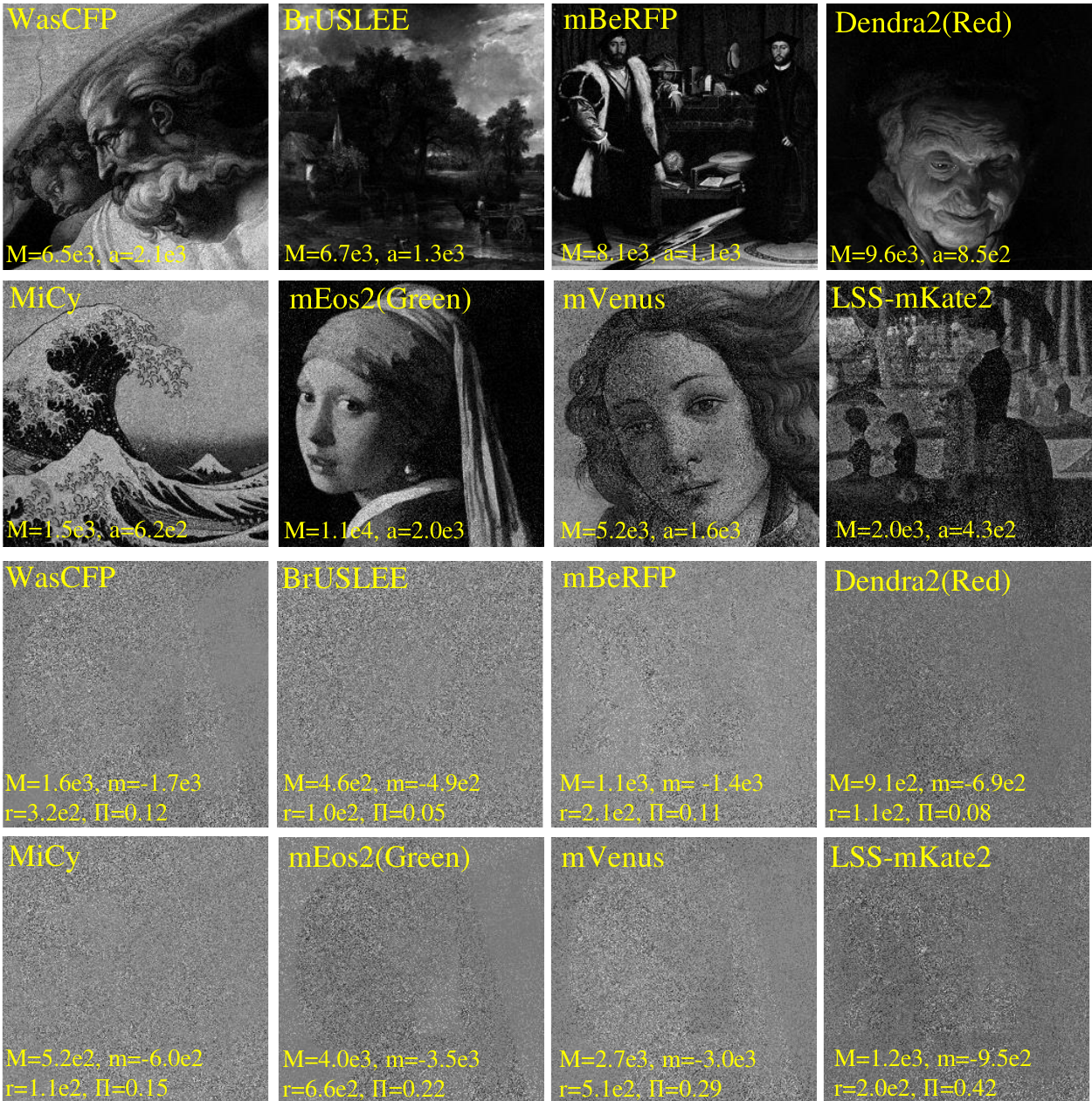}
	\caption{Spatial distributions obtained by applying uFLIM to sFLIM synthetic data generated with 8 FPs (see labels), having spatial patterns given by selected paintings, and detected by a two-channel FLIM set-up (see text). uFLIM\add{,} in this case\add{,} assumes prior knowledge on the FP dynamics, i.e. uses fixed $\bT_f$. Greyscale is from $m$ to $M$. Top rows: Retrieved \bS\, with $m=0$ and the maximum ($M$) as indicated. The spatially averaged pixel values ($a$) are also given, for comparison with the nominal values $\hat{c}_f\Itot$, see \Tab{tab:8FPs}. Bottom rows: Difference between the nominal and retrieved distributions corresponding to the top rows, with $M$ and $m$ as given.}
	\label{Fig_UnmixingFixed_fastNMF_H}
\end{figure}

This synthetic data is then analysed by uFLIM according to the method described in \Sec{sec:uFLIM_method} with $\xi=0$ whitening threshold for the spatial and time averages. As a first test, for direct comparison with \Onlinecite{NiehorsterNM16},  we retrieved the spatial distribution $\bSw$ in a single step NMF, with the dynamics $\bTw$ fixed by $\bT_f$, i.e. assuming prior knowledge on the dynamics. The resulting $\bS$ are shown in \Fig{Fig_UnmixingFixed_fastNMF_H} for $\Itot=10^4$. To quantify the retrieval performance\add{,} we \delete{have} calculated the root-mean-squares (rms) $r$ of the distribution differences $\bS_f-\hat{c}_{f}\Itot\bF{f}$ for each FP $f$ and its relative counterpart $\Pi$ obtained by dividing $r$ with the rms of $\hat{c}_{f}\Itot\bF{f}$. The spatial distributions of all \edit{8}{eight} FPs are well retrieved, with an average root-mean-square error of about $300$ photons and a relative error $\Pi\sim20\%$. We emphasise that this was achieved using only \edit{2}{two} channels in excitation and detection, compared to 3 and 32 channels in \Onlinecite{NiehorsterNM16}. FPs with properties differing significantly from each other are well recovered, while more error is visible for FPs with similar properties, for example\add{,} for mEos2 and mVenus, and for FPs with weak emission, such as LSmKate2. Even for a much smaller photon budget $\Itot=100$ (see \Fig{FigSM_UnmixingFixed_fastNMF_L}), spatial distributions are recovered, albeit with accordingly larger noise and reconstruction error.

To evaluate if the retrieval could be improved by maximizing the likelihood for the Poisson statistics, we have implemented a gradient descent minimising the KLD. We have used a multiplicative update rule\,\cite{LeeBook01}, and, as \add{an} initial guess of $\bS$, either the result of the NMF, or the solution of the linear system $\bD=\bS\bT$ (see SI \Sec{sec:graddescent}). In both cases, we did not observe a relevant improvement of the results compared to the fast NMF algorithm (see \Fig{FigSM_UnmixingFixed_fastNMFGD_H} and \Fig{FigSM_UnmixingFixed_GD_H} for $\Itot=10^4$ and \Fig{FigSM_UnmixingFixed_fastNMFGD_L} and \Fig{FigSM_UnmixingFixed_GD_L} for $\Itot=100$), despite a  15--50 times longer computational time. Using the fast NMF, the uFLIM computational time was 5\,\textmu s/pixel. This indicates that the whitening transformation applied by us, combined with fast NMF algorithm, is a suitable alternative to the computationally expensive gradient descent method.

Next, we applied uFLIM to retrieve the spatial distribution and the FP spectral and dynamic properties directly from the simulated photon counting data, with no prior knowledge. Since determining the FP properties additionally to the spatial distribution is more taxing on the data information content, we have removed the three FPs with the smallest differences in their properties from the \edit{8}{eight} previously used (see \Tab{tab:5FPs}). To introduce unknown variations from the nominal FP properties, often encountered in the cellular environment, the sFLIM data are generated with a $\pm20\%$ relative variation in $c_{def}$ and $1/\gamma_f$, and an additional $\pm5\%$ on the resulting $\hat{c}_f$, taken at random from a uniform distribution. We use the iterative uFLIM method\add{,} where both $\bS$ and $\bT$ are  calculated. The nominal FP properties, before parameter variation, are used to generate the initial value of $\bT$, while the guesses for $\bS$ are obtained by solving the system $\bD=\bS\bT$ and then setting negative values to zero. We constrain the dynamics of a given FP to be the same for all excitation and detection channels\delete{,} by replacing at each NMF iteration step the dynamics calculated for the different channels with their average. The iteration is stopped if the factorisation error has not improved for three consecutive steps, allowing for a maximum of 100 iterations. Here, a single iteration step took about 2\,\textmu s/pixel, and typically 10-25 steps were used.

\begin{figure*}
	\includegraphics[width=\textwidth]{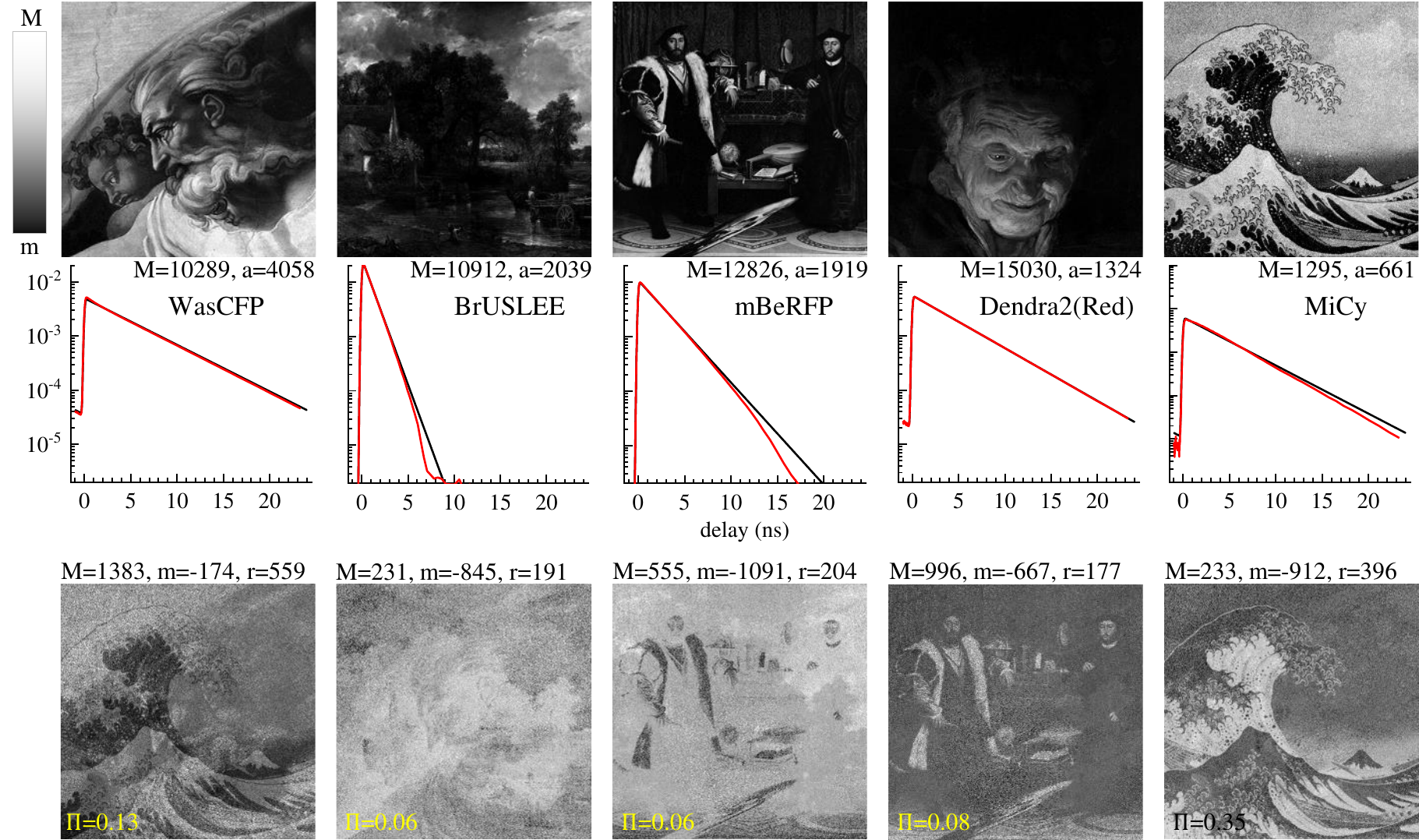}
	\caption{Spatial distributions and properties of 5 FPs retrieved by uFLIM from sFLIM synthetic data with $\Itot=10^4$, generated as described in the text, on a greyscale from $m$ to $M$. \edit{uFLIM in this cases}{In these cases, uFLIM} is applied with no prior knowledge on the FP properties. Top row: Retrieved \bS\, with $m=0$ and the maximum $M$ as indicated. The spatially averaged pixel values  ($a$) are given, having the nominal values $\hat{c}_f \Itot$, see \Tab{tab:5FPs}. Middle row: Retrieved dynamics $\bT_f$ (red), and corresponding original dynamics (black). Bottom row: Difference between the retrieved and original distributions, using $M$ and $m$ as given.}
	\label{Fig_UnmixingFree_fastNMF_H}
\end{figure*}

\begin{table*}[t]
	\caption{\label{tab:5FPs} Spectral properties and lifetimes of the 5 FPs used in the synthetic sFLIM data analysed in \Fig{Fig_UnmixingFree_fastNMF_H}. The values retrieved from a single data realization by uFLIM for $\Itot=10^4$ are given in red, where the lifetimes are the first moment of the retrieved dynamics for positive times. The standard deviations of the retrieved parameters due to photon shot noise are given in green.}
	\begin{center}	
		\begin{tabular}{lc|c|c|c|c|c|c} 
			Name of FP / & $f$ & $\hat{c}_{11f}$ & $\hat{c}_{12f}$ & $\hat{c}_{21f}$ & $\hat{c}_{22f}$ & $\hat{c}_f$ & $\tau$ (ns) \\
			 {\it painting} && & & & & &\\
			\hline\hline
			{WasCFP} / & 1 & 0.27 & 0.07 & 0.53 & 0.13 & {0.36} & 5.05  \\
			{\it The creation of} && {\color{red} 0.28} & {\color{red} 0.08} & {\color{red} 0.51} & {\color{red} 0.13} & {\color{red} 0.41} & {\color{red} 4.73} \\
			{\it Adam} && {\color{green} 2.2e-4} & {\color{green} 1.4e-4} & {\color{green} 2.8e-4} & {\color{green} 1.2e-4} & {\color{green} 3.0e-4} & {\color{green} 0.017} \\			 
			\hline
			{BrUSLEE} / & 2 & 0.26 & 0.09 & 0.52 & 0.14 & {0.22} & 0.94 \\
			{\it The Hay Wain} && {\color{red} 0.26} & {\color{red} 0.07} & {\color{red} 0.53} & {\color{red} 0.14} & {\color{red} 0.20} & {\color{red} 0.95} \\
	{} && {\color{green} 1.3e-4} & {\color{green} 1.6e-4} & {\color{green} 1.6e-4} & {\color{green} 0.9e-4} & {\color{green} 1.6e-4} & {\color{green} 4.8e-4} \\
			\hline
			{mBeRFP} / & 3 & 0.01 & 0.64 & 0.01 & 0.34 & {0.19} & 2.31  \\
			{\it The ambassadors}&& {\color{red} 0.01} & {\color{red} 0.65} & {\color{red} 0.01} & {\color{red} 0.33} & {\color{red} 0.19} & {\color{red} 2.24} \\
			{} && {\color{green} 1.4e-4} & {\color{green} 2.9e-4} & {\color{green} 2.9e-4} & {\color{green} 3.8e-4} & {\color{green} 2.1e-4} & {\color{green} 0.0024} \\
			\hline 
			{Dendra2(Red)} / & 4 & 0.01 & 0.44 & 0.01 & 0.54 & {0.13} & 4.46 \\
			{\it Old woman and} && {\color{red} 0.01} & {\color{red} 0.44} & {\color{red} 0.01} & {\color{red} 0.54} & {\color{red} 0.13} & {\color{red} 4.38} \\
			{\it boy with candles} && {\color{green} 3.8e-4} & {\color{green} 2.4e-4} & {\color{green} 4.0e-4} & {\color{green} 3.0e-4} & {\color{green} 2.8e-3} & {\color{green} 0.0024} \\
			\hline
			{MiCy} / & 5 & 0.63 & 0.11 & 0.22 & 0.04 & {0.10} & 3.90 \\
			{\it  The great wave} && {\color{red} 0.80} & {\color{red} 0.12} & {\color{red} 0.08} & ${\color{red} 0.00}$ & {\color{red} 0.07} & {\color{red} 3.72}\\
			{\it off Kanagawa} && {\color{green} 6.4e-4} & {\color{green} 6.2e-4} & {\color{green} 7.8e-4} & {\color{green} 2.8e-4} & {\color{green} 1.3e-4} & {\color{green} 0.0027} \\
		\end{tabular}
				
	\end{center}
\end{table*}

\Fig{Fig_UnmixingFree_fastNMF_H} shows the retrieved spatial distributions and FP properties obtained for $\Itot=10^4$. The spectral and temporal properties extracted from the retrieved quantities are given in red in \Tab{tab:5FPs}, showing a good agreement between the retrieved and original $\bS$ and $\bT$, with  \bT\ being slightly faster. Even for $\Itot=10^2$ (see \Fig{FigSM_UnmixingFree_fastNMF_L}), the retrieval works reasonably, showing only some crostalk between the FPs with most similar properties, mBeRFP and Dendra2(Red). Results can be slightly improved by subsequently minimizing the KLD (see \Fig{FigSM_UnmixingFree_fastNMFGD_H} and \Fig{FigSM_UnmixingFree_fastNMFGD_L})\edit{, however}{. However,} this takes \edit{2 to 5}{two to five} times longer than the fast NMF \add{,} depending on \Itot\, and the choice of initial guesses. 

We note that the number of FPs retrievable within a certain error depends in a complex way on their properties, especially on their differences, as well as the signal strength \Itot, and the FP spatial distributions. Therefore, for a given experiment, a reliable determination of the retrieval error should be obtained via repeated retrievals using new realizations of the photon counts \bD\ from probability distributions determined by the measured counts. To give an example, for the parameters shown in \Tab{tab:5FPs}, we evaluated \edit{10}{ten} realisations of the photon shot noise, and found that the absolute deviations for $\hat{c}_{def}$ and $\hat{c}_f$ and the relative deviation for $\tau$ are below 1\%, as shown in  \Tab{tab:5FPs}. 

To exemplify the benefits of using retrieved properties versus fixed properties, we show in \Fig{FigSM_UnmixingFree_fastNMFFixed_H} the FP distributions obtained from the data of \Fig{Fig_UnmixingFree_fastNMF_H} fixing the FP properties to the nominal ones, not including the variations introduced. Significant systematic errors are found for weak FPs, e.g. Dendra2(Red) and MiCy. With decreasing \Itot, the noise in the data is increasing and the relative importance of the systematic error decreases, so that for $\Itot=100$ (see \Fig{FigSM_UnmixingFree_fastNMFFixed_L}), these systematic errors are less relevant.

We emphasize that while we have chosen here exponential dynamics allowing to use known FP parameters, the method is applicable for any dynamics -- as example we show in \Sec{sec:unmixlogn} results for a log-normal distribution. The retrieval quality\add{,} even when using a broad distribution $\sigma=0.8$\add{,} is similar to the case of exponential dynamics, confirming that the method is suited for a wide range of FP dynamics.

We stress that retrieving both the spatial distribution and the FP spectral and dynamic properties from the measured data eliminates the need for separate measurements on reference samples with individual FPs. Notably, the spectral and dynamic properties of FPs \edit{are varying}{vary} with their environment, and thus can be different between pure solutions and cellular samples. Furthermore, \add{a} long-term drift of the instrument response can \edit{add to}{introduce} systematic deviations between the FP properties used and the ones present in the sample of interest. Removing the need for such prior knowledge is\add{,} therefore\add{,} a major advantage of uFLIM.

\subsection{uFLIM-FRET application I: Analysis of synthetic data}\label{sec:resultsFRET}

To verify the uFLIM-FRET method, we first use synthetic data. We consider two detectors which are mostly detecting the donor and acceptor emission, respectively, given by $\R{d}=\R{a}=0.9$. The FLIM system is the same as in \Sec{sec:sFLIM}. We consider that \add{the} donor and \add{the} acceptor fluorescence have \delete{an} exponential dynamics, with decay rates of \gD=0.33/ns and \gA=0.385/ns, respectively, corresponding to the decay lifetimes of mNeonGreen and \edit{mCherry}{mRuby}. We vary the spatially averaged time-integrated photon counts of the donor emission $\Id$, proportional to the one of the acceptor emission, $\Ia$, using $\Ia=0.8\Id$ throughout. The dynamics of the DAP detected by the two channels \edit{is}{are} calculated according to \Eq{eq:fretint}, considering a log-normal distribution of FRET rates. 


We generated data with \gb\ taking values of $\gbs=0.1$/ns, 0.5/ns and 0.9/ns, and \sg\ given by $\sgs=0.5$. The relative detection efficiency between donor and acceptor was taken to be $\qs=1$.
In the following, symbols without subscript refer to the parameter values changed by the algorithm, while symbols with the subscript s refer to the values used to generate the data, and symbols with the subscript r are values resulting from the algorithm.

Various relative strength\add{s} of DAP and donor emission, $\Idap/\Id$, are considered, where $\Idap$ are the spatially averaged time-integrated photon counts of the DAP emission. As spatial distributions of donor, acceptor, and  DAP\add{,} we used Monet's {\it Nymph\'eas}, Van Gogh's {\it Starry Night}, and  Leonardo's {\it La Gioconda}, respectively. The drawings \cite{Wikipedia} were cropped, resized to $256\times256$ pixels, and converted to greyscale. The synthetic data $\bD^{\rm s}=\bSp{s}\bTp{s}+b$ are then created by multiplying each pixel of the images with the corresponding decay curve\delete{,} and adding the dark counts $b$, which we characterize by their equivalent intensity $\Ib=b\Nt$. For the data shown, we have considered $b=0$, 0.001, and 0.01, corresponding to $\Ib=0$, 2, and 20.

The photon counting data $\bD$ is generated from $\bD^{\rm s}$ using Poissonian statistics as before, and we repeated the analysis for 10 realizations of $\bD$. To reduce the analysis time, we apply a time binning with \tb=25\,ps and \rb=0.05 (see SI \Sec{sec:binning}). 
The data are then factorised using the donor (\bTd), acceptor (\bTa), and FRET (\bbT) components over a grid of the FRET parameters \gb, \sg, and $q$. Donor and acceptor dynamics without FRET are taken as known --  in experiments\add{,} these would have been measured and retrieved by uFLIM. No free components are used\delete{,} so that the factorization is a single step NMF for the spatial distributions $\bSw$\add{,} which minimise the residual $E$. The initial guesses for $\bSw$ are random.

\begin{figure}[h]
	\includegraphics[width=\columnwidth]{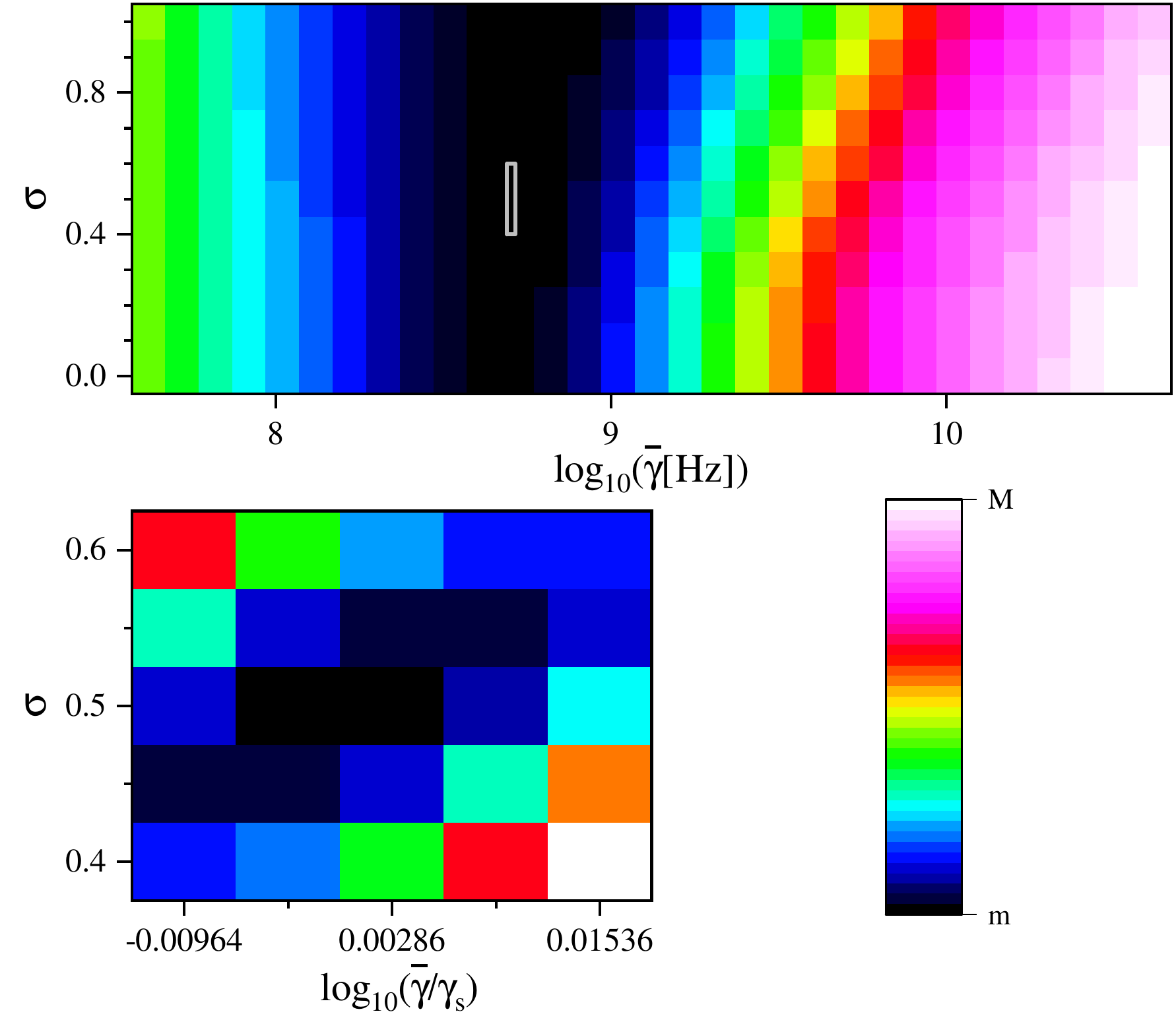}
	\caption{Factorization error $E$ as a function of the FRET distribution parameters \gb\ and \sg\ for $q=1$ for data generated using $\Id=\Idap=10^4$, \gbs=0.5/ns, \sgs=0.5, \qs=1, $\kappa=1$ and $\Ib=2$.  Top: Coarse grid, $m=399$, $M=537$. Bottom: refined grid (see \Sec{sec:FRETparmin}), $m=399.163$, $M=399.256$. The refinement domain is defined by the grey rectangle in the top panel. \edit{Colorscale}{Colour scale} as given from $m$ to $M$.} 
	\label{FigSM_FRET_Error}
\end{figure}

The dependence of the factorisation error $E$ over the parameter space is shown in \Fig{FigSM_FRET_Error}.
The top panel of \Fig{FigSM_FRET_Error} shows $E$ over the coarse grid of FRET parameters \gb\ and \sg\ for $q=1$, and $\Id=\Idap=10^4$. The bottom image shows $E$ calculated during the grid refinement step, within the finer grid range indicated by the grey rectangle in the top panel. The residual is minimized to a relative change better than $10^{-5}$. Note that the non-zero residual is entirely due to the shot noise in the photon counts. The estimated parameter values are close to the ground truth of the simulated data (the relative errors are 0.07\%, -0.72\%, and 0.17\% for \gb, $\sigma$ and $q$, respectively), with remaining deviations due to the photon shot noise.

\begin{figure*}[h]
	\begin{center}
	\includegraphics[width=0.7\textwidth]{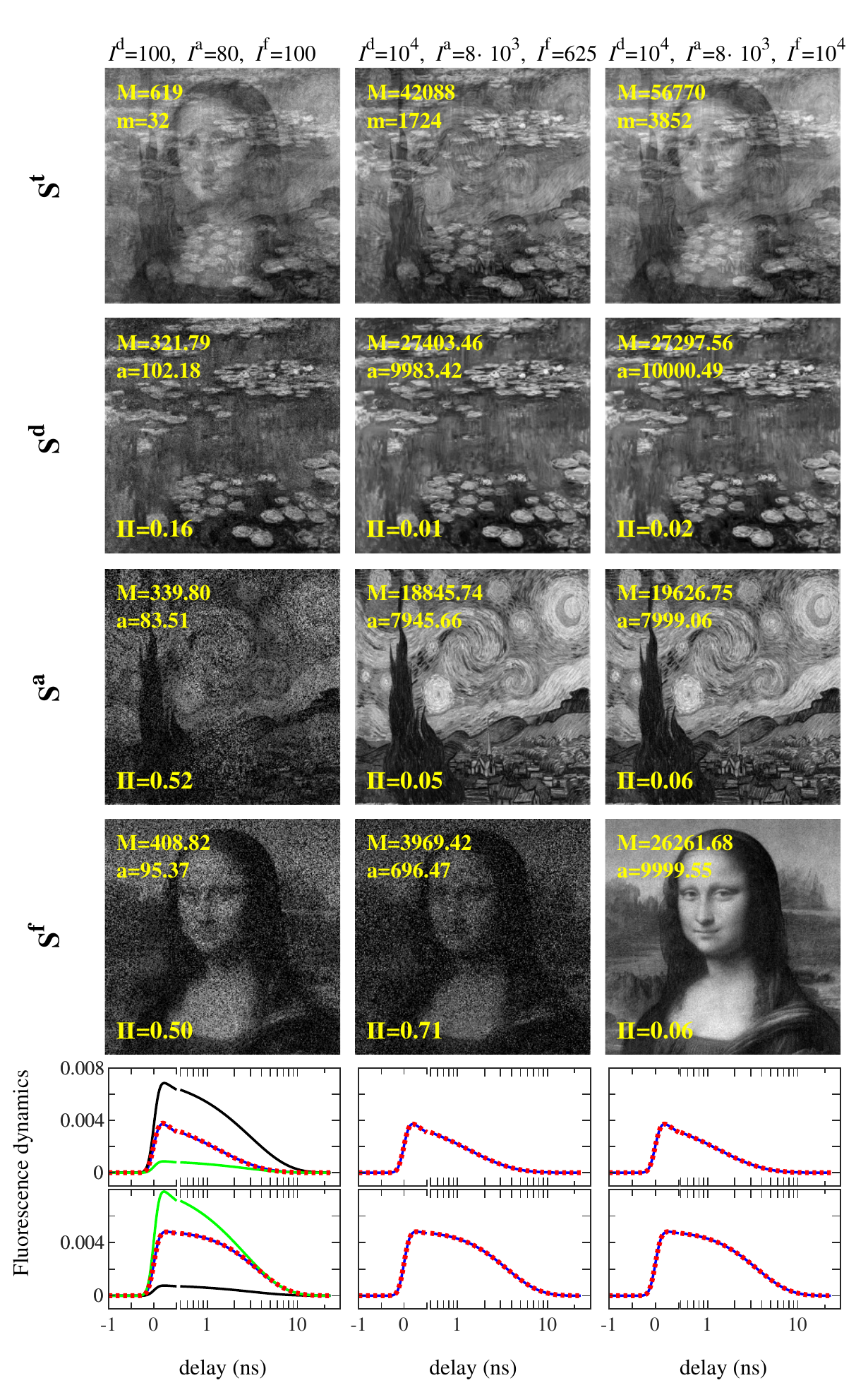}
	\caption{Results of uFLIM-FRET for synthetic data \add{generated} using $\gbs=0.5$/ns, $\sgs=0.5$, $\qs=1$, $\Ib=2$ and $\kappa=1$. The three columns refer to different intensities\add{,} as indicated \add{on the top}. Top row: the time summed data $\St=\Nt\Sav$, on a linear grey scale from a minimum $m$ (black) to a maximum $M$ (white) as indicated. The second to fourth rows show the retrieved spatial distributions of the donor \Sd, acceptor \Sa, and FRET \Sf. Here $m=0$\add{,} and $a$ is the average pixel value over the image. The bottom panels show the \edit{synthetic original}{ground truth} dynamics of donor \bTd (black), acceptor \bTa (green), and DAPs undergoing FRET \bbT (blue), with the retrieved FRET dynamics given as red dashed lines. The signal\add{s} \edit{of}{acquired at} the donor (acceptor) detector are given in the top (bottom) panel, respectively. The dynamics are normalized to have a sum of unity over the 2000 temporal points of both detectors. The retrieved FRET rate distribution parameters are $\gbf=(480.52\pm 0.03)/\mu$s, $\sgf=0.0732 \pm 0.0013$ and $\qf=0.98143\pm 0.00007$ for the first column, $\gbf=(502.55\pm 0.39)/\mu$s, $\sgf=0.5201 \pm 0.0019$ and $\qf=0.99961 \pm 0.00006$ for the second column, and $\gbf=(499.72\pm 0.11)/\mu$s, $\sgf=0.499 \pm 0.00055$, and $\qf=1.00042 \pm 0.00002$ for the third column. The errors given are the uncertainty of the minimum position of the second order polynomial fit to the reconstruction error (see \Fig{FigSM_FRET_Error}).}
	\label{Fig_FRET_g=5_k=1_Ib=2}
\end{center}
\end{figure*}

\Fig{Fig_FRET_g=5_k=1_Ib=2} shows uFLIM-FRET results for the values of \gbf, \sgf, and \qf\ minimizing the residual, for a specific data realisation with $\gbs=0.5$/ns, $\Ib=2$ and $\kappa=1$. Results for small intensity and strong FRET ($\Id=\Idap=100$) are given on the left, for large intensity and weak FRET ($\Id=16\Idap=10000$) in the middle, and for large intensity and strong FRET ($\Id=\Idap=10000$) on the right. The first row shows the data summed over the temporal channels, $\Nt\Sav$, where the images of donor and  acceptor are visible, and the FRET image is discernible for strong FRET. The synthetic data dynamics \bTp{s,d}, \bTp{s,a}, and \bTp{s,f}, are given as solid lines in \Fig{Fig_FRET_g=5_k=1_Ib=2} (bottom).  The second, third and fourth rows from the top show the spatial distributions \Sd, \Sa, \Sf\ retrieved by NMF, recovering the corresponding images well, also in conditions of small intensity (left) and weak FRET (middle). The difference between the original and retrieved data is quantified using the relative error $\epsilon=||\bD^{\rm s}-\bS\bT||_2/||\bD^{\rm s}||_2$, and similarly the reconstruction of the individual components is quantified by the relative errors $\epsilon_i=||\bSp{s,\it i}\bTp{s,\it i}-\bSp{\it i}\bTp{\it i}||_2/||\bSp{s,\it i}\bTp{s,\it i}||_2$,  where $i\in\{{\rm d,a,f}\}$. The mean values ($\langle.\rangle$) and the standard deviations ($].[$) of the reconstruction errors calculated over the data realizations are shown in \Fig{FigSM_FRET_RecError_g=5_k=1_Ib=2}. As expected, the reconstruction error decreases with increasing intensities. We find that the error scales approximately as $1/\sqrt{\Id}$ (see \Fig{FigSM_FRET_IdIfDep}). In general, $\epsilon$, $\epsilon_{\rm d}$, and $\epsilon_{\rm a}$ depend mostly on $\Id$, while $\epsilon_{\rm f}$ is affected by both $\Id$ and $\Idap$. We note that all errors are below 10\% for the \edit{high intensity}{high-intensity} case, and that they are always much larger than the parameter retrieval error, since they are dominated by the shot noise in the realizations.
 
The uFLIM-FRET analysis is largely superior to the phasor analysis approach, as we show in the SI \Sec{sec:phasor} using the same data. Specifically, to extract quantitative information, a phasor analysis needs to assume a simple model of the dynamics, and the abundance of the donor-acceptor pairs undergoing FRET and the FRET efficiency are hardly disentangled. Furthermore, the spatial distributions of the donor-only and DAPs obtained with the phasor analysis poorly reflect the original distributions (see \Fig{FigSM_Phasor}).

The retrieved dynamics of the FRET component \bbT\ (dashed lines in \Fig{Fig_FRET_g=5_k=1_Ib=2}) agree well with the \edit{original synthetic dynamics}{ground truth}, which is confirmed by the close match of \edit{original}{true} and retrieved values of the parameters \gbf, \sgf, and \qf\ given in the caption. Their mean values and standard deviations over the ensemble of realizations are given in \Fig{FigSM_FRET_ParamError_g=5_k=1_Ib=2}. The errors decrease as the intensities increase, showing that the method is correctly retrieving the FRET parameters. The standard deviation, which is due to the photon shot noise in each realization, is rather similar for the different parameters, with \sg\ being retrieved with \delete{somewhat} less accuracy as its influence on the dynamics \bbT\ is lower. However, we also see some systematics for low intensities, in particular \sg\ is underestimated. To verify if this could be due to the remaining non-whiteness of the noise in the analysed data \bDw\add{,} we \delete{have} repeated the factorisation using the gradient descent minimizing the KLD, with the fast NMF results as initial guesses. The retrieved spatial distributions and FRET parameters obtained with the two methods are generally very similar (see \Fig{FigSM_FRET_g=5_k=1_Ib=2_GD}), confirming the suitability of the fast NMF algorithm on whitened data for the analysis of data showing Poisson noise. Additionally, the gradient descent comes with \add{more than} two orders of magnitude longer computational time \add{for a single CPU core} of about \edit{50\,\textmu s}{1\,ms}/pixel for given FRET parameters, and of the order of 5000 evaluations are used to find the parameters \edit{minimizing}{that minimise} the error, which makes it unsuitable for real\add{-}time analysis.

The difference in the accuracy among the different parameters can be understood by looking at the curvature of the reconstruction error along the directions defined by the parameters. The curvature is much smaller along the \sg\ direction, resulting in a lower accuracy in the determination of this parameter (see \Sec{sec:fretquad}). 

Further results for different \gbs\ and \Ib\ are given in the SI \Sec{sec:fretpara}.  In the case of a small FRET rate $\gbs=0.1$/ns, donor and DAP dynamics are similar, making the retrieval more challenging, so that for small intensities\add{,} the FRET image bleeds through to the \Sd\ component, \bTp{f}\ differs from \bTp{s,f}, and the value of \sg\ is underestimated. For a higher rate $\gbs=0.9$/ns instead, the DAP dynamics  and spatial distribution \edit{is}{are} recovered with higher accuracy. The obtained average parameters for the two cases of $\gb=0.1$/ns and $\gb=0.9$/ns are also given in the SI \Sec{sec:fretpara}. The dark count rate \edit{is adding}{adds} uncertainty to the retrieval. Without dark rate ($\Ib=0$), the method is able to retrieve the correct parameters of the FRET distribution with smaller error than for $\Ib=2$ (see \Fig{FigSM_FRET_g=5_k=1_Ib=0} and \Fig{FigSM_FRET_ParamError_g=5_k=1_Ib=0}), and \add{the} retrieval is possible \edit{down to lower}{for} intensities \add{as small as} $\Id=32$ and mean FRET rates of 0.5 and 0.9/ns. Conversely\add{,} for large dark rate ($\Ib=20$), higher \Id\ and \Idap\ are required for retrieval, see SI \Fig{FigSM_FRET_g=5_k=1_Ib=20} and \Fig{FigSM_FRET_ParamError_g=5_k=1_Ib=20}.

The dependence of the reconstruction and FRET parameter retrieval errors on the image size is analysed in the SI \Fig{FigSM_FRET_SizeDependence_k=1}. The systematic errors of the mean FRET parameters are not significantly affected by the number of pixels \Ns. The standard deviation\edit{ instead is scaling}{, instead, scales} as $1/\Ns$, \add{which is} steeper than the $1/\sqrt{\Ns}$ dependence expected for the shot noise. We note that each pixel comes with its own concentration in \bS, so that the number of photons per retrieved information is independent of \Ns, as long as the number of spatial points is much larger than the number of FRET parameters.

We have repeated the analysis in the case of negligible direct excitation of the acceptor molecules, choosing $\kappa=0$ in \Eq{eq:Ta}\edit{, and accordingly not including}{. Accordingly, we do not include} a pure acceptor component with \delete{a} dynamics $\bTa$ in the NMF. The corresponding increase in contrast\delete{,} and reduction in free parameters\delete{,} results in smaller errors of both reconstruction and retrieved parameters, as shown in the SI \Fig{FigSM_FRET_g=1_k=0_Ib=0} to \Fig{FigSM_FRET_g=9_k=0_Ib=20}.
\edit{Finally, we}{We also} show the ability of uFLIM-FRET to retrieve the FRET parameters and the DAP spatial distribution in the presence of an additional component, such as autofluorescence, in the SI \Sec{sec:resultsFRETAF}. \add{The method performs well even in the presence of multiple autofluorescent species, such as bound and unbound NADH and FAD, when taking data for additional excitation and detection channels, as shown in \Sec{sec:resultsFRETNADHFAD}.}

\add{We have also considered the case of environmental conditions which could alter the dynamics, such as a spatial dependent pH, resulting in a modification of the unquenched donor dynamics similar to a FRET process. By providing two donor and two acceptor dynamics, corresponding to the end points of the pH dependence present in the data (such dynamics could be extracted from uFLIM analysis), and including a constrain given by a single spatially dependent environmental parameter, we show in \Sec{sec:resultsFRETpH} that such environment effects can be disentangled from the FRET process and quantified by the uFLIM-FRET method.}

\clearpage
\subsection{uFLIM-FRET application II: Analysis of experimental data}
To show that uFLIM-FRET works well also with experimental data, we analysed FLIM-FRET {\it in vivo} experiments \delete{)}using the data published in \,\cite{SmithPNAS19}\delete{)}, where four Matrigel plugs containing different donor (AF700)-acceptor (AF750) ratios (ROI$_1$: D:A=1:0, ROI$_2$: D:A=1:1, ROI$_3$: D:A=1:2, ROI$_4$: D:A=1:3) are implanted subcutaneously into a mouse and imaged \,\cite{SmithPNAS19,SinsuebphonJB18}. Only one channel, centered at the donor emission, has been acquired in the FLIM measurements. In the analysis, the data \add{of the regions} corresponding to the four Matrigel plugs were \edit{considered}{used}. Before performing the uFLIM-FRET, we \delete{have} compensated for the possible pixel-dependent variation of the laser pulse arrival time. For each pixel, we \delete{have} defined the pulse arrival time as the time when the measured intensity is half of the maximum recorded signal. To align the time axis, data were interpolated, and we used linear extrapolation to take into account \edit{for}{the} truncated dynamics. Only data with \add{a} delay larger than -0.22ns were used to limit the contribution of \add{the} signal at negative time delays.

After \edit{this}{these} pre-processing steps, we have used the pixels in the Matrigel region with D:A=1:0 (ROI$_1$) to obtain the dynamics of the free donor applying uFLIM with one component. The data were time binned ($\tb=0.04$\,ns and $\rb=0.05$) to improve \add{the single pixel} signal-to-noise ratio and reduce computational time. \add{We note that such a temporal binning step might be useful also for other analysis methodologies.}
uFLIM-FRET was then used to estimate the distribution of the DAP undergoing FRET\add{,} including all pixels in the four ROIs. Since the data were acquired using only a single channel resonant with the donor emission, and the acceptor bleed-through was not characterised, we have performed our analysis assuming $\R{a}=1$ and $\eta=0$ and searching for the combination of \gb\ and \sg\ minimising the NMF error. We did not apply partial whitening as the noise in the data did not show a significant intensity dependence, which may be due to dominating read noise or other classical noise.

\begin{figure}[h]
	\includegraphics[width=\columnwidth]{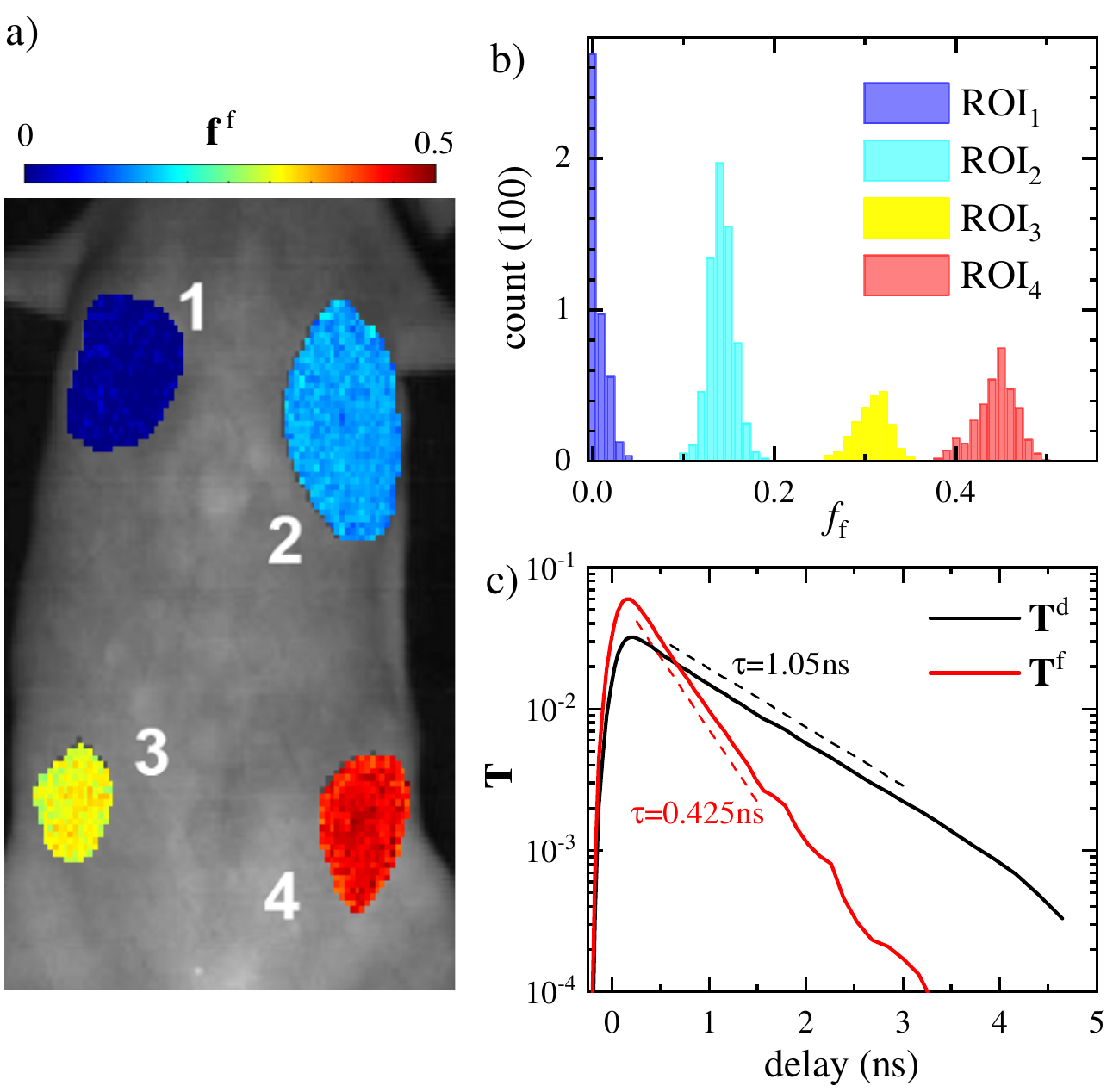}
	\caption{Results of the uFLIM-FRET analysis on data from \Onlinecite{SmithPNAS19}. The different Matrigel plugs contain different donor-acceptor ratios (ROI$_1$: D:A=1:0, ROI$_2$: D:A=1:1, ROI$_3$: D:A=1:2, ROI$_4$: D:A=1:3). a) Spatial distribution of the quenched donor fraction \ff\ in the different ROIs. b) Histograms of \ff\ measured in the four ROIs. c) Dynamics of the uFLIM-FRET components for the unquenched (\bTp{d}, black) and quenched (\bTp{f}, red) donor.}
	\label{Fig_FRET_invivo}
\end{figure}

\Fig{Fig_FRET_invivo} shows the results of the uFLIM-FRET analysis. The retrieved FRET rate distribution has a mean rate \gb\ of $\sim1.2$\,GHz with a negligible width ($\sg\sim0$). We \delete{have} calculated the fraction of photons emitted by the donor undergoing FRET as point-wise $\ff=\Sf/\left(\Sd+\Sf\right)$. The spatial distribution of \ff\ is shown in \Fig{Fig_FRET_invivo}a. The different ROIs present rather uniform values of \ff\ quantified by the histograms in \Fig{Fig_FRET_invivo}b. The retrieved dynamics of the unquenched (\bTp{d}) and quenched (\bTp{f}) donor (\Fig{Fig_FRET_invivo}c) show approximately mono-exponential decays with lifetimes of 1.05\,ns and 0.425\,ns, respectively. Our results are in agreement with least-square fitting and deep-learning approaches (see SI of \Onlinecite{SmithPNAS19}). Importantly, uFLIM-FRET retrieves a more uniform distribution of \ff\ in the different ROIs (narrower histograms), which is closer to the uniform distribution expected from the experiment.

\begin{figure}[h]
	\includegraphics[width=\columnwidth]{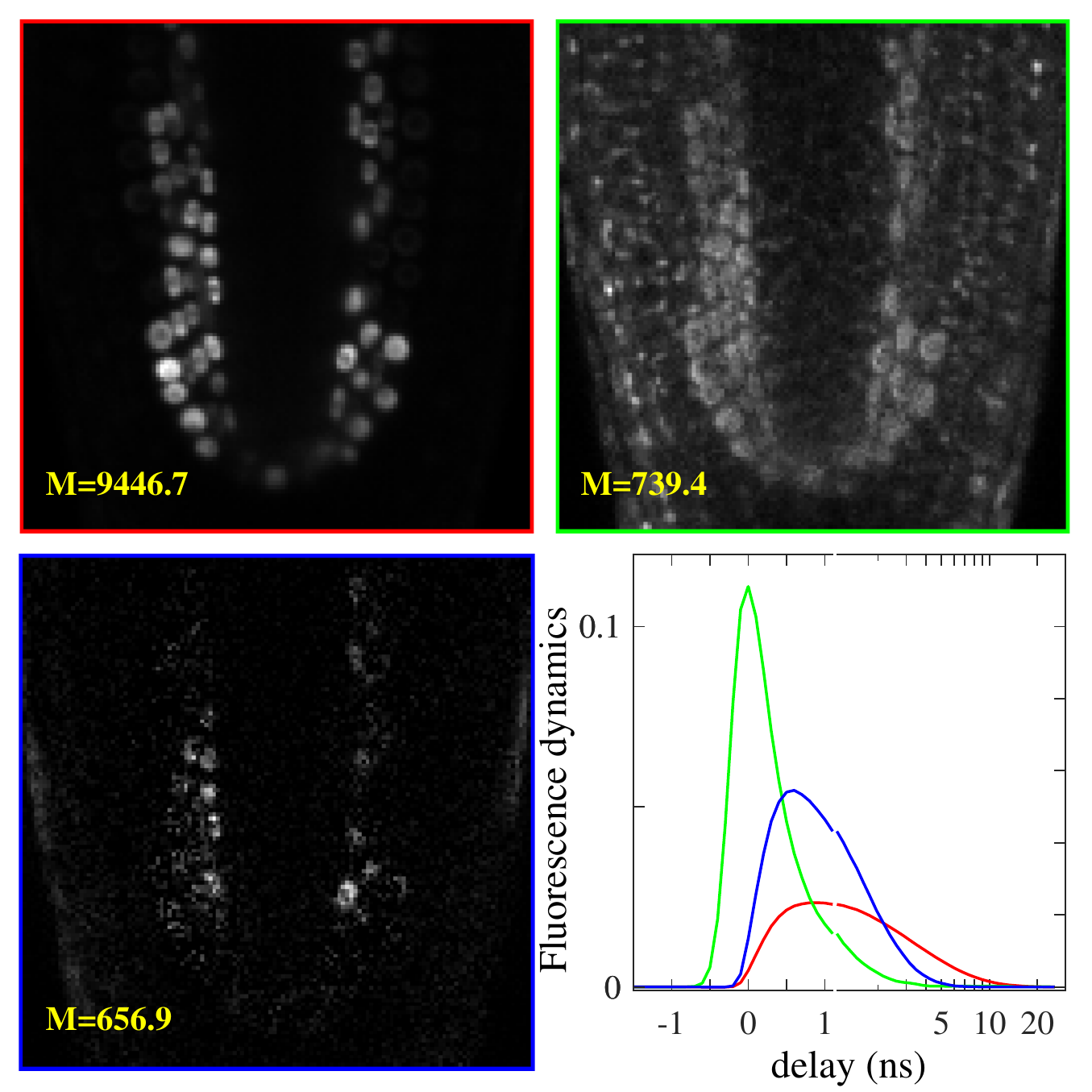}
	\caption{uFLIM-FRET analysis of pSCR expressed SCR and SHR in the Arabidopsis root endodermis. The images show the distribution of the three components used in the uFLIM-FRET analysis on a grayscale as in \Fig{Fig_UnmixingFree_fastNMF_H} with $m=0$. Red: donor ({\it pSCR::SCR:YFP}), green: autofluorescence, Blue: DAP. The corresponding dynamics \bTd\ (red), \bbT\ (blue), and \bTAF\ (green) are shown in the graph.}
	\label{Fig_Arabidopsis}
\end{figure}

Additional unknown fluorescence components, such as autofluorescence, can be included in uFLIM-FRET, as we demonstrate here using FLIM-FRET experiments reported in \Onlinecite{LongN17} on Arabidopsis roots co-expressing two tagged interacting transcription factors, SHORT-ROOT (SHR) and SCARECROW (SCR). The levels of both proteins are elevated in the endodermis controlled by the SCR promoter (pSCR). The SCR factor is tagged with YFP acting as donor, while the SHR protein is tagged with the RFP acting as acceptor. Only one channel, centered at the donor emission, has been acquired in the FLIM measurements. The data \edit{was}{were} binned both spatially ($2\times 2$) and temporally (\tb=100\,ps, \rb=0). We \delete{have} used uFLIM on images of roots expressing only {\it pSCR::SCR:YFP} to retrieve the donor (\bTd) and \delete{an} autofluorescence (\bTAF) dynamics. Using \edit{this}{these} donor dynamics, we have applied uFLIM-FRET on data from roots co-expressing {\it pSCR::SCR:YFP} and {\it pSCR::RFP:SHR}, using a binning \tb=100\,ps and \rb=0.1, and the time zero set to the peak of the autofluorescence component. Since only the donor was measured, we \edit{use}{used} $R^d=1$ and $R^a=0$ and the FRET dynamics \edit{simplifies}{simplified} to\add{:}
\be \bbT{}(\gb,\sg)=\int P(\gamma;\gb,\sg) \btT{d}(\gamma)d\gamma. \ee
\Fig{Fig_Arabidopsis} shows the results of uFLIM-FRET, yielding $\gbf=0.57/$ns and $\sgf\sim0$. This corresponds to a quenched donor decay time of $\tau_{\rm f}=\tau_{\rm d}/(1+\gbf\tau_{\rm d})=1.2$\,ns and a FRET efficiency of $E=1-(\gbf\tau_{\rm d}+1)^{-1}=0.66$, where a donor lifetime of $\tau_{\rm d}= 3.57$\,ns has been estimated from the first moment of \bTp{d}.

The analysis reveals an accumulation of DAPs in the endodermis of the root, where both donor and acceptor are expressed, in line with \add{the} reported single-pixel lifetime analysis of \delete{these data} \Onlinecite{LongN17} (see also the distribution of the retrieved average lifetime in the SI \Sec{sec:HeatMap}). The computational time was about  \edit{0.5}{2}\,\textmu s/pixel for a single iteration \add{performed by a single CPU core}, and $\sim700$ iterations were used to find the parameters \edit{minimizing}{that minimise} the error. \add{With our CPU, the total analysis time was ~10\,s, including fitting. The computational time can be significantly reduced if a GPU is used, allowing more parallel calculations.} As with the synthetic data, using the gradient descent minimizing the KLD instead of fast NMF does not lead to significant changes in the parameters ($\gbf=0.47/$ns and $\sgf\sim0$, see \Fig{FigSM_Arabidopsis_GD}). 
\section{Conclusion}
We have demonstrated a data analysis method, which we call uFLIM, to analyze FLIM data in an unsupervised way. It employs a fast non-negative factorization algorithm on partially whitened data to infer the emission dynamics and the spatial distribution of emitting molecules. The method offers several advantages compared to other approaches in the analysis of FLIM data available in \add{the} literature. Firstly, it does not make assumptions on the shape of the dynamics, which is instead the starting point of standard fitting techniques. Secondly, the algorithm does not require reference patterns, which are the component dynamics, as input. It can unmix spectrally resolved FLIM images where several spectrally overlapping fluorescing probes are present, extending the multiplexing capabilities of FLIM. Furthermore, the method uses a fast NMF algorithm, capable of analysing data in real\add{-}time on desktop computers. This speed comes with an approximate treatment of the noise in the data, but we have verified that the resulting systematic errors in the retrieval are not significant by comparing with a gradient descent algorithm \edit{which}{that} uses the exact noise model of the data, at the cost of orders of magnitude longer computational time.

Based on uFLIM, we developed uFLIM-FRET\add{,} which extracts FRET rates and spatial distributions. Here, the individual donor (and acceptor if detected) emission dynamics, which can be determined by uFLIM, are used to calculate the DAP dynamics for a distribution of FRET rates. uFLIM-FRET determines the values of the FRET distribution parameters which minimize the residual of the NMF, at the same time as determining the spatial distribution of donor, acceptor, and DAPs. The distribution parameters characterize the fluctuations in the separation and orientation of the donor and acceptor in the DAP, going beyond the approximation of a single FRET rate. Additional known or unknown components can be added to the retrieval. uFLIM-FRET can estimate the FRET parameters even in the presence of unknown autofluorescence. The method can be adapted to retrieve donor, acceptor, and DAP dynamics without separate donor and acceptor data. Generally, the more information is available, the more parameters can be retrieved. The precision of retrieval depends on the corresponding effect on the data -- the larger the difference between, for example, donor, acceptor, and FRET dynamics over the detected channels, the higher the precision. 

\add{Both uFLIM and uFLIM-FRET have been demonstrated on synthetic data with known ground truth and realistic photon shot-noise, as well as on experimental data taken from a range of applications, showing its wide suitability and performance. FRET could be retrieved even in presence of spatially varying donor and acceptor lifetimes due to e.g. pH dependencies, and in the presence of strong autofluorescence with multiple components, such as bound and free FAD and NADH.}  

Notably, the method also offers the possibility to compress the data of FLIM experiments into the spatial distributions of few components, which facilitates the usage of FLIM-FRET as a high-throughput tool for cell biology.

In order to enable widespread adoption of uFLIM-FRET as \add{a} method of choice to analyze FLIM data, the corresponding software is provided (http://langsrv.astro.cf.ac.uk/uFLIM/uFLIM.html). Information on the data underpinning the results presented here, including how to access them, can be found in the
Cardiff University data catalogue at http://doi.org/10.17035/d.2020.0115661402.
\clearpage	
\acknowledgements{
This work was supported by the Cardiff University Data Innovation URI Seedcorn Fund. F.M. acknowledges the Ser Cymru II programme (Case ID 80762-CU-148) which is part-funded by Cardiff University and the European Regional Development Fund through the Welsh Government. P.B. acknowledges the Royal Society for her Wolfson research merit award (Grant WM140077). The authors thank Christopher Dunsby (Imperial College London) and Ikram Blilou (King Abdullah University of Science and Technology) for kindly providing the experimental data used in the manuscript.
Discussions with Peter Watson and Camille Blakebrough-Fairbairn are gratefully acknowledged.}

\section*{Contribution statement}
F.M. and W.L. developed the method, with input from P.B. and W.D.. F.M. implemented the method and analyzed the data, with input from W.L.. All authors contributed to the manuscript writing.

\let\addcontentsline\oldaddcontentsline

\clearpage

\setcounter{section}{0}
\setcounter{table}{0}
\setcounter{figure}{0}
\setcounter{page}{0}
\setcounter{equation}{0}
\setcounter{NAT@ctr}{0}
	
\title{uFLIM -- Unsupervised analysis of FLIM-FRET microscopy data - Supplementary Information}
\maketitle{}
\onecolumngrid

\renewcommand{\thesection}{S\arabic{section}}
\renewcommand{\thefigure}{S\arabic{figure}}
\renewcommand{\thetable}{S\arabic{table}}
\renewcommand{\thesubsection}{\roman{subsection}}
\renewcommand{\thepage}{S-\arabic{page}}
\renewcommand{\theequation}{S\arabic{equation}}
\renewcommand{\citenumfont}[1]{S#1} 
\renewcommand{\bibnumfmt}[1]{[S#1]} 

\tableofcontents
\clearpage

\section{Temporal binning} \label{sec:binning}
To increase the number of counts in the tail of the fluorescence dynamics, resulting in a Poisson distribution which is more similar to the Gaussian distribution assumed for the fast NMF, and to reduce computational time, we bin the temporal points, using both an absolute time resolution \tb\ and a relative time resolution \rb, whichever is larger, yielding a binning size ${\rm ceil}({\rm max}(\rb t(i),\tb)/\Delta)$, with the time from excitation at point $i$ given by $t(i)$, and the time step $\Delta$. We start the binning at the first time step analyzed and then sequentially apply the binning to all points, skipping the final incomplete bin. The pseudocode to produce the binning is given by

\begin{algorithmic}
	\State	$c \gets 1$
	\State	$i_c \gets 1$
	\State	$n_c  \gets {\rm ceil}({\rm max}(\rb t(i_c),\tb)/\Delta)$
	\While {$i_c+n_c-1\leq l$} 
	\State	$\tilde{I}_{c}  \gets \mathlarger{\mathlarger{\sum}}_{i=i_c}^{i_c+n_c-1} I_i$
	\State	$i_{c+1}  \gets i_c+n_{c}$
	\State $c  \gets c+1$
	\State	$n_{c}  \gets {\rm ceil}({\rm max}(\rb t(i_c),\tb)/\Delta)$
	\EndWhile
\end{algorithmic}
resulting in the binned data $\tilde{I}_c$ and binning $n_c$, at times $t_c=t(i_c)+(n_c-1)\Delta/2$. 
The dynamics \bT\ shown in the manuscript are normalized by the channel width $n_i$ to represent the un-binned intensity. 
In the partial whitening transformation of \Eq{eq:pw}, the background removed from the binned data is $b n_c$.

\section{uFLIM applied to FLIM data T2-AMPKAR-991 compound.}
\label{sec:AMPCAR}
\begin{figure}
	\includegraphics[width=\textwidth]{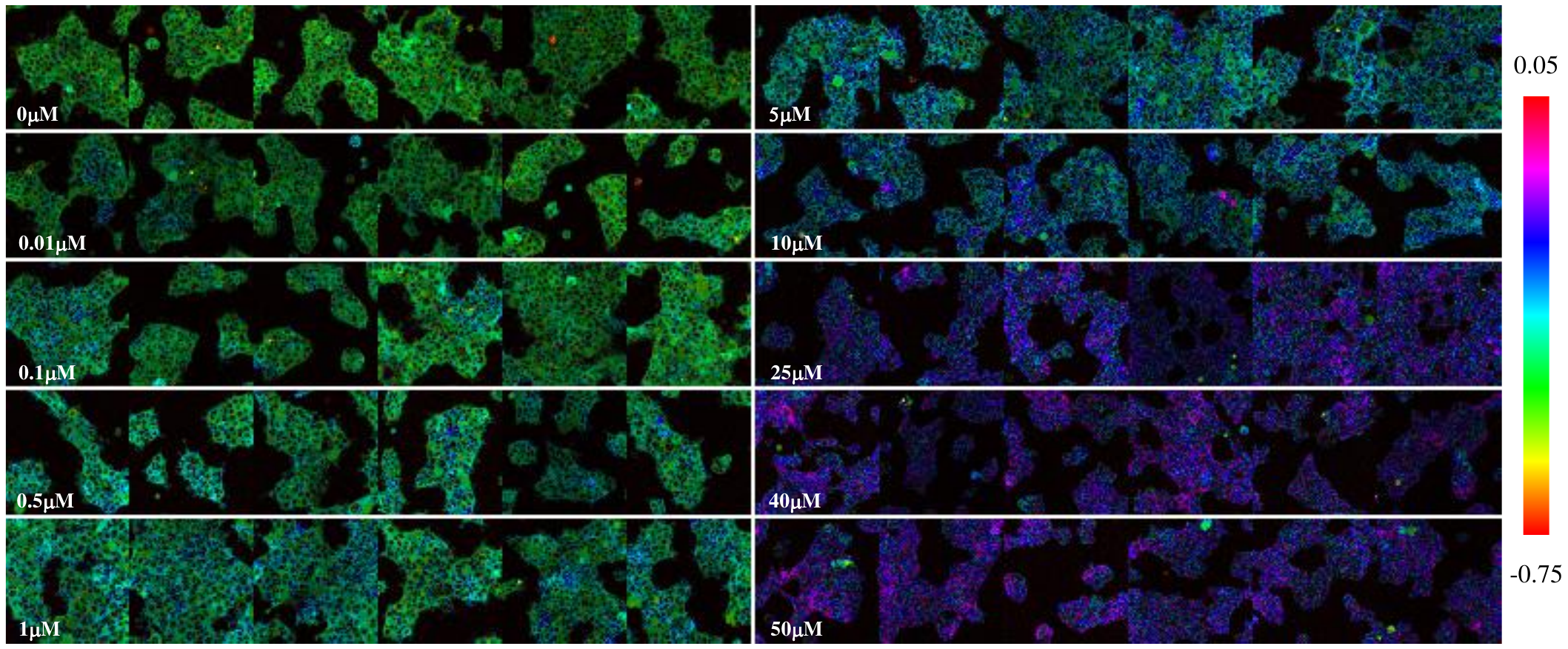}
	\caption{Concentrations of the factorisation components associated with the emission of T2-AMPKAR with and without 991, displayed with a HSV mapping as discussed in the main text and \Fig{Fig_T2AMPKAR}. All available 991 activator concentrations are shown.}
	\label{FigSM_T2AMPKAR}
\end{figure}

\Fig{FigSM_T2AMPKAR} shows the results of the uFLIM algorithm applied on FLIM dataset on HepG2 cells expressing the T2-AMPKAR compound for all concentrations of the 991 activator available in the data\,\cite{ChennelSen16S}, using the same formatting as in the main text \Fig{Fig_T2AMPKAR}.

\begin{figure}
	\includegraphics[width=0.5\textwidth]{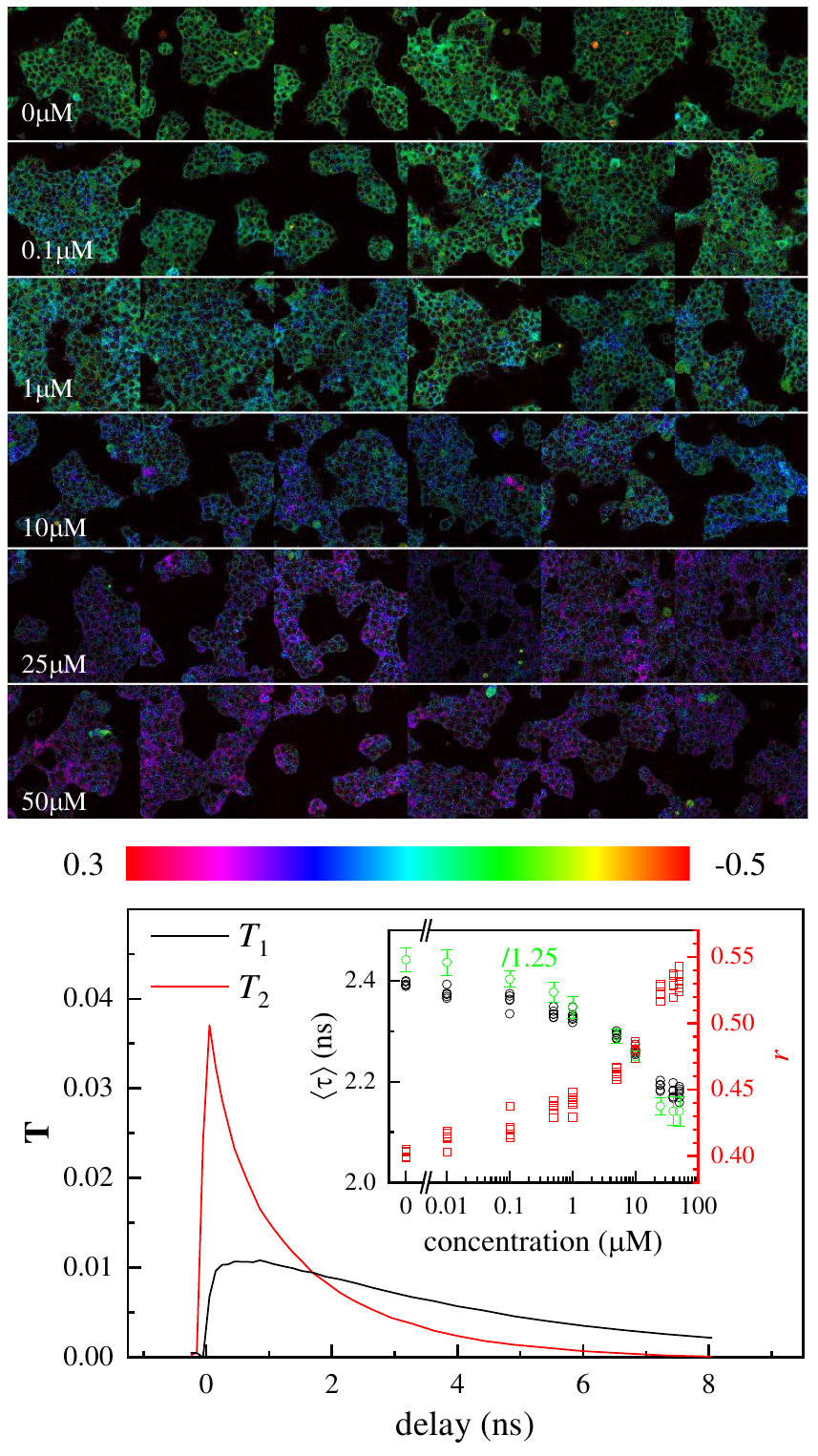}
	\caption{Same as \Fig{Fig_T2AMPKAR} but using a $2\times 2$ spatial binning: \add{Results of the uFLIM algorithm applied on a dataset from \Onlinecite{ChennelSen16} of TCSPC FLIM on HepG2 expressing the T2-AMPKAR compound as a function of the concentration of the 991 activator using a $2\times 2$ spatial binning.}}
	\label{FigSM_T2AMPKAR_Bin2}
\end{figure}

\begin{figure}
	\includegraphics[width=0.5\textwidth]{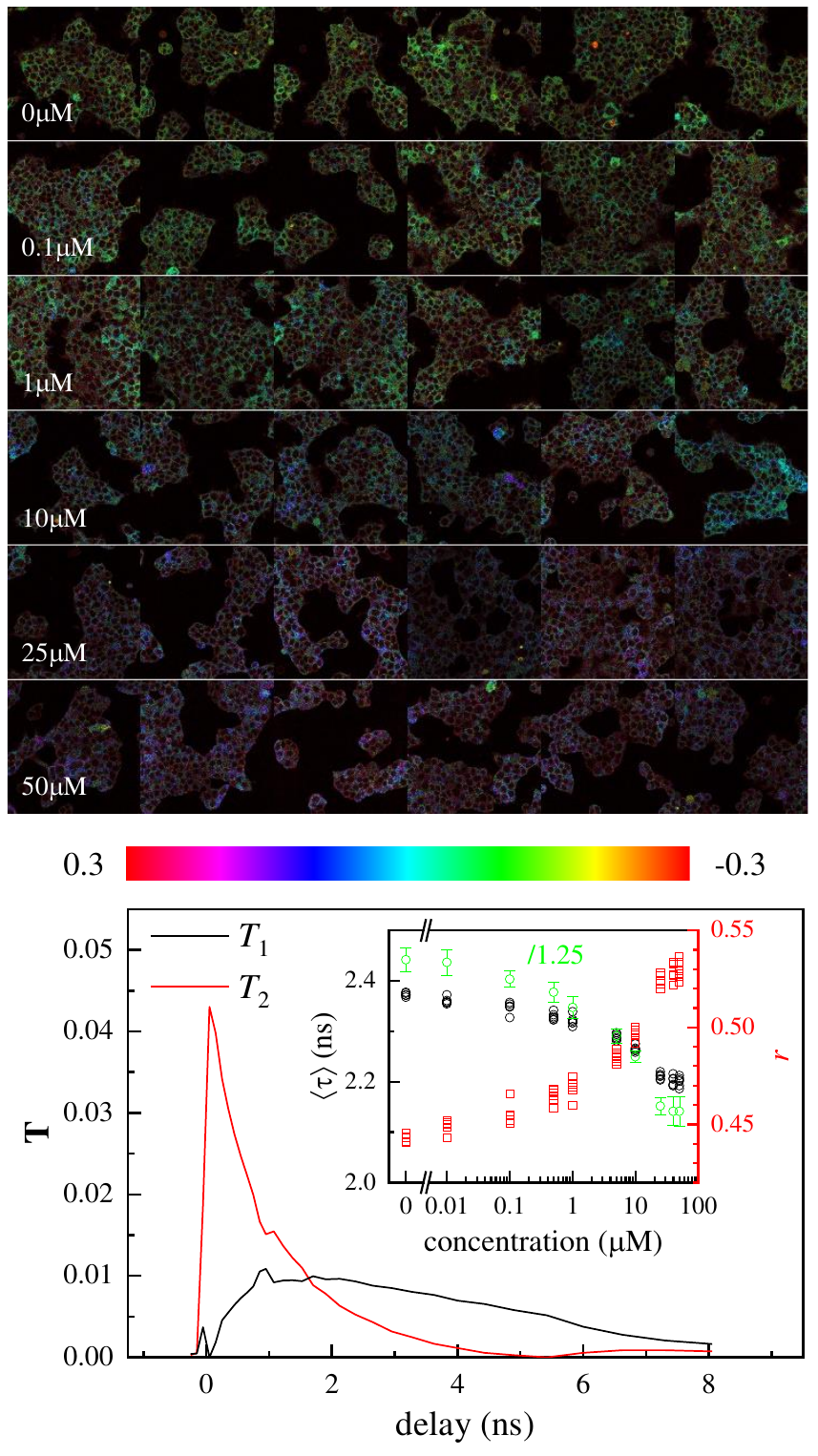}
	\caption{Same as \Fig{Fig_T2AMPKAR} but using no spatial binning: \add{Results of the uFLIM algorithm applied on a dataset from \Onlinecite{ChennelSen16} of TCSPC FLIM on HepG2 expressing the T2-AMPKAR compound as a function of the concentration of the 991 activator using a $2\times 2$ spatial binning.}}
	\label{FigSM_T2AMPKAR_Bin1}
\end{figure}

\begin{figure}
	\includegraphics[width=0.5\textwidth]{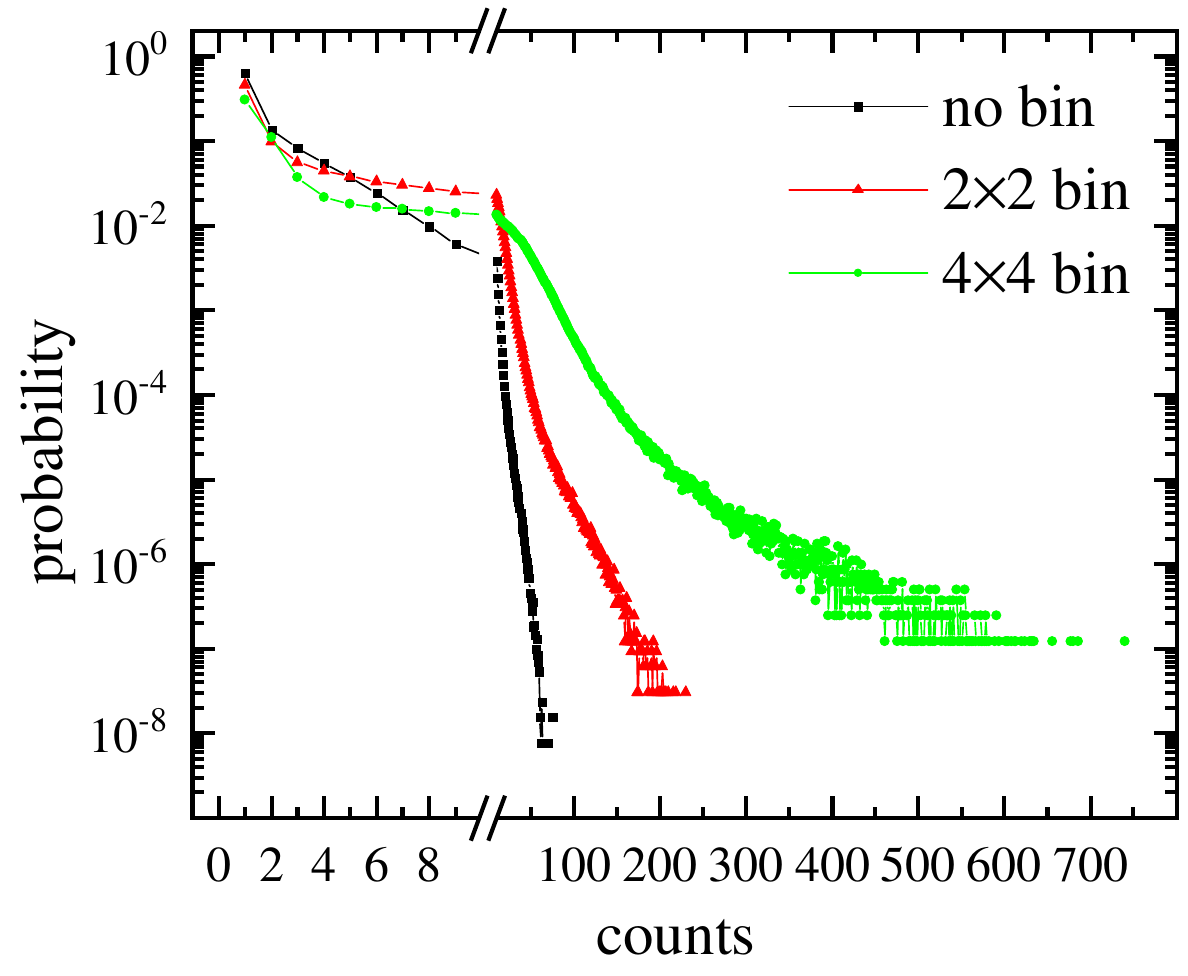}
	\caption{Histogram of counts per pixel \add{of a dataset from \Onlinecite{ChennelSen16} of TCSPC FLIM on HepG2 expressing the T2-AMPKAR compound for different spatial binning: Black squares-- no binning, red triangles -- $2\times 2$ binning, green diamonds -- $4\times 4$ binning.}}
	\label{FigSM_T2AMPKAR_Hist}
\end{figure}

\Fig{FigSM_T2AMPKAR_Bin2} and \Fig{FigSM_T2AMPKAR_Bin1} shows the results of uFLIM applied on the same data as \Fig{Fig_T2AMPKAR} with a $2\times 2$ spatial binning or no binning, respectively. \Fig{FigSM_T2AMPKAR_Hist} shows the histogram of the pixel counts for the different spatial binnings. 
Using $2\times 2$ binning (see \Fig{FigSM_T2AMPKAR_Bin2}) the results are similar to the case of the $4\times 4$ binning, but in the case of no binning (see \Fig{FigSM_T2AMPKAR_Bin1}) the two components are less separated as indicated by the slower rise time of $\bT_1$, a signature of cross-talk with $\bT_2$.

\begin{figure}
	\includegraphics[width=0.75\textwidth]{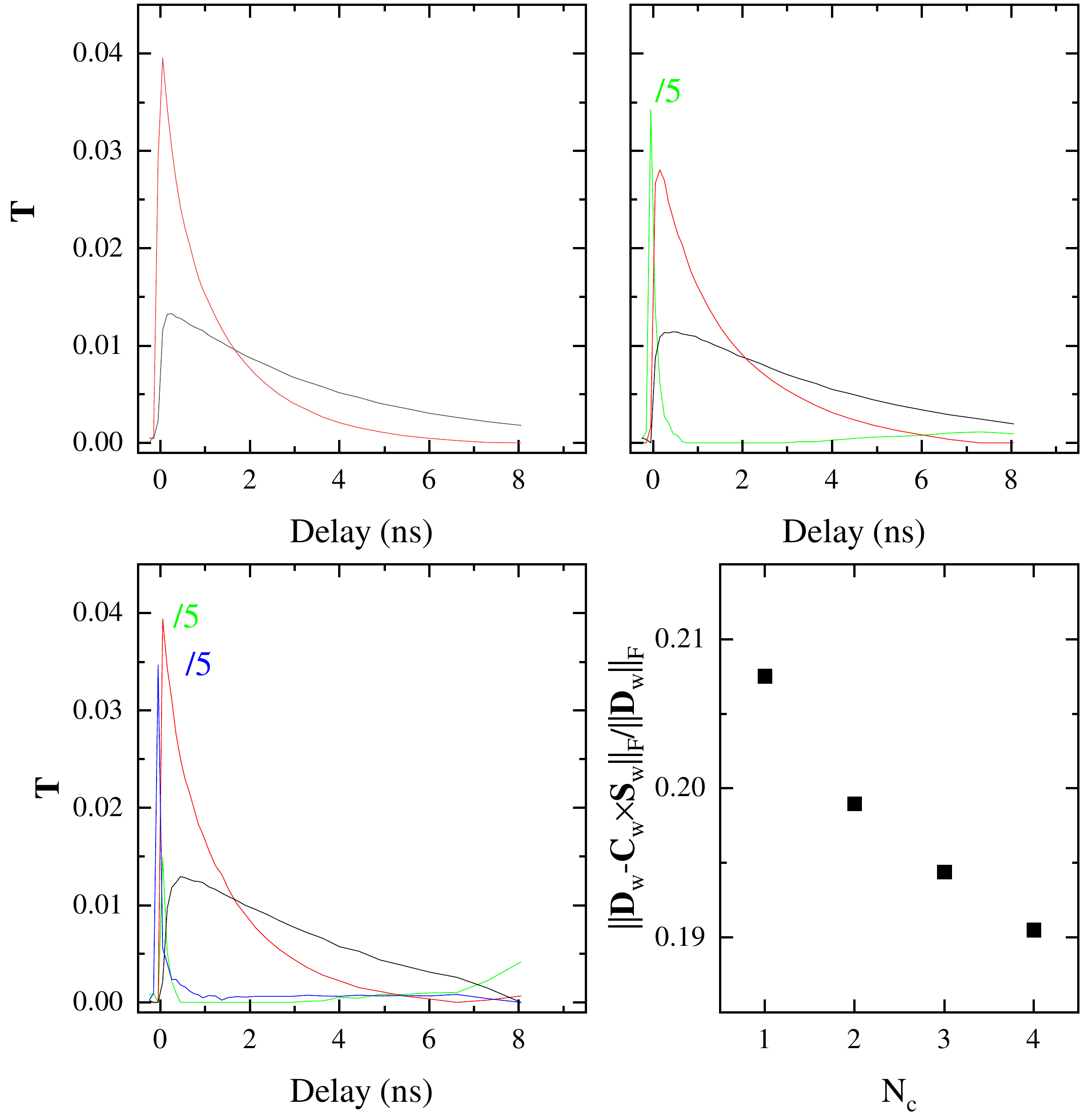}
	\caption{uFLIM \bT\ dynamics obtained varying the number of components $N_{\rm c}$. a) $N_{\rm c}=2$, b) $N_{\rm c}=3$, c) $N_{\rm c}=4$. d) Factorisation error as a function of the number of components. The black component in b) and the red component in c) have been divided by 2 as indicated.}
	\label{FigSM_T2AMPKARvNc}
\end{figure}

We have also investigated the effect of the number of components $N_{\rm c}$ on the reconstruction. \Fig{FigSM_T2AMPKARvNc} shows the resulting dynamics obtained by uFLIM on the \Onlinecite{ChennelSen16} datasets as a function of the number of components for the case of $4\times4$ binning. The relative reconstruction error $||\bDw-\bSw\bTw||_2/||\bDw||_2$ decreases with increasing $N_{\rm c}$, and for $N_{\rm c}$ above 2, the factorisation retrieves additional components with fast and somewhat erratic dynamics (see \Fig{FigSM_T2AMPKARvNc}). If we consider only the components with dynamics similar to the $N_{\rm c}=2$ case we observe a similar dependence on activator concentration of the weighted average lifetime and the short component fraction $r$ (not shown here). We note that the relatively large reconstruction error around 20\% is dominated by the shot noise in the data, which even after binning still has more than 30\% of pixels at zero counts, as can be seen in \Fig{FigSM_T2AMPKAR_Hist}.

\clearpage
\section{uFLIM applied to FLIM data from controlled mixtures}\label{sec:resultsFLIM}
\begin{figure*}
	\includegraphics*[width=\textwidth]{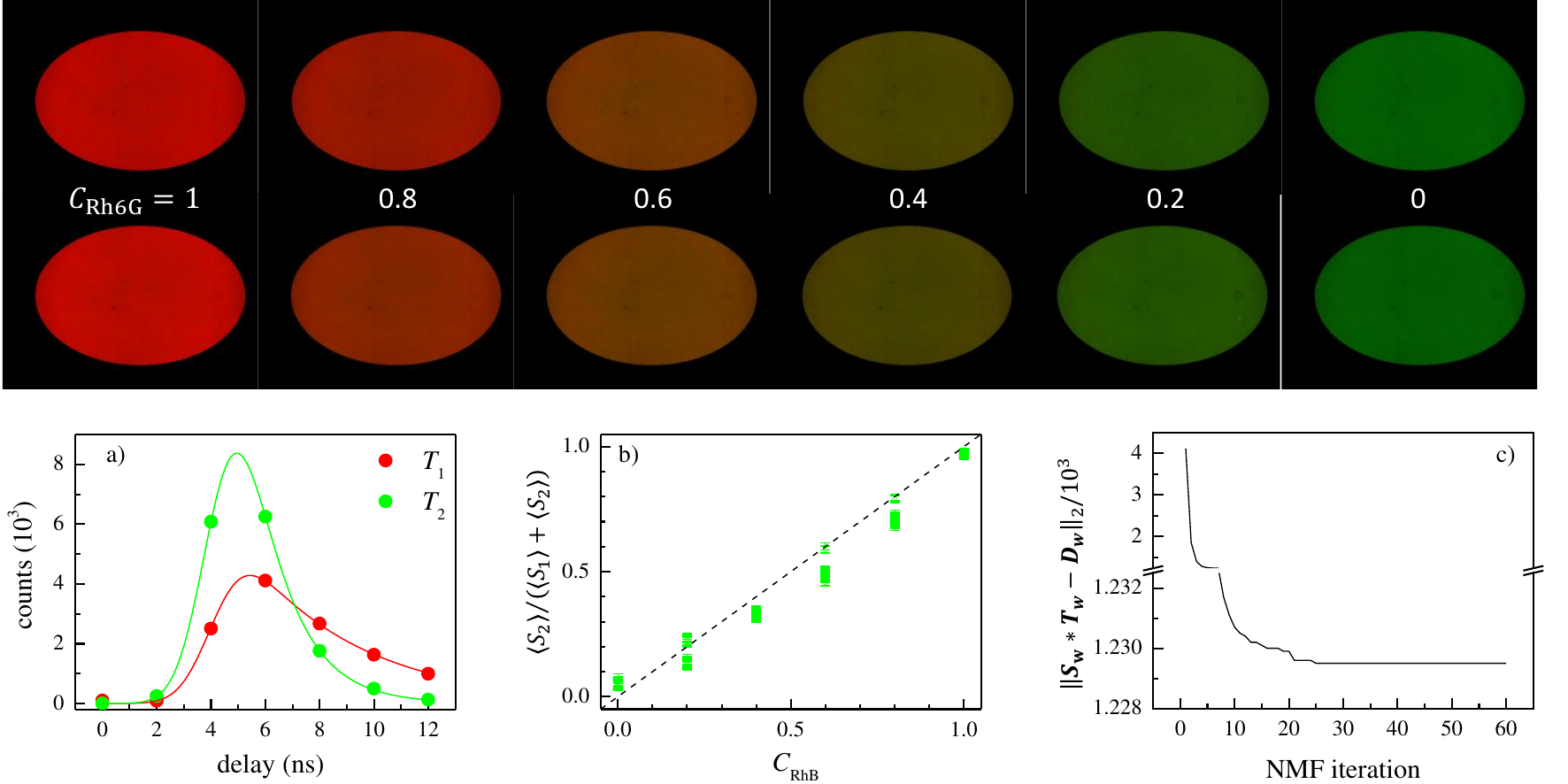}
	\caption{uFLIM applied to time-gated FLIM on Rhodamine 6G and Rhodamine B mixtures from \Onlinecite{WarrenPO13}. Top: Spatial concentration \bS\ displayed as a colour overlay with the red (green) given by the first (second) component. a) Temporal dynamics \bT, with red (green) symbols represent the first (second) component. The solid lines are a fit to the data as discussed in the text. b) Relative concentration \bS\ of the second components averaged over the different fields of view as a function of the nominal concentration of Rhodamine B. The dashed line shows the nominal concentration. c) NMF error versus number of iterations.}
	\label{FigSM_Rh_fNMF}
\end{figure*}
Here we demonstrate that uFLIM can correctly factorize FLIM data, determining the dynamics and spatial distribution of two molecules in a mixture by using random initial guesses. For this purpose, we analysed the data of experiment 1 in \Onlinecite{WarrenPO13}, representing FLIM on mixtures of Rhodamine 6G and Rhodamine B in water at six different relative concentrations acquired with time-gated wide-field imaging of wells containing the mixtures. We removed a background given by averaging the data outside the wells for each dataset separately. We then selected an oval central region of the wells (similar to \Onlinecite{WarrenPO13}) to avoid the inhomogeneous regions in the proximity of the walls of the wells. We corrected for any residual spatial inhomogeneity in the delay-integrated signal by fitting the fluorescence with a two-dimensional Gaussian $\exp(f_0+f(x,y))$ where $f(x,y)$ is a second order polynomial without constant term, and then multiplying the data by $\exp(-f(x,y))$. Finally we constructed \bD\ by reshaping the data including all wells, having a total number of 5.1e7 spatial points and 7 temporal points. \Fig{FigSM_Rh_fNMF} shows the results of the NMF analysis on \bD\ using random initial guesses for \bT\ and \bS\ using two components, and $\xi=1$. Here we present the results using the components in \bS\ normalized to minimise the deviation from unity of the sum over the components in \bS, while applying a corresponding normalization of the dynamics of the components in \bT\ to retain the factorization \bS\bT.
The top row of \Fig{FigSM_Rh_fNMF} shows the dependence of \bS\ for a selection of the 324 fields of view analyzed, showing different nominal concentrations of dyes as indicated. The two components of \bS\ are encoded as red and green channel respectively. A visual inspection of \bS\ suggests that the first component (red channel) decreases as the concentration of Rhodamine 6G increases, with the opposite happening for the second component (green channel). The resulting \bT\ is shown in \Fig{FigSM_Rh_fNMF}a (solid symbols). The two components show different dynamics, with the first being slower than the second. To verify that the two components are consistent with the concentrations and dynamics of Rhodamine 6G and Rhodamine B respectively, we fitted the obtained dynamics with an exponential decay convoluted with a Gaussian IRF given by $\exp(-t^2/w^2)$ with the width $w$, resulting in
\be
T\propto \exp\left(\frac{w^2}{4\tau^2}-\frac{t-t_0}{\tau}\right)\left(1+{\rm erf}\left(\frac{t-t_0}{w}-\frac{w}{2\tau}\right)\right),
\ee
where $t_0$ is the time of excitation, and $\tau$ the decay lifetime. A fit for the two components in \bT, with common $t_0$ and $w$ and two lifetimes $\tau_{1,2}$ (solid lines in \Fig{FigSM_Rh_fNMF}a), results in $\tau_1=(4.196\pm0.087)$\,ns and $\tau_2=(1.467\pm0.031)$\,ns, in good agreement with the lifetimes of 4.08\,ns and 1.52\,ns for the two dyes reported in \Onlinecite{MagdePP99S}. We found a IRF width $w$ of 1.35\,ns, corresponding to a full-width-at-half-maximum of 2.25\,ns, in good agreement with the experimental gate width of 2\,ns. We also find a proportionality between the normalized average concentration of the second component over each well, and the nominal Rhodamine B concentration in the well, as shown in \Fig{FigSM_Rh_fNMF}b. The factorization takes about 40\,s per NMF iteration using an Intel® Xeon® Processor E5-2630 v4 (2.20 GHz), and sufficient convergence (corresponding to a change in the error smaller of $1\%$ within two iterations) was reached within $6$ iterations (see \Fig{FigSM_Rh_fNMF}c). This result verifies that uFLIM can recover quantitatively the dynamics and spatial distribution of mixtures of components. For comparison, we note that the CPU time required by the global fit analysis described in \Onlinecite{WarrenPO13} performed on the same dataset was reported to be 45s on a Intel Core i7 870 (2.93 GHz).

\clearpage

\section{Gradient descent algorithm}\label{sec:graddescent}
The use of the fast NMF algorithm comes with an approximate treatment of the noise. For Poisson noise, the correct estimator to minimise is the KLD \,\cite{LeeBook01S,LaurenceNMet10S}
\be
\sum_{i,j}\left(\left(\bS\bT\right)_{i,j}+b_{i,j}-D_{i,j}\right) - \sum_{\substack{i,j \\ \left(\bS\bT\right)_{i,j}+b_{i,j}>0,\,  D_{i,j}>0}}D_{i,j}\log\left(\frac{\left(\bS\bT\right)_{i,j}+b_{i,j}}{D_{i,j}}\right),\label{eq:mle}
\ee
where $i(j)$ is the index of the spatial (temporal) pixels, and $b_{i,j}$ are the expected dark counts. To minimise the estimator, a gradient descent method can be applied where the elements of the matrices \bS, $S_{i,k}$, and \bT, $T_{i,k}$, can be found iteratively by subtracting a quantity proportional to the gradient of the estimator function with respect to $S_{i,k}$ and  $T_{i,k}$, respectively. Rather than using a constant fraction of the gradient, we have used the multiplicative update rule\,\cite{LeeBook01S}
\be\label{eq:gd}
S_{i,k} \gets S_{i,k}\frac{\sum_j \frac{T_{k,j} D_{i,j}}{\sum_\kappa S_{i,\kappa}T_{\kappa,j}+b_{i,j}}}{\sum_j T_{k,j}}\,,\quad
T_{k,j} \gets T_{k,j}\frac{\sum_i \frac{S_{i,k} D_{i,j}}{\sum_\kappa S_{i,\kappa}T_{\kappa,j}+b_{i,j}}}{\sum_i S_{i,k}}.
\ee

In case the dynamics is fixed, the update of $T_{k,j}$ is simply omitted. We iterate the update rules until either the estimator value has converged, or it has not improved for three consecutive iterations, or 1000 iterations are reached.

We define convergence in the following way. Once $\iota_m$ iterations are reached, we fit the dependence of the estimator on the iteration index $\iota$ using a exponential law, i.e. $\alpha+\beta\phi^{\iota-0.75\iota_m}$, and considering only the second half of the iterations. Initially we consider, $\iota_m=50$. If the difference between the last estimator and the extrapolated minimum $\alpha$ is smaller than $10^{-3}$ times the total number of data points (elements of the matrix $\bD$) we consider the estimator converged. Otherwise we run the update rules for 10 additional iteration and repeat the fit, always including the second half of the total iterations.  The convergence condition implies that the extrapolated probability for each measured data point in $D_{i,j}$ to be representing a Poisson distribution of expectation value $\sum_k S_{i,k} T_{k,j}$ is on average by less than a factor of $10^{-3}$ higher than for the factorization of the last iteration. After the iteration has completed, we keep the factorization corresponding to the minimum obtained estimator over all iteration steps.
We have considered different initial guess options for \bS\ and \bT\, to investigate their influence on the convergence and final results.
For the unmixing with fixed spectra and dynamics, we have used 
\begin{enumerate}
	\item inversion: \bS\ from solving the linear system $\bS\bT=\bD$ using a QR solver.
	\item single fast NMF: \bS\ obtained with a single step of the fast NMF.
\end{enumerate}
For the unmixing with free spectra and dynamics, we have used
\begin{enumerate}
	\item inversion: \bT\ from the nominal parameters and \bS\ from solving the linear system $\bS\bT=\bD$ using a QR solver.
	\item single fast NMF: the \bT\ from the nominal parameters and \bS\ obtained with a single step of the fast NMF.
	\item fast NMF: \bT\ and \bS\ obtained with the fast NMF.
\end{enumerate}

For the FRET data, we have considered the dynamics obtained from the fast NMF algorithm.
We need to avoid that the initial guesses have elements close to zero as the multiplicative update rule will take many iterations to lift them if needed. In the limit case of elements equal to 0, they will not be updated, and the results can be affected by this. 
For all the above options we have replaced the elements of the initial guess matrices which were smaller than a threshold (defined as 1\% of the average value for the corresponding component) with the threshold itself, i.e.

 \begin{algorithmic}

 	\ForAll{$1\leq i\leq N_s$}
 	\State	$\hat{S}\gets \Sigma_{k=1}^{N_c}S_{i,k}/N_c$
 	\ForAll{$1\leq k\leq N_c$}
 	\If{$S_{i,k}<0.1\hat{S}$}
 	\State $S_{i,k}\gets\hat{S}$
 	\EndIf
 	\EndFor
 	\EndFor
 	
 \ForAll{$1\leq j\leq N_t$}
 \State	$\hat{T}\gets \Sigma_{k=1}^{N_c}T_{k,j}/N_c$
 \ForAll{$1\leq k\leq N_c$}
 \If{$T_{k,j}<0.1\hat{T}$}
 \State $T_{k,j}\gets0.1\hat{T}$
 \EndIf
 \EndFor
 \EndFor
 \end{algorithmic}

\clearpage
\section{Spectral unmixing of multiple fluorescent proteins}
\subsection{Generation of synthetic data} \label{subsec:gensyn}
In the manuscript we have introduced the proportionality factor $c_{def}$ between the distribution matrix \bF{f}\ of the FP $f$ and the scaled spatial distribution for a certain excitation wavelength (index $e$) and detection channel $d$, i.e. $\bF{def}=c_{def}\bF{f}$. $c_{def}$ is given by
\be c_{def}=\varepsilon_{ef}\phi_{df}\nu_f\,,\label{eq:cdef}\ee
where $\nu_f$ is the quantum efficiency of the FP, $\varepsilon_{ef}$ is the extinction coefficient of the FP $f$ at the excitation wavelength \lex{e} (\lex{1}=460\,nm, \lex{2}=490\,nm), and $\phi_{df}$ is the fraction of photon emission of FP $f$ detected in channel $d$,
\be \phi_{df}=\int_{\lambda^-_d}^{\lambda^+_d}E_f(\lambda)d\lambda, \, \mbox{with}\, \int_0^\infty E_f(\lambda) d\lambda=1\,, \ee
where $\lambda^-_d (\lambda^+_d)$ represent the minimum (maximum) wavelength detected by channel $d$ ($\lambda^-_1=500$\,nm, $\lambda^+_1=550$\,nm, $\lambda^-_2=550$\,nm and $\lambda^+_2=700$\,nm), and $E$ is a normalized emission rate per wavelength. 
The FP temporal dynamics is given by 
\bea\label{eq:decay}
\bT_f&=&A\exp\left(\frac{\gamma_f^2w^2}{4} - t\gamma_f\right) \times \\\nonumber&&\left(\frac{2}{e^{\gamma_f/r}-1}+{\rm erf}\left( \frac{t}{w}-\frac{\gamma_f w}{2} \right) + 1\right)\,
\eea
which includes the pile-up due to the periodic excitation with repetition rate $r$. The normalization constant $A$ ensures $\{\bT_f\}=1$. The decay rate $\gamma_f$ is given by the inverse of the FP lifetime $\tau$.
\subsection{Fixed spectro-temporal FP properties}

\Tab{tab:8FPs} summarises the spectral and temporal properties of the FPs used to create the synthetic data for the unmixing using known dynamics and spectra.  The spectral properties and lifetimes are obtained from {\it www.fpbase.org}. The available figures showing results of the unmixing using fixed FP spectra and dynamics are tabulated in \Tab{tab:simfigsunmixfixed}.

\begin{table}[b]
	\begin{center}		
		\begin{tabular}{l|c|c|c|c|c|c|c|l} 
			\hline
Name of FP & $f$ & $\hat{c}_{11f}$ & $\hat{c}_{12f}$ & $\hat{c}_{21f}$ & $\hat{c}_{22f}$ & $\hat{c}_f$ & $\tau$ (ns) & name of painting\\
			\hline\hline
			WasCFP & 1 & 0.31 & 0.09 & 0.47 & 0.13 & 0.21 & 5.1 & {\it The creation of Adam}\\
			\hline
			BrUSLEE & 2 & 0.29 & 0.07 & 0.52 & 0.12 &  0.13 & 0.82 & {\it The Hay Wain}\\
			\hline
			mBeRFP & 3 & 0.01 & 0.67 & 0.01 & 0.31 & 0.12 & 2.0 & {\it The ambassadors}\\
			\hline
			Dendra2(Red) & 4 & 0.01 & 0.49 & 0.01 & 0.49 & 0.08 & 4.4 & {\it Old woman and boy with candles}\\
			\hline
			MiCy & 5 & 0.67 & 0.11 & 0.19 & 0.03 & 0.06 & 3.4 & {\it The great wave off Kanagawa}\\
			\hline
			mEos2 (Green) & 6 & 0.2 & 0.06 & 0.57 & 0.17 & 0.19 & 3.5 & {\it Girl with a pearl earring}\\
			\hline
			mVenus & 7 & 0.13 & 0.07 & 0.52 & 0.28 & 0.16 & 2.7 & {\it The birth of Venus}\\
			\hline
			LSS-mKate2 & 8 & 0.02 & 0.54 & 0.02 & 0.42 & 0.04 & 1.4 & {\it A Sunday on la grande Jatte}\\
			\hline
		\end{tabular}
		\caption{Spectral properties and lifetimes of the 8 FPs used to generate the synthetic data with the corresponding painting used as spatial distribution. Definitions of $\hat{c}_{f}$ and $\hat{c}_{def}$ are given in \Sec{sec:sFLIM}. $\tau$ is the lifetime. \label{tab:8FPs}}		
	\end{center}
\end{table}

\clearpage
\begin{table}
	\begin{tabular}{l|l|l|l|l}
		\hline
		Figure & \Itot\ & Method & iterations & mle \\
		\hline
		\Fig{Fig_UnmixingFixed_fastNMF_H} & $10000$ & single fast NMF & 1 & 1.408e7\\
		\Fig{FigSM_UnmixingFixed_GD_H} & $10000$ & GD KLD from inversion & 50 & 1.410e7\\
		\Fig{FigSM_UnmixingFixed_fastNMFGD_H} & $10000$ & GD KLD from single fast NMF & 50 & 1.407e7\\ 
		\Fig{FigSM_UnmixingFixed_fastNMF_L} & $100$ & single fast NMF & 1 & 8.424e6\\
		\Fig{FigSM_UnmixingFixed_GD_L} & $100$ & GD KLD from inversion & 50  & 8.428e6\\
		\Fig{FigSM_UnmixingFixed_fastNMFGD_L} & $100$ & GD KLD from single fast NMF & 50 & 8.416e6\\ 
		\hline
	\end{tabular}
	\caption{Overview over simulation results shown for unmixing of 8 FPs of fixed dynamics and spectra, with the number of iterations and the achieved mle.  The mle of the simulated data is 1.432e7 for $\Itot=10^4$ and 8.608e6 for $\Itot=10^2$.}
	\label{tab:simfigsunmixfixed}
\end{table}

\begin{figure}
	\includegraphics[width=\textwidth]{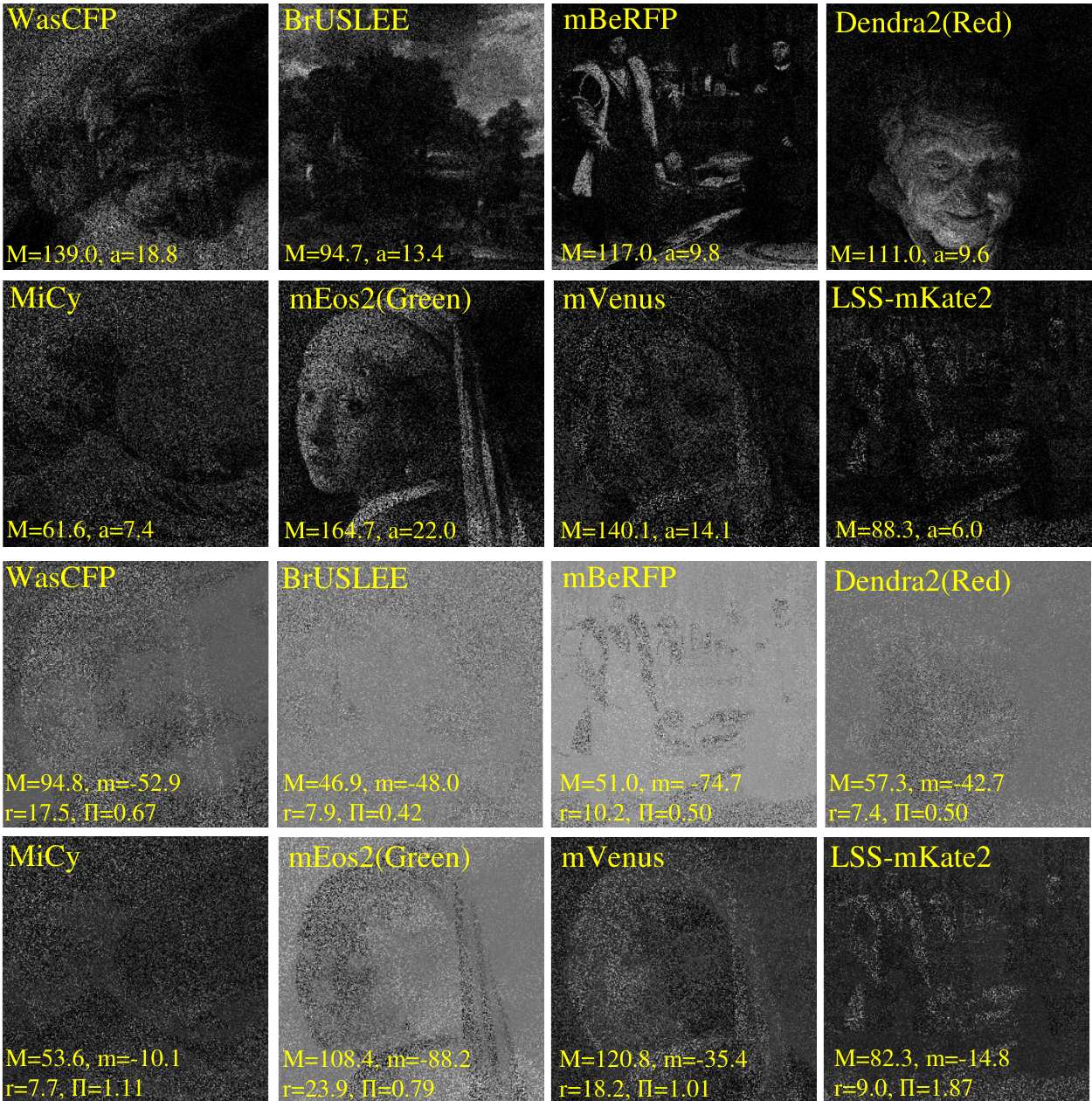}
	\caption{Same as \Fig{Fig_UnmixingFixed_fastNMF_H} but for $I_{\rm tot}=100$ photons: \add{Results of the uFLIM analysis on sFLIM synthetic data generated with 8 FPs with $I_{\rm tot}=100$ photons using the fast NMF algorithm. The FP dynamics are fixed in the analysis. Top: Retrieved \bS\, with $m=0$ and the maximum ($M$) as indicated. The spatially averaged pixel values ($a$) are also given, for comparison with the nominal values $\hat{c}_f\Itot$, see \Tab{tab:8FPs}. Bottom rows: Difference between the nominal and retrieved distributions corresponding to the top rows, with $M$ and $m$ as given.}}
	\label{FigSM_UnmixingFixed_fastNMF_L}
\end{figure}

\begin{figure}
	\includegraphics[width=\textwidth]{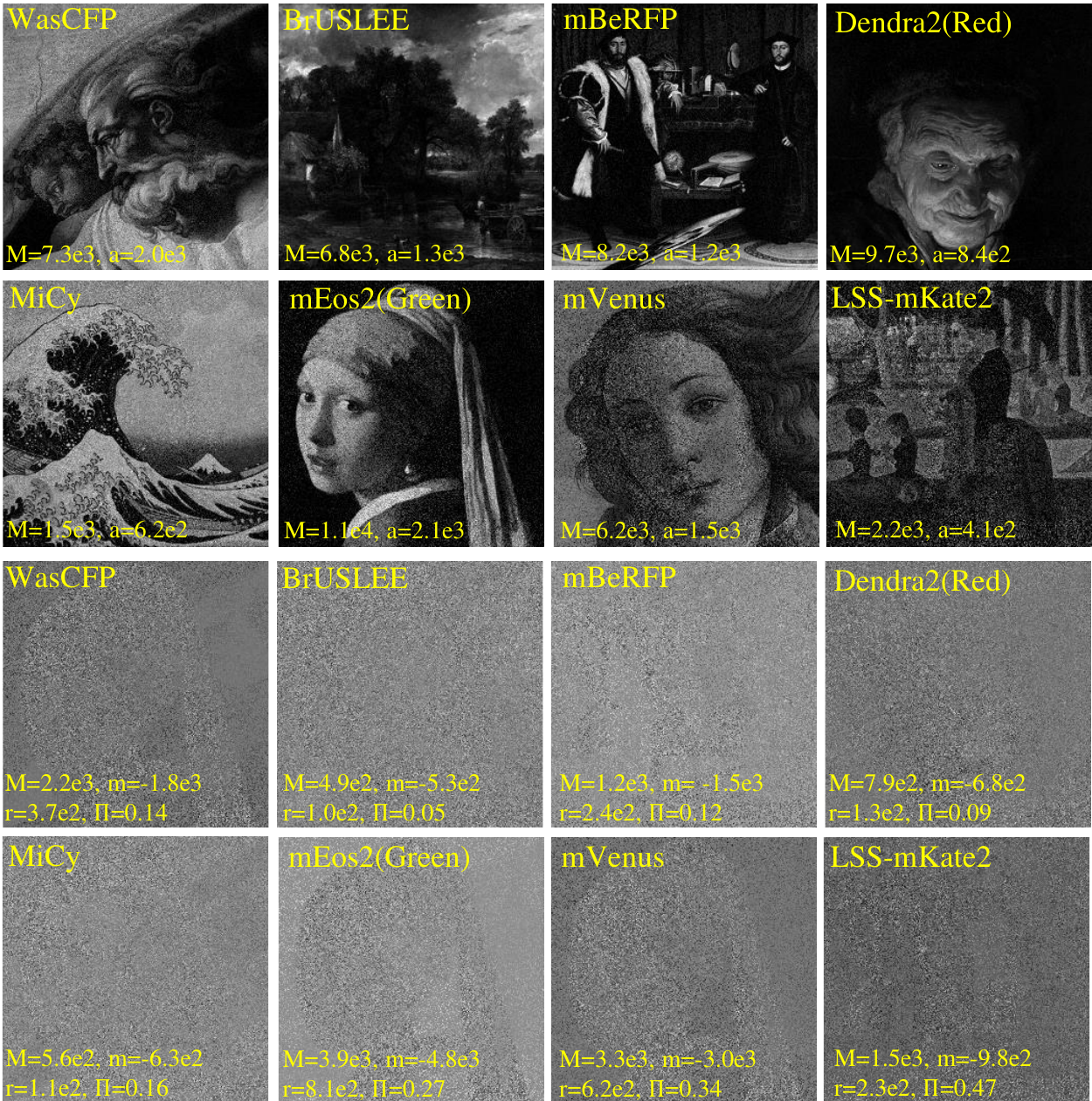}
	\caption{Same as \Fig{Fig_UnmixingFixed_fastNMF_H} but for the gradient descent method using inversion for the initial guess: \add{Results of the uFLIM analysis on sFLIM synthetic data generated with 8 FPs with $I_{\rm tot}=10000$ photons using the gradient descent method with the initial guesses obtained by inversion. The FP dynamics are fixed in the analysis. Top: Retrieved \bS\, with $m=0$ and the maximum ($M$) as indicated. The spatially averaged pixel values ($a$) are also given, for comparison with the nominal values $\hat{c}_f\Itot$, see \Tab{tab:8FPs}. Bottom rows: Difference between the nominal and retrieved distributions corresponding to the top rows, with $M$ and $m$ as given.}}
	\label{FigSM_UnmixingFixed_GD_H}
\end{figure}

\begin{figure}
	\includegraphics[width=\textwidth]{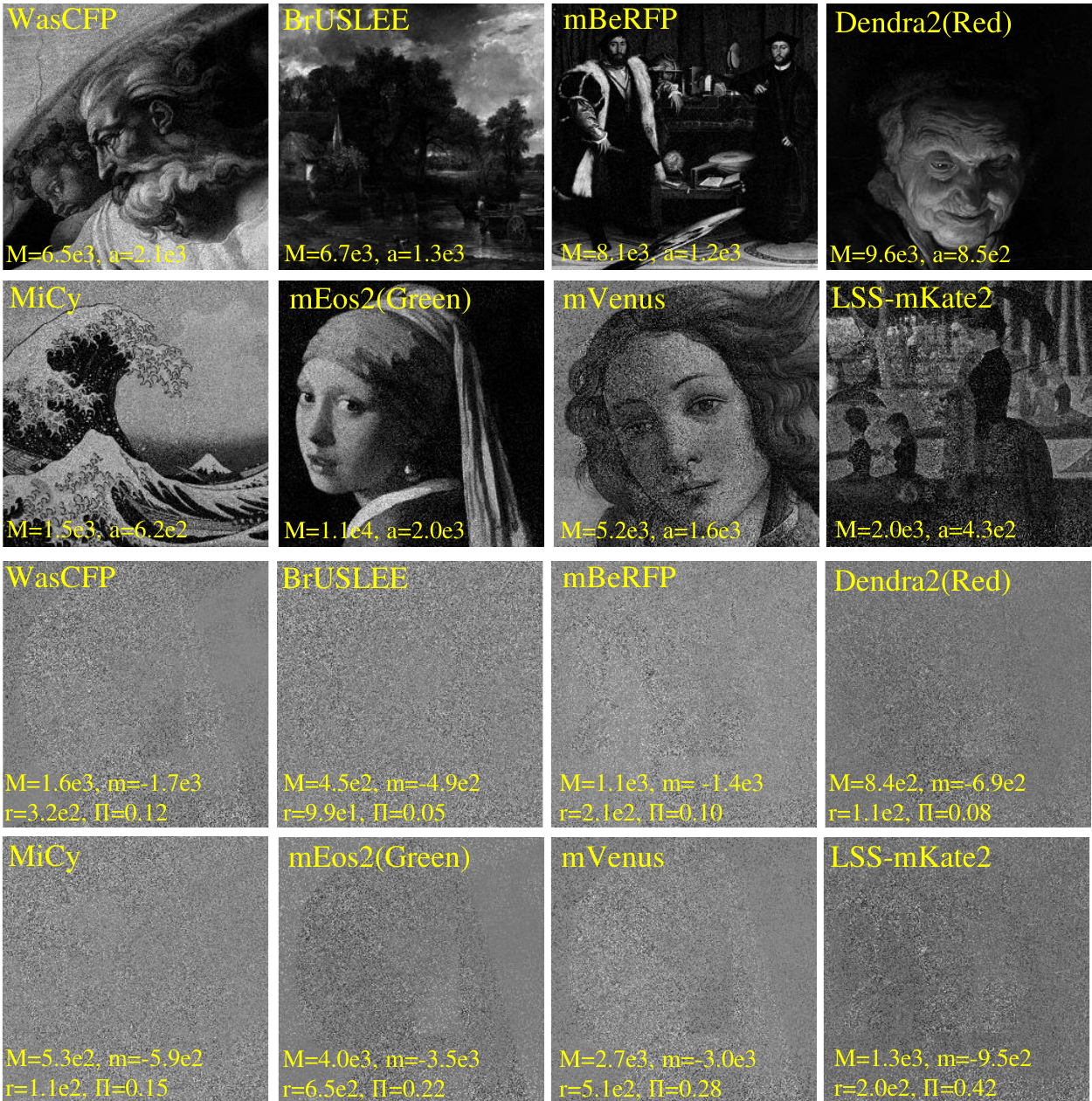}
	\caption{Same as \Fig{FigSM_UnmixingFixed_GD_H} but using a single step fast NMF for the initial guess: \add{Results of the uFLIM analysis on sFLIM synthetic data generated with 8 FPs with $I_{\rm tot}=10000$ photons using the gradient descent method with the initial guesses obtained by the fast NMF algorithm. The FP dynamics are fixed in the analysis. Top: Retrieved \bS\, with $m=0$ and the maximum ($M$) as indicated. The spatially averaged pixel values ($a$) are also given, for comparison with the nominal values $\hat{c}_f\Itot$, see \Tab{tab:8FPs}. Bottom rows: Difference between the nominal and retrieved distributions corresponding to the top rows, with $M$ and $m$ as given.}}
	\label{FigSM_UnmixingFixed_fastNMFGD_H}
\end{figure}

\begin{figure}
	\includegraphics[width=\textwidth]{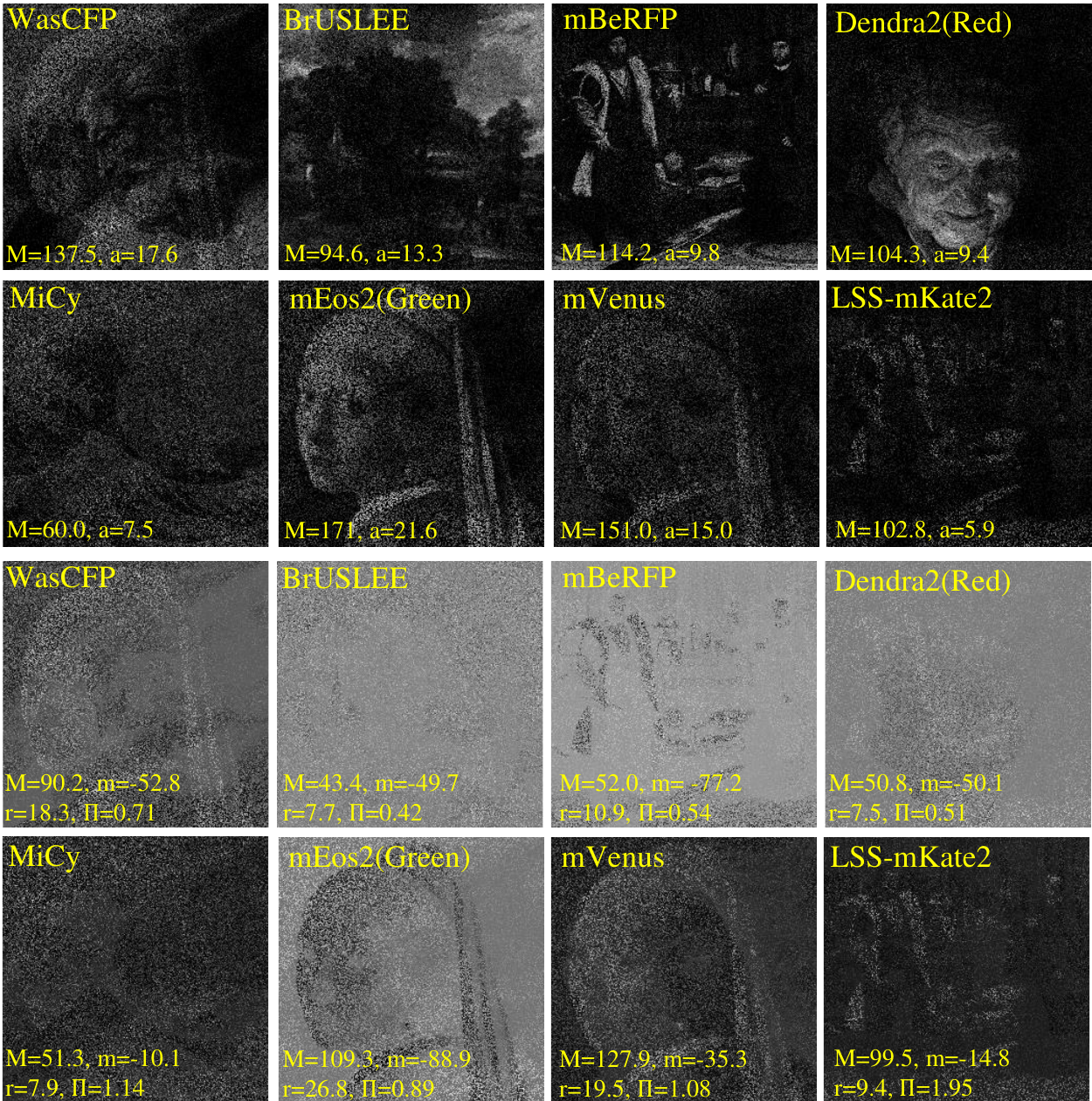}
	\caption{Same as \Fig{FigSM_UnmixingFixed_GD_H} but for $I_{\rm tot}=100$ photons: \add{Results of the uFLIM analysis on sFLIM synthetic data generated with 8 FPs with $I_{\rm tot}=100$ photons using the gradient descent method with the initial guesses obtained by inversion. The FP dynamics are fixed in the analysis. Top: Retrieved \bS\, with $m=0$ and the maximum ($M$) as indicated. The spatially averaged pixel values ($a$) are also given, for comparison with the nominal values $\hat{c}_f\Itot$, see \Tab{tab:8FPs}. Bottom rows: Difference between the nominal and retrieved distributions corresponding to the top rows, with $M$ and $m$ as given.}}
	\label{FigSM_UnmixingFixed_GD_L}
\end{figure}

\begin{figure}
	\includegraphics[width=\textwidth]{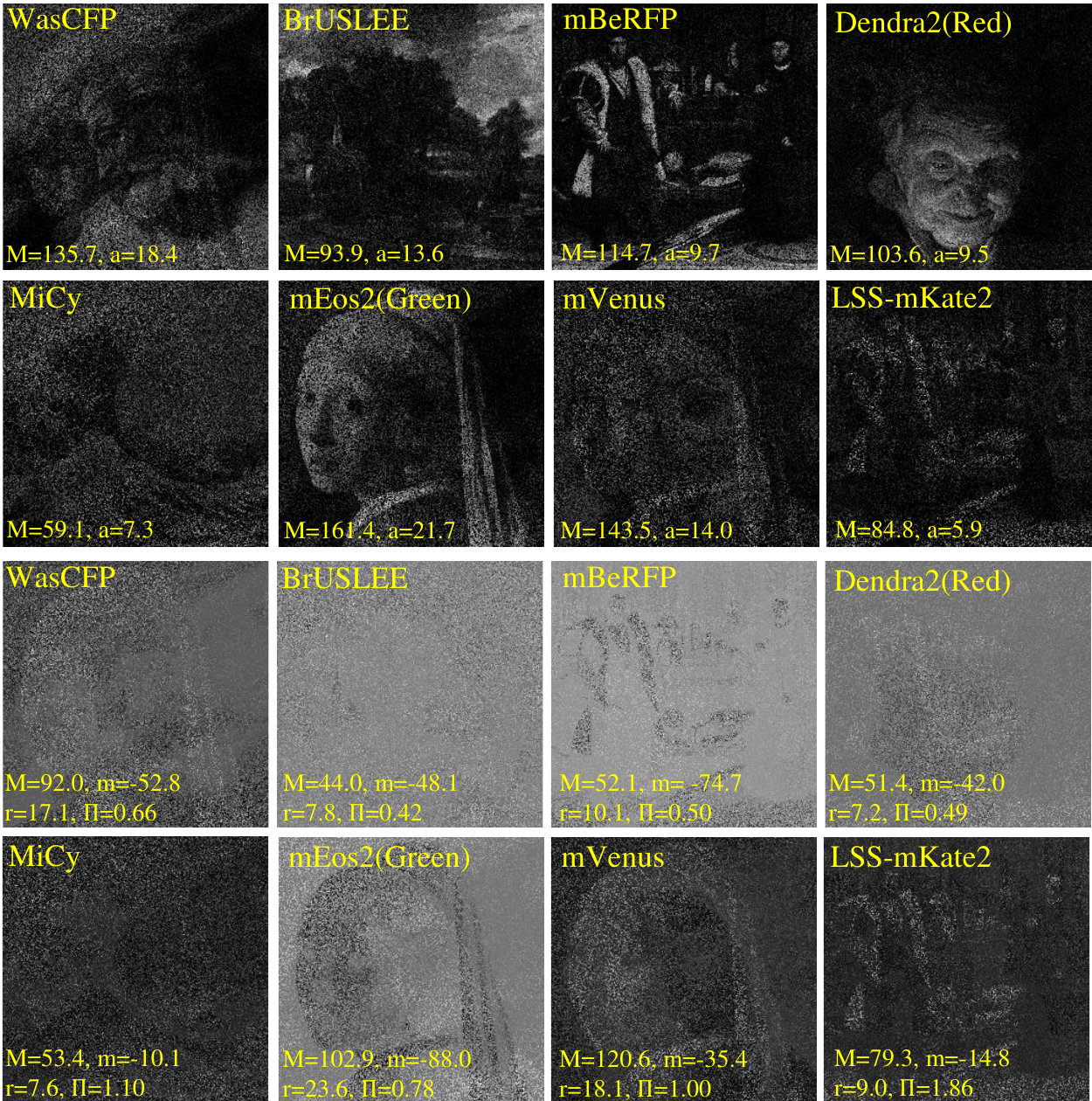}
	\caption{Same as \Fig{FigSM_UnmixingFixed_fastNMFGD_H} but for $I_{\rm tot}=100$ photons: \add{Results of the uFLIM analysis on sFLIM synthetic data generated with 8 FPs with $I_{\rm tot}=100$ photons using the gradient descent method with the initial guesses obtained by the fast NMF algorithm. The FP dynamics are fixed in the analysis. Top: Retrieved \bS\, with $m=0$ and the maximum ($M$) as indicated. The spatially averaged pixel values ($a$) are also given, for comparison with the nominal values $\hat{c}_f\Itot$, see \Tab{tab:8FPs}. Bottom rows: Difference between the nominal and retrieved distributions corresponding to the top rows, with $M$ and $m$ as given.}}
	\label{FigSM_UnmixingFixed_fastNMFGD_L}
\end{figure}

\begin{figure}
	\includegraphics[width=\textwidth]{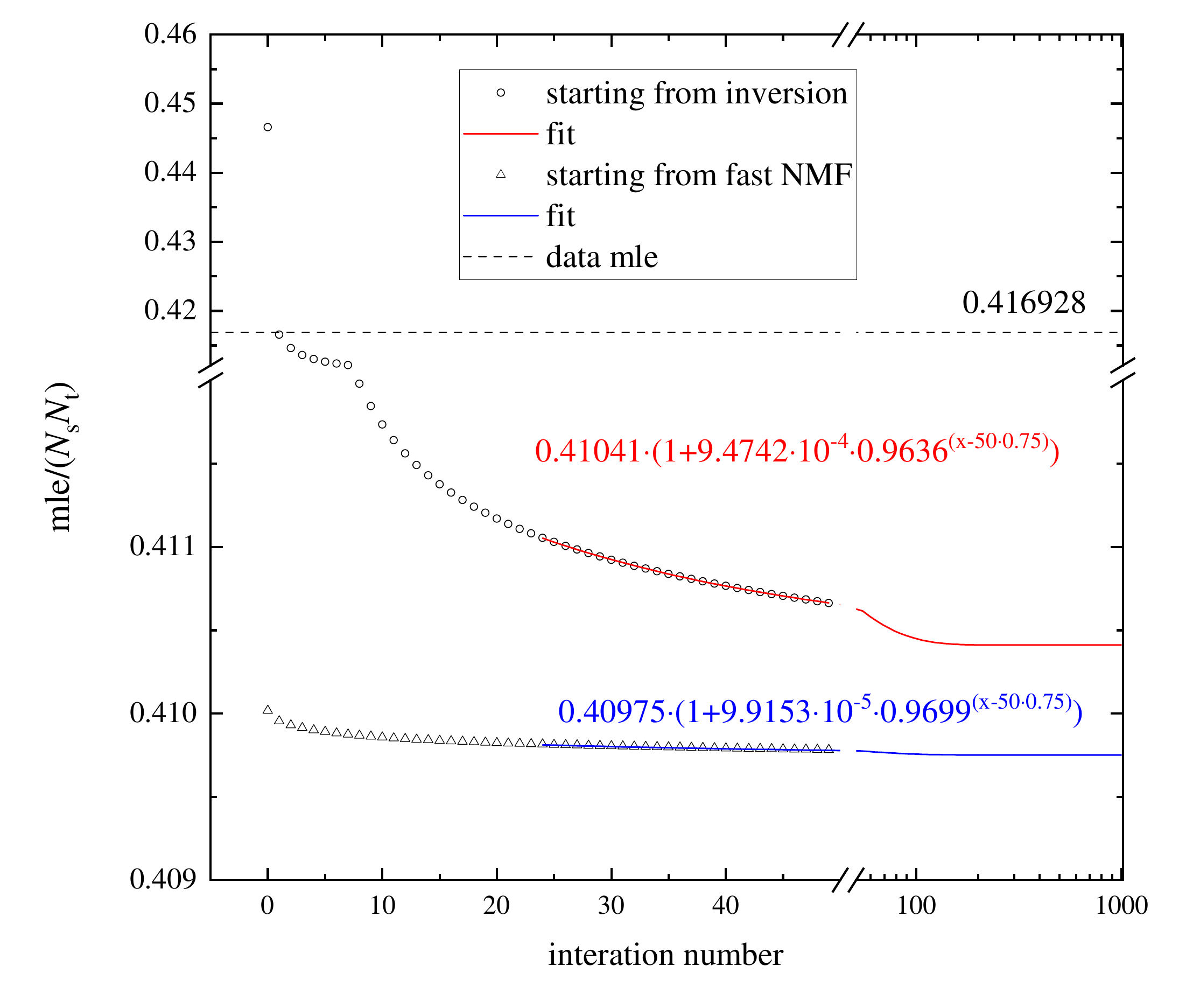}
	\caption{Convergence of the KLD during gradient descent applied to the data of \Fig{Fig_UnmixingFixed_fastNMF_H} using inversion (see \Fig{FigSM_UnmixingFixed_GD_H}) or fast NMF (see \Fig{FigSM_UnmixingFixed_fastNMFGD_H}) as initial guesses . The evaluated mle (symbols) is the KLD maximum likelihood estimator (\Eq{eq:mle}) normalised by the number of elements of \bD. The  red (blue) lines are the fits used to determine convergence, also given as analytic expressions.}
	\label{FigSM_UnmixingFixed_conv}
\end{figure}

\Fig{FigSM_UnmixingFixed_fastNMF_L} shows the results of uFLIM applied to the unmixing of sFLIM data with $I_{\rm tot}=100$ and fixed FP properties as given in \Tab{tab:8FPs}. The uFLIM results obtained using the gradient descent instead of the fast NMF are shown in \Fig{FigSM_UnmixingFixed_GD_H}-\ref{FigSM_UnmixingFixed_fastNMFGD_L}, for different $I_{\rm tot}$ and initial guesses as given in the captions. The gradient descent using the fast NMF results as initial guess produces a slightly improved factorization to the fast NMF method, specifically at low photon count. Instead, we observe marginal worse results using option 1 for the initial guess instead. This is confirmed from the convergence plots (\Fig{FigSM_UnmixingFixed_conv}) which show a final maximum likelihood estimator for option 1 larger than the value calculated for the factorization obtained by the fast NMF. Also, the convergence is about twice faster if option 2 is used. The mle calculated using the ground truth spatial and temporal distribution is few percent larger than the values obtained after fast NMF or gradient descent. This is interesting, as it shows that the retrieved distributions are already fitting the specific realization including its random noise better that the ground-truth, showing that the algorithm has converged to a residual below the effect due to noise.

\clearpage

\subsection{Free spectro-temporal FP properties}

\begin{figure}
	\includegraphics[width=\textwidth]{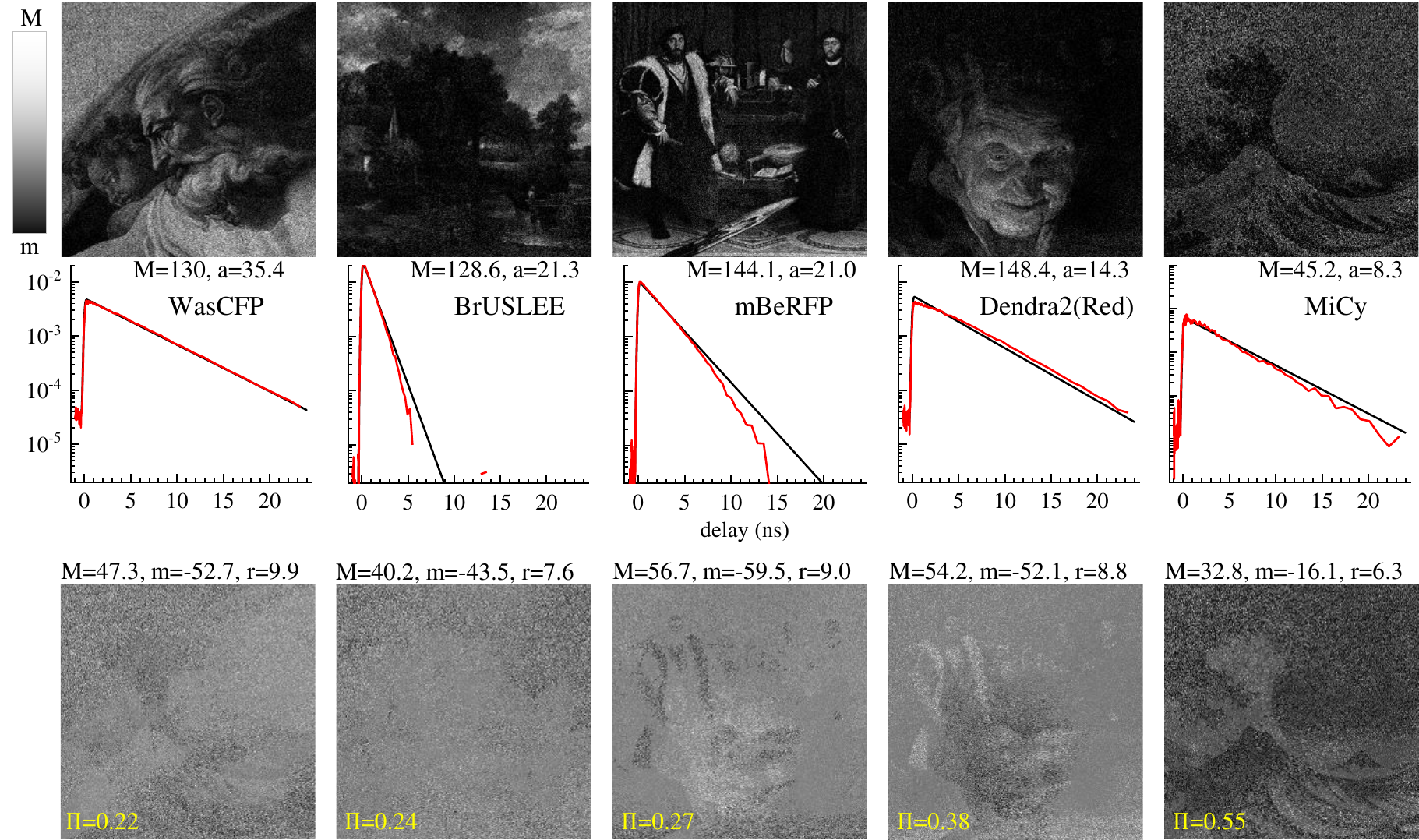}
	\caption{Same as \Fig{Fig_UnmixingFree_fastNMF_H} but for $I_{\rm tot}=100$ photons: \add{Results of the uFLIM analysis on sFLIM synthetic data generated with five FPs with $I_{\rm tot}=100$ photons using the fast NMF algorithm. The FP dynamics are obtained by the algorithm in the analysis. Top: Retrieved \bS\, with $m=0$ and the maximum ($M$) as indicated. The spatially averaged pixel values ($a$) are also given, for comparison with the nominal values $\hat{c}_f\Itot$, see \Tab{tab:5FPs}. Middle row: Retrieved dynamics $\bT_f$ (red), and corresponding original dynamics (black). Bottom rows: Difference between the nominal and retrieved distributions corresponding to the top rows, with $M$ and $m$ as given.}}
	\label{FigSM_UnmixingFree_fastNMF_L}
\end{figure}

\Fig{FigSM_UnmixingFree_fastNMF_L} shows the results of the uFLIM method applied to the unmixing of sFLIM data with $\Itot=100$ and FPs according to \Tab{tabSM:5FPs}, for the case of $I_{\rm tot}=100$ photons, retrieving also the FP properties. The resulting retrieved FP parameters are given in \Tab{tabSM:5FPs} in red.

\begin{table}[t]
	\begin{center}	
		\begin{tabular}{lc|c|c|c|c|c|c} 
			Name of FP / & $f$ & $\hat{c}_{11f}$ & $\hat{c}_{12f}$ & $\hat{c}_{21f}$ & $\hat{c}_{22f}$ & $\hat{c}_f$ & $\tau$ (ns) \\
			{\it painting} && & & & & &\\
			\hline\hline
			{WasCFP} / & 1 & 0.27 & 0.07 & 0.53 & 0.13 & {0.36} & 5.05  \\
			{\it The creation of} && {\color{red} 0.25} & {\color{red} 0.06} & {\color{red} 0.57} & {\color{red} 0.11} & {\color{red} 0.35} & {\color{red} 4.96} \\
			{\it Adam} && {\color{green} 2.0e-3} & {\color{green} 1.0e-3} & {\color{green} 3.6e-3} & {\color{green} 1.2e-3} & {\color{green} 2.9e-3} & {\color{green} 0.013} \\			 
			\hline
			{BrUSLEE} / & 2 & 0.26 & 0.09 & 0.52 & 0.14 & {0.22} & 0.94 \\
			{\it The Hay Wain} && {\color{red} 0.26} & {\color{red} 0.06} & {\color{red} 0.55} & {\color{red} 0.13} & {\color{red} 0.21} & {\color{red} 0.88} \\
			{} && {\color{green} 8e-4} & {\color{green} 6.1e-4} & {\color{green} 8.9e-4} & {\color{green} 4.1e-4} & {\color{green} 1.2e-3} & {\color{green} 3.8e-3} \\
			\hline
			{mBeRFP} / & 3 & 0.01 & 0.64 & 0.01 & 0.34 & {0.19} & 2.31  \\
			{\it The ambassadors}&& {\color{red} 0.01} & {\color{red} 0.67} & {\color{red} 0.01} & {\color{red} 0.31} & {\color{red} 0.21} & {\color{red} 2.11} \\
			{} && {\color{green} 3.2e-4} & {\color{green} 8.6e-4} & {\color{green} 8.7e-4} & {\color{green} 1.5e-3} & {\color{green} 1.8e-3} & {\color{green} 6.8e-3} \\
			\hline 
			{Dendra2(Red)} / & 4 & 0.01 & 0.44 & 0.01 & 0.54 & {0.13} & 4.46 \\
			{\it Old woman and} && {\color{red} 0.01} & {\color{red} 0.39} & {\color{red} 0.01} & {\color{red} 0.60} & {\color{red} 0.14} & {\color{red} 4.91} \\
			{\it boy with candles} && {\color{green} 8.1e-4} & {\color{green} 1.6e-3} & {\color{green} 3.9e-4} & {\color{green} 1.4e-3} & {\color{green} 7.4e-4} & {\color{green} 2.1e-2} \\
			\hline
			{MiCy} / & 5 & 0.63 & 0.11 & 0.22 & 0.04 & {0.10} & 3.90 \\
			{\it  The great wave} && {\color{red} 0.91} & {\color{red} 0.07} & {\color{red} 0.01} & {\color{red} 0.01} & {\color{red} 0.08} & {\color{red} 3.52}\\
			{\it off Kanagawa} && {\color{green} 4.4e-3} & {\color{green} 4.9e-3} & {\color{green} 1.7e-3} & {\color{green} 6.4e-4} & {\color{green} 8.5e-4} & {\color{green} 2.4e-2} \\
		\end{tabular}
		\caption{Same as \Tab{tab:5FPs} but for $\Itot=10^2$: \add{Spectral properties and lifetimes of the 5 FPs used in the synthetic sFLIM data. The values retrieved from a single data realization by uFLIM for $\Itot=10^2$ are given in red, where the lifetimes are the first moment of the retrieved dynamics for positive times. The standard deviations of the retrieved parameters due to photon shot noise are given in green.}}
		\label{tabSM:5FPs}		
	\end{center}
\end{table}

\begin{figure}
	\includegraphics[width=\textwidth]{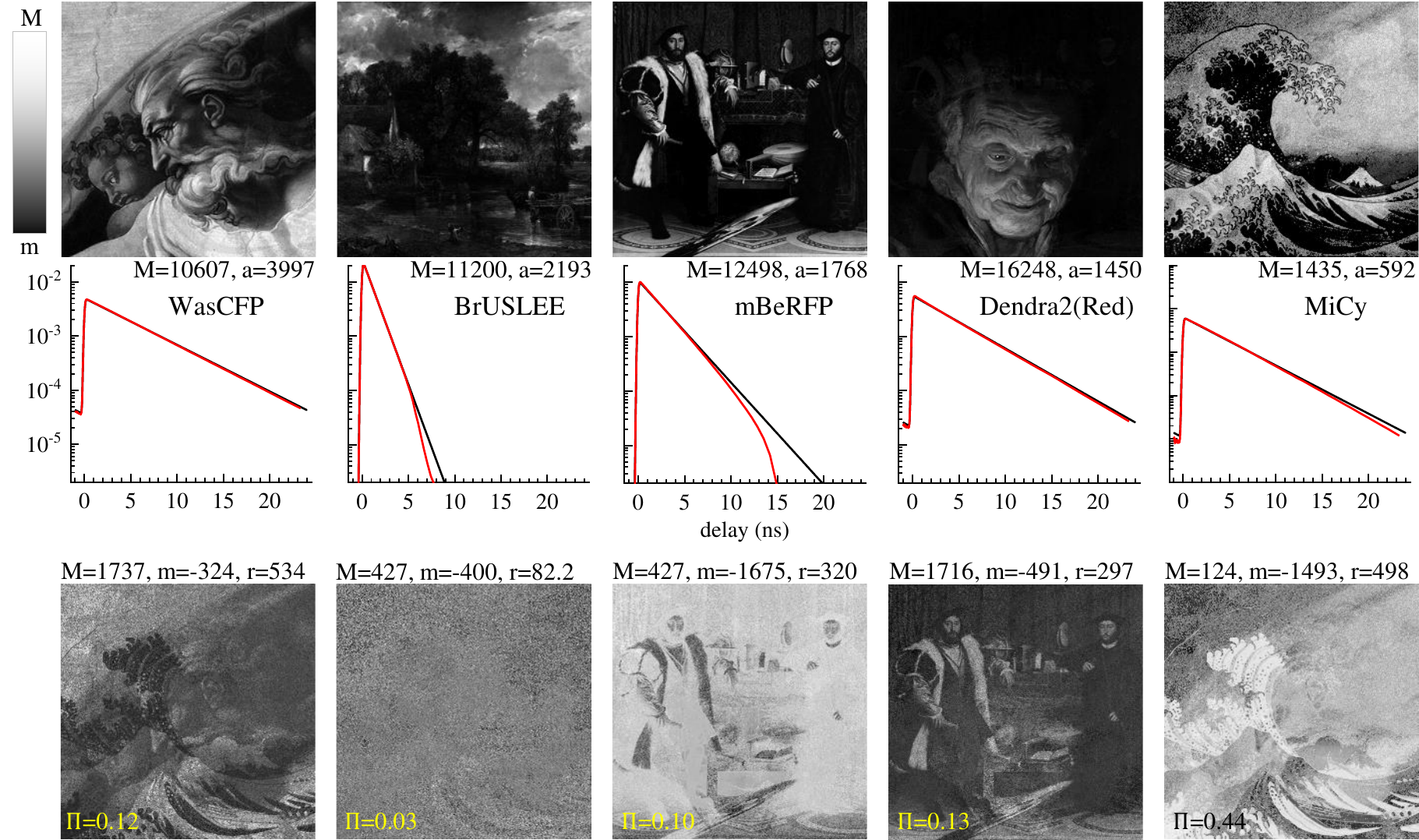}
	\caption{Same as \Fig{Fig_UnmixingFree_fastNMF_H} but for the gradient descent method using inversion for the initial guess: \add{Results of the uFLIM analysis on sFLIM synthetic data generated with five FPs with $I_{\rm tot}=10000$ photons using the gradient descent method where the initial guesses are obtained by inversion. The FP dynamics are obtained by the algorithm in the analysis. Top: Retrieved \bS\, with $m=0$ and the maximum ($M$) as indicated. The spatially averaged pixel values ($a$) are also given, for comparison with the nominal values $\hat{c}_f\Itot$, see \Tab{tab:5FPs}. Middle row: Retrieved dynamics $\bT_f$ (red), and corresponding original dynamics (black). Bottom rows: Difference between the nominal and retrieved distributions corresponding to the top rows, with $M$ and $m$ as given.}}
	\label{FigSM_UnmixingFree_GD0_H}
\end{figure}

\begin{figure}
	\includegraphics[width=\textwidth]{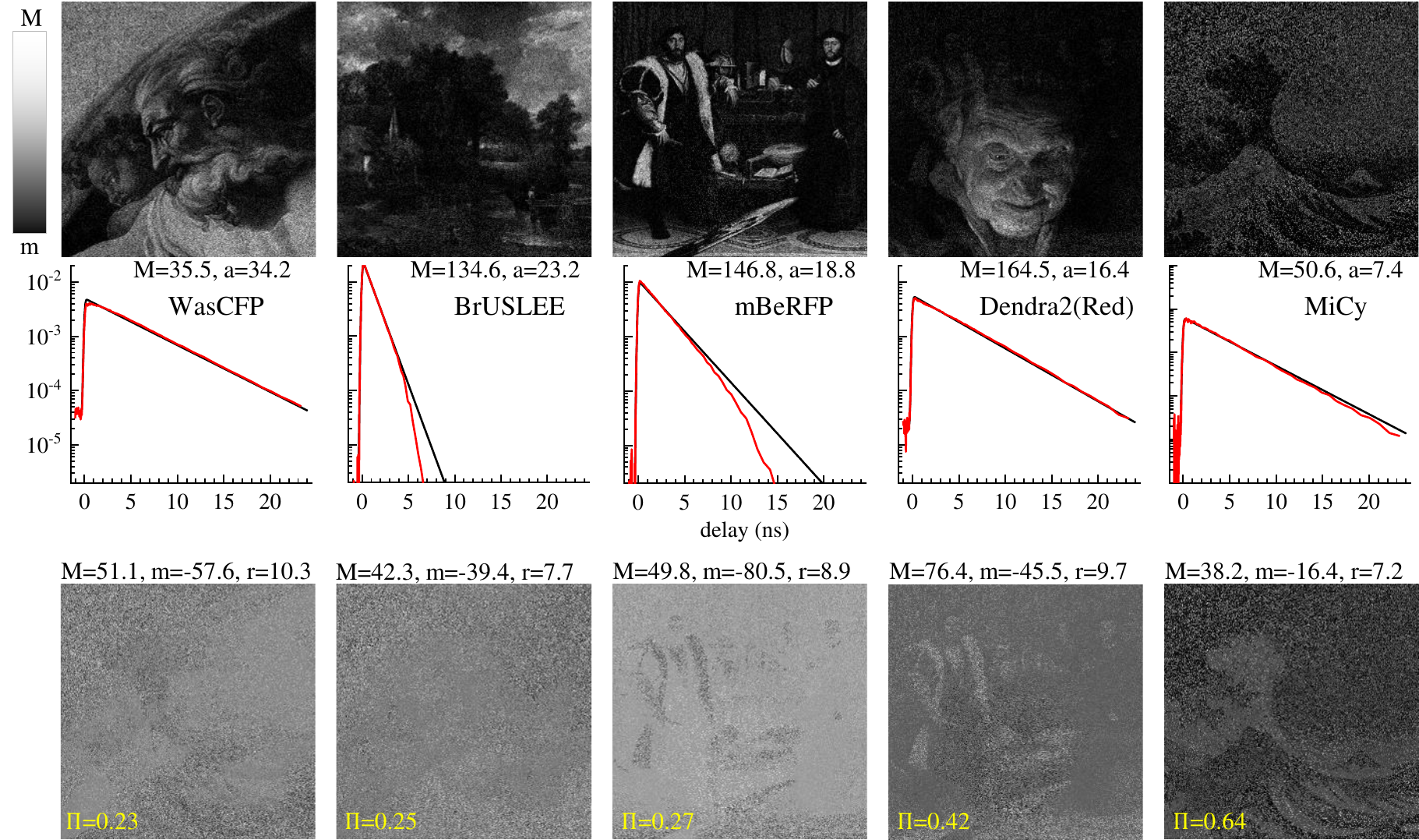}
	\caption{Same as \Fig{FigSM_UnmixingFree_GD0_H} but for $I_{\rm tot}=100$ photons: \add{Results of the uFLIM analysis on sFLIM synthetic data generated with five FPs with $I_{\rm tot}=100$ photons using the gradient descent method where the initial guesses are obtained by inversion. The FP dynamics are obtained by the algorithm in the analysis. Top: Retrieved \bS\, with $m=0$ and the maximum ($M$) as indicated. The spatially averaged pixel values ($a$) are also given, for comparison with the nominal values $\hat{c}_f\Itot$, see \Tab{tab:5FPs}. Middle row: Retrieved dynamics $\bT_f$ (red), and corresponding original dynamics (black). Bottom rows: Difference between the nominal and retrieved distributions corresponding to the top rows, with $M$ and $m$ as given.}}
	\label{FigSM_UnmixingFree_GD0_L}
\end{figure}

\begin{figure}
	\includegraphics[width=\textwidth]{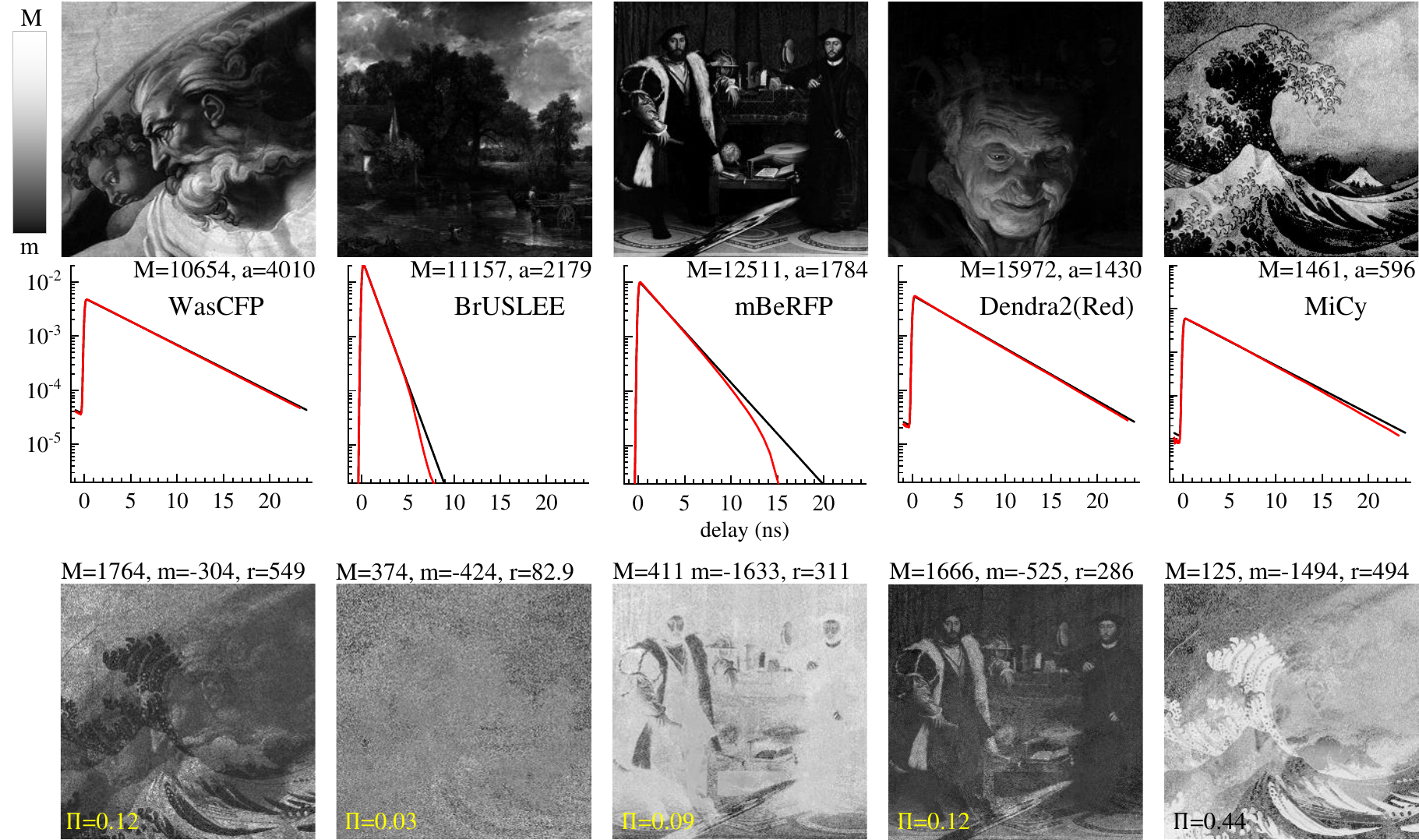}
	\caption{Same as \Fig{FigSM_UnmixingFree_GD0_H} but using inversion for the initial guess: \add{Results of the uFLIM analysis on sFLIM synthetic data generated with five FPs with $I_{\rm tot}=10000$ photons using the gradient descent method where the initial guesses are obtained by a single fast NMF step. The FP dynamics are obtained by the algorithm in the analysis. Top: Retrieved \bS\, with $m=0$ and the maximum ($M$) as indicated. The spatially averaged pixel values ($a$) are also given, for comparison with the nominal values $\hat{c}_f\Itot$, see \Tab{tab:5FPs}. Middle row: Retrieved dynamics $\bT_f$ (red), and corresponding original dynamics (black). Bottom rows: Difference between the nominal and retrieved distributions corresponding to the top rows, with $M$ and $m$ as given.}}
	\label{FigSM_UnmixingFree_GD_H}
\end{figure}

\begin{figure}
	\includegraphics[width=\textwidth]{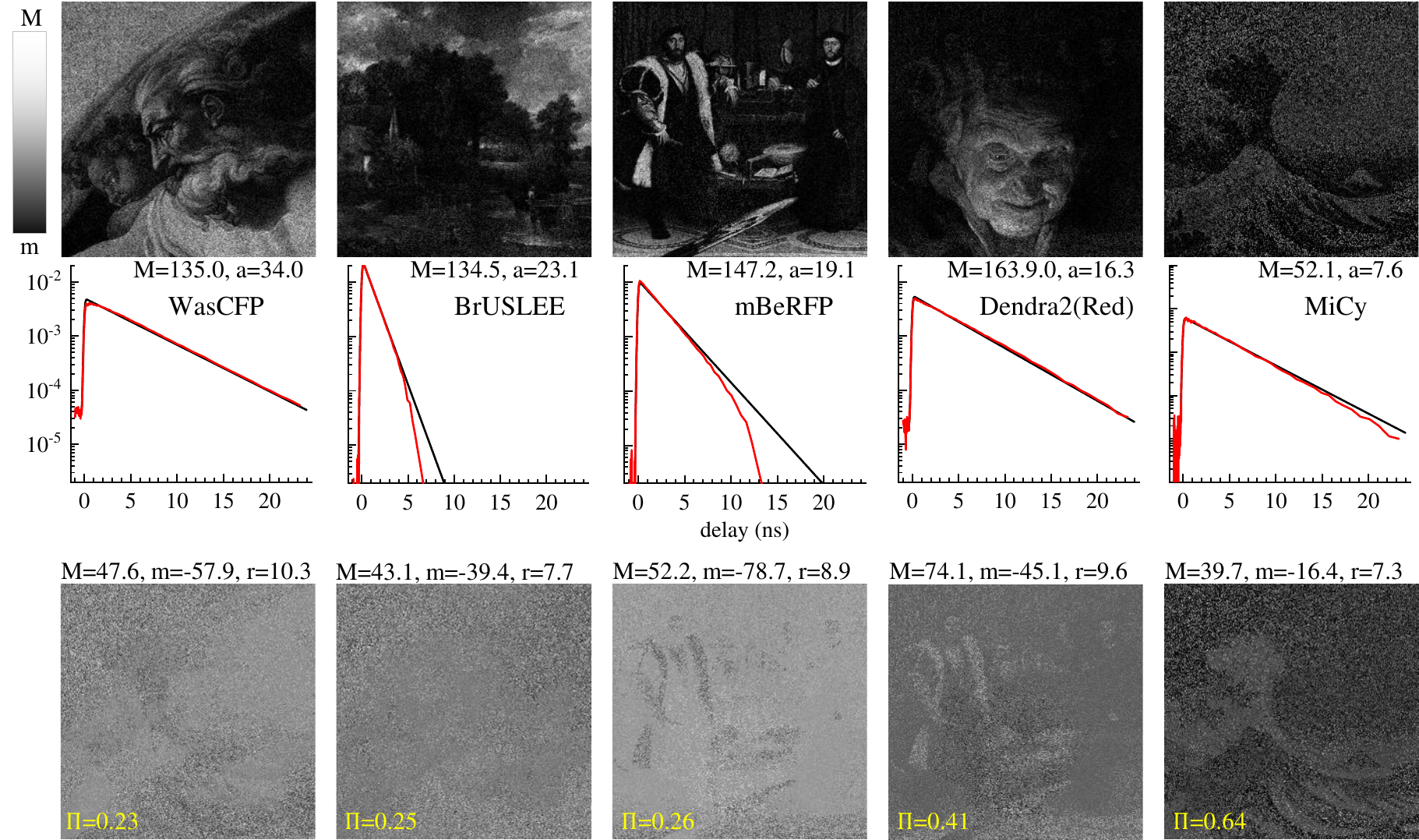}
	\caption{Same as \Fig{FigSM_UnmixingFree_GD_H} but for $I_{\rm tot}=100$ photons: \add{Results of the uFLIM analysis on sFLIM synthetic data generated with five FPs with $I_{\rm tot}=100$ photons using the gradient descent method where the initial guesses are obtained by a single fast NMF step. The FP dynamics are obtained by the algorithm in the analysis. Top: Retrieved \bS\, with $m=0$ and the maximum ($M$) as indicated. The spatially averaged pixel values ($a$) are also given, for comparison with the nominal values $\hat{c}_f\Itot$, see \Tab{tab:5FPs}. Middle row: Retrieved dynamics $\bT_f$ (red), and corresponding original dynamics (black). Bottom rows: Difference between the nominal and retrieved distributions corresponding to the top rows, with $M$ and $m$ as given.}}
	\label{FigSM_UnmixingFree_GD_L}
\end{figure}

\begin{figure}
	\includegraphics[width=\textwidth]{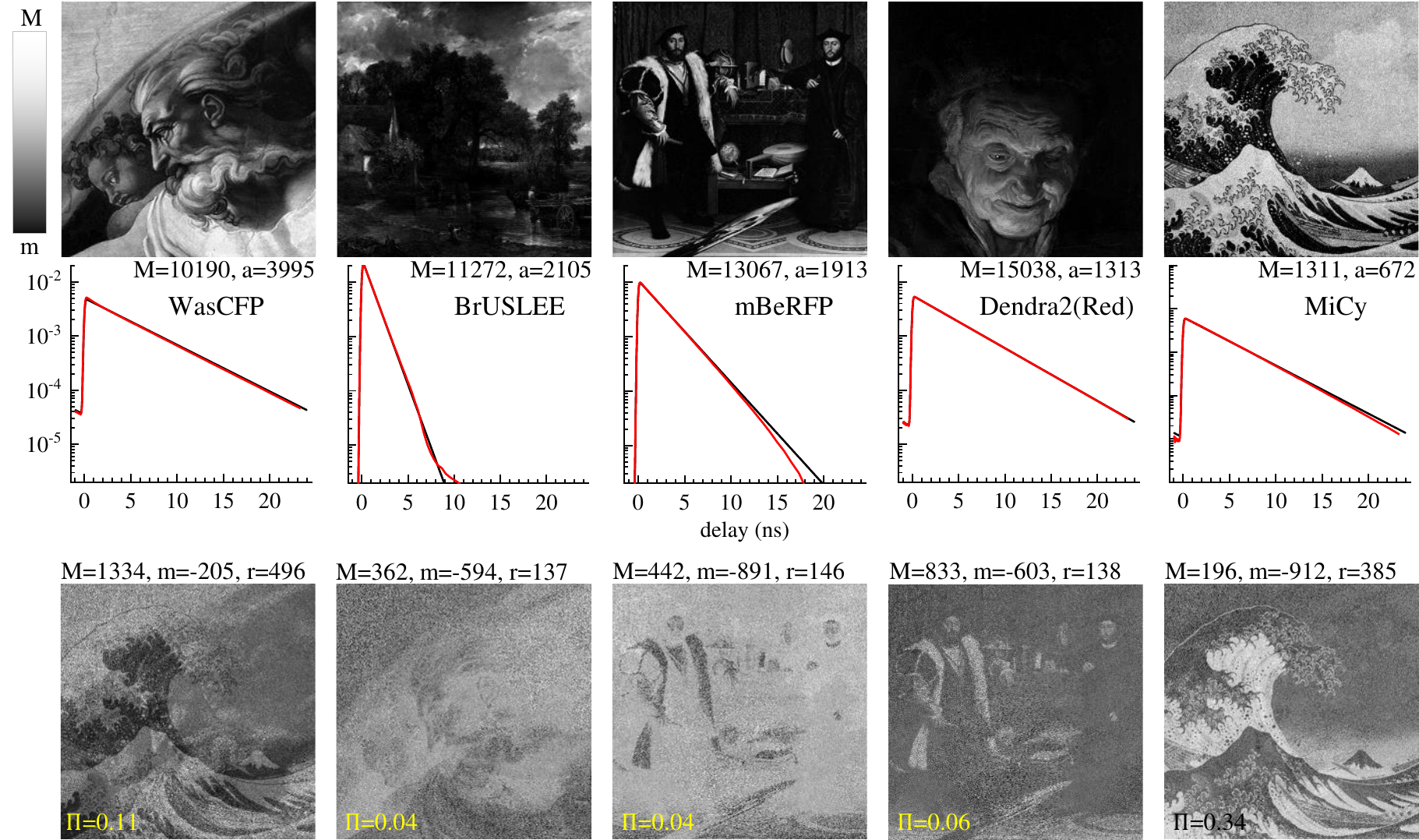}
	\caption{Same as \Fig{FigSM_UnmixingFree_GD0_H} but using fast NMF for the initial guess: \add{Results of the uFLIM analysis on sFLIM synthetic data generated with five FPs with $I_{\rm tot}=10000$ photons using the gradient descent method where the initial guesses are obtained by fast NMF. The FP dynamics are obtained by the algorithm in the analysis. Top: Retrieved \bS\, with $m=0$ and the maximum ($M$) as indicated. The spatially averaged pixel values ($a$) are also given, for comparison with the nominal values $\hat{c}_f\Itot$, see \Tab{tab:5FPs}. Middle row: Retrieved dynamics $\bT_f$ (red), and corresponding original dynamics (black). Bottom rows: Difference between the nominal and retrieved distributions corresponding to the top rows, with $M$ and $m$ as given.}}
	\label{FigSM_UnmixingFree_fastNMFGD_H}
\end{figure}

\begin{figure}
	\includegraphics[width=\textwidth]{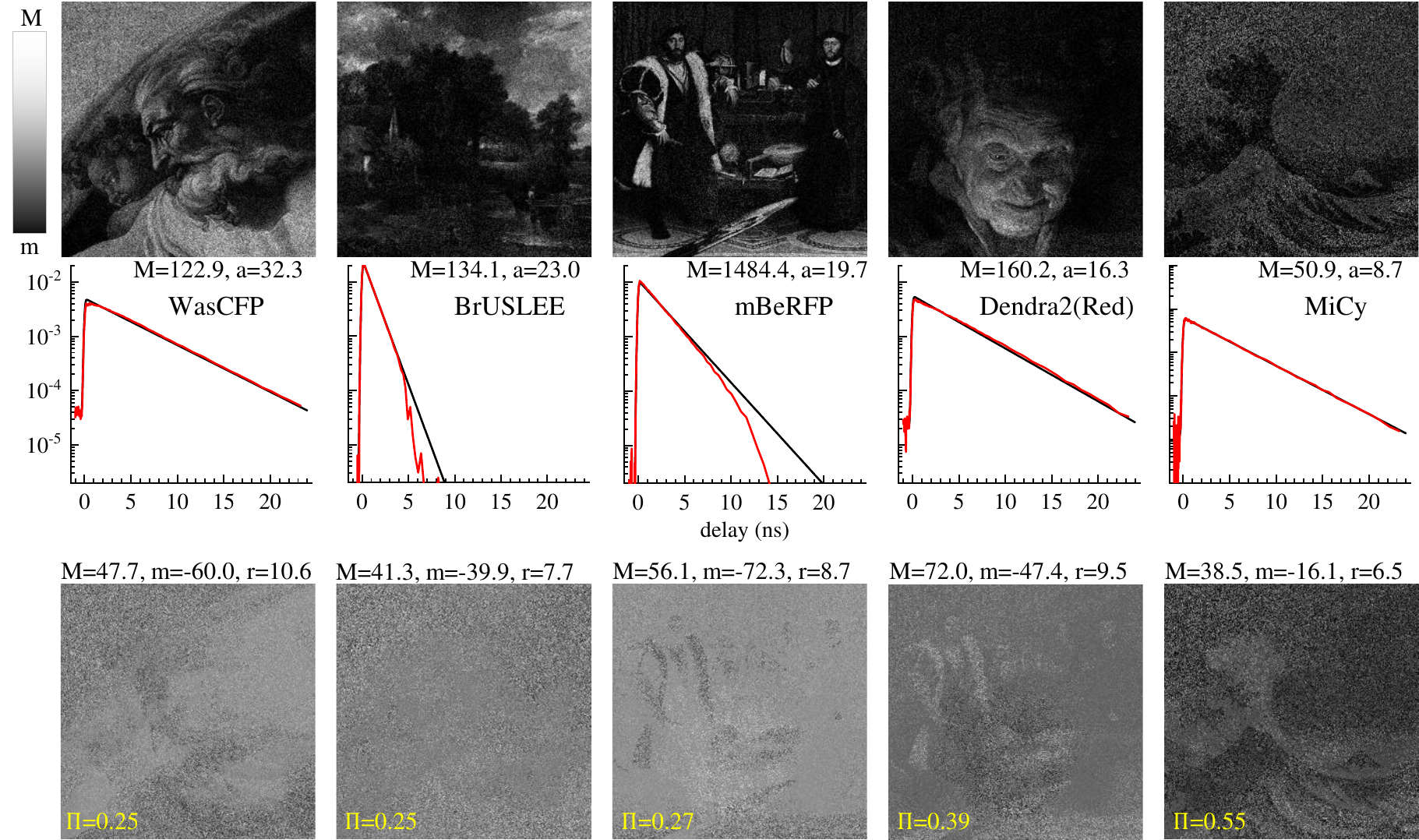}
	\caption{Same as \Fig{FigSM_UnmixingFree_fastNMFGD_H} but for $I_{\rm tot}=100$ photons: \add{Results of the uFLIM analysis on sFLIM synthetic data generated with five FPs with $I_{\rm tot}=100$ photons using the gradient descent method where the initial guesses are obtained by fast NMF. The FP dynamics are obtained by the algorithm in the analysis. Top: Retrieved \bS\, with $m=0$ and the maximum ($M$) as indicated. The spatially averaged pixel values ($a$) are also given, for comparison with the nominal values $\hat{c}_f\Itot$, see \Tab{tab:5FPs}. Middle row: Retrieved dynamics $\bT_f$ (red), and corresponding original dynamics (black). Bottom rows: Difference between the nominal and retrieved distributions corresponding to the top rows, with $M$ and $m$ as given.}}
	\label{FigSM_UnmixingFree_fastNMFGD_L}
\end{figure}

\begin{figure}
	\includegraphics[width=\textwidth]{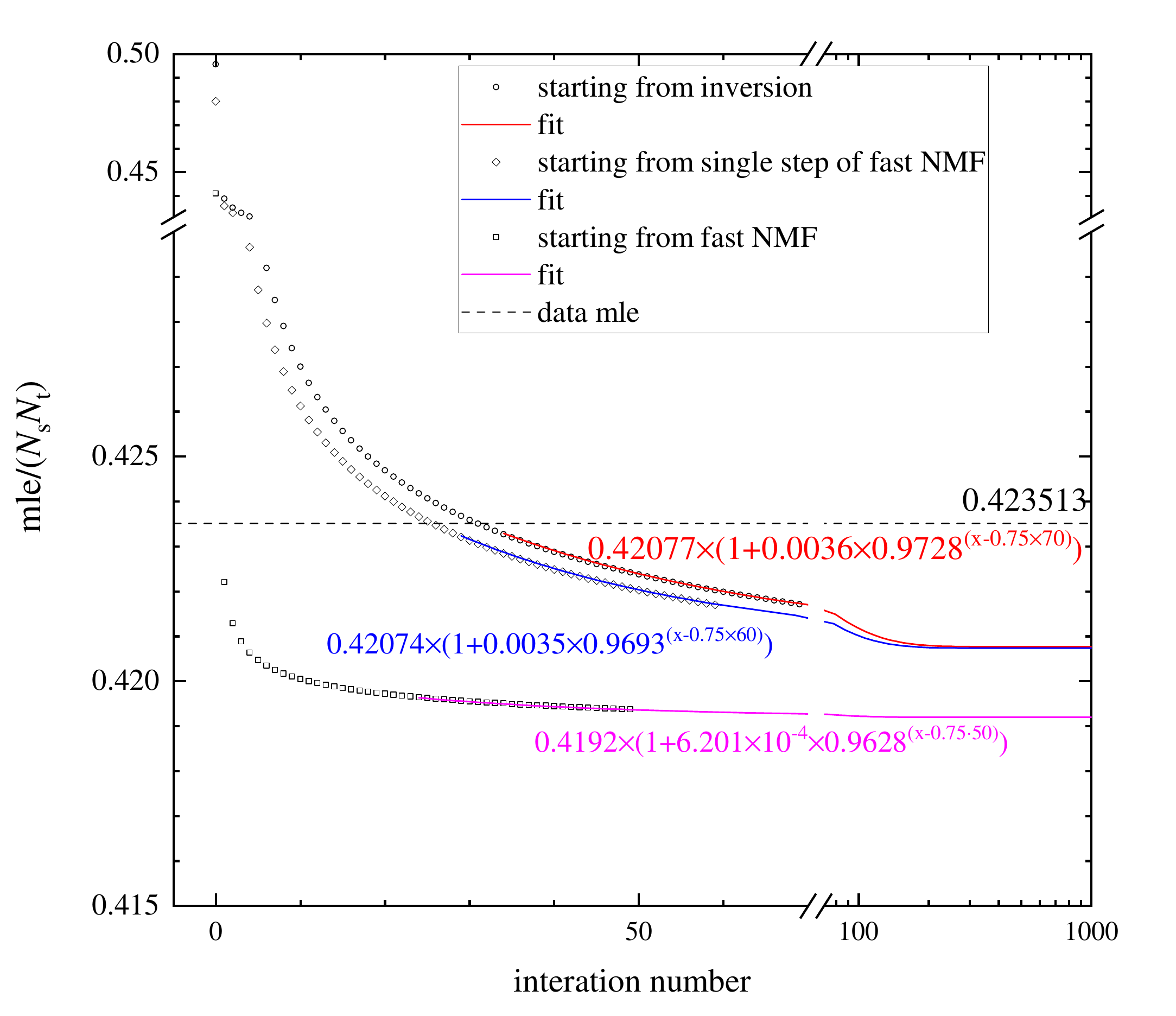}
	\caption{Same as \Fig{FigSM_UnmixingFixed_conv} for the analysis of the data of \Fig{Fig_UnmixingFree_fastNMF_H} using different initial guesses as labelled: \add{Convergence of the KLD during gradient descent applied to the sFLIM data of \Fig{Fig_UnmixingFixed_fastNMF_H} using different initial guesses as labelled. The evaluated mle (symbols) is the KLD maximum likelihood estimator (\Eq{eq:mle}) normalised by the number of elements of \bD. The  coloured lines are the fits used to determine convergence, also given as analytic expressions.}}
	\label{FigSM_UnmixingFree_conv}
\end{figure}

The gradient descent method can be applied also in this situation by iterating between the two update rules of \Eq{eq:mle}. Here we have imposed that the dynamics does not depend on the excitation and detection channel, by replacing \bT\ calculated by \Eq{eq:gd} with its average across the different channels. The results of the analysis are shown in \Fig{FigSM_UnmixingFree_GD0_H}-\ref{FigSM_UnmixingFree_fastNMFGD_L} for the different initial guesses used and number of photons detected, as given in \Tab{tab:simfigsunmixfree}.
The convergence of the estimator during the gradient descent depends on the choice of the initial guess, and we find that choosing the fast NMF results provides the fastest convergence and the lowest estimator (see \Fig{FigSM_UnmixingFree_conv}).

\begin{table}
	\begin{tabular}{l|l|l|l|l}
		\hline
		Figure & \Itot\ & Method & iterations & mle \\
		\hline
		\Fig{Fig_UnmixingFree_fastNMF_H} & $10000$ & fast NMF & 25 & $1.514\times10^7$\\
		\Fig{FigSM_UnmixingFree_GD0_H} & $10000$ & GD KLD from inversion  & 70 & $1.448\times10^7$\\
		\Fig{FigSM_UnmixingFree_GD_H} & $10000$ & GD KLD from single fast NMF step  & 60 & $1.448\times10^7$\\
		\Fig{FigSM_UnmixingFree_fastNMFGD_H} & $10000$ & GD KLD from fast NMF & 50 & $1.440\times10^7$\\ 
		\Fig{FigSM_UnmixingFree_fastNMF_L} & $100$ & fast NMF & 9 & $8.340\times10^6$\\
		\Fig{FigSM_UnmixingFree_GD0_L} & $100$ & GD KLD from inversion & 50 & $8.321\times10^6$\\
		\Fig{FigSM_UnmixingFree_GD_L} & $100$ & GD KLD from single fast NMF step  & 50 & $8.321\times10^6$\\
		\Fig{FigSM_UnmixingFree_fastNMFGD_L} & $100$ & GD KLD from fast NMF & 50 & $8.316\times10^6$\\ 
		\hline
	\end{tabular}
	\caption{Overview over simulation results shown for unmixing of 5 FPs of free dynamics and spectra, with the number of iterations and achieved mle. The mle of the simulated data is $1.454\times10^7$ for $\Itot=10^4$ and $8.477\times10^6$ for $\Itot=10^2$.}
	\label{tab:simfigsunmixfree}
\end{table}


\clearpage
\subsection{Systematic errors due to unmixing with inaccurate fixed spectro-temporal properties}
\begin{figure}[b]
	\includegraphics[width=\textwidth]{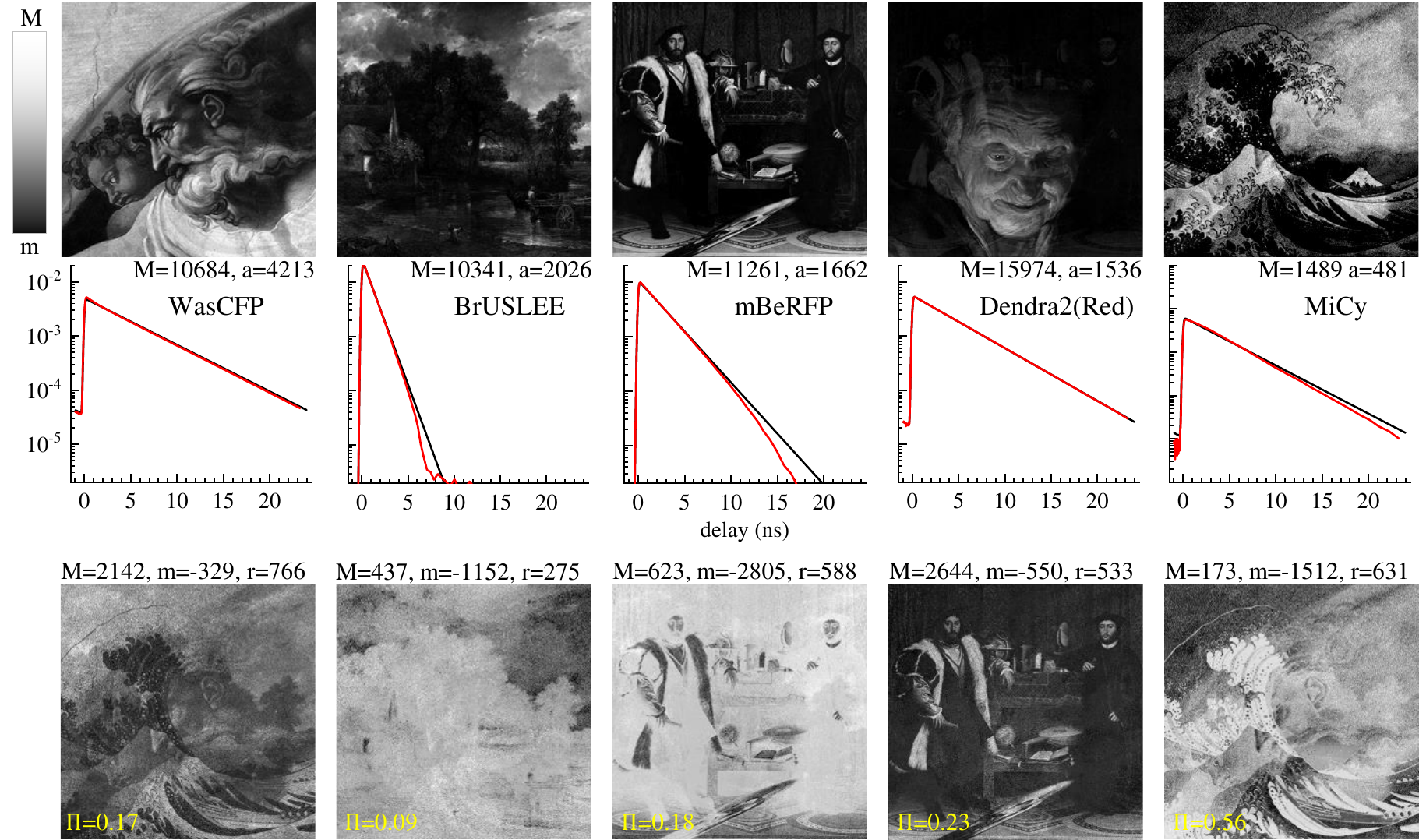}
	\caption{\edit{Factorisation results of the data generated using the FP properties in \Tab{tab:5FPs} and  $\Itot=10^4$ photons, but analysed using the corresponding FP properties of \Tab{tab:8FPs}.}{uFLIM factorisation results of the sFLIM data generated using the FP properties in \Tab{tab:5FPs} and  $\Itot=10^4$ photons, but analysed using the corresponding FP properties of \Tab{tab:8FPs}.}}
	\label{FigSM_UnmixingFree_fastNMFFixed_H}
\end{figure}

\begin{figure}
	\includegraphics[width=\textwidth]{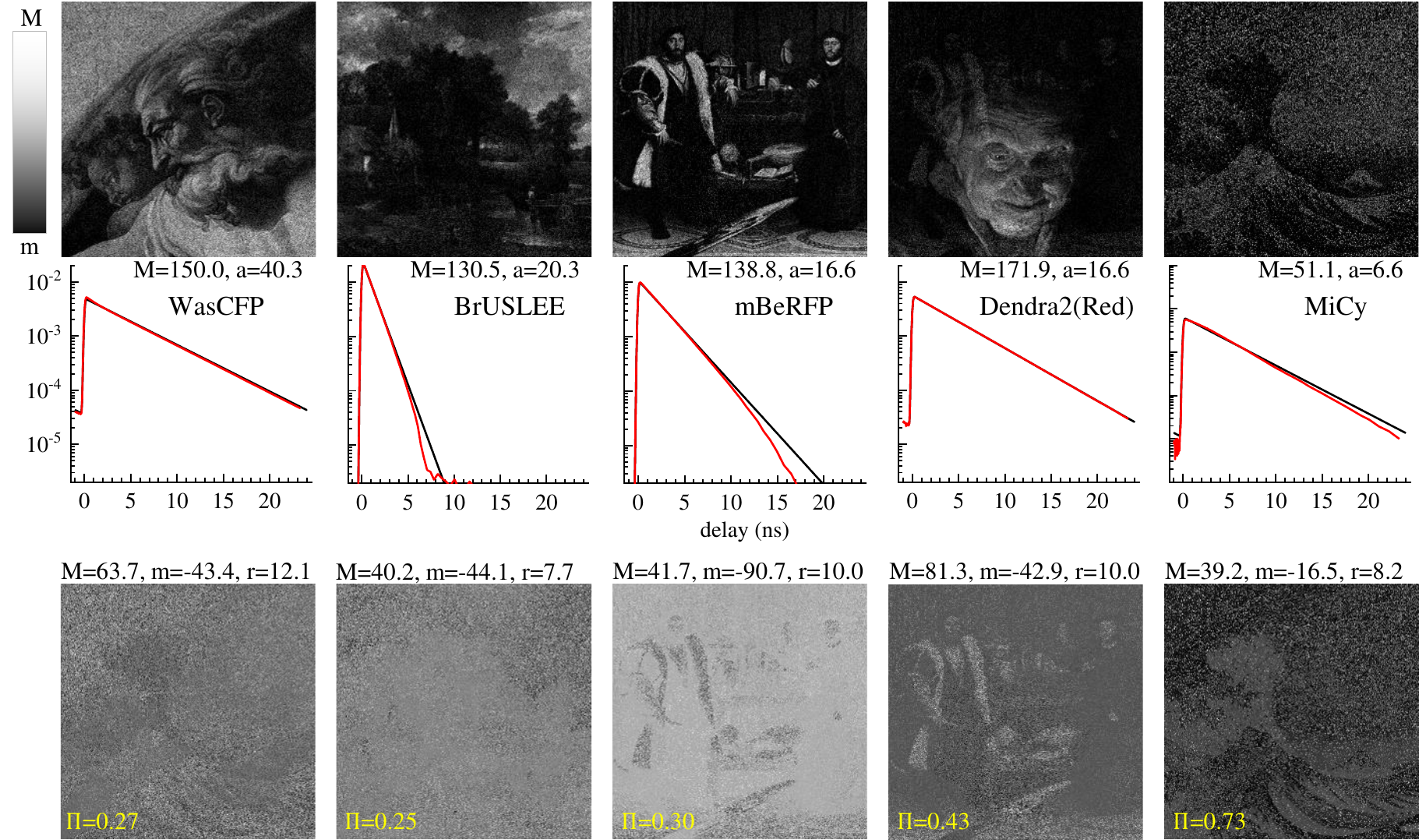}
	\caption{Same as \Fig{FigSM_UnmixingFree_fastNMFFixed_H} but for $\Itot=100$ photons: \add{uFLIM factorisation results of the sFLIM data generated using the FP properties in \Tab{tab:5FPs} and  $\Itot=100$ photons, but analysed using the corresponding FP properties of \Tab{tab:8FPs}.}}
	\label{FigSM_UnmixingFree_fastNMFFixed_L}
\end{figure}

Systematic errors arise when unmixing the distribution of FPs using fixed  spectro-temporal properties somewhat deviating from the correct ones, which is likely the case when they are measured on different samples. \Fig{FigSM_UnmixingFree_fastNMFFixed_H}-\ref{FigSM_UnmixingFree_fastNMFFixed_L} shows the result of the factorisation of the data generated using the properties of \Tab{tab:5FPs} and applying the fast NMF method with fixed FP properties taken as corresponding values listed in \Tab{tab:8FPs}. The resulting systematic error is clearly visible for higher signal to noise ratio ($\Itot=10^4$, \Fig{FigSM_UnmixingFree_fastNMFFixed_H}), where the retrieved distributions of the weaker FPs are significantly perturbed.  For low signal-to-noise-ratio ($\Itot=100$, \Fig{FigSM_UnmixingFree_fastNMFFixed_L}), where the shot noise dominates, the effect of the systematic error is less relevant.

\clearpage
\subsection{Unmixing of FPs having non-exponential dynamics}\label{sec:unmixlogn}

To calculate the FP dynamics with non-exponential dynamics given by a decay rate distribution $P(\gamma)$ efficiently, we have discretized the rate distribution using a logarithmic grid of rates, plus zero and infinity, i.e.  $\gamma_k\in[0, 2r, 4r, .., 2^n r, \infty]$, where $r$ is the inverse of the measured time range $l \Delta$, and $n$ is the smallest integer for which $2^n>l$. We then determine for each interval $\gamma_k$ to $\gamma_{k+1}$ the probability 
\be P_k=\int_{\gamma_k}^{\gamma_{k+1}}P(\gamma)d\gamma\label{eq:discprob}\ee
and the average decay rate $\bar{\gamma}_k$ as the first moment  
\be \bar{\gamma}_k=\frac{1}{P_k}\int_{\gamma_k}^{\gamma_{k+1}}\gamma P(\gamma)d\gamma\,,\label{eq:avdecrate} \ee 
and then calculate the dynamics as 
\be \mathbf{T}_{P}=\sum_{k}P_k\mathbf{T}_k\,,\ee
using the single exponential dynamics $\mathbf{T}_k$ given by \Eq{eq:decay} with $\gamma_f=\gamma_k$. 

\begin{figure}
	\includegraphics[width=0.7\textwidth]{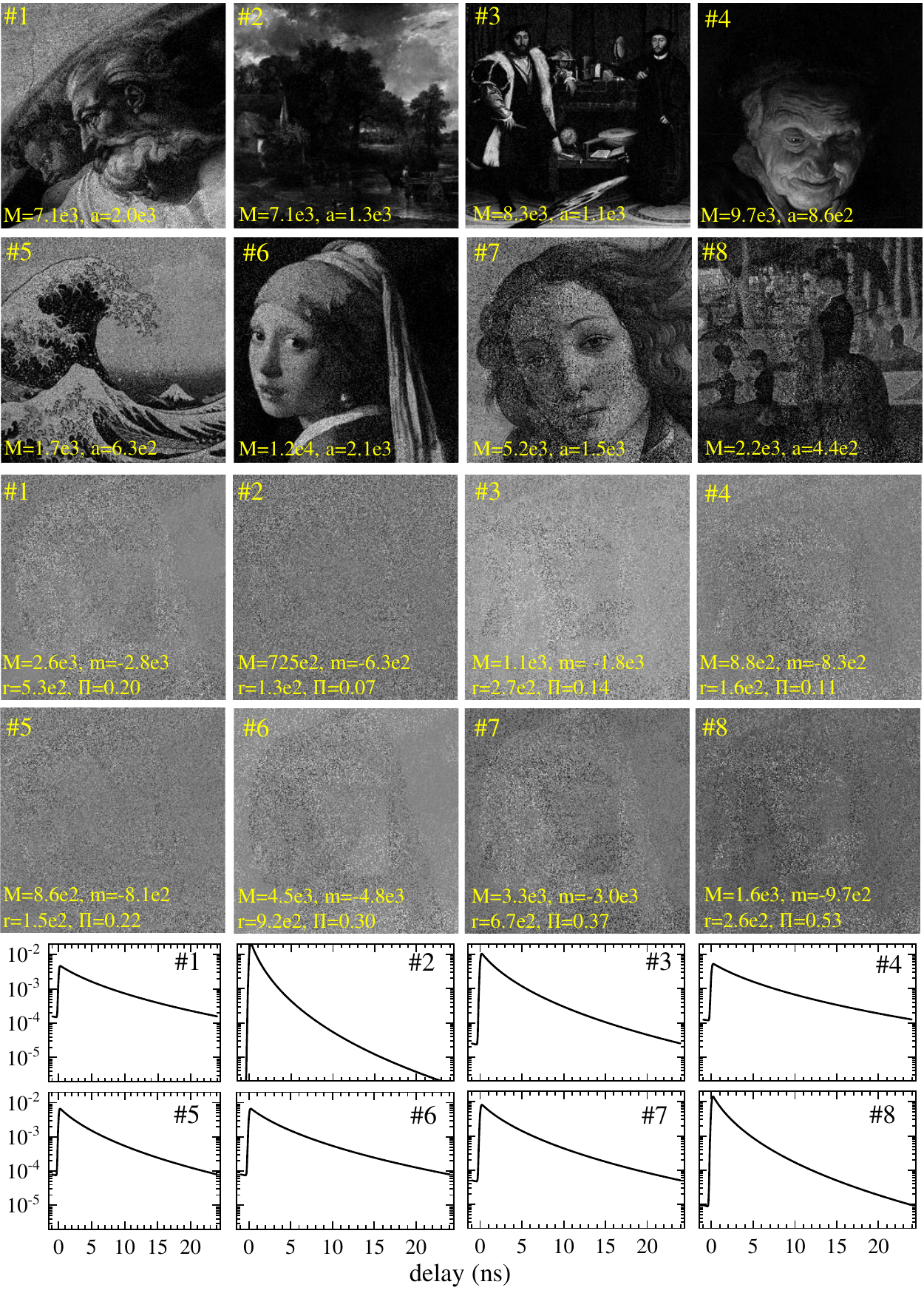}
	\caption{Same as \Fig{Fig_UnmixingFixed_fastNMF_H} for the case of non-exponential dynamics: \add{Results of the uFLIM analysis on sFLIM synthetic data generated with 8 FPs with $I_{\rm tot}=100$ photons and non-exponential dynamics. The FP dynamics are fixed in the analysis which uses the fast NMF algorithm. Top: Retrieved \bS\, with $m=0$ and the maximum ($M$) as indicated. The spatially averaged pixel values ($a$) are also given, for comparison with the nominal values $\hat{c}_f\Itot$, see \Tab{tab:8FPs}. Middle rows: Difference between the nominal and retrieved distributions corresponding to the top rows, with $M$ and $m$ as given.} The last two rows show the non-exponential FP fluorescence dynamics used, resulting from the log-normal distribution \Eq{eq:lognorm} with $\gb=1/\tau$ and $\sigma=0.8$.}
	\label{FigSM_UnmixingFixed_LogNorm}
\end{figure}

\begin{figure}
	\includegraphics[width=\textwidth]{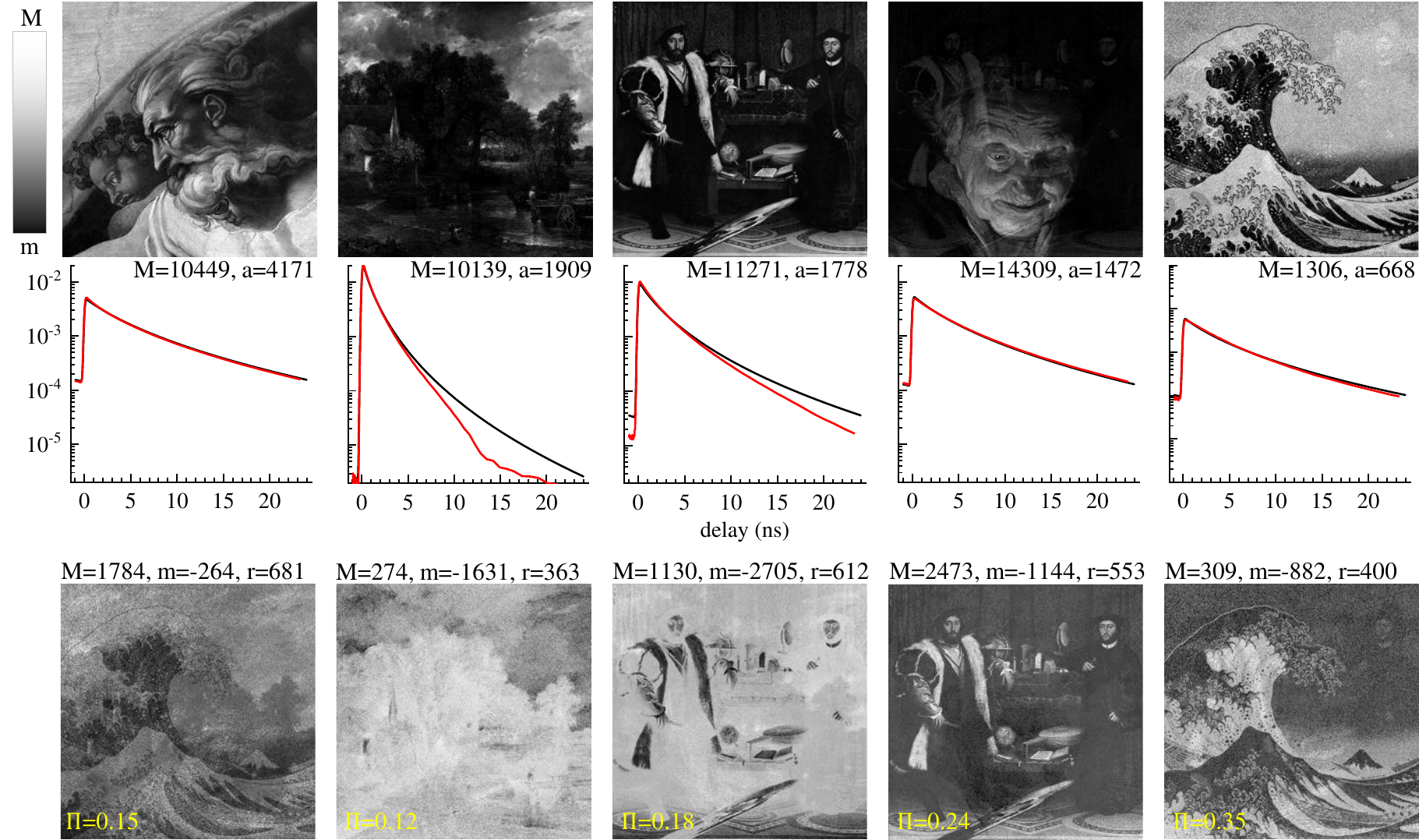}
	\caption{Same as \Fig{Fig_UnmixingFree_fastNMF_H} for the case of non-exponential dynamics: \add{Results of the uFLIM analysis on sFLIM synthetic data generated with five FPs with $I_{\rm tot}=100$ photons for the case of non-exponential dynamics. The FP dynamics are obtained by the fast NMF algorithm in the analysis. Top: Retrieved \bS\, with $m=0$ and the maximum ($M$) as indicated. The spatially averaged pixel values ($a$) are also given, for comparison with the nominal values $\hat{c}_f\Itot$, see \Tab{tab:5FPs}. Middle row: Retrieved dynamics $\bT_f$ (red), and corresponding original dynamics (black). Bottom rows: Difference between the nominal and retrieved distributions corresponding to the top rows, with $M$ and $m$ as given.}}
	\label{FigSM_UnmixingFree_LogNorm}
\end{figure}

For the simulations we took $P(\gamma)$ as the log-normal distribution \Eq{eq:lognorm} with $\gb$ given by the inverse of the FP lifetime $\tau$, and \sg=0.8. \Fig{FigSM_UnmixingFixed_LogNorm} and \Fig{FigSM_UnmixingFree_LogNorm} show the results of the unmixing analysis for these datasets where the dynamics are known or retrieved, respectively.

\clearpage
\section{Analytical solution of donor and acceptor dynamics undergoing FRET}
\label{sec:analytic}
\begin{figure}[b]
	\includegraphics[width=0.5\textwidth]{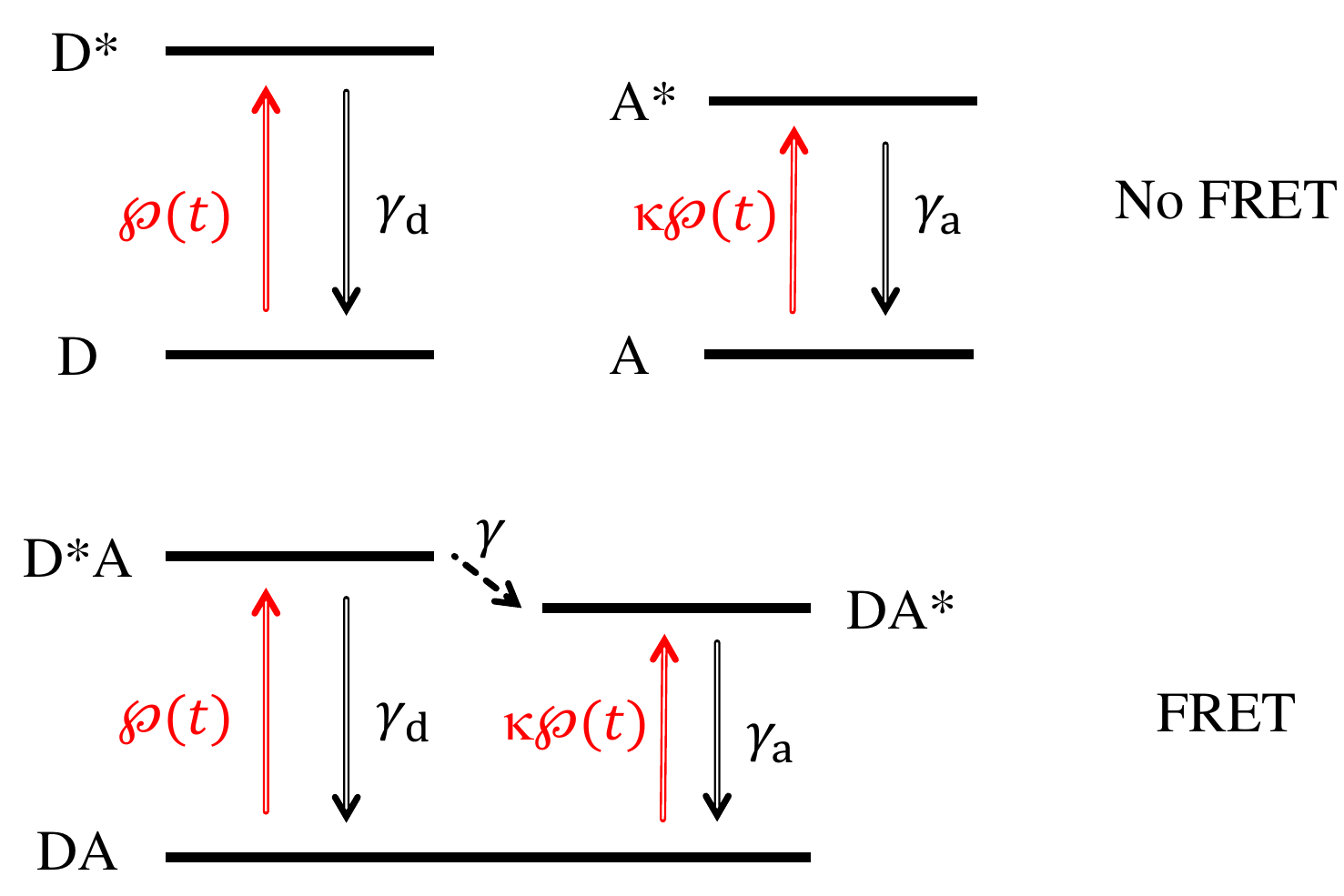}
	\includegraphics[width=0.6\textwidth]{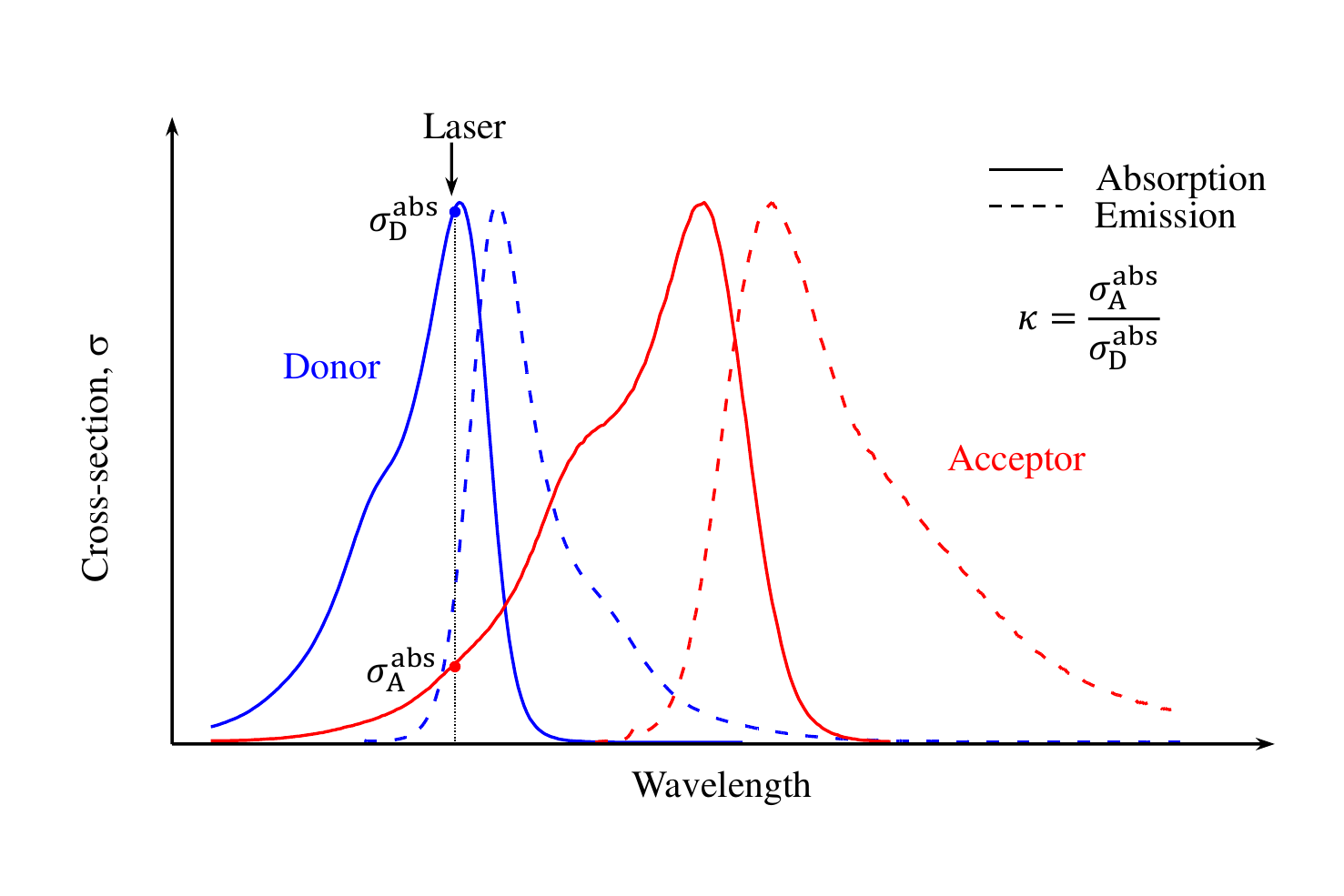}
	\caption{Top: Level scheme showing the dynamic processes in absence (top) and presence (bottom) of a FRET process. Bottom: Sketch of the absorption and emission cross-section. The parameter $\kappa$ is the ratio of acceptor to donor absorption cross-sections at the excitation wavelength.}\label{FigSM_FRETLevelScheme}
\end{figure}

The change in the fluorescence dynamics of the donor and acceptor in presence of FRET can be analytically calculated in the case of a monoexponential decay for the pure species and a Gaussian excitation pulse. The level scheme used to determine the rate equations is shown in \Fig{FigSM_FRETLevelScheme}. We are calulating here the dynamics after excitation of the donor only, as the dynamics of the directly excited acceptor in the DAP is unaltered.

The rate equations governing the resulting dynamics can be written as
\be
\frac{{\rm d}}{{\rm d} t}N_{\rm D^*A}=\wp(t)-(\gamma+\gamma_{\rm d})N_{\rm D^*A}\label{eq:DonorRateEquation}
\ee
and
\be
\frac{{\rm d} }{{\rm d} t}N_{\rm DA^*}=\gamma N_{\rm D^*A}-\gamma_{\rm a} N_{\rm DA^*},\label{eq:AcceptorRateEquation}
\ee
where $N_{\rm D^*A}$ ($N_{\rm DA^*}$) is the number of donor-acceptor pairs with the excitation localised on the donor (acceptor) site, respectively, $\gamma_{\rm d}(\gamma_{\rm a})$ is the recombination rate of the donor (acceptor), and $\gamma$ is the FRET rate.
If we assume a Gaussian excitation pulse $\wp(t)=\frac{1}{\sqrt{2\pi}s}e^{-\frac{t^2}{2s^2}}$ of root-mean square width $s$ and unity integral, the equations above, for an initially unexcited system, have the analytical solution
\be
N_{\rm D^*A}=\frac{e^{-(\gamma_{\rm d}+\gamma)}e^{\left(\gamma_{\rm d}+\gamma\right)^2s^2/2}}{2}\left[1+{\rm erf}\left(\frac{t-\left(\gamma_{\rm d}+\gamma\right)s^2}{\sqrt{2}s}\right)\right]\,,\label{eq:DonorAnalyticalEquation}
\ee
and

\begin{multline}	
N_{\rm DA^*}(t)=\frac{\gamma}{2\Gamma}e^{-\gamma_{\rm a}t}e^{\left(\gamma_{\rm d}+\gamma\right)^2s^2/2}\Biggl[e^{\Gamma s^2/2}e^{-\Gamma\left(\gamma_{\rm d}+\gamma\right)s^2}\left[1+{\rm erf}\left(\frac{\Gamma s^2+t-\left(\gamma_{\rm d}+\gamma\right)s^2}{\sqrt{2}s}\right)\right]\\
-e^{-\Gamma t}\left[1+{\rm erf}\left(\frac{t-\left(\gamma_{\rm d}+\gamma\right)s^2}{\sqrt{2}s}\right)\right]\Biggr],\label{eq:AcceptorAnalyticalEquation}
\end{multline}
where $\Gamma=\gamma_{\rm d}+\gamma-\gamma_{\rm a}$. Setting $\gamma=0$,  \Eq{eq:DonorAnalyticalEquation} describes the donor or acceptor dynamics in absence of a FRET process using the corresponding $\gamma_{\rm d}$ or $\gamma_{\rm a}$.

\begin{figure}
	\includegraphics[width=0.6\textwidth]{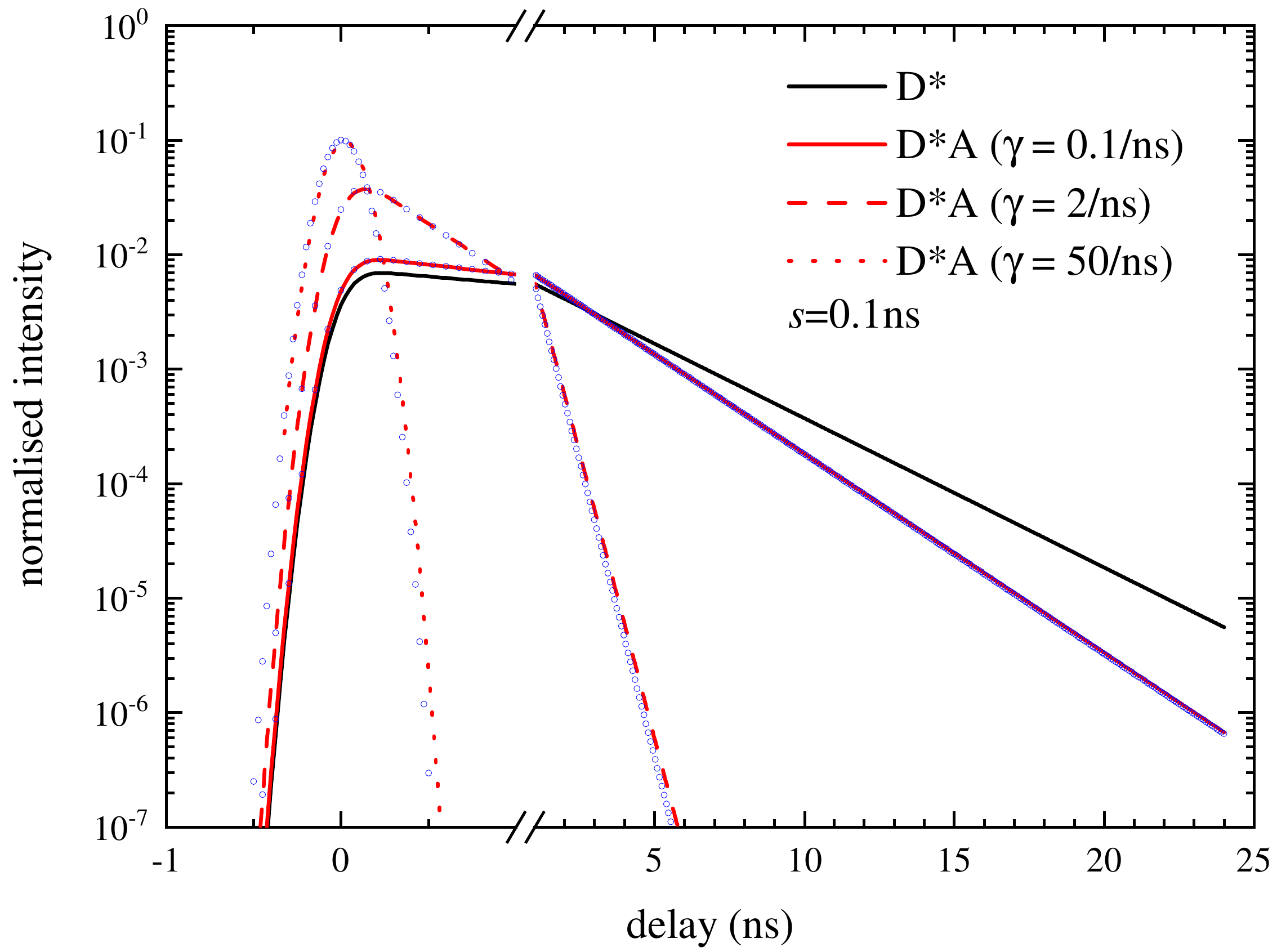}
	\caption{Fluorescence dynamics for the pure donor ($N_{\rm D^*}$, black line) and for a donor undergoing FRET ($N_{\rm D^*A}$, red lines) as calculated using \Eq{eq:DonorAnalyticalEquation} for different values of $\gamma$, and the corresponding dynamics (blue symbols) determined using \Eq{eq:Td} for $\Delta=0.025$\,ns.}\label{FigSM_fitAnalyticDonorOnly}
\end{figure}

We have used the model developed in the main manuscript to calculate the dynamics of a donor molecule undergoing FRET as a function of the FRET rate and the temporal sampling $\Delta$. In the left panel of \Fig{FigSM_fitAnalyticDonorOnly} we shows the increase in the relative error in the determination of $\gamma$ for increasing values of $\gamma$ and $\Delta$. 
As discussed in the main manuscript, the model is retrieving the simulated FRET rate with good accuracy for $\Delta\gamma \ll 1$, that is for a weak dynamics within the sampling interval, as expected. The good agreement between the dynamics calculated using the analytical solution (red lines) and the one obtained by the model (blue symbols) are shown in \Fig{FigSM_fitAnalyticDonorOnly}.

\begin{figure}
	\includegraphics[width=0.6\textwidth]{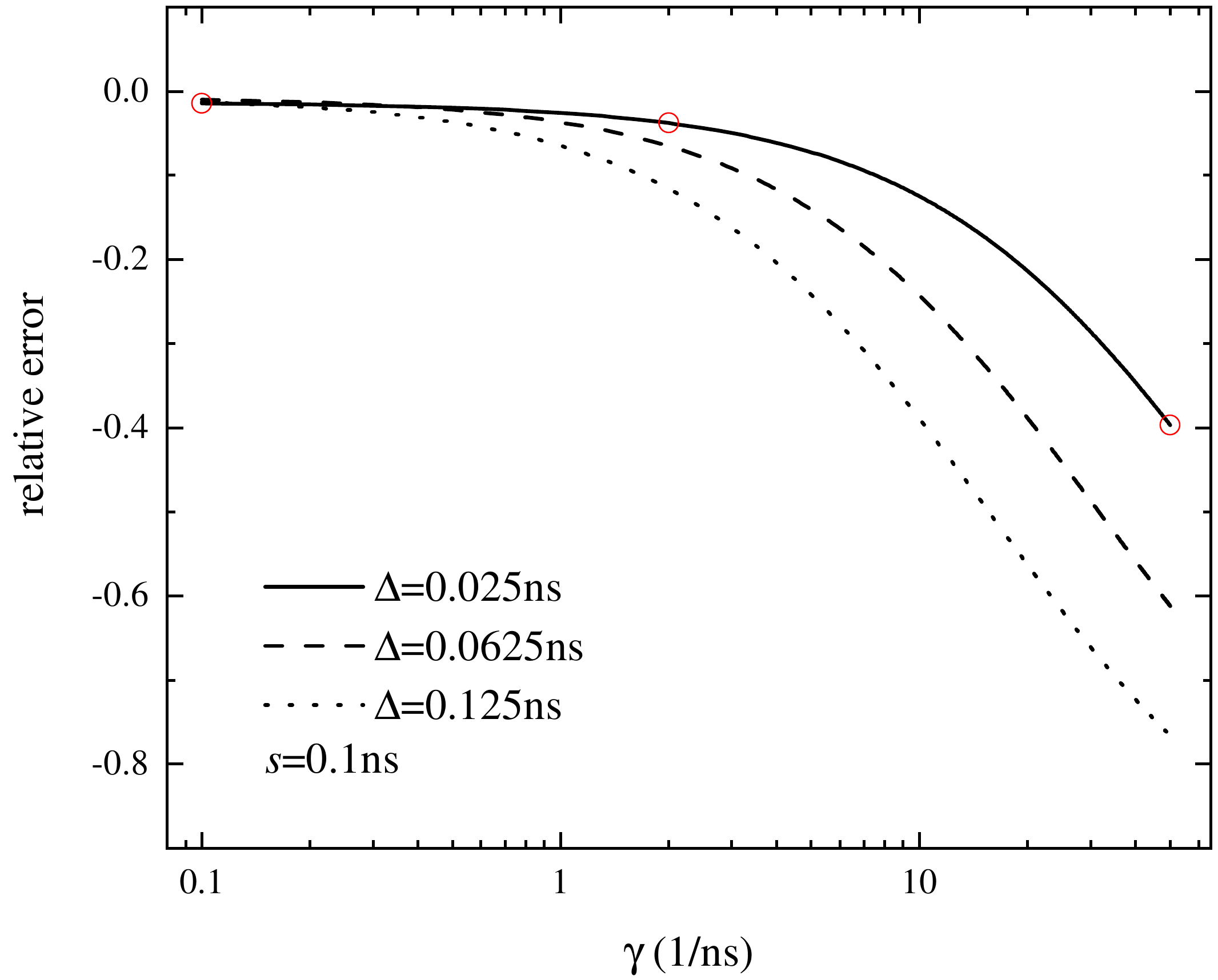}
	\caption{Relative error of the FRET rate $\gamma$ obtained by the fit using \Eq{eq:Td} as a function of $\gamma$ and for different values of the temporal sampling interval $\Delta$. The red circles indicate the $\gamma$ values used for the plots in \Fig{FigSM_fitAnalyticDonorOnly}.}\label{FigSM_RelError_fitAnalyticDonorOnly}
\end{figure}

\begin{figure}
	\includegraphics[width=0.8\textwidth]{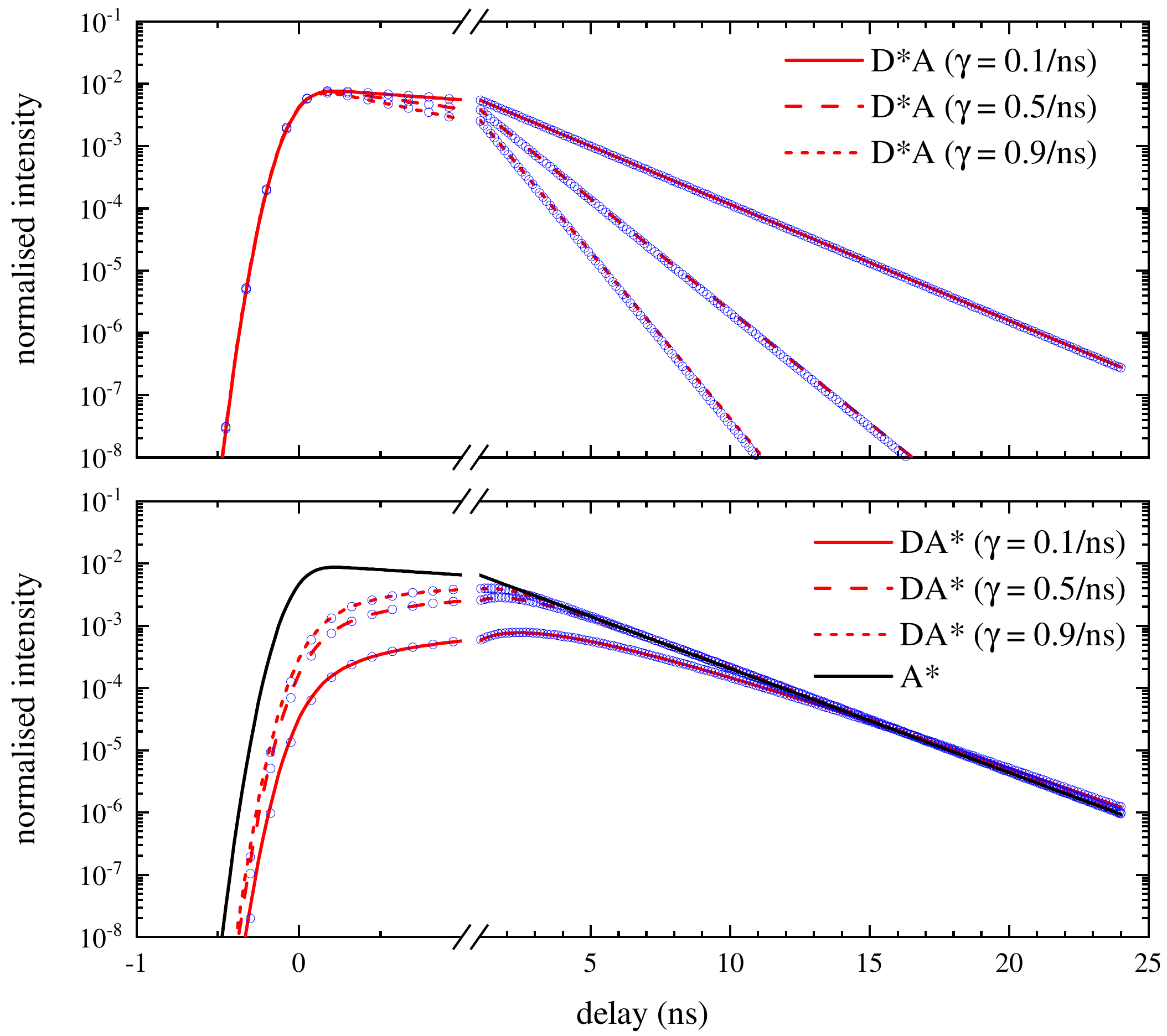}
	\caption{Dynamics of the donor $N_{\rm D^*A}$ (top) and acceptor $N_{\rm DA^*}$ (bottom) undergoing FRET for different transfer rates. Red lines are the analytical solutions, while the blue symbols are determined using \Eq{eq:Td} (top) and \Eq{eq:Ta} (bottom) for $\Delta=0.025$\,ns and $s=0.1$\,ns. The black line shows the dynamics of a directly excited acceptor $N_{\rm A^*}$.}\label{FigSM_fitAnalytic}
\end{figure}

\Fig{FigSM_fitAnalytic} shows the results of the fits once the acceptor dynamics is also taken into account. For this analysis we used $s=0.1$\,ns and $\Delta=0.025$\,ns. The FRET rates determined by the fits using the model developed in the main text are within 1\% of the values used in \Eq{eq:DonorAnalyticalEquation} and \Eq{eq:AcceptorAnalyticalEquation} to calculate the dynamics.

\clearpage
\section{Computationally efficient implementation of FRET parameter determination minimizing the NMF error}\label{sec:FRETparmin}

To implement the required minimization of the residual computationally efficiently, we discretized the rate distribution using \Eq{eq:discprob} and calculate the corresponding average decay rate $\bar{\gamma}_k$ from \Eq{eq:avdecrate} to approximate \Eq{eq:fretint} by the discrete sum
\be \bbT{}(\gb,\sg,q)=\sum_{k}P_k(\gb,\sg)\btT{}(\bar{\gamma}_k,q)\,.\ee
The obtained dynamics is rescaled to have $\{\bbT{}(\gb,\sg,q)\}=1$.
We then define a coarse grid of values for \gb, \sg\ and $q$ over a wide parameter range to identify the \edit{global minimum region}{the minimum region within the physically meaningful range of parameters}. For the results shown later for the simulated data,  we use a logarithmically spaced grid in \gb\ spanning from 1/8 of the donor decay rate, taken as the inverse of the first moment of \bTd, to half of the inverse of the temporal IRF of the data, with 31 steps, corresponding to a factor of about 1.26 per step. To calculate the first moment of \bTd\ we use the excitation maximum as time zero. For $\sg$, we use a linear grid from -0.2 to 1, with a step size of 0.1, while for the relative efficiency $q$ we use a linear grid between -0.4 and 2, with a step size of 0.2. The grid points at negative values are included to allow finding minima close to zero. We evaluate the NMF residual for each mesh point, and denote the indices of the mesh point of minimum error with $(i_0, j_0, k_0)$.

\begin{figure}
	\includegraphics[width=0.7\columnwidth]{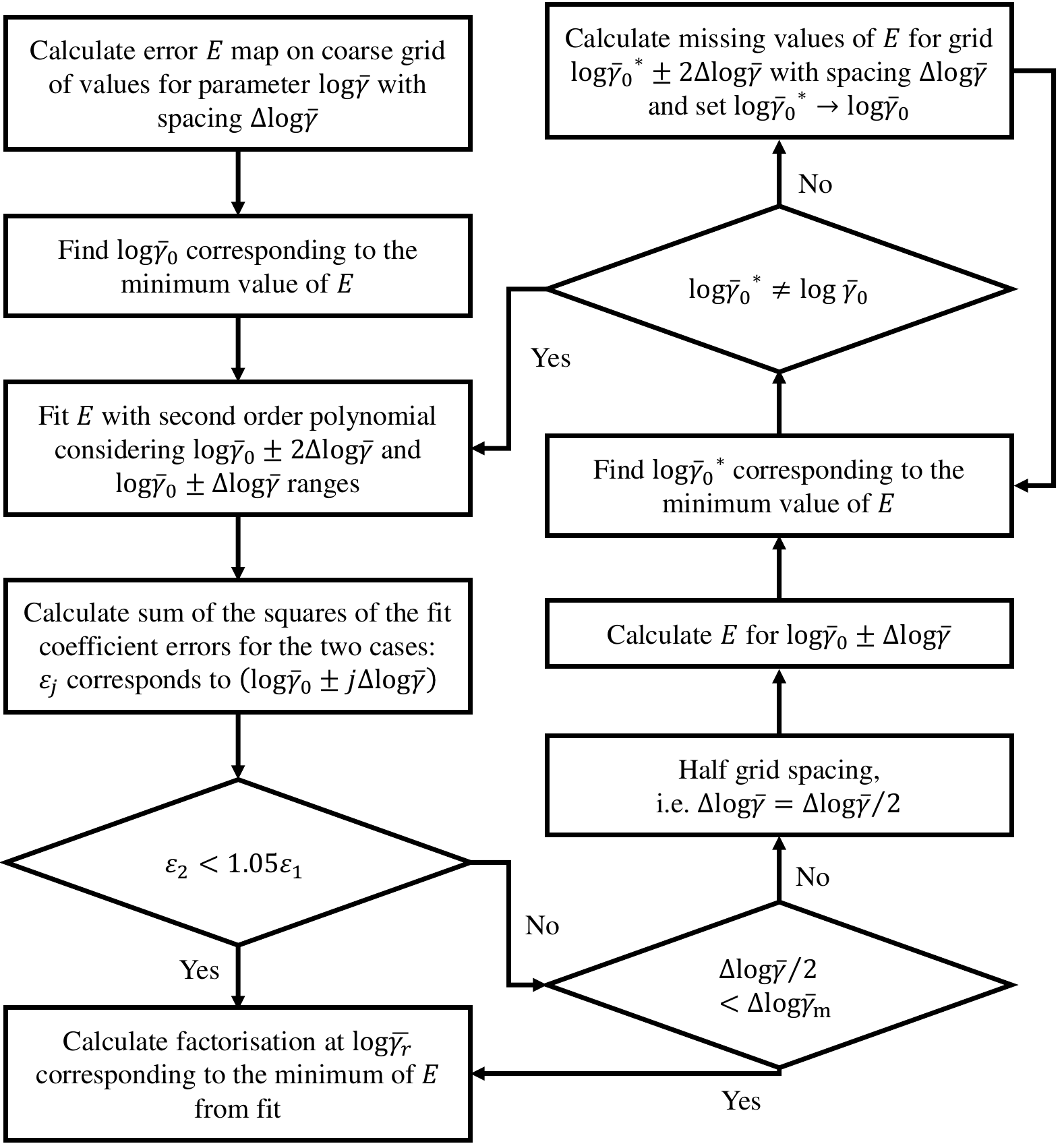}
	\caption{Flow diagram exemplifying the algorithm steps for the refinement of the parameter domain. Here we show the case of a one dimensional space for parameter $\log\gb$, with initial grid spacing $\Delta\log\gb$. $\Delta\log\gb_{\rm m}$ represents the minimum spacing allowed for the refined grid and it is set to $0.005$. For \sg\ and $q$ we have used a minimum grid spacing of $0.01$.}
	\label{FigSM_Method}
\end{figure}

To refine the calculation around this minimum, we fit the error in the local region $(i_0\pm1, j_0\pm1, k_0\pm1)$  using a second order polynomial. The fit returns the minimum position $(\gbf,\sgf,\qf)$, and its error $(\epsilon_{\gbf},\epsilon_{\sgf},\epsilon_{\qf})$. We consider the root-sum-square (RSS) of the errors in the minimum position, $\sqrt{\epsilon_{\gbf}^2+\epsilon_{\sgf}^2+\epsilon_{\qf}^2}$. We then increase the range sequentially in each parameter by one point as long as RSS decreases with respect to the previous step. To increase the speed in the analysis we require that the relative change in the RSS is larger than a given threshold, chosen to be 5\% for the results shown. This condition defines a initial range of the parameters.  We require that at least 5 points are included in each parameter range, i.e. $(i_0 \pm 2, j_0 \pm 2, k_0 \pm 2)$. If this is not the case for a given parameter, we half its grid step size, and repeat the procedure until either the RSS increases above the threshold for all parameters when increasing the number of points in the fit from 3 to 5 or the step size in the grid reaches a lower limit (for the analysis below we have used 0.005 for log$(\gb)$, 0.01 for \sg\ and q). Finally, we recalculate the NMF at the resulting minimum $(\gbf,\sgf,\qf)$. A simplified flow diagram of the algorithm used to refine the grid of FRET parameters and obtain their final values is shown in \Fig{FigSM_Method}. The flow diagram refers only to the refinement of a single parameter ($\log\gb$ in this case), while the algorithm takes into account the multidimensional parameter space.

\clearpage 

\clearpage
\section{Phasor Analysis}\label{sec:phasor}
In phasor analysis, the decay $I(t)$ at each pixel is represented by a pair of values $(g,s)$, the phasor $p$, corresponding to its Fourier cosine and sine components, respectively, given by
\be \label{eq:phasor} p=g+is=\frac{\int_0^{\infty} I(t) \exp{(i\omega t)} dt}{\int_0^{\infty} I(t) dt}, \ee
where $\omega$ is a multiple of the excitation modulation angular frequency. The phasor $p$ is analyzed in the complex plane. For a mono-exponential decay, $I(t)\propto \exp(-t/\tau)$,  $p$ forms a half-circle given by $|2p-1|=1$. Since \Eq{eq:phasor} is a linear transform, superpositions of decays have phasors given by the weighted average of the component phasors, in particular the phasors of a two-component mixture lie on the straight line connecting the phasors of the components. Adding additional components with other decay constants, for example due to autofluorescence, the phasors lie in the area spanned by all connecting lines. In FRET experiments, the lifetime of the quenched donor is calculated from the FRET efficiency $E$, $\tauf=(1-E)\taud$, where $\tauf(\taud)$ is the lifetime of the quenched (unquenched) donor, and the superposition of both components has a phasor given by
\be \pf=(1-\ff) \frac{1+i\omega\taud}{1+(\omega\taud)^2}+\ff\frac{1+i\omega\tauf}{1+(\omega\tauf)^2},\label{eq:Phasor} \ee
where $\ff$ is the fraction of quenched donor signal to total donor signal in the focal volume. The fraction of quenched to total donor concentration is accordingly given by $1/(1+(1/\ff+1)\tauf/\taud)$. The phasors of the unquenched donor and background can be determined in a control sample. Notably, different combinations of fractions and efficiency can lead to close-lying phasors, making it difficult to disentangle the two quantities. FRET trajectories are calculated according to the definition of FRET efficiency. The phasor of the unquenched donor is obtained from a FLIM image where the acceptor is absent, while the FRET phasor is calculated from \Eq{eq:Phasor}. All possible FRET phasors with different efficiencies describe a trajectory in the
phasor plot. For each pixel, the position of the phasor along the trajectory determines the FRET efficiency.\,\cite{DigmanBJ08S}

\begin{figure*}
	\includegraphics[width=\textwidth]{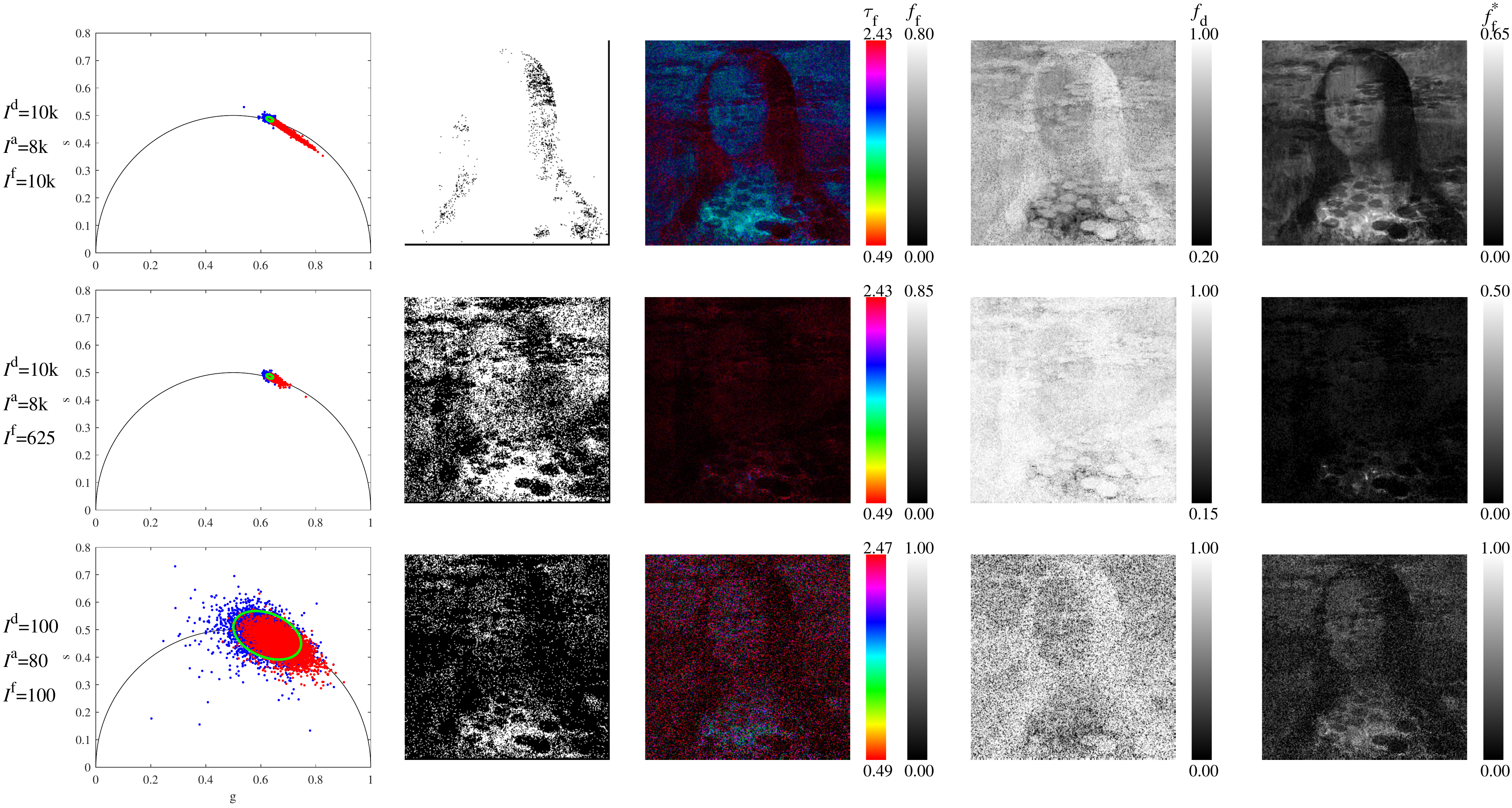}
	\caption{Results of the phasor analysis on the data of \Fig{Fig_FRET_g=5_k=1_Ib=2}. First column: Scatter plot of phasors. Blue symbols refer to control (donor only image). Red symbols correspond to data with all species included. Green lines are ellipses centred at mean values of $g$ and $s$ for the control image, with semi-axis aligned along the eigenvector of the covariance matrix and their length equal to two times the square root of the corresponding eigenvalues, that is a two standard deviations range. Second column: Binary image where white pixels correspond to phasors of the data with all species included outside the ellipses. Third column: HSV image of retrieved $\tau_{\rm f}$ and $\ff$ encoded as hue and value, respectively. Fourth column: Distribution of the estimated $f_{\rm d}$. Fifth column: Distribution of $\ff$ retrieved fixing the FRET rate to the ground truth.}
	\label{FigSM_Phasor}
\end{figure*}

\Fig{FigSM_Phasor} show the results of the phasor analysis for the data of \Fig{Fig_FRET_g=5_k=1_Ib=2}. Only the signal detected by the donor channel is considered, and we assume that the acceptor emission into the donor channel is negligible. The first column shows scatter plots of the phasors for two FLIM images, one including the mixed population (red points) and one with only the unquenched donor (blue points). While the phasors from the image with the unquenched donor are forming a localised cluster on the circle, the mixed population image phasors show an elongated distribution. As the number of detected photons decreases, the distributions spread due to the increasing effect of shot noise. 

To analyze the data using the phasor plot, we have defined a region associated to pure donor molecules. We have calculated the covariance matrix $C$ of the $(g,s)$ pairs for the donor only image, and determined the ellipse with semi-axis length equal to 2 times the square root of the eigenvalues of $C$ and directions defined by its eigenvectors, i.e. defining a contour at two standard deviations. We have then created a binary images from the phasors of the image showing the FRET component, assigning 1 (0) to the pixels outside (inside) the ellipse. The second column of \Fig{FigSM_Phasor} shows the resulting images. For the case of high photon counts for both the pure donor and FRET component (top row), most of the pixels show the presence of the FRET process. As the relative strength of the signal of the FRET component is reduced (middle row), or for small signals (bottom row), the number of pixels associated with the FRET process reduces. This approach gives only a binary information of the distribution of the FRET pairs. 

\begin{figure}[b]
	\includegraphics[width=\textwidth]{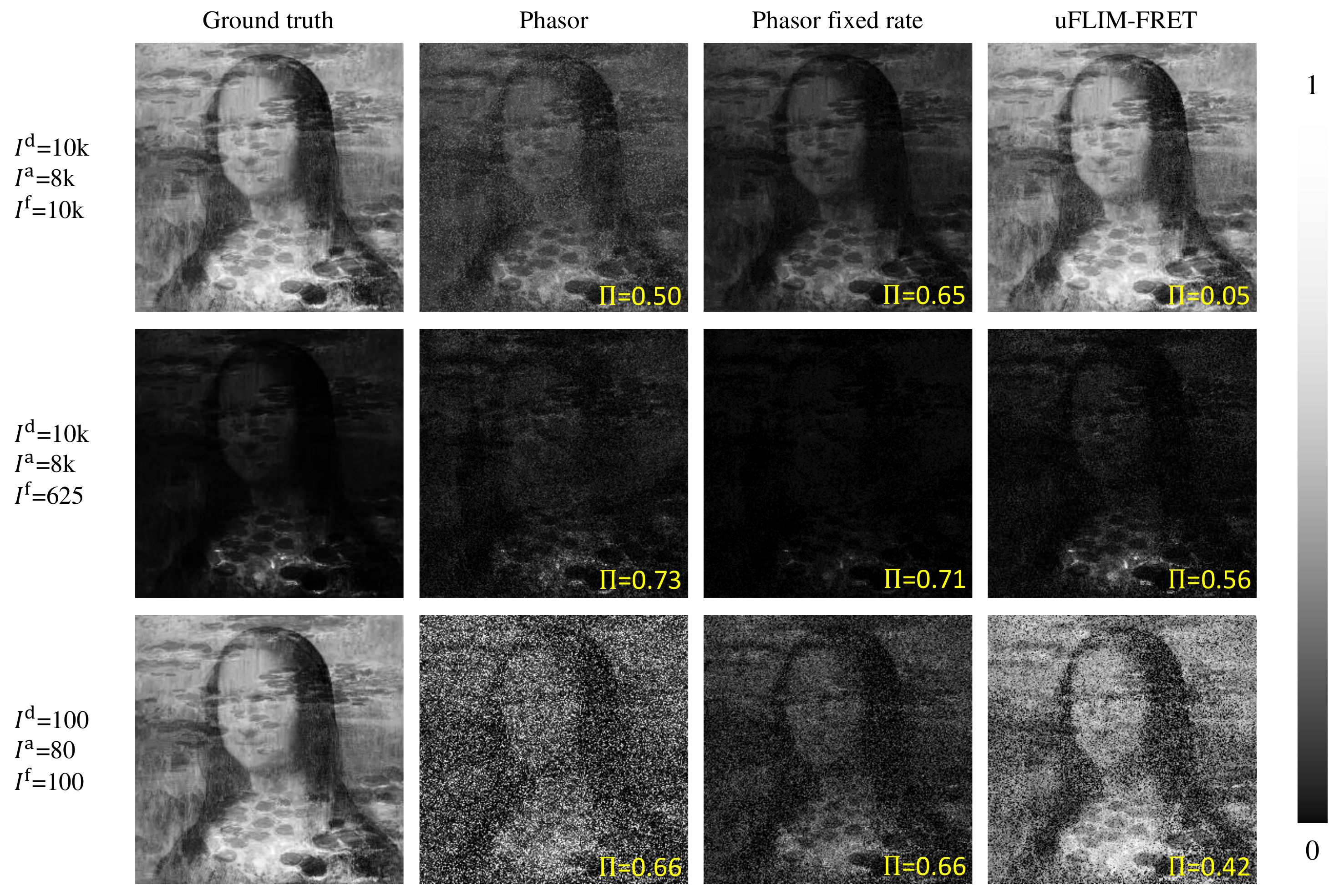}
	\caption{Distribution of the fraction \ff\ of donor undergoing FRET for different values of \Id\ and \Idap. First column: ground truth obtained from the distributions used to calculate the synthetic data. Second column: phasor analysis. Third column: phasor analysis fixing the FRET rate to the ground truth. Fourth column: uFLIFRET method.}
	\label{FigSM_Ph_vs_uFLIM}
\end{figure}

To extract more quantitative information, we have calculated the phasors  $\pf(\ff,\tauf)$ from \Eq{eq:Phasor}, by equidistant sampling of the parameter space considering $0.2\leq E\leq1$ (17 steps) and $0\leq \ff \leq1$ (21 steps) and assigned to each data phasor $p$ the parameter pair $(\ff,\tauf)$ corresponding to the minimum distance $|p-\pf(\ff,\tauf)|$. The donor lifetime is determined from the mean phasor $\bar{p}$ of the control (donor only) image by minimising $|\bar{p}-\pf(0,\taud)|$, resulting in $\tau_d=3.04$\,ns, in good agreement with the nominal value used (3\,ns).
The third column of \Fig{FigSM_Phasor} show the results of the analysis with $\tau_{\rm f}$  encoded as hue and the \ff\ as value, and full saturation. The fourth column shows the retrieved distribution of \fd\ and \ff\ in the image. The obtained distributions of \fd\ and \ff\ poorly reflect the expected distributions of donors and DAP signals. For comparison, we have calculated the signal fraction of donor molecules undergoing FRET using the uFLIFRET results shown in \Fig{Fig_FRET_g=5_k=1_Ib=2} as point-wise $\Sf/\left(\Sd+\Sf\right)$ (Note that \bS\ is refering to normalized dynamics and thus is proportional to the signal, as in the phasor analysis definition of \ff. The FRET pair fractions obtained with the different methods are shown in \Fig{FigSM_Ph_vs_uFLIM} for different values of \Id\ and \Idap\ and compared with the nominal distribution. Clearly, the uFLIFRET results show a better agreement with the ground truth, as confirmed from a lower value of the relative error $\Pi$ (here calculated using the distribution of \ff).

We have also estimated $\ff$ using the method above but fixing the FRET efficiency $E$ to the ground truth, shown in \Fig{FigSM_Phasor} on the right. Despite using additional prior knowledge, the relative error $\Pi$is much higher than the uFLIFRET results. The high error can be ascribed to the contribution of the modified acceptor dynamics into the DAP signal. We note that the phasor analysis is computationally light, with a CPU time of about 1s needed for the analysis of a single image.

To further compare our method with phasor analysis we have analysed data from FRET-FLIM experiments {\it in vivo}\,\cite{ChenJB18S}. In this experiment, donor- and acceptor- labeled transferrin (Tf) are injected in anaesthetized mice to monitor engagement of Tf with receptor (TfR) and non-specific accumulation of Tf in the liver (as this is a major site of iron
homeostasis regulation and displays high levels of TfR expression) and its elimination via the bladder (where the free dyes are accumulated before excretion). AF700 and AF750 were used as donor and acceptor, respectively with a 1:2 relative concentration. A control experiment where the mouse was injected only with Tf-AF700 was also performed. The donor- and acceptor- labelled Tf were injected at distinct time points separated by 15 minutes. FLIM acquisition was performed every ~45 seconds over ~2 hours. More details on the experimental procedure can be found in \Onlinecite{ChenJB18S}. 

In the analysis, only the data corresponding to the liver and the bladder were considered. Before performing the uFLIFRET, we have compensated the possible pixel-dependent variation of the laser pulse arrival time. For each pixel, we have defined the pulse arrival time as the time when the measured intensity is half of the maximum recorded signal\,\cite{ChenJB18S}. To align the time axis data were interpolated, and we used linear extrapolation to take into account for truncated dynamics. Only data with delay larger than -0.22ns were used to limit the contribution of signal at negative time delays.

After this pre-processing steps, we have used the first 5 FLIM images of the bladder to obtain the dynamics of the free donor using uFLIM with one component. The data were successively time binned ($\tb=0.04$\,ns and $\rb=0.05$) to improve signal-to-noise ratio and reduce computational time. uFLIFRET was then used to estimate the distribution of the DAP undergoing FRET. Since the data were acquired only using a single channel resonant with the donor emission and the acceptor bleed-through was not characterised we have performed our analysis assuming $\R{a}=1$ and $\eta=0$ and searching for the combination of \gb\ and \sg\ minimising the NMF error. We have decided not to apply the partial whitening step as the data were presenting high systematic fluctuations which concealed the Poisson noise.

\begin{figure}[b]
	\includegraphics[width=\textwidth]{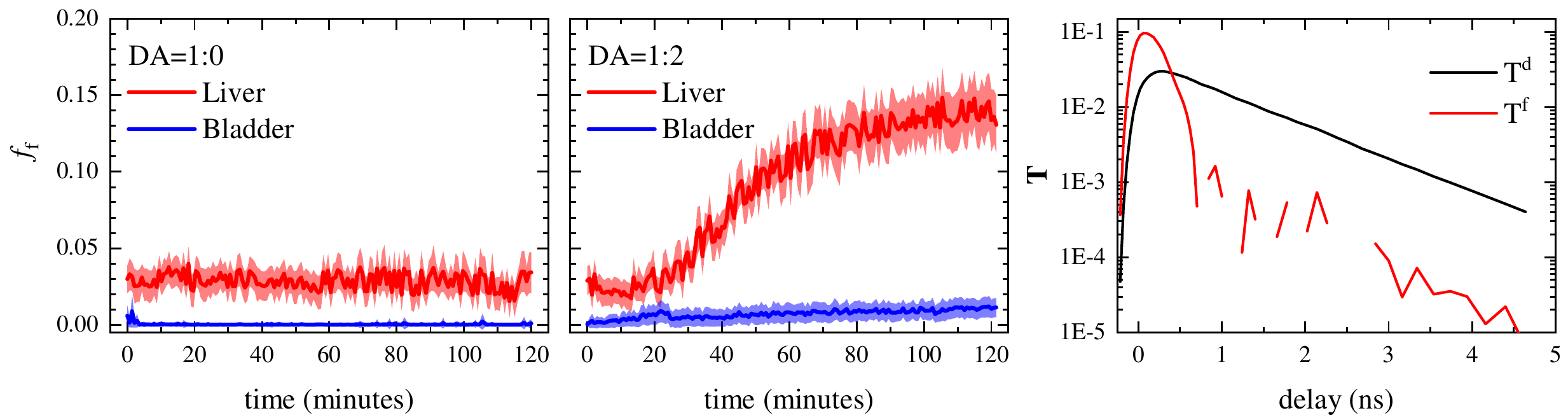}
	\caption{uFLIFRET analysis of FLIFRET data on Tf-TfR engagement {\it in vivo}. Left panel shows the evolution of the quenched donor fraction \ff\ at different acquisition time in the control sample (D:A=1:0) in the liver (red line) and bladder (blue). Solid line shows the average \ff\ calculated over the organ region of interest, while the shaded area corresponds to the standard deviation of the values. Central panel: same as left for the sample where acceptor-labelled Tf is also injected (D:A=1:2). Right panel: Dynamics of the free donor (black) used in the calculation of the DAP dynamics (red) The minimum error is reached for $\gb=6.8$\,GHz and $\sg\sim0$.}
	\label{FigSM_FRET_invivo}
\end{figure}

\Fig{FigSM_FRET_invivo} shows the results of the uFLIFRET analysis. The retrieved FRET rate distribution has a mean rate \gb\ of $\sim6.8$\,GHz with a negligible width ($\sg\sim0$). The dynamics of the two component show a mono-exponential behaviour with lifetime of $\sim1$\,ns for the unquenched donor and $\sim0.15$\,ns for the DAP. We have inferred the fraction of donor molecules undergoing FRET (\ff) by calculating the point-wise ratio $\Sf/\left(\Sd+\Sf\right)$. In \Fig{FigSM_FRET_invivo} we have displayed the average \ff\ (solid line) and standard deviation (shaded area) calculated over the liver (red) and bladder (blue) region-of interests.

For the control experiment (D:A=1:0), the bladder shows a constant $\ff\sim 0$ indicating that only free donor are accumulating there. The liver shows instead a $\ff\sim 0.025$ even in absence of acceptor, probably due to a influence of the different chemical environment and/or different autofluorescence. When the acceptor is injected (D:A=1:2), the engagement of TfR can be observed in an increase of \ff\ in the liver reaching a saturation value of $\sim~0.14$. Notably, the dynamics of \ff\ obtained by uFLIFRET is in agreement with what observed with phasor analysis\,\cite{ChenJB18S}. The smaller absolute value of the uFLIFRET result is due to the faster retrieved quenched donor dynamics (0.15\,ns)  compared to the one used (0.53\,ns) in the phasor analysis \cite{ChenJB18S}. Interestingly, this is a result of using the bladder data as donor dynamics to analyze the liver data, which also results in the reconstructed finite \ff\ in absence of the acceptor.

\begin{figure}[b]
	\includegraphics[width=0.8\textwidth]{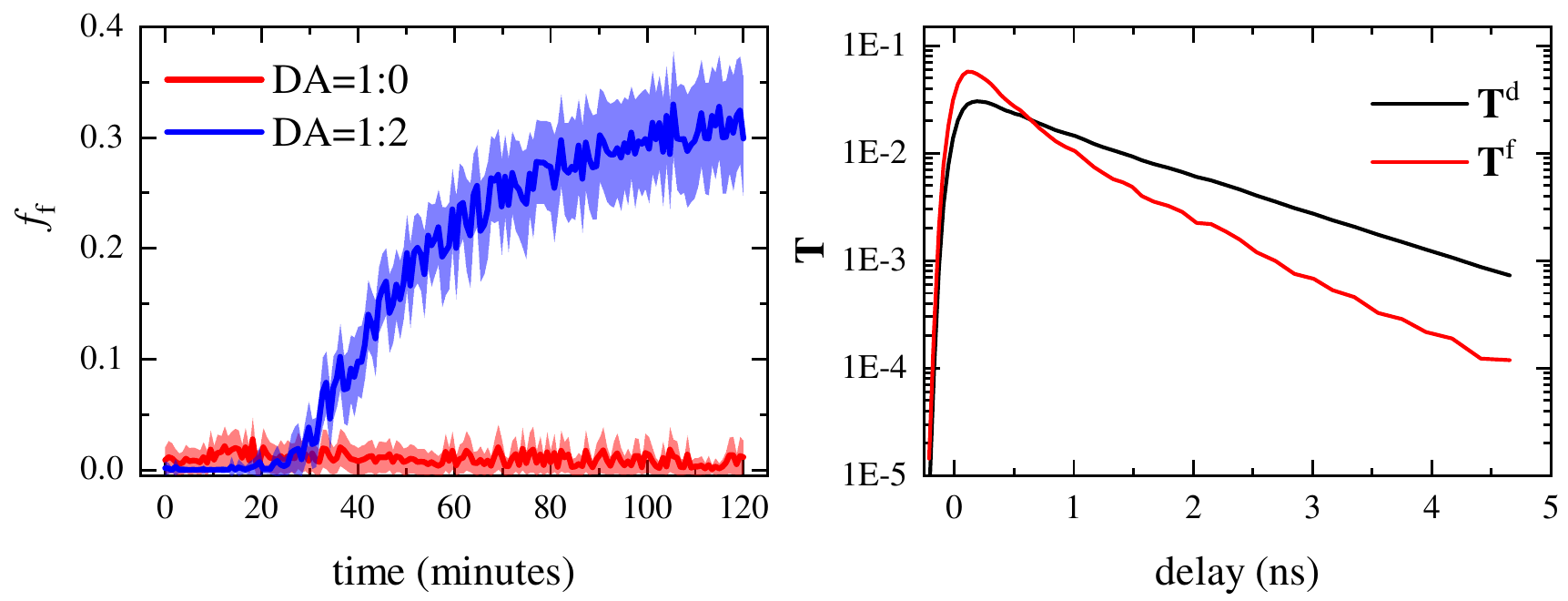}
	\caption{Same as \Fig{FigSM_FRET_invivo} but considering only the liver ROI for the analysis, including the
		estimation of the unquenched donor dynamics: \add{uFLIFRET analysis of FLIFRET data on Tf-TfR engagement {\it in vivo}, considering only the liver ROI for the analysis, including the estimation of the unquenched donor dynamics.} Left panel shows the evolution of the quenched donor fraction \ff\ at different acquisition time in the control sample (D:A=1:0, red) and the sample with acceptor (D:A=1:2, blue). Solid line shows the average \ff\ calculated over the liver region of interest, while the shaded area corresponds to the standard deviation of the values. Right panel: Dynamics of the free donor (black) used in the calculation of the DAP dynamics (red) The minimum error is reached for $\gb=2.3$\,GHz and $\sg\sim1.3$.}
	\label{FigSM_FRET_invivo_liver}
\end{figure}

We have therefore re-analyzed only the data from the liver ROI using the early times to extract the donor dynamics. The result is given in \Fig{FigSM_FRET_invivo_liver} showing zero initial \ff\ and also a slower DAP dynamics with a initial decay lifetime of $\sim0.4$\,ns, in good agreement with the one measured in reference samples (see \Fig{Fig_FRET_invivo}).

\clearpage
\section{Dependences of reconstruction errors on \Id\ and \Idap}\label{sec:err_Id_If}

\begin{figure}[b]
	\includegraphics[width=0.6\textwidth]{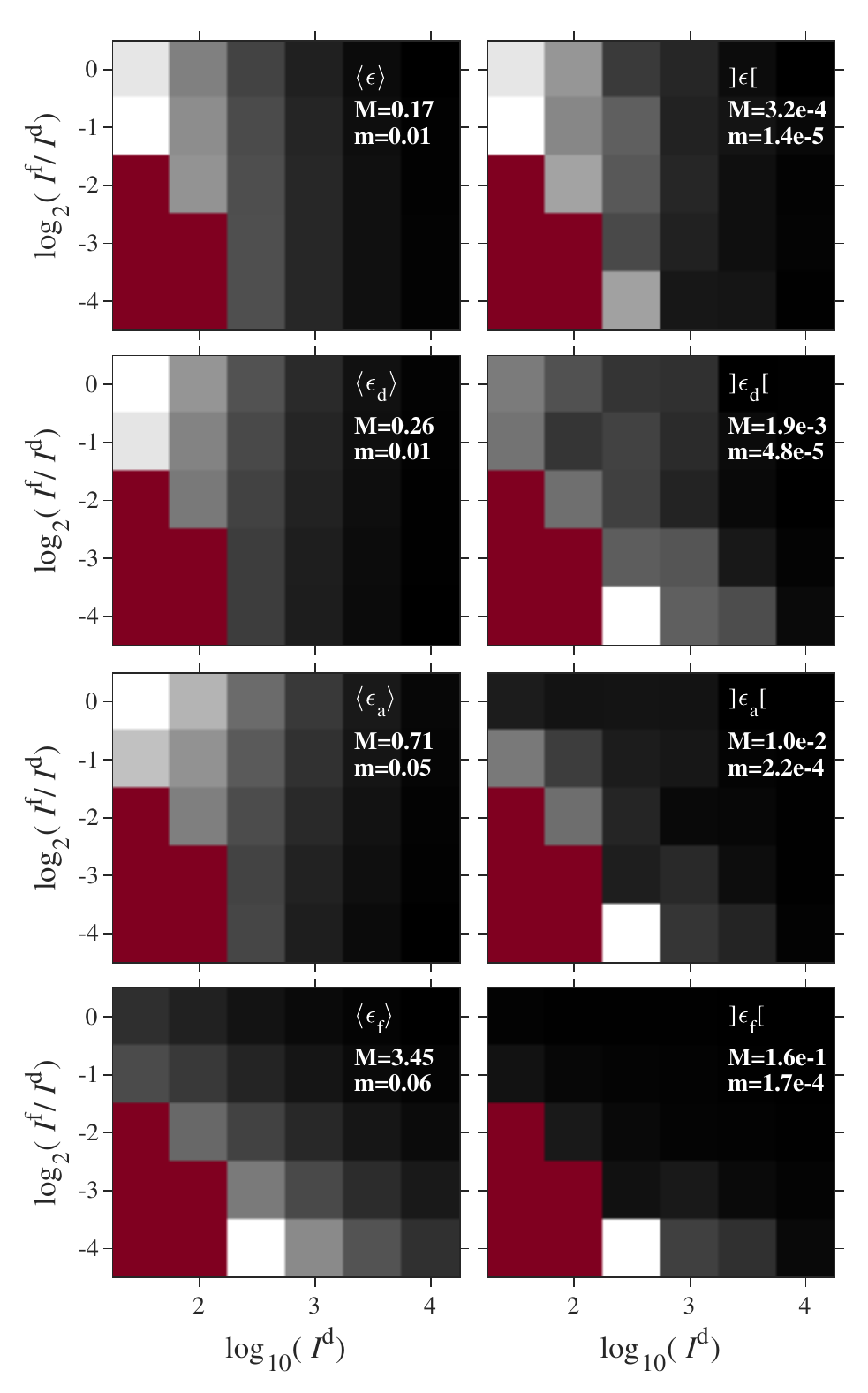}
	\caption{Relative errors of the image reconstruction versus intensities \Id\ and \Idap, for $\Ia=0.8\Id$,  $\gbs=0.5$/ns, $\sgs=0.5$, and $\Ib=2$. First row: sum data $\epsilon$; Second to fourth row: individual components, $\epsilon_{\rm d}$, $\epsilon_{\rm a}$, and $\epsilon_{\rm f}$. The left (right) column refers to the parameter mean value (standard deviation), respectively, over the data realisations. Red marks points resulting in FRET parameters outside the coarse grid.}
	\label{FigSM_FRET_RecError_g=5_k=1_Ib=2}
\end{figure}

\begin{figure}
	\includegraphics[width=0.6\textwidth]{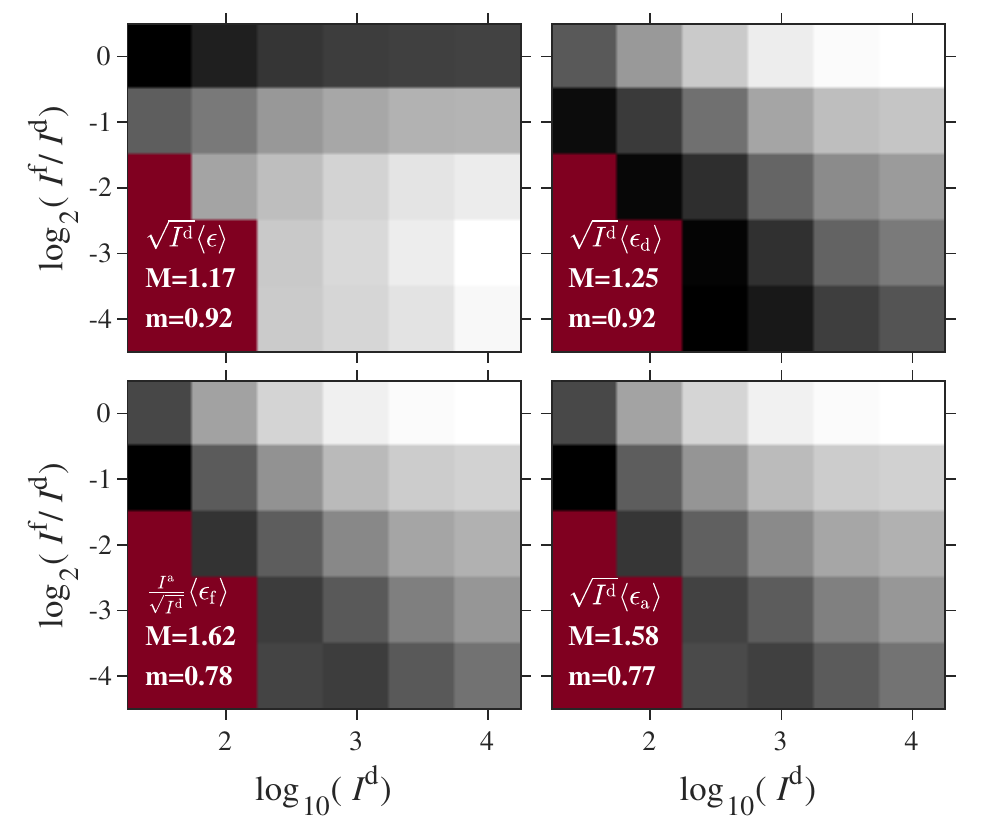}
	\caption{Average reconstruction error shown in \Fig{FigSM_FRET_RecError_g=5_k=1_Ib=2}, normalised by their approximate scaling with \Id\ and \Idap, as indicated. Grey-scale from m (black) to M (white) as given.}
	\label{FigSM_FRET_IdIfDep}
\end{figure}

The uFLIFRET reconstruction errors for $\Idap \leq \Id$, shown in \Fig{FigSM_FRET_RecError_g=5_k=1_Ib=2}, exhibit different dependencies on \Id\ and \Idap. The total error \Ert, the donor error \Erd, and the acceptor error \Era, are mostly affected by changes in \Id\ (where $\Ia=0.8\Id$), scaling as $1/\sqrt{\Id}$, which is the case since \Id\ and \Ia\ are dominating the shot noise, scaling as $\sqrt{\Id}$. On the other hand, the FRET error \Erf\ shows a combined dependence on \Id\ and \Idap, scaling approximately as $\sqrt{\Id}/{\Idap}$, so roughly the ratio between shot noise and FRET signal.  \Fig{FigSM_FRET_IdIfDep} shows that the reconstruction errors of \Fig{FigSM_FRET_RecError_g=5_k=1_Ib=2} normalised by these dependences are approximately constant.

\clearpage
\section{Dependence of reconstruction error and FRET parameter retrieval error on image size}\label{sec:err_Ns}
\begin{figure}
	\includegraphics[width=\textwidth]{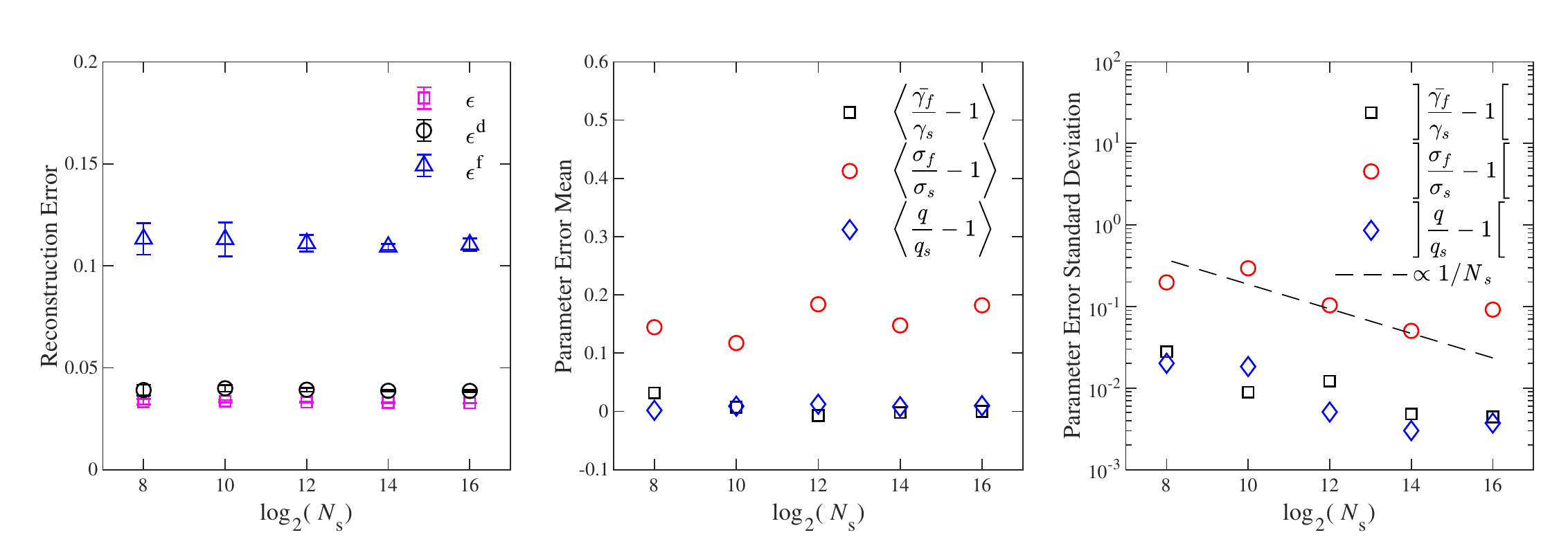}
	\caption{Retrieval errors as a function of the number \Ns\ of spatial pixels in the image.  Left: Total (\Ert, magenta squares), donor (\Erd, black circles) and FRET (\Erf, blue diamonds) reconstruction errors. The points (error bars) refer to the average (standard deviation) over 10 noise realisations. Mean (middle), and standard deviation (right) of the relative FRET parameter errors.  Data are generated using $\Id=1000$, $\Idap=250$, \gbs=0.5/ns, \sgs=0.5, \qs=1, $\Ib=2$ and $\kappa=0$.}
	\label{FigSM_FRET_SizeDependence_k=0}
\end{figure}

\begin{figure}
	\includegraphics[width=\textwidth]{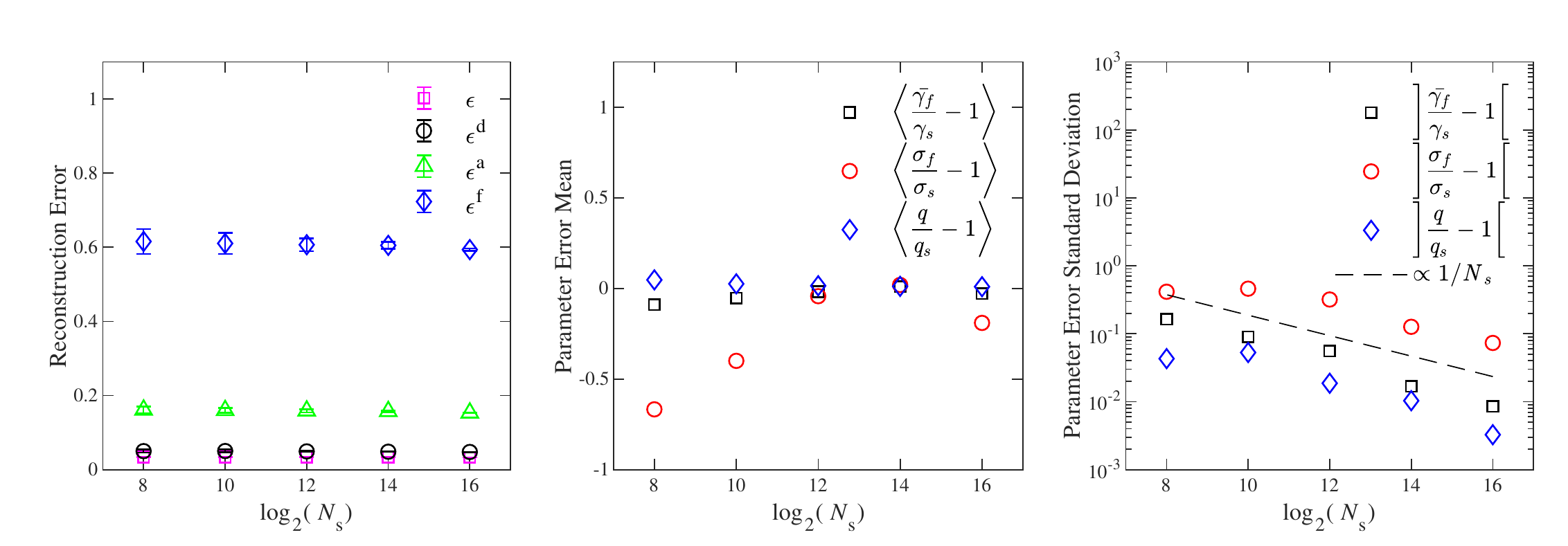}
	\caption{Same as \Fig{FigSM_FRET_SizeDependence_k=0}, but for $\kappa=1$: \add{Retrieval errors as a function of the number \Ns\ of spatial pixels in the image.  Left: Total (\Ert, magenta squares), donor (\Erd, black circles), acceptor(\Era, green triangles) and FRET (\Erf, blue diamonds) reconstruction errors. The points (error bars) refer to the average (standard deviation) over 10 noise realisations. Mean (middle), and standard deviation (right) of the relative FRET parameter errors.  Data are generated using $\Id=1000$, $\Idap=250$, \gbs=0.5/ns, \sgs=0.5, \qs=1, $\Ib=2$ and $\kappa=1$.}}
	\label{FigSM_FRET_SizeDependence_k=1}
\end{figure}

We have investigated the performance of uFLIFRET as a function of the number of spatial points \Ns\ in the image. We start with the case of no direct excitation of the acceptor ($\kappa=0$), and show the resulting reconstruction and FRET parameter errors in \Fig{FigSM_FRET_SizeDependence_k=0}. 
The reconstruction errors are limited by the shot noise in the data, which depends on the intensities \Id, \Idap\ and \Ia, and thus does not change with the number of spatial points.
The mean FRET parameter error is also hardly affected by \Ns. When a strong pure acceptor component ($\kappa=1$) is introduced in the model, the retrieved parameters have somewhat larger errors (see \Fig{FigSM_FRET_SizeDependence_k=1}). Notably though, the FRET parameter errors are changing significantly with image size. In both cases, the variation of the reconstruction error over the different noise realisations reduces with increasing \Ns, approximately with a $1/\Ns$ dependence.

\clearpage

\clearpage
\section{Extraction of FRET parameters for different FRET rate distributions, background intensities and relative acceptor excitation}\label{sec:fretpara}
In this section we investigate uFLIFRET for different average FRET rates \gbs, background intensities \Ib, and the relative excitation of the acceptor $\kappa$. In general, an  average FRET rate smaller than the donor and acceptor rates reduces the contrast of the FRET process in the overall dynamics, and thus also the reconstruction and parameter retrieval fidelity for given intensities. The background intensity \Ib\ is adversely affecting the retrieval as previously noted. For $\Ib=0$, the FRET parameters can be determined correctly over a large range of \Id\ and \Idap, while for $\Ib=20$, only data generated with large \Id\ and \Idap\ can be factorized correctly.
Reducing the relative excitation of the acceptor ($\kappa$) increases the contrast and thus the fidelity of retrieval, which we show by comparing $\kappa=1$ and $\kappa=0$. The results are shown in \Fig{FigSM_FRET_g=1_k=1_Ib=0} to \Fig{FigSM_FRET_ParamError_g=9_k=0_Ib=20} where the above discussed trends are exemplified for various values of \Id, \Idap, \Ib, $\kappa$, \gbs\ as given in the respective captions. All simulations used \sgs=0.5 and \qs=1. An overview of the figures and the parameters used is given in \Tab{tab:simfigs}.

\begin{table}[h]
	\begin{tabular*}{\textwidth}{@{\extracolsep{\fill}}ccccc}
		\hline
		Images & Errors & $\kappa$ & $\gbs$ (ns$^{-1}$) & \Ib \\
		\hline
		\Fig{FigSM_FRET_g=1_k=1_Ib=0} & \Fig{FigSM_FRET_ParamError_g=1_k=1_Ib=0} & $1$ & $0.1$	& $0$ \\
		\Fig{FigSM_FRET_g=1_k=1_Ib=2} & \Fig{FigSM_FRET_ParamError_g=1_k=1_Ib=2} & $1$ & $0.1$	& $2$ \\
		\Fig{FigSM_FRET_g=1_k=1_Ib=20} & \Fig{FigSM_FRET_ParamError_g=1_k=1_Ib=20} & $1$ & $0.1$	& $20$ \\
		\Fig{FigSM_FRET_g=5_k=1_Ib=0} & \Fig{FigSM_FRET_ParamError_g=5_k=1_Ib=0} & $1$ & $0.5$	& $0$ \\
		\Fig{Fig_FRET_g=5_k=1_Ib=2} & \Fig{FigSM_FRET_ParamError_g=5_k=1_Ib=2} & $1$ & $0.5$	& $2$ \\
		\Fig{FigSM_FRET_g=5_k=1_Ib=20} & \Fig{FigSM_FRET_ParamError_g=5_k=1_Ib=20} & $1$ & $0.5$	& $20$ \\
		\Fig{FigSM_FRET_g=9_k=1_Ib=0} & \Fig{FigSM_FRET_ParamError_g=9_k=1_Ib=0} & $1$ & $0.9$	& $0$ \\
		\Fig{FigSM_FRET_g=9_k=1_Ib=2} & \Fig{FigSM_FRET_ParamError_g=9_k=1_Ib=2} & $1$ & $0.9$	& $2$ \\
		\Fig{FigSM_FRET_g=9_k=1_Ib=20} & \Fig{FigSM_FRET_ParamError_g=9_k=1_Ib=20} & $1$ & $0.9$	& $20$ \\
		\Fig{FigSM_FRET_g=1_k=0_Ib=0} & \Fig{FigSM_FRET_ParamError_g=1_k=0_Ib=0} & $0$ & $0.1$	& $0$ \\
		\Fig{FigSM_FRET_g=1_k=0_Ib=2} & \Fig{FigSM_FRET_ParamError_g=1_k=0_Ib=2} & $0$ & $0.1$	& $2$ \\
		\Fig{FigSM_FRET_g=1_k=0_Ib=20} & \Fig{FigSM_FRET_ParamError_g=1_k=0_Ib=20} & $0$ & $0.1$	& $20$ \\
		\Fig{FigSM_FRET_g=5_k=0_Ib=0} & \Fig{FigSM_FRET_ParamError_g=5_k=0_Ib=0} & $0$ & $0.5$	& $0$ \\
		\Fig{FigSM_FRET_g=5_k=0_Ib=2} & \Fig{FigSM_FRET_ParamError_g=5_k=0_Ib=2} & $0$ & $0.5$	& $2$ \\
		\Fig{FigSM_FRET_g=5_k=0_Ib=20} & \Fig{FigSM_FRET_ParamError_g=5_k=0_Ib=20} & $0$ & $0.5$	& $20$ \\
		\Fig{FigSM_FRET_g=9_k=0_Ib=0} & \Fig{FigSM_FRET_ParamError_g=9_k=0_Ib=0} & $0$ & $0.9$	& $0$ \\
		\Fig{FigSM_FRET_g=9_k=0_Ib=2} & \Fig{FigSM_FRET_ParamError_g=9_k=0_Ib=2} & $0$ & $0.9$	& $2$ \\
		\Fig{FigSM_FRET_g=9_k=0_Ib=20} & \Fig{FigSM_FRET_ParamError_g=9_k=0_Ib=20} & $0$ & $0.9$	& $20$ \\
		\hline
	\end{tabular*}
	\caption{Overview over simulation results shown.}
\label{tab:simfigs}
\end{table}

\begin{figure}
	\includegraphics[width=0.9\textwidth]{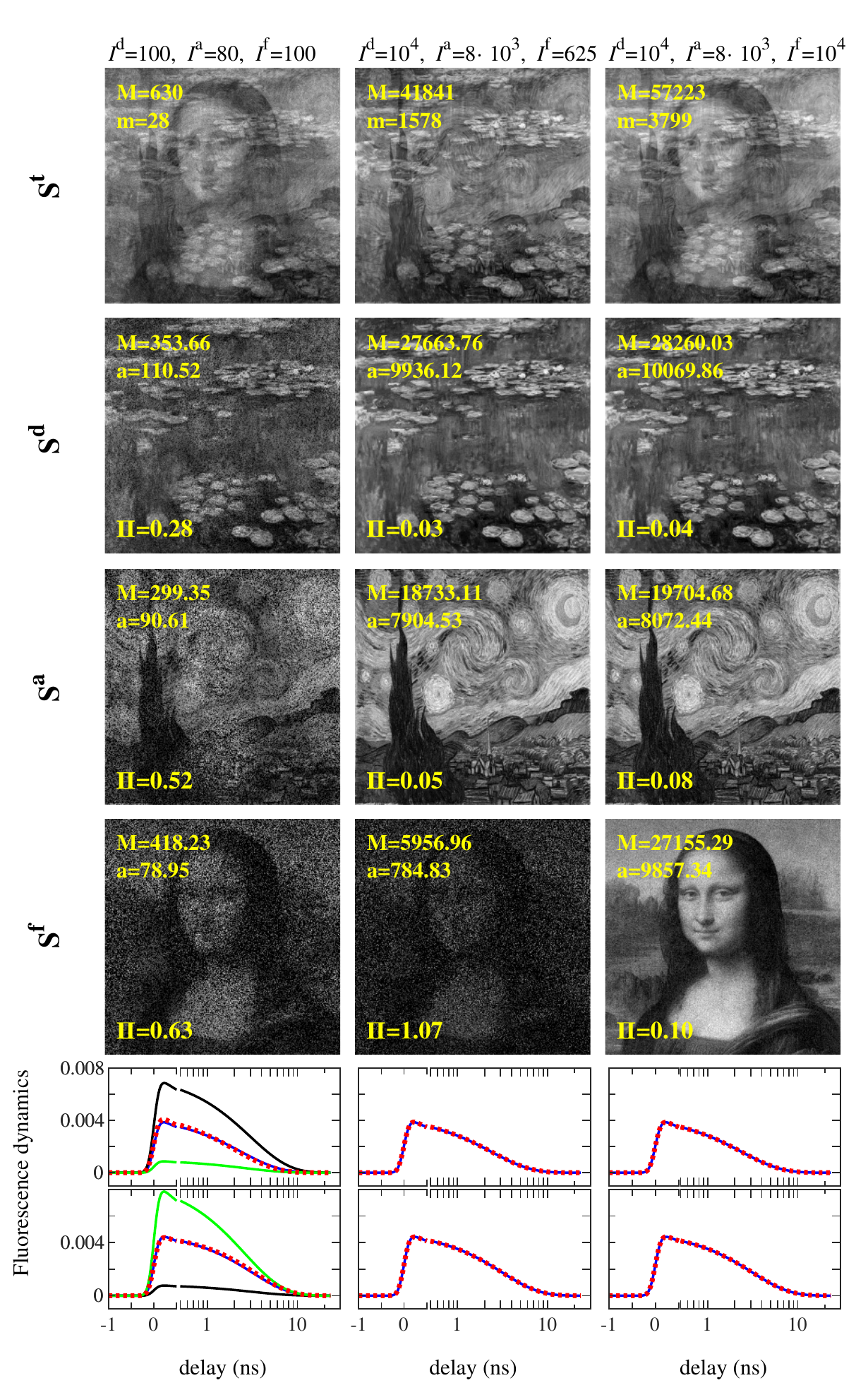}
\end{figure}
\begin{figure}[t]
	\caption{Same as \Fig{Fig_FRET_g=5_k=1_Ib=2} but for $\gbs=0.1/$\,ns, $\Ib=0$ and $\kappa=1$: \add{Results of uFLIFRET for synthetic data generated using $\gbs=0.1$/ns, $\sgs=0.5$, $\qs=1$, $\Ib=0$ and $\kappa=1$. The three columns refer to different intensities as indicated. Top row: the time summed data $\St=\Nt\Sav$, on a linear grey scale from a minimum $m$ (black) to a maximum $M$ (white) as indicated. The second to fourth rows show the retrieved spatial distributions of the donor \Sd, acceptor \Sa, and FRET \Sf. Here $m=0$ and $a$ is the average pixel value over the image. The bottom panels show the synthetic original dynamics of donor \bTd (black), acceptor \bTa (green), and DAPs undergoing FRET \bbT (blue), with the retrieved FRET dynamics given as red dashed lines. The signal acquired at the donor (acceptor) detector are given in the top (bottom) panel, respectively. The dynamics are normalized to have a sum of unity over the 2000 temporal points of both detectors.}}\label{FigSM_FRET_g=1_k=1_Ib=0}
\end{figure}

\begin{figure}
	\includegraphics[width=0.6\textwidth]{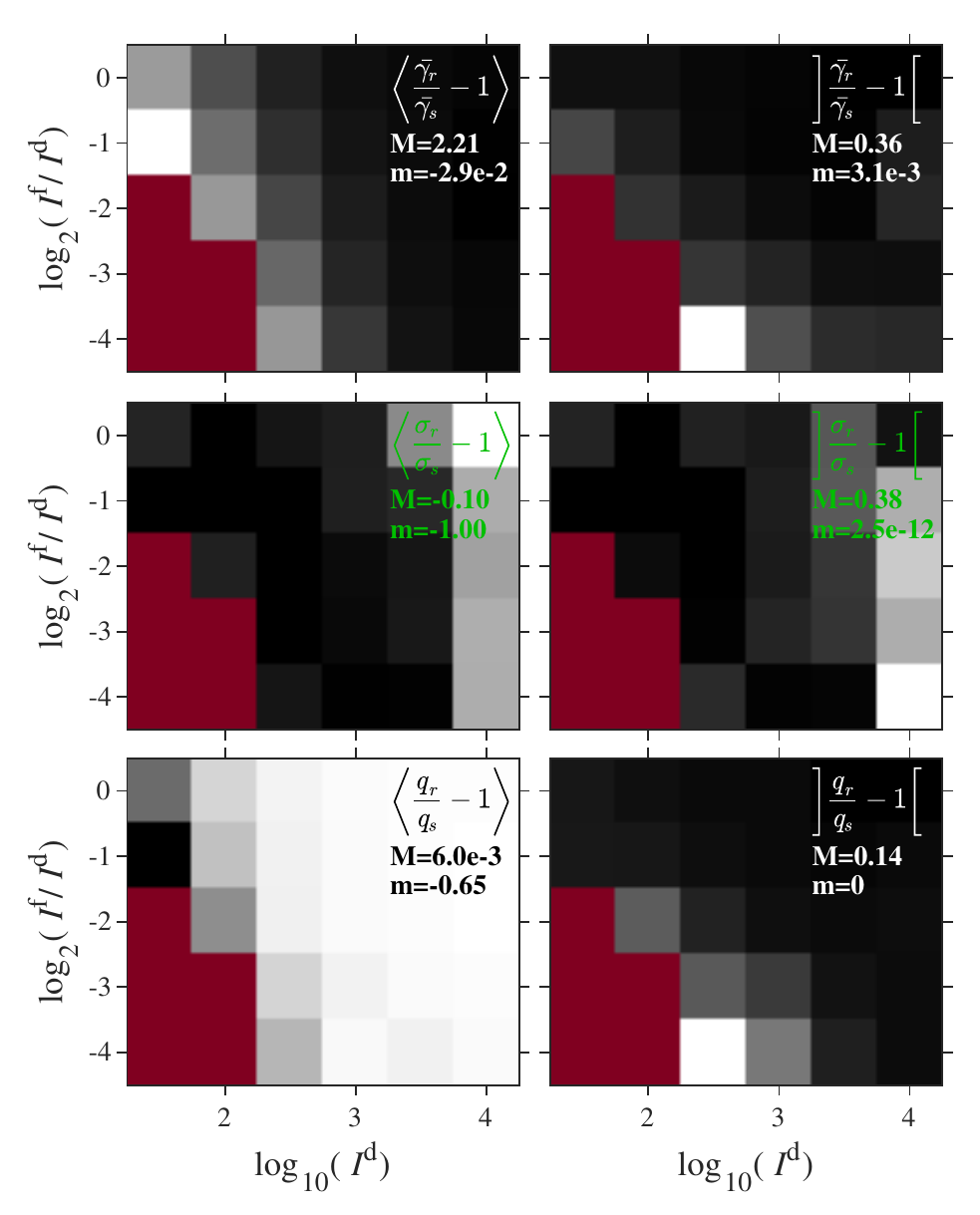}
	\caption{Relative error of retrieved FRET parameters, versus \Id\ and \Idap, using the same simulation parameters as in \Fig{FigSM_FRET_g=1_k=1_Ib=0} \add{($\gbs=0.1$/ns, $\sgs=0.5$, $\qs=1$, $\Ib=0$ and $\kappa=1$}). The mean values over data realisations are shown on the left ($\langle\gbf/\gbs-1\rangle$, $\langle\sgf/\sgs-1\rangle$ and $\langle\qf/\qs-1\rangle$ from top to bottom), while the standard deviations are shown on the right.}\label{FigSM_FRET_ParamError_g=1_k=1_Ib=0}
\end{figure}

\begin{figure}
	\includegraphics[width=0.9\textwidth]{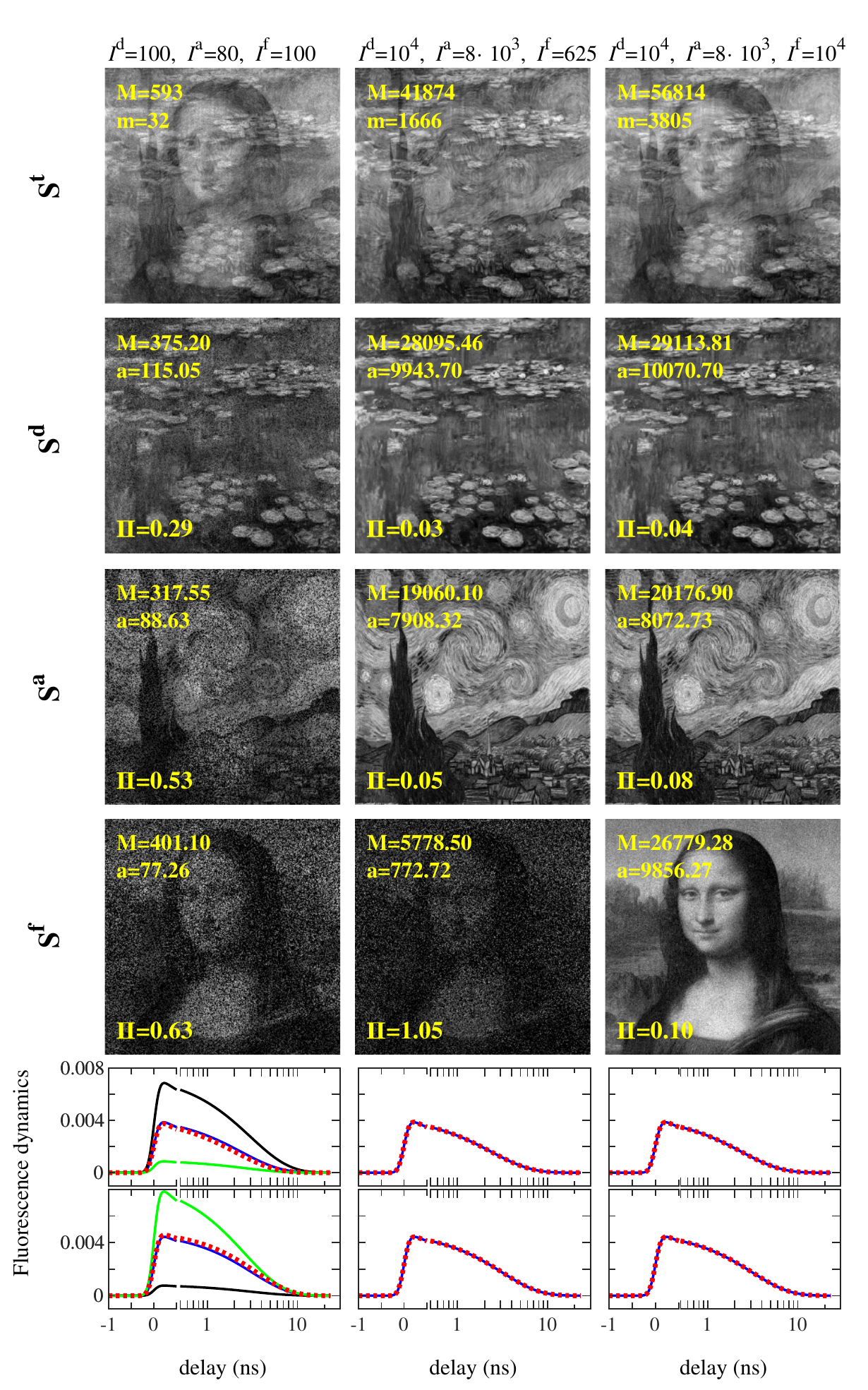}
\end{figure}
\begin{figure}
	\caption{Same as \Fig{Fig_FRET_g=5_k=1_Ib=2} but for $\gbs=0.1/$\,ns, $\Ib=2$ and $\kappa=1$: \add{Results of uFLIFRET for synthetic data generated using $\gbs=0.1$/ns, $\sgs=0.5$, $\qs=1$, $\Ib=0$ and $\kappa=1$. The three columns refer to different intensities as indicated. Top row: the time summed data $\St=\Nt\Sav$, on a linear grey scale from a minimum $m$ (black) to a maximum $M$ (white) as indicated. The second to fourth rows show the retrieved spatial distributions of the donor \Sd, acceptor \Sa, and FRET \Sf. Here $m=0$ and $a$ is the average pixel value over the image. The bottom panels show the synthetic original dynamics of donor \bTd (black), acceptor \bTa (green), and DAPs undergoing FRET \bbT (blue), with the retrieved FRET dynamics given as red dashed lines. The signal acquired at the donor (acceptor) detector are given in the top (bottom) panel, respectively. The dynamics are normalized to have a sum of unity over the 2000 temporal points of both detectors.}\label{FigSM_FRET_g=1_k=1_Ib=2}}
\end{figure}

\begin{figure}
	\includegraphics[width=0.6\textwidth]{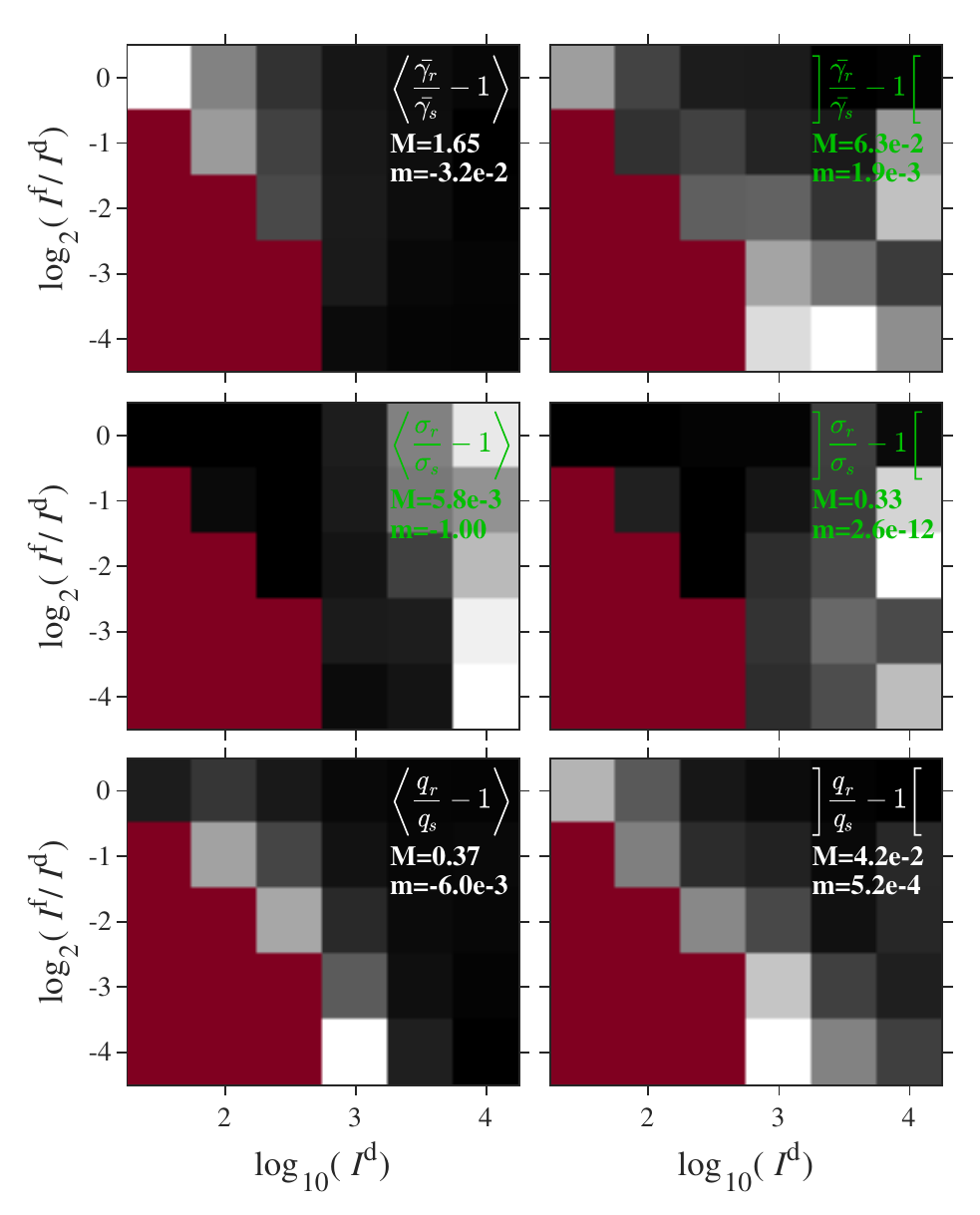}
	\caption{Same as \Fig{FigSM_FRET_ParamError_g=1_k=1_Ib=0} but for $\gbs=0.1/$\,ns, $\Ib=2$ and $\kappa=1$: \add{Relative error of retrieved FRET parameters, versus \Id\ and \Idap, for the analysis of synthetic data generated with $\gbs=0.1$/ns, $\sgs=0.5$, $\qs=1$, $\Ib=2$ and $\kappa=1$. The mean values over data realisations are shown on the left ($\langle\gbf/\gbs-1\rangle$, $\langle\sgf/\sgs-1\rangle$ and $\langle\qf/\qs-1\rangle$ from top to bottom), while the standard deviations are shown on the right.}}\label{FigSM_FRET_ParamError_g=1_k=1_Ib=2}
\end{figure}

\begin{figure}
	\includegraphics[width=0.38\textwidth]{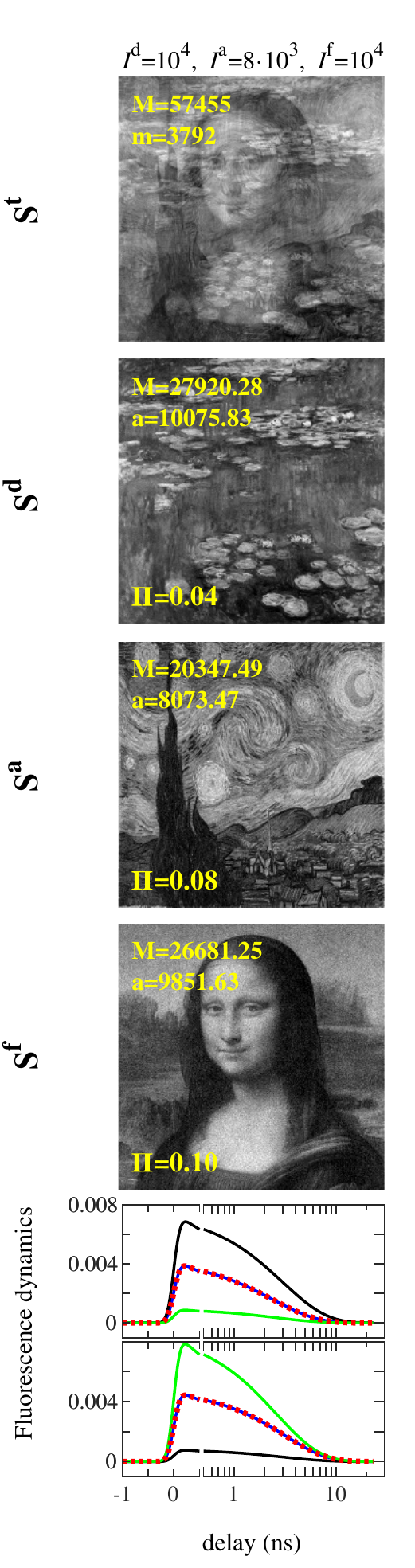}
\end{figure}
\begin{figure}
	\caption{Same as \Fig{Fig_FRET_g=5_k=1_Ib=2} but for $\gbs=0.1/$\,ns, $\Ib=20$ and $\kappa=1$: \add{Results of uFLIFRET for synthetic data generated using $\gbs=0.1$/ns, $\sgs=0.5$, $\qs=1$, $\Ib=20$ and $\kappa=1$. Top row: the time summed data $\St=\Nt\Sav$, on a linear grey scale from a minimum $m$ (black) to a maximum $M$ (white) as indicated. The second to fourth rows show the retrieved spatial distributions of the donor \Sd, acceptor \Sa, and FRET \Sf. Here $m=0$ and $a$ is the average pixel value over the image. The bottom panels show the synthetic original dynamics of donor \bTd (black), acceptor \bTa (green), and DAPs undergoing FRET \bbT (blue), with the retrieved FRET dynamics given as red dashed lines. The signal acquired at the donor (acceptor) detector are given in the top (bottom) panel, respectively. The dynamics are normalized to have a sum of unity over the 2000 temporal points of both detectors.}\label{FigSM_FRET_g=1_k=1_Ib=20}}
\end{figure}
\begin{figure}
	\includegraphics[width=0.6\textwidth]{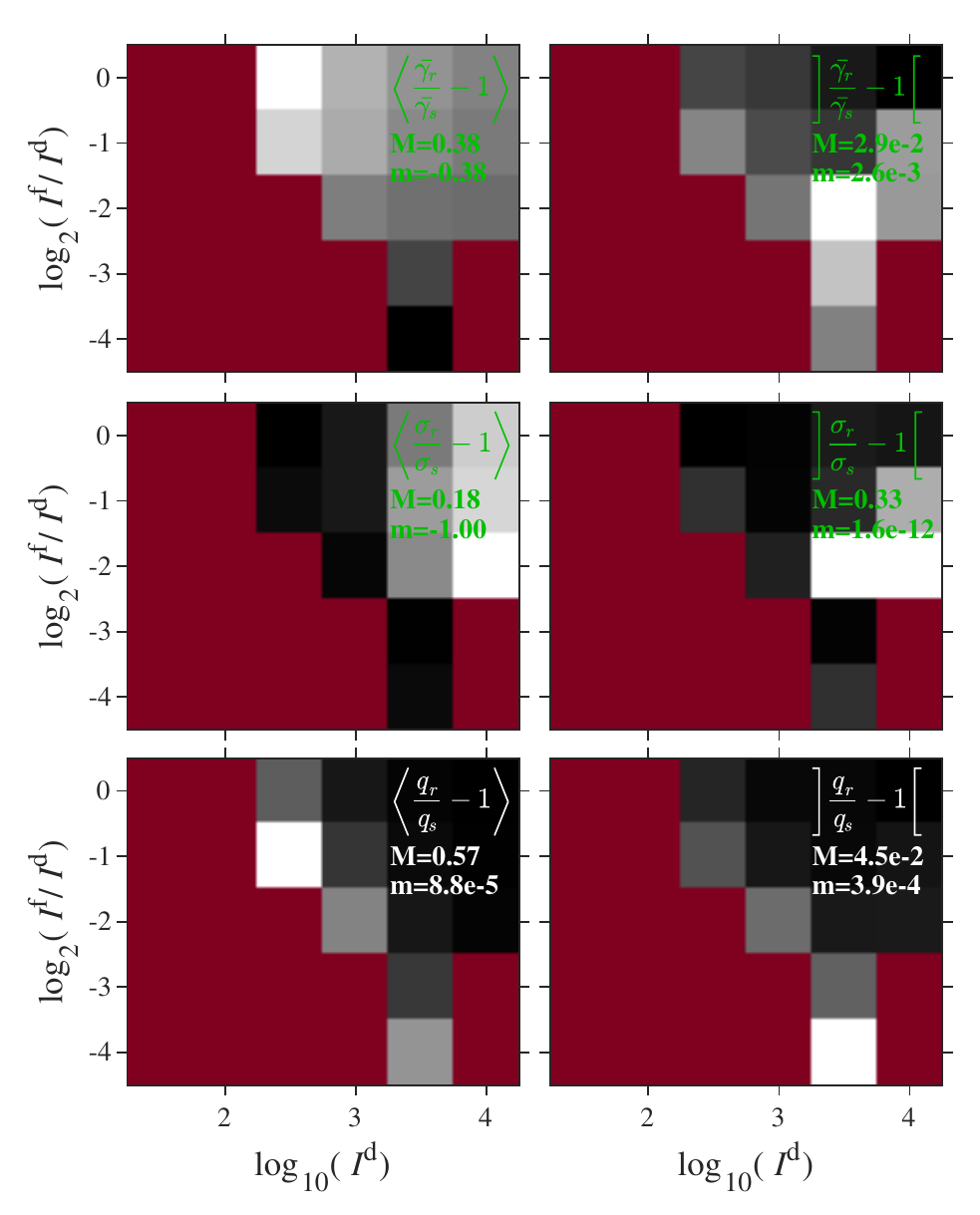}
	\caption{Same as \Fig{FigSM_FRET_ParamError_g=1_k=1_Ib=0} but for $\gbs=0.1/$\,ns, $\Ib=20$ and $\kappa=1$: \add{Relative error of retrieved FRET parameters, versus \Id\ and \Idap, for the analysis of synthetic data generated with $\gbs=0.1$/ns, $\sgs=0.5$, $\qs=1$, $\Ib=20$ and $\kappa=1$. The mean values over data realisations are shown on the left ($\langle\gbf/\gbs-1\rangle$, $\langle\sgf/\sgs-1\rangle$ and $\langle\qf/\qs-1\rangle$ from top to bottom), while the standard deviations are shown on the right.}}\label{FigSM_FRET_ParamError_g=1_k=1_Ib=20}
\end{figure}

\begin{figure}
	\includegraphics[width=0.9\textwidth]{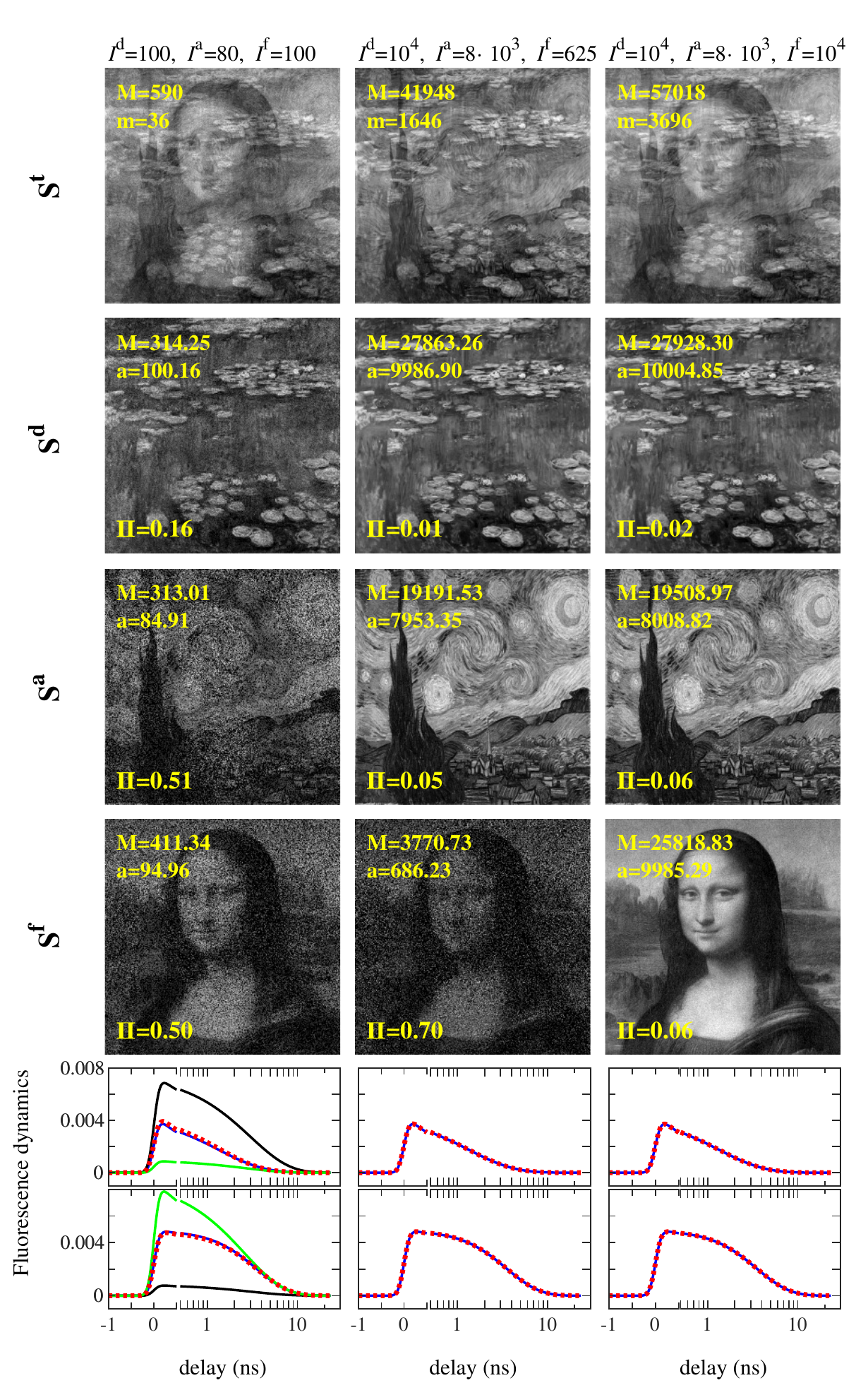}
\end{figure}
\begin{figure}
	\caption{Same as \Fig{Fig_FRET_g=5_k=1_Ib=2}) but for $\gbs=0.5/$\,ns, $\Ib=0$ and $\kappa=1$: \add{Results of uFLIFRET for synthetic data generated using $\gbs=0.5$/ns, $\sgs=0.5$, $\qs=1$, $\Ib=0$ and $\kappa=1$. The three columns refer to different intensities as indicated. Top row: the time summed data $\St=\Nt\Sav$, on a linear grey scale from a minimum $m$ (black) to a maximum $M$ (white) as indicated. The second to fourth rows show the retrieved spatial distributions of the donor \Sd, acceptor \Sa, and FRET \Sf. Here $m=0$ and $a$ is the average pixel value over the image. The bottom panels show the synthetic original dynamics of donor \bTd (black), acceptor \bTa (green), and DAPs undergoing FRET \bbT (blue), with the retrieved FRET dynamics given as red dashed lines. The signal acquired at the donor (acceptor) detector are given in the top (bottom) panel, respectively. The dynamics are normalized to have a sum of unity over the 2000 temporal points of both detectors.}\label{FigSM_FRET_g=5_k=1_Ib=0}}
\end{figure}
\begin{figure}
	\includegraphics[width=0.6\textwidth]{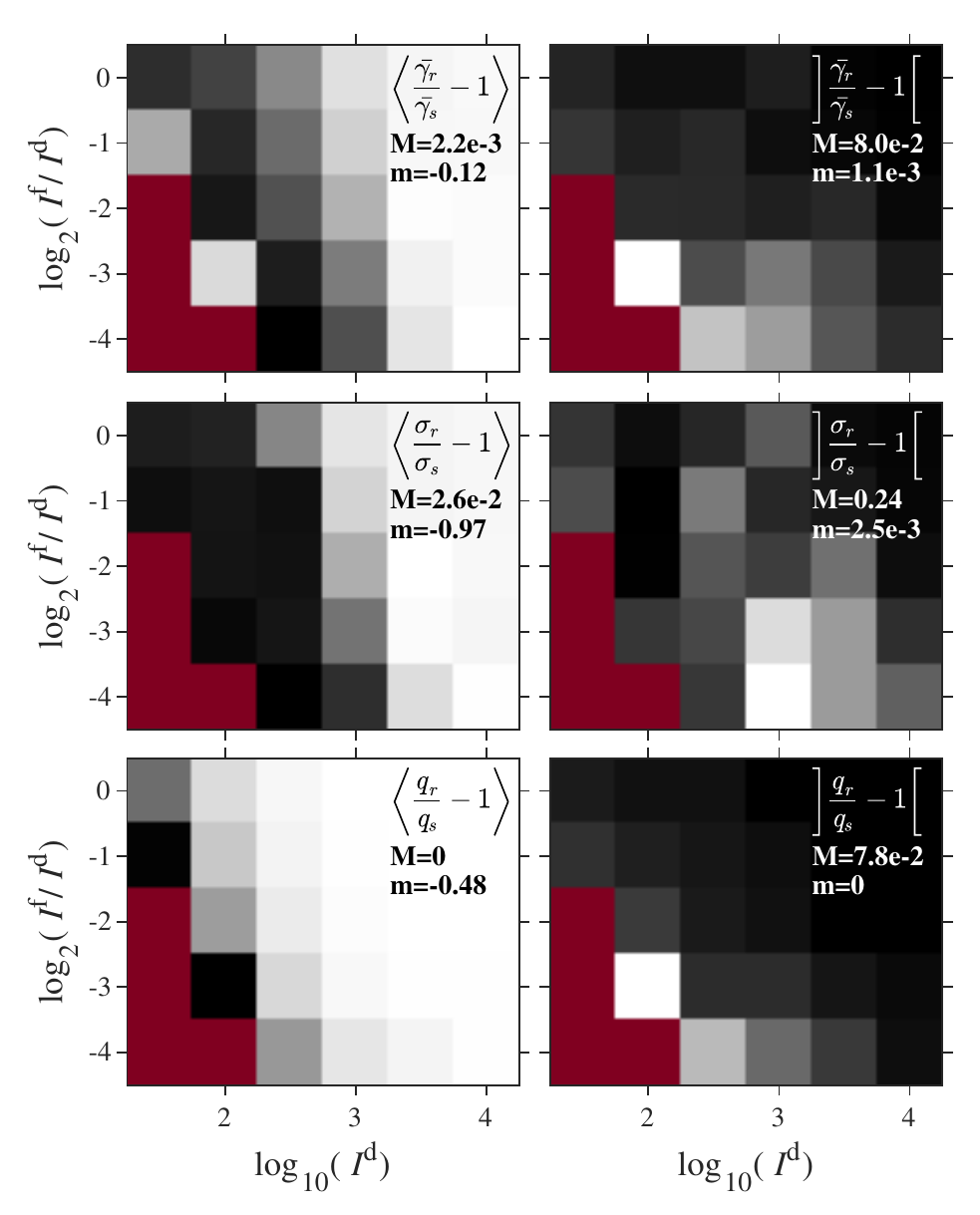}
	\caption{Same as \Fig{FigSM_FRET_ParamError_g=1_k=1_Ib=0} but for $\gbs=0.5/$\,ns, $\Ib=0$ and $\kappa=1$: \add{Relative error of retrieved FRET parameters, versus \Id\ and \Idap, for the analysis of synthetic data generated with $\gbs=0.5$/ns, $\sgs=0.5$, $\qs=1$, $\Ib=0$ and $\kappa=1$. The mean values over data realisations are shown on the left ($\langle\gbf/\gbs-1\rangle$, $\langle\sgf/\sgs-1\rangle$ and $\langle\qf/\qs-1\rangle$ from top to bottom), while the standard deviations are shown on the right.}}\label{FigSM_FRET_ParamError_g=5_k=1_Ib=0}
\end{figure}

\begin{figure}
	\includegraphics[width=0.6\textwidth]{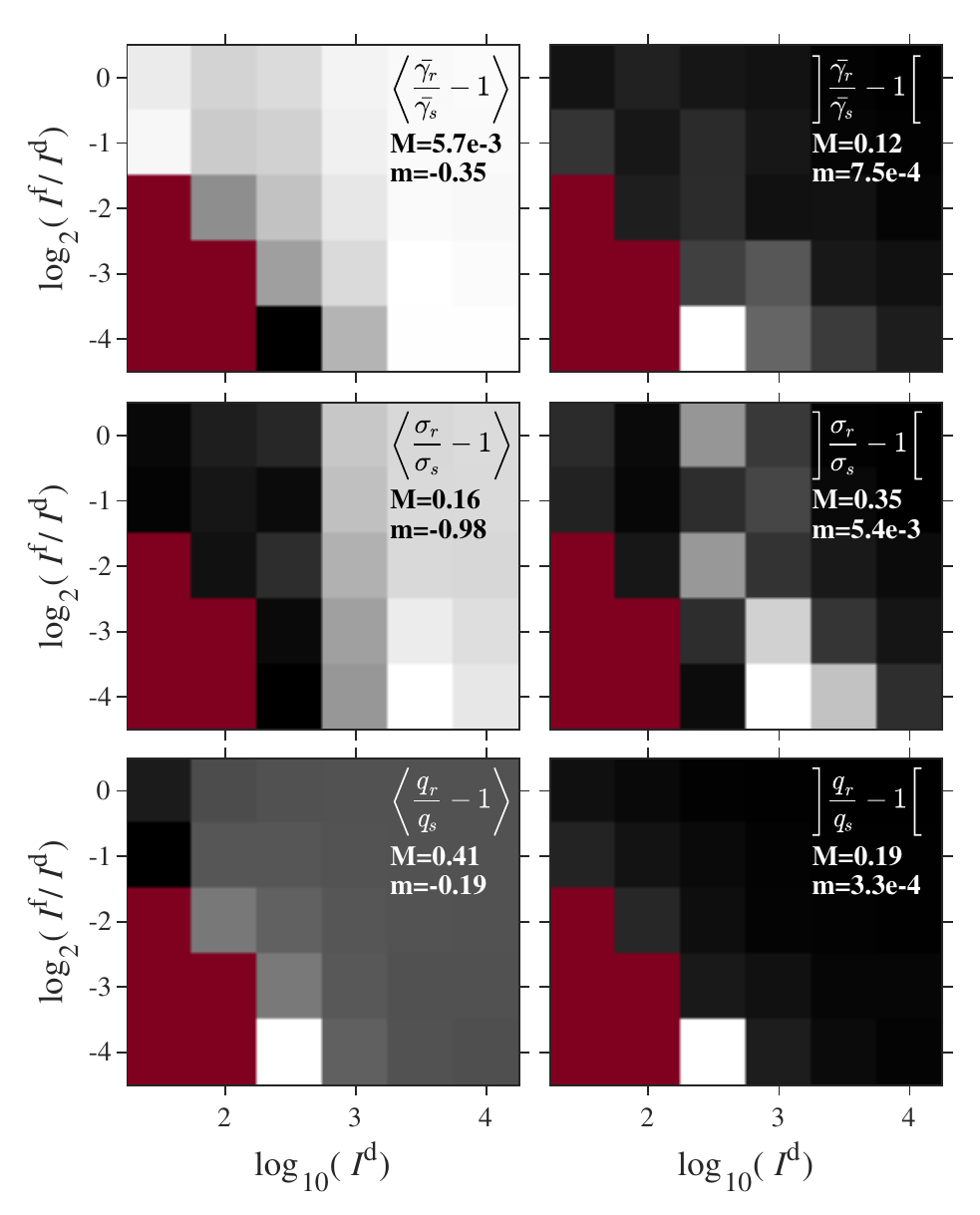}
	\caption{Same as \Fig{FigSM_FRET_ParamError_g=1_k=1_Ib=0} but for $\gbs=0.5/$\,ns, $\Ib=2$ and $\kappa=1$: \add{Relative error of retrieved FRET parameters, versus \Id\ and \Idap, for the analysis of synthetic data generated with $\gbs=0.5$/ns, $\sgs=0.5$, $\qs=1$, $\Ib=2$ and $\kappa=1$. The mean values over data realisations are shown on the left ($\langle\gbf/\gbs-1\rangle$, $\langle\sgf/\sgs-1\rangle$ and $\langle\qf/\qs-1\rangle$ from top to bottom), while the standard deviations are shown on the right.}}
	\label{FigSM_FRET_ParamError_g=5_k=1_Ib=2}
\end{figure}

\begin{figure}
	\includegraphics[width=0.9\textwidth]{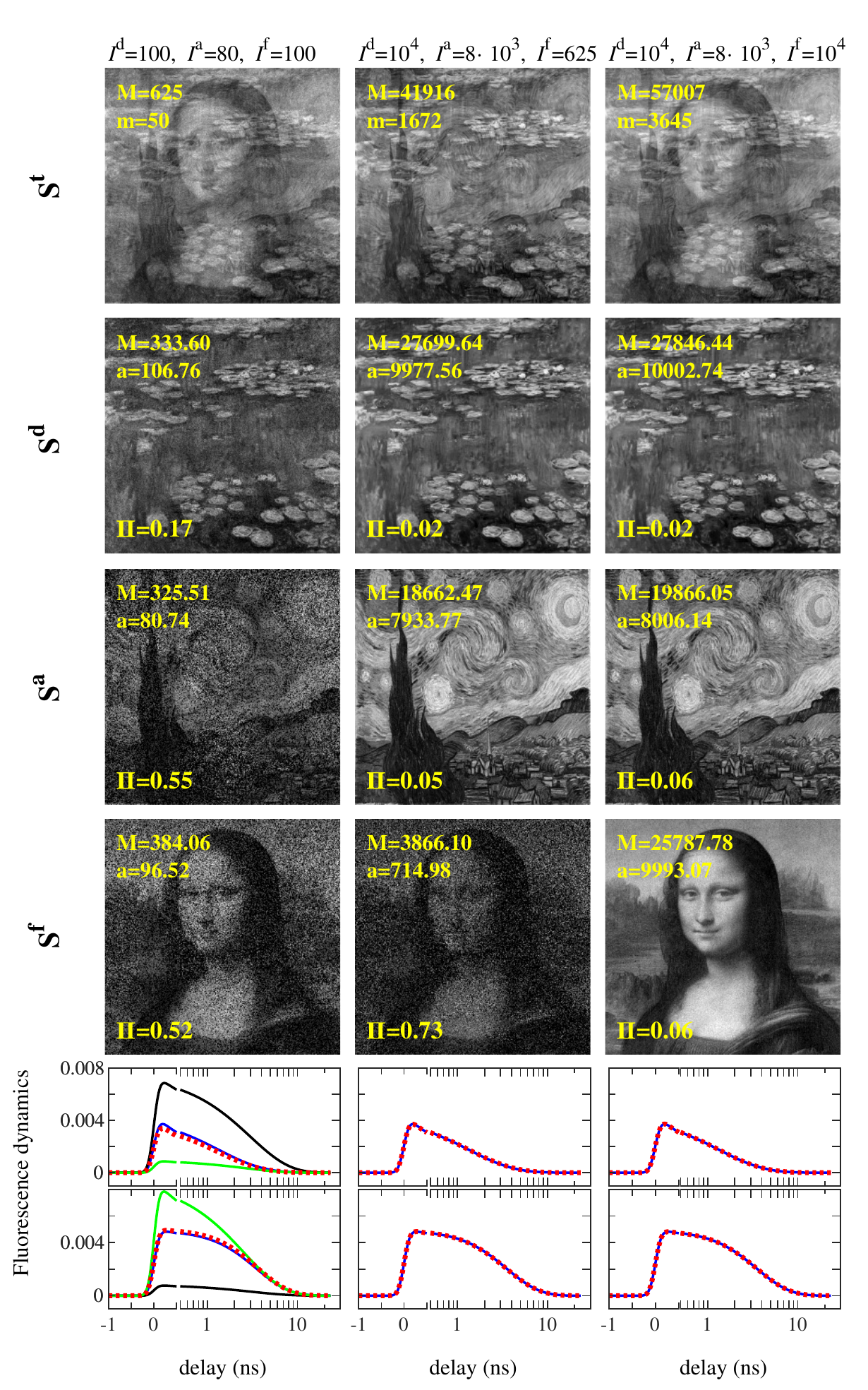}
\end{figure}
\begin{figure}
	\caption{Same as \Fig{Fig_FRET_g=5_k=1_Ib=2} but for $\gbs=0.5/$\,ns, $\Ib=20$ and $\kappa=1$: \add{Results of uFLIFRET for synthetic data generated using $\gbs=0.5$/ns, $\sgs=0.5$, $\qs=1$, $\Ib=20$ and $\kappa=1$. The three columns refer to different intensities as indicated. Top row: the time summed data $\St=\Nt\Sav$, on a linear grey scale from a minimum $m$ (black) to a maximum $M$ (white) as indicated. The second to fourth rows show the retrieved spatial distributions of the donor \Sd, acceptor \Sa, and FRET \Sf. Here $m=0$ and $a$ is the average pixel value over the image. The bottom panels show the synthetic original dynamics of donor \bTd (black), acceptor \bTa (green), and DAPs undergoing FRET \bbT (blue), with the retrieved FRET dynamics given as red dashed lines. The signal acquired at the donor (acceptor) detector are given in the top (bottom) panel, respectively. The dynamics are normalized to have a sum of unity over the 2000 temporal points of both detectors.}\label{FigSM_FRET_g=5_k=1_Ib=20}}
\end{figure}
\begin{figure}
	\includegraphics[width=0.6\textwidth]{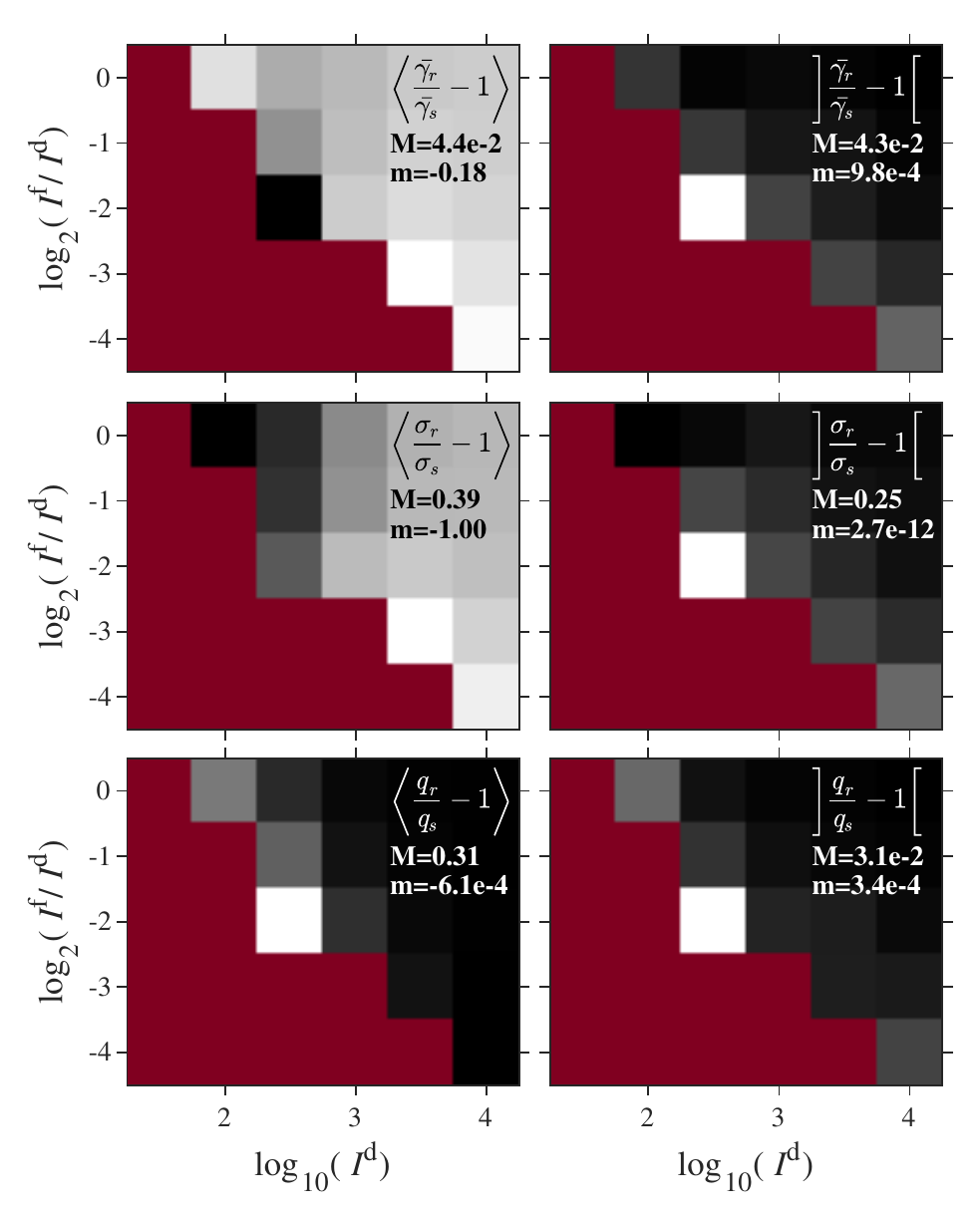}
	\caption{Same as \Fig{FigSM_FRET_ParamError_g=1_k=1_Ib=0} but for $\gbs=0.5/$\,ns, $\Ib=20$ and $\kappa=1$: \add{Relative error of retrieved FRET parameters, versus \Id\ and \Idap, for the analysis of synthetic data generated with $\gbs=0.1$/ns, $\sgs=0.5$, $\qs=1$, $\Ib=20$ and $\kappa=1$. The mean values over data realisations are shown on the left ($\langle\gbf/\gbs-1\rangle$, $\langle\sgf/\sgs-1\rangle$ and $\langle\qf/\qs-1\rangle$ from top to bottom), while the standard deviations are shown on the right.}}\label{FigSM_FRET_ParamError_g=5_k=1_Ib=20}
\end{figure}

\begin{figure}
	\includegraphics[width=0.9\textwidth]{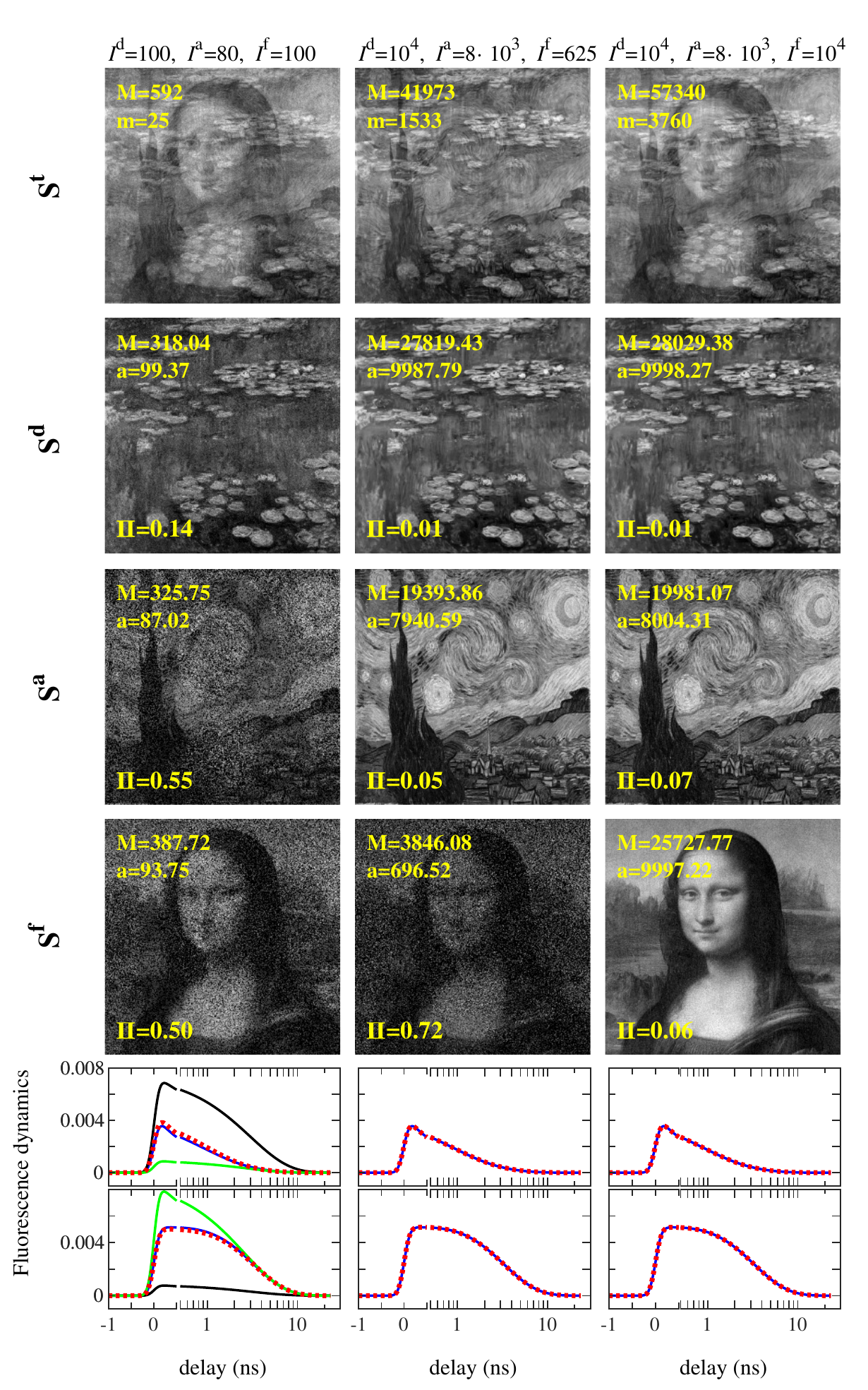}
\end{figure}
\begin{figure}
	\caption{Same as \Fig{Fig_FRET_g=5_k=1_Ib=2} but for $\gbs=0.9/$\,ns, $\Ib=0$ and $\kappa=1$: \add{Results of uFLIFRET for synthetic data generated using $\gbs=0.9$/ns, $\sgs=0.5$, $\qs=1$, $\Ib=0$ and $\kappa=1$. The three columns refer to different intensities as indicated. Top row: the time summed data $\St=\Nt\Sav$, on a linear grey scale from a minimum $m$ (black) to a maximum $M$ (white) as indicated. The second to fourth rows show the retrieved spatial distributions of the donor \Sd, acceptor \Sa, and FRET \Sf. Here $m=0$ and $a$ is the average pixel value over the image. The bottom panels show the synthetic original dynamics of donor \bTd (black), acceptor \bTa (green), and DAPs undergoing FRET \bbT (blue), with the retrieved FRET dynamics given as red dashed lines. The signal acquired at the donor (acceptor) detector are given in the top (bottom) panel, respectively. The dynamics are normalized to have a sum of unity over the 2000 temporal points of both detectors.}\label{FigSM_FRET_g=9_k=1_Ib=0}}
\end{figure}
\begin{figure}
	\includegraphics[width=0.6\textwidth]{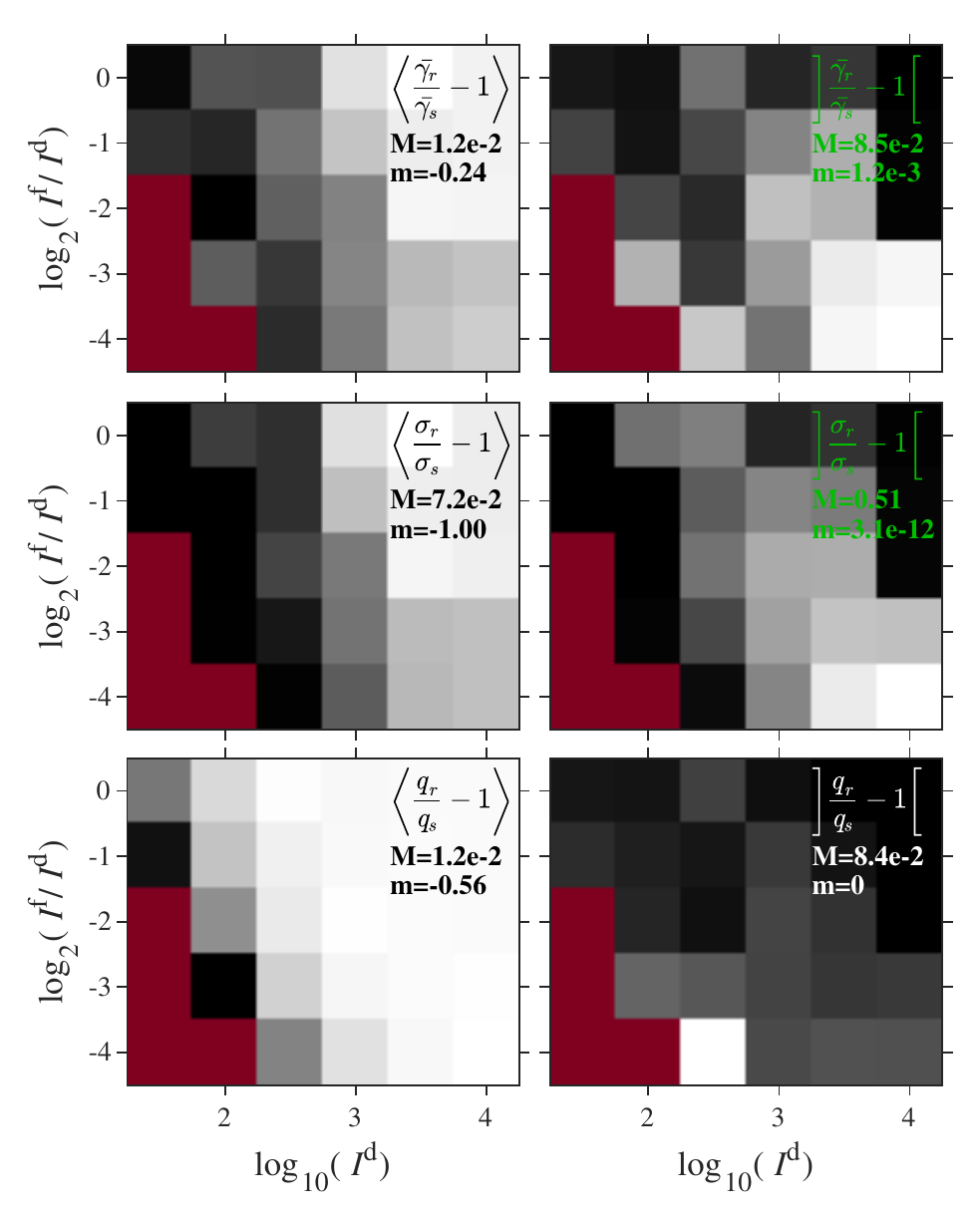}
	\caption{Same as \Fig{FigSM_FRET_ParamError_g=1_k=1_Ib=0} but for $\gbs=0.9/$\,ns, $\Ib=0$ and $\kappa=1$: \add{Relative error of retrieved FRET parameters, versus \Id\ and \Idap, for the analysis of synthetic data generated with $\gbs=0.9$/ns, $\sgs=0.5$, $\qs=1$, $\Ib=0$ and $\kappa=1$. The mean values over data realisations are shown on the left ($\langle\gbf/\gbs-1\rangle$, $\langle\sgf/\sgs-1\rangle$ and $\langle\qf/\qs-1\rangle$ from top to bottom), while the standard deviations are shown on the right.}}\label{FigSM_FRET_ParamError_g=9_k=1_Ib=0}
\end{figure}

\begin{figure}
	\includegraphics[width=0.9\textwidth]{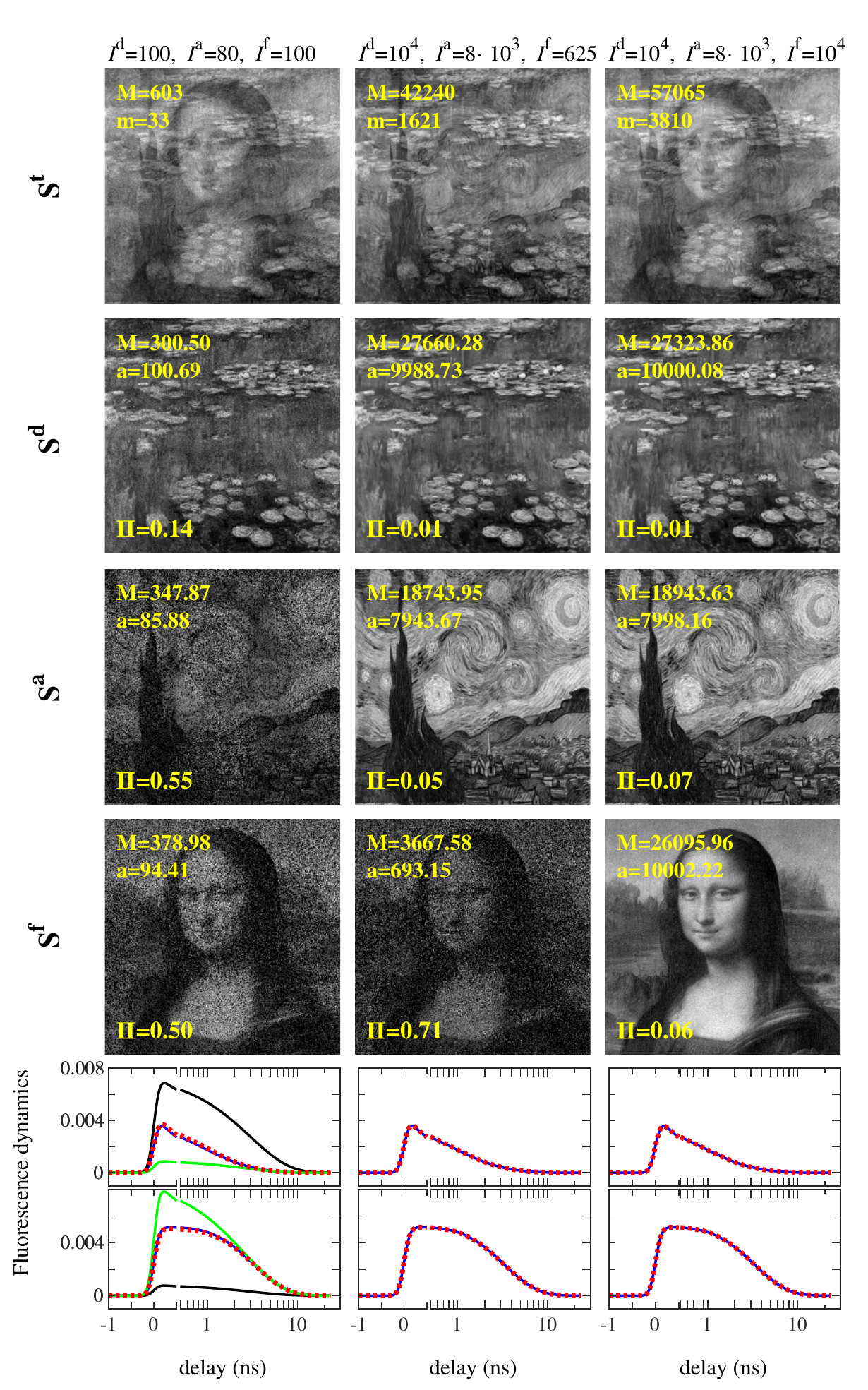}
\end{figure}
\begin{figure}
	\contcaption{Same as \Fig{Fig_FRET_g=5_k=1_Ib=2} but for $\gbs=0.9/$\,ns, $\Ib=2$ and $\kappa=1$: \add{Results of uFLIFRET for synthetic data generated using $\gbs=0.9$/ns, $\sgs=0.5$, $\qs=1$, $\Ib=2$ and $\kappa=1$. The three columns refer to different intensities as indicated. Top row: the time summed data $\St=\Nt\Sav$, on a linear grey scale from a minimum $m$ (black) to a maximum $M$ (white) as indicated. The second to fourth rows show the retrieved spatial distributions of the donor \Sd, acceptor \Sa, and FRET \Sf. Here $m=0$ and $a$ is the average pixel value over the image. The bottom panels show the synthetic original dynamics of donor \bTd (black), acceptor \bTa (green), and DAPs undergoing FRET \bbT (blue), with the retrieved FRET dynamics given as red dashed lines. The signal acquired at the donor (acceptor) detector are given in the top (bottom) panel, respectively. The dynamics are normalized to have a sum of unity over the 2000 temporal points of both detectors.}\label{FigSM_FRET_g=9_k=1_Ib=2}}
\end{figure}
\begin{figure}
	\includegraphics[width=0.6\textwidth]{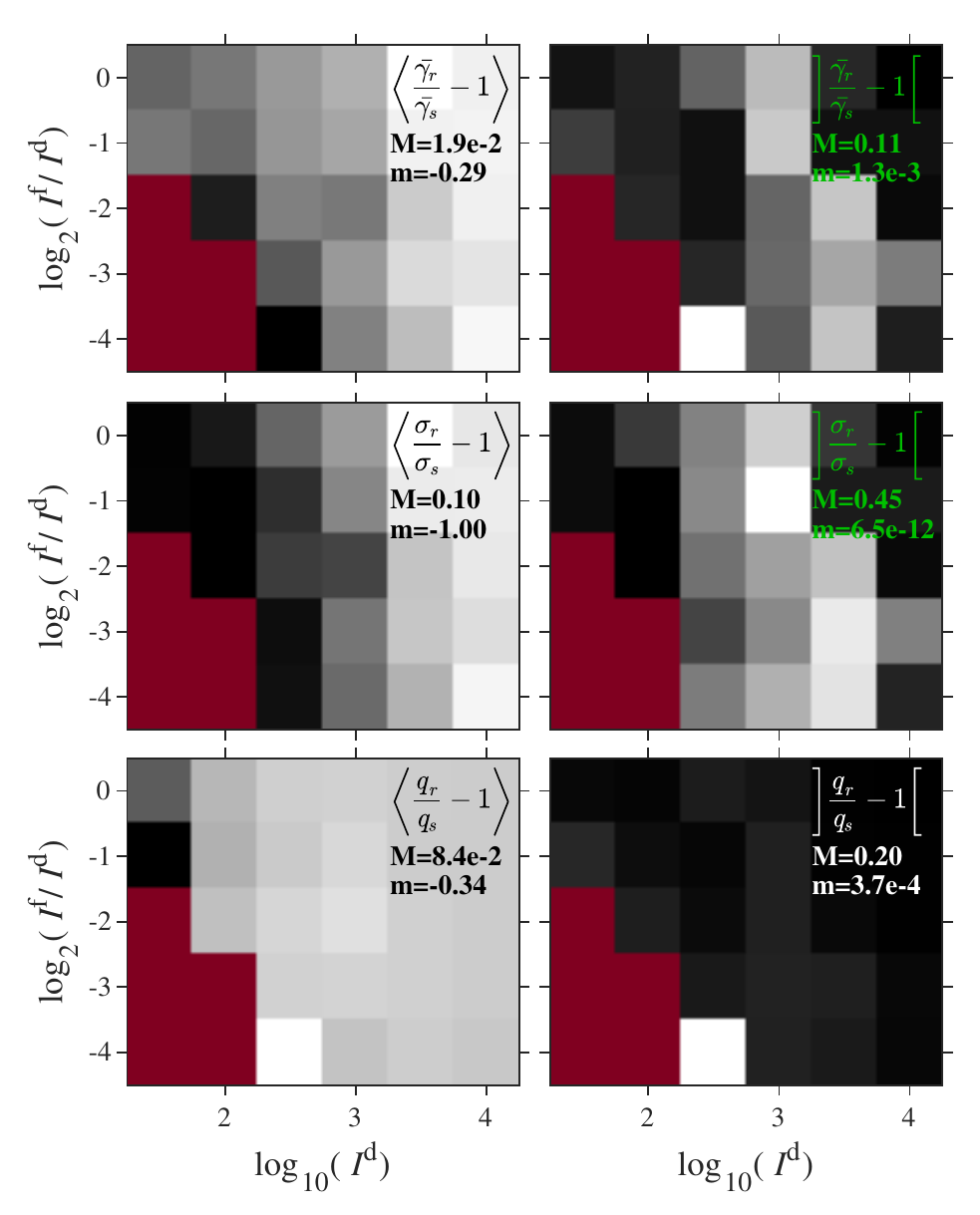}
	\caption{Same as \Fig{FigSM_FRET_ParamError_g=1_k=1_Ib=0} but for $\gbs=0.9/$\,ns, $\Ib=2$ and $\kappa=1$: \add{Relative error of retrieved FRET parameters, versus \Id\ and \Idap, for the analysis of synthetic data generated with $\gbs=0.9$/ns, $\sgs=0.5$, $\qs=1$, $\Ib=2$ and $\kappa=1$. The mean values over data realisations are shown on the left ($\langle\gbf/\gbs-1\rangle$, $\langle\sgf/\sgs-1\rangle$ and $\langle\qf/\qs-1\rangle$ from top to bottom), while the standard deviations are shown on the right.}}\label{FigSM_FRET_ParamError_g=9_k=1_Ib=2}
\end{figure}

\begin{figure}
	\includegraphics[width=0.9\textwidth]{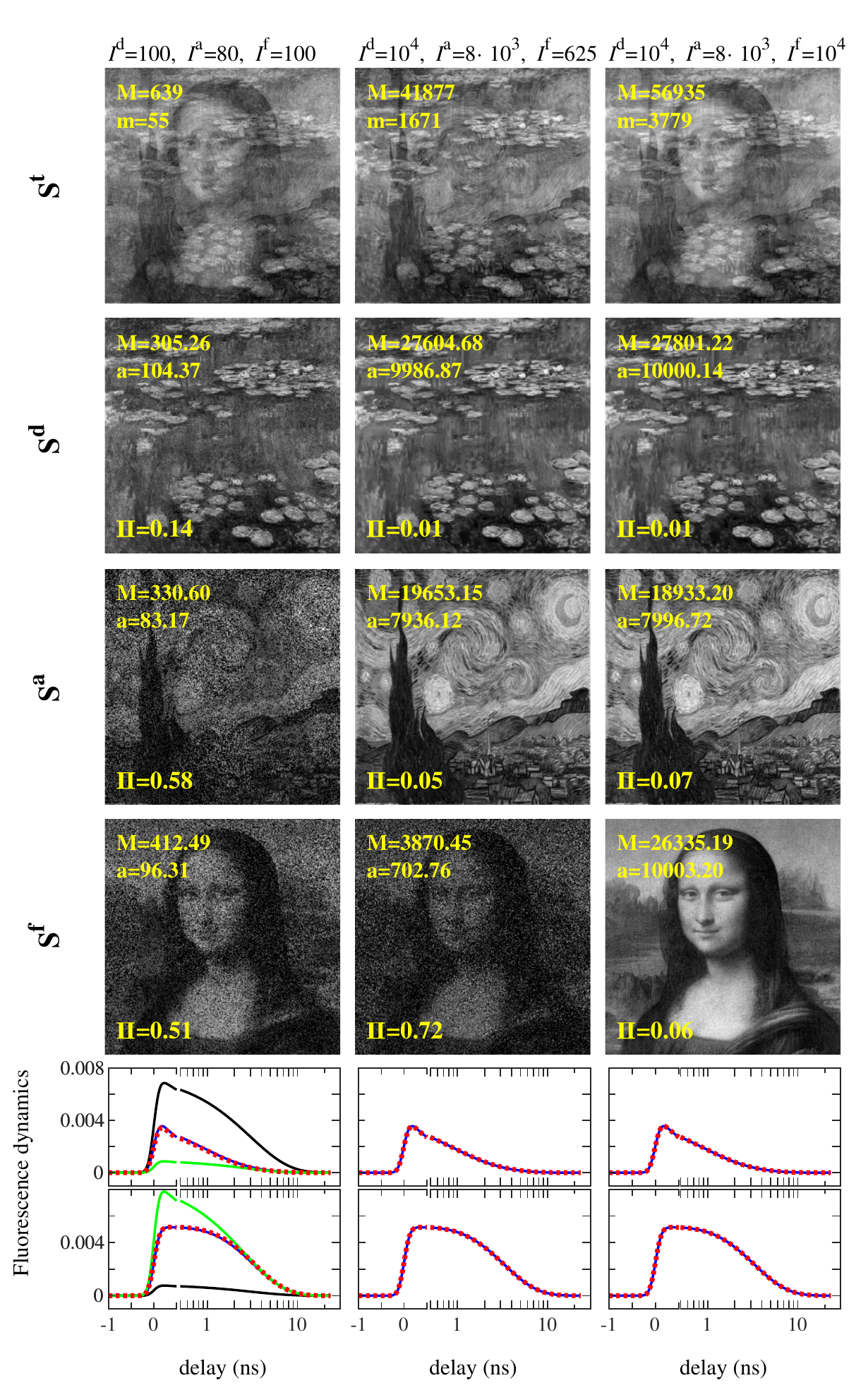}
\end{figure}
\begin{figure}
	\caption{Same as \Fig{Fig_FRET_g=5_k=1_Ib=2} but for $\gbs=0.9/$\,ns, $\Ib=20$ and $\kappa=1$: \add{Results of uFLIFRET for synthetic data generated using $\gbs=0.9$/ns, $\sgs=0.5$, $\qs=1$, $\Ib=20$ and $\kappa=1$. The three columns refer to different intensities as indicated. Top row: the time summed data $\St=\Nt\Sav$, on a linear grey scale from a minimum $m$ (black) to a maximum $M$ (white) as indicated. The second to fourth rows show the retrieved spatial distributions of the donor \Sd, acceptor \Sa, and FRET \Sf. Here $m=0$ and $a$ is the average pixel value over the image. The bottom panels show the synthetic original dynamics of donor \bTd (black), acceptor \bTa (green), and DAPs undergoing FRET \bbT (blue), with the retrieved FRET dynamics given as red dashed lines. The signal acquired at the donor (acceptor) detector are given in the top (bottom) panel, respectively. The dynamics are normalized to have a sum of unity over the 2000 temporal points of both detectors.}\label{FigSM_FRET_g=9_k=1_Ib=20}}
\end{figure}
\begin{figure}
	\includegraphics[width=0.6\textwidth]{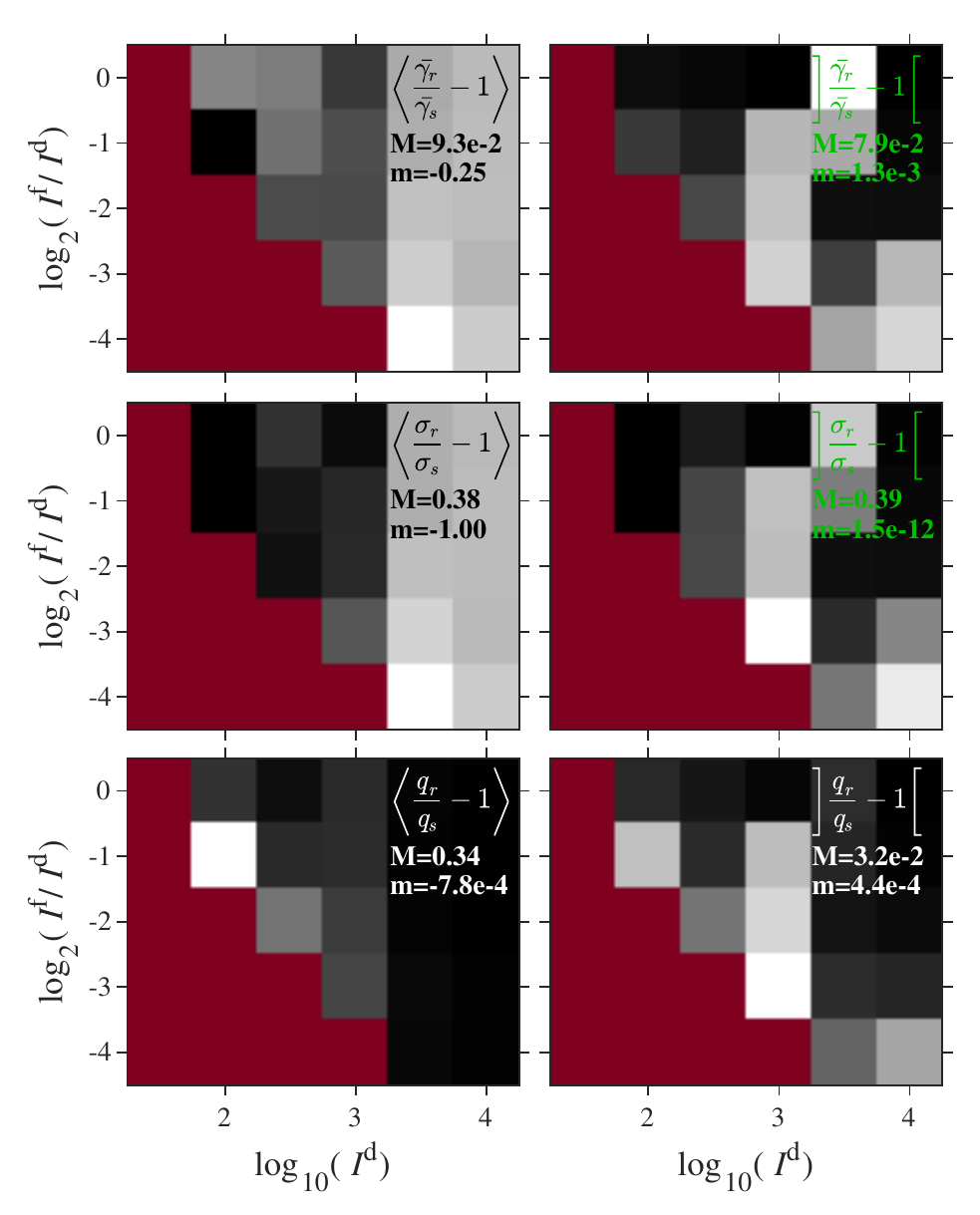}
	\caption{Same as \Fig{FigSM_FRET_ParamError_g=1_k=1_Ib=0} but for $\gbs=0.9/$\,ns, $\Ib=20$ and $\kappa=1$: \add{Relative error of retrieved FRET parameters, versus \Id\ and \Idap, for the analysis of synthetic data generated with $\gbs=0.9$/ns, $\sgs=0.5$, $\qs=1$, $\Ib=20$ and $\kappa=1$. The mean values over data realisations are shown on the left ($\langle\gbf/\gbs-1\rangle$, $\langle\sgf/\sgs-1\rangle$ and $\langle\qf/\qs-1\rangle$ from top to bottom), while the standard deviations are shown on the right.}}\label{FigSM_FRET_ParamError_g=9_k=1_Ib=20}
\end{figure}

\begin{figure}
	\includegraphics[width=0.9\textwidth]{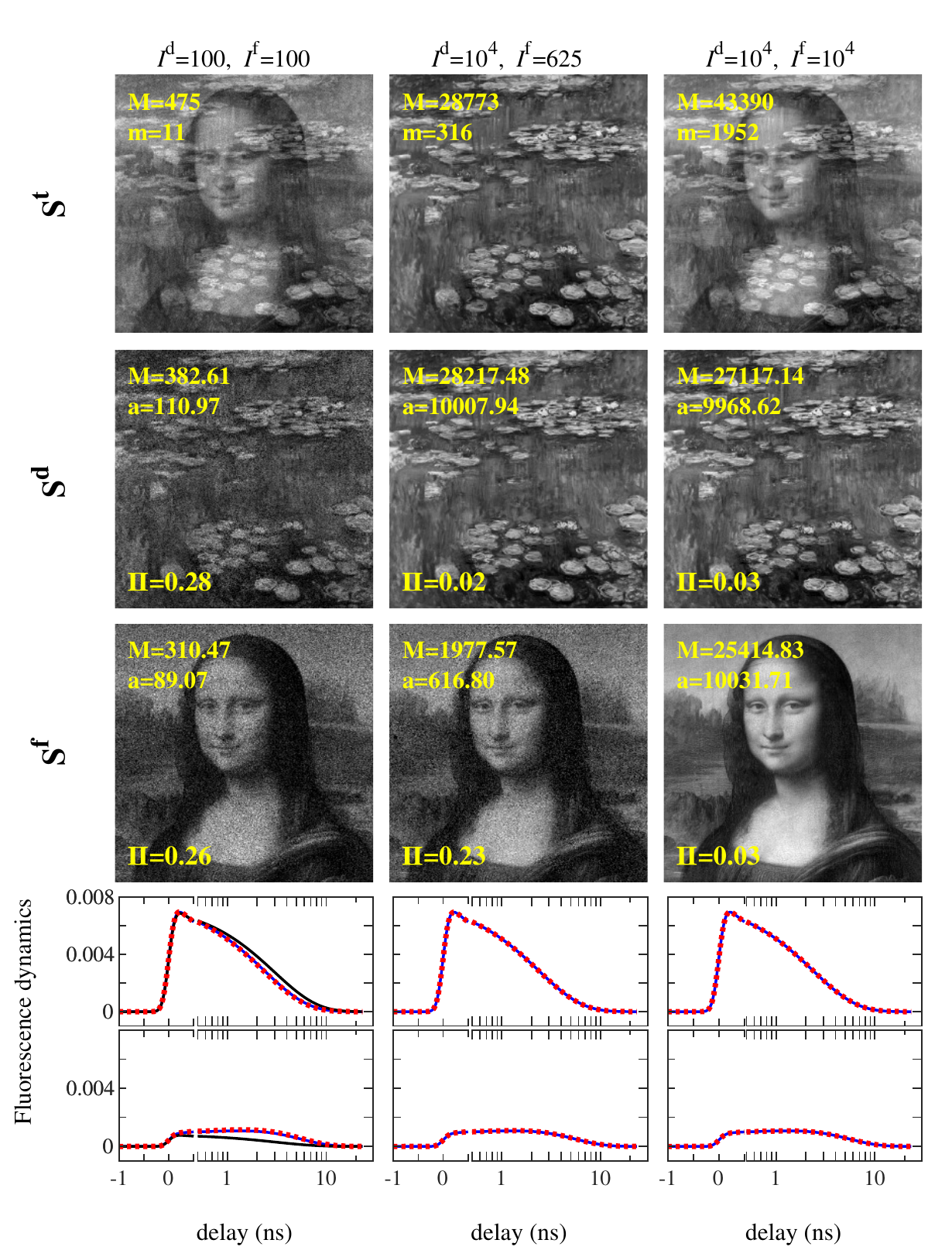}
\end{figure}
\begin{figure}
	\caption{Same as \Fig{Fig_FRET_g=5_k=1_Ib=2} but for $\gbs=0.1/$\,ns, $\Ib=0$ and $\kappa=0$: \add{Results of uFLIFRET for synthetic data generated using $\gbs=0.1$/ns, $\sgs=0.5$, $\qs=1$, $\Ib=0$ and $\kappa=0$. The three columns refer to different intensities as indicated. Top row: the time summed data $\St=\Nt\Sav$, on a linear grey scale from a minimum $m$ (black) to a maximum $M$ (white) as indicated. The second to fourth rows show the retrieved spatial distributions of the donor \Sd, acceptor \Sa, and FRET \Sf. Here $m=0$ and $a$ is the average pixel value over the image. The bottom panels show the synthetic original dynamics of donor \bTd (black), acceptor \bTa (green), and DAPs undergoing FRET \bbT (blue), with the retrieved FRET dynamics given as red dashed lines. The signal acquired at the donor (acceptor) detector are given in the top (bottom) panel, respectively. The dynamics are normalized to have a sum of unity over the 2000 temporal points of both detectors.}\label{FigSM_FRET_g=1_k=0_Ib=0}}
\end{figure}
\begin{figure}
	\includegraphics[width=0.6\textwidth]{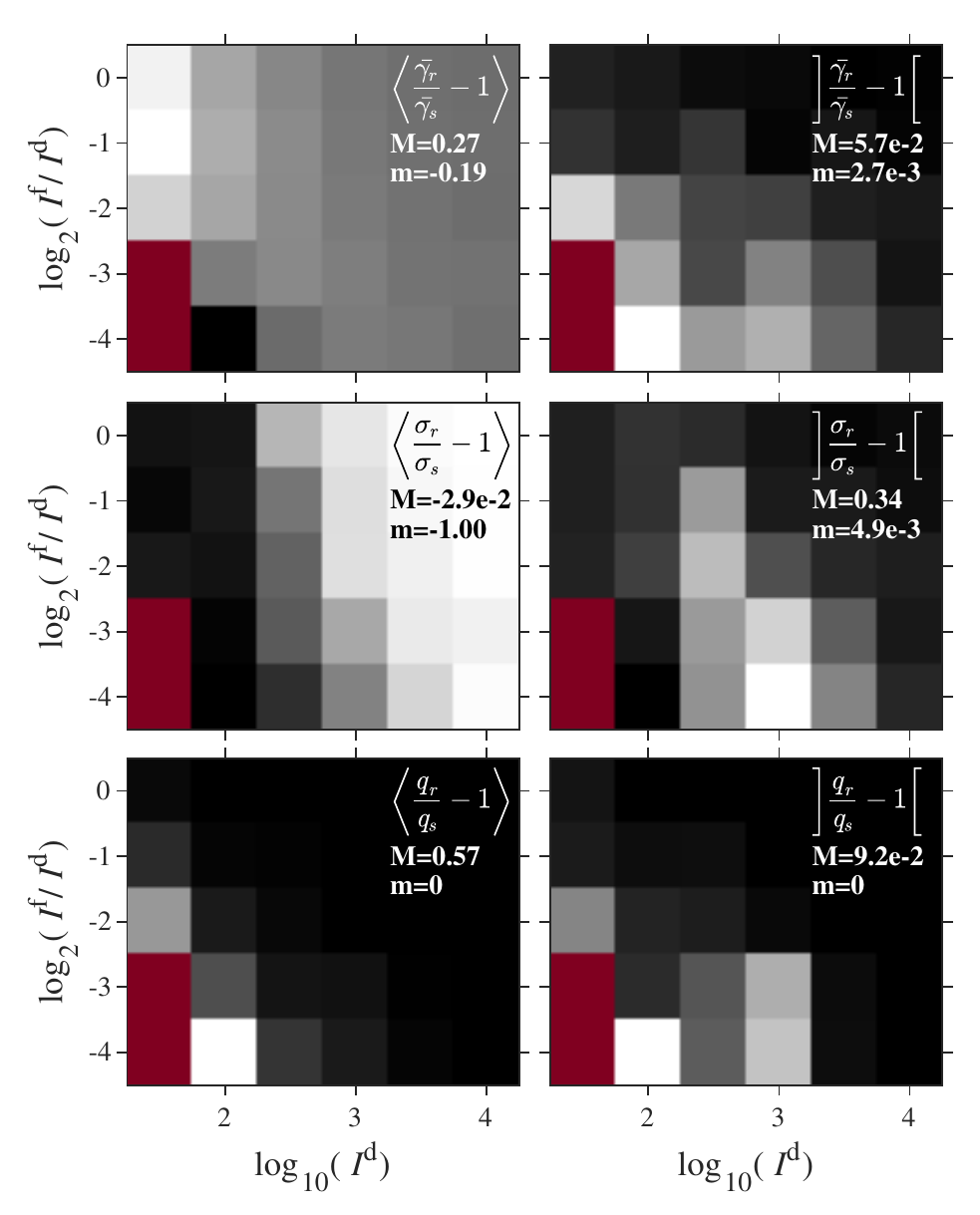}
	\caption{Same as \Fig{FigSM_FRET_ParamError_g=1_k=1_Ib=0} for $\gbs=0.1/$\,ns, $\Ib=0$ and $\kappa=0$: \add{Relative error of retrieved FRET parameters, versus \Id\ and \Idap, for the analysis of synthetic data generated with $\gbs=0.1$/ns, $\sgs=0.5$, $\qs=1$, $\Ib=0$ and $\kappa=0$. The mean values over data realisations are shown on the left ($\langle\gbf/\gbs-1\rangle$, $\langle\sgf/\sgs-1\rangle$ and $\langle\qf/\qs-1\rangle$ from top to bottom), while the standard deviations are shown on the right.}}\label{FigSM_FRET_ParamError_g=1_k=0_Ib=0}
\end{figure}

\begin{figure}
	\includegraphics[width=0.9\textwidth]{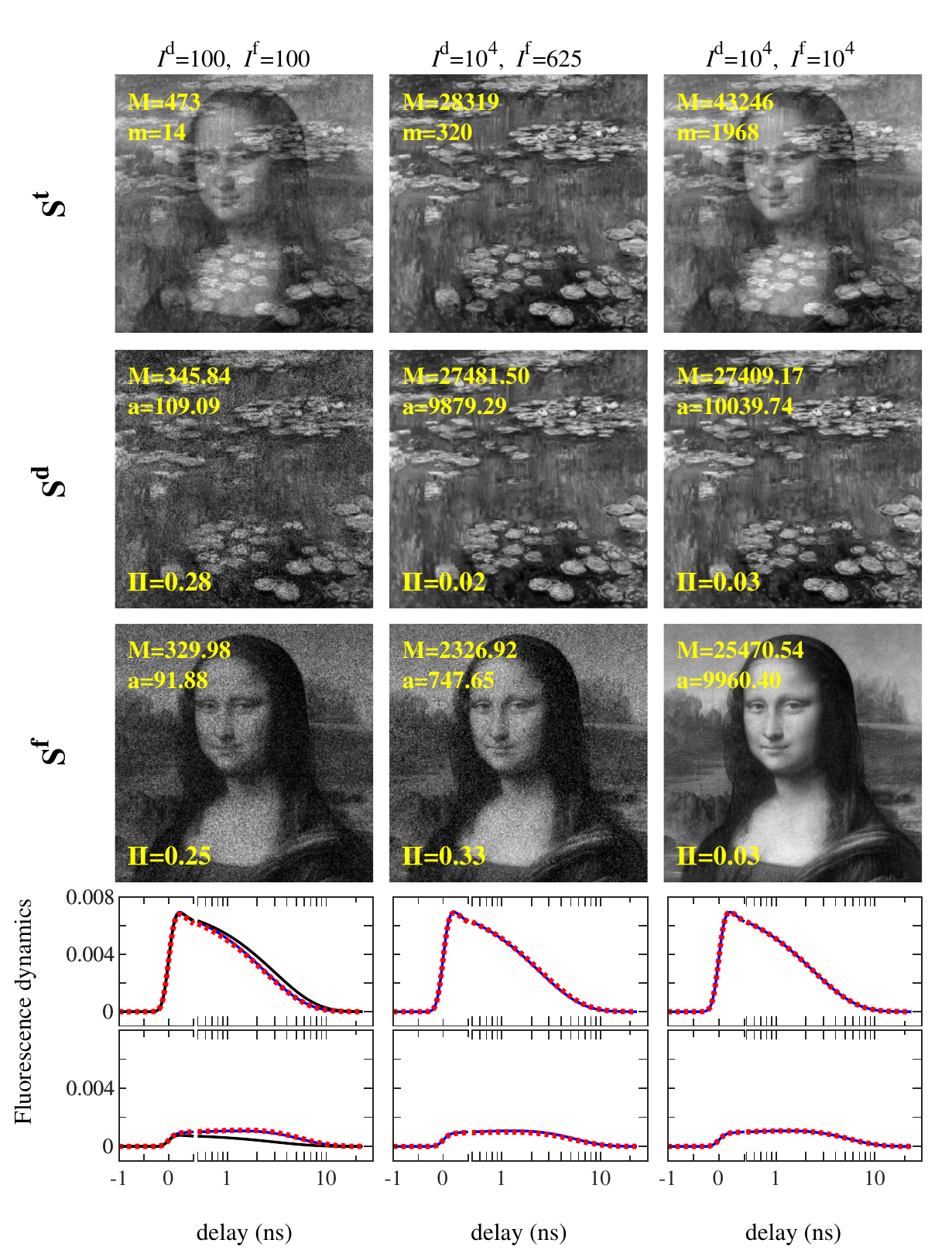}
\end{figure}
\begin{figure}
	\caption{Same as \Fig{Fig_FRET_g=5_k=1_Ib=2} for $\gbs=0.1/$\,ns, $\Ib=2$ and $\kappa=0$: \add{Results of uFLIFRET for synthetic data generated using $\gbs=0.1$/ns, $\sgs=0.5$, $\qs=1$, $\Ib=2$ and $\kappa=0$. The three columns refer to different intensities as indicated. Top row: the time summed data $\St=\Nt\Sav$, on a linear grey scale from a minimum $m$ (black) to a maximum $M$ (white) as indicated. The second to fourth rows show the retrieved spatial distributions of the donor \Sd, acceptor \Sa, and FRET \Sf. Here $m=0$ and $a$ is the average pixel value over the image. The bottom panels show the synthetic original dynamics of donor \bTd (black), acceptor \bTa (green), and DAPs undergoing FRET \bbT (blue), with the retrieved FRET dynamics given as red dashed lines. The signal acquired at the donor (acceptor) detector are given in the top (bottom) panel, respectively. The dynamics are normalized to have a sum of unity over the 2000 temporal points of both detectors.}\label{FigSM_FRET_g=1_k=0_Ib=2}}
\end{figure}
\begin{figure}
	\includegraphics[width=0.6\textwidth]{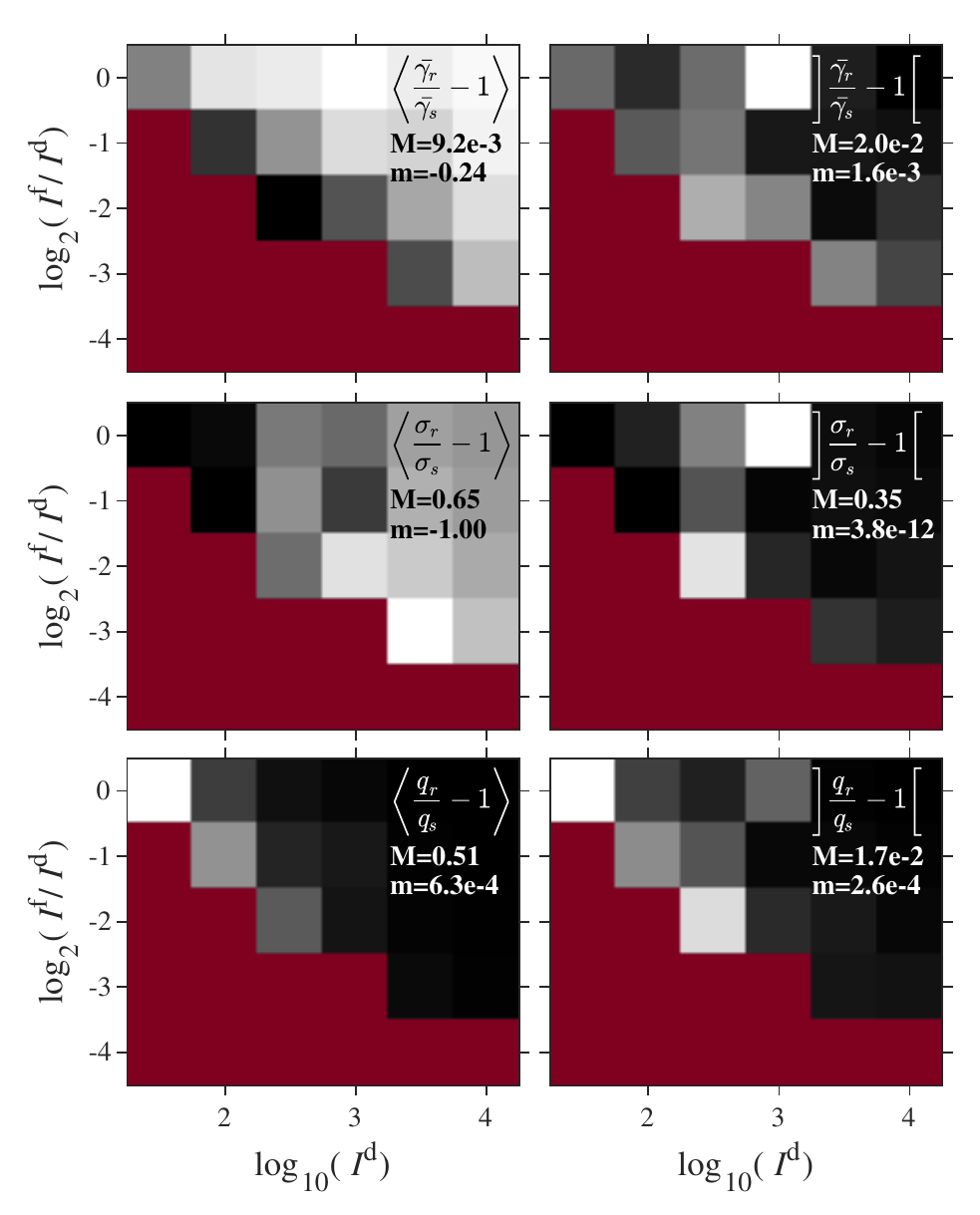}
	\caption{Same as \Fig{FigSM_FRET_ParamError_g=1_k=1_Ib=0} for $\gbs=0.1/$\,ns, $\Ib=2$ and $\kappa=0$: \add{Relative error of retrieved FRET parameters, versus \Id\ and \Idap, for the analysis of synthetic data generated with $\gbs=0.1$/ns, $\sgs=0.5$, $\qs=1$, $\Ib=2$ and $\kappa=0$. The mean values over data realisations are shown on the left ($\langle\gbf/\gbs-1\rangle$, $\langle\sgf/\sgs-1\rangle$ and $\langle\qf/\qs-1\rangle$ from top to bottom), while the standard deviations are shown on the right.}}\label{FigSM_FRET_ParamError_g=1_k=0_Ib=2}
\end{figure}

\begin{figure}
	\includegraphics[width=0.38\textwidth]{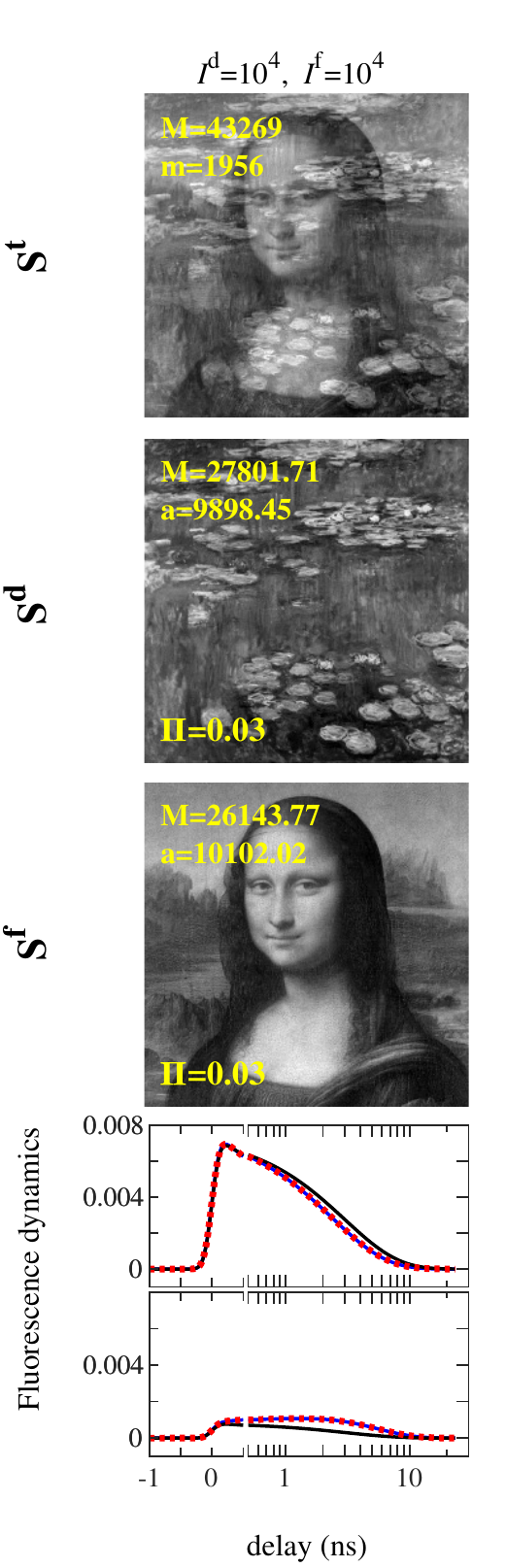}
\end{figure}
\begin{figure}
	\caption{Same as \Fig{Fig_FRET_g=5_k=1_Ib=2} but for $\gbs=0.1/$\,ns, $\Ib=20$ and $\kappa=0$: \add{Results of uFLIFRET for synthetic data generated using $\gbs=0.1$/ns, $\sgs=0.5$, $\qs=1$, $\Ib=20$ and $\kappa=0$. Top row: the time summed data $\St=\Nt\Sav$, on a linear grey scale from a minimum $m$ (black) to a maximum $M$ (white) as indicated. The second to fourth rows show the retrieved spatial distributions of the donor \Sd, acceptor \Sa, and FRET \Sf. Here $m=0$ and $a$ is the average pixel value over the image. The bottom panels show the synthetic original dynamics of donor \bTd (black), acceptor \bTa (green), and DAPs undergoing FRET \bbT (blue), with the retrieved FRET dynamics given as red dashed lines. The signal acquired at the donor (acceptor) detector are given in the top (bottom) panel, respectively. The dynamics are normalized to have a sum of unity over the 2000 temporal points of both detectors.}\label{FigSM_FRET_g=1_k=0_Ib=20}}
\end{figure}
\begin{figure}
	\includegraphics[width=0.6\textwidth]{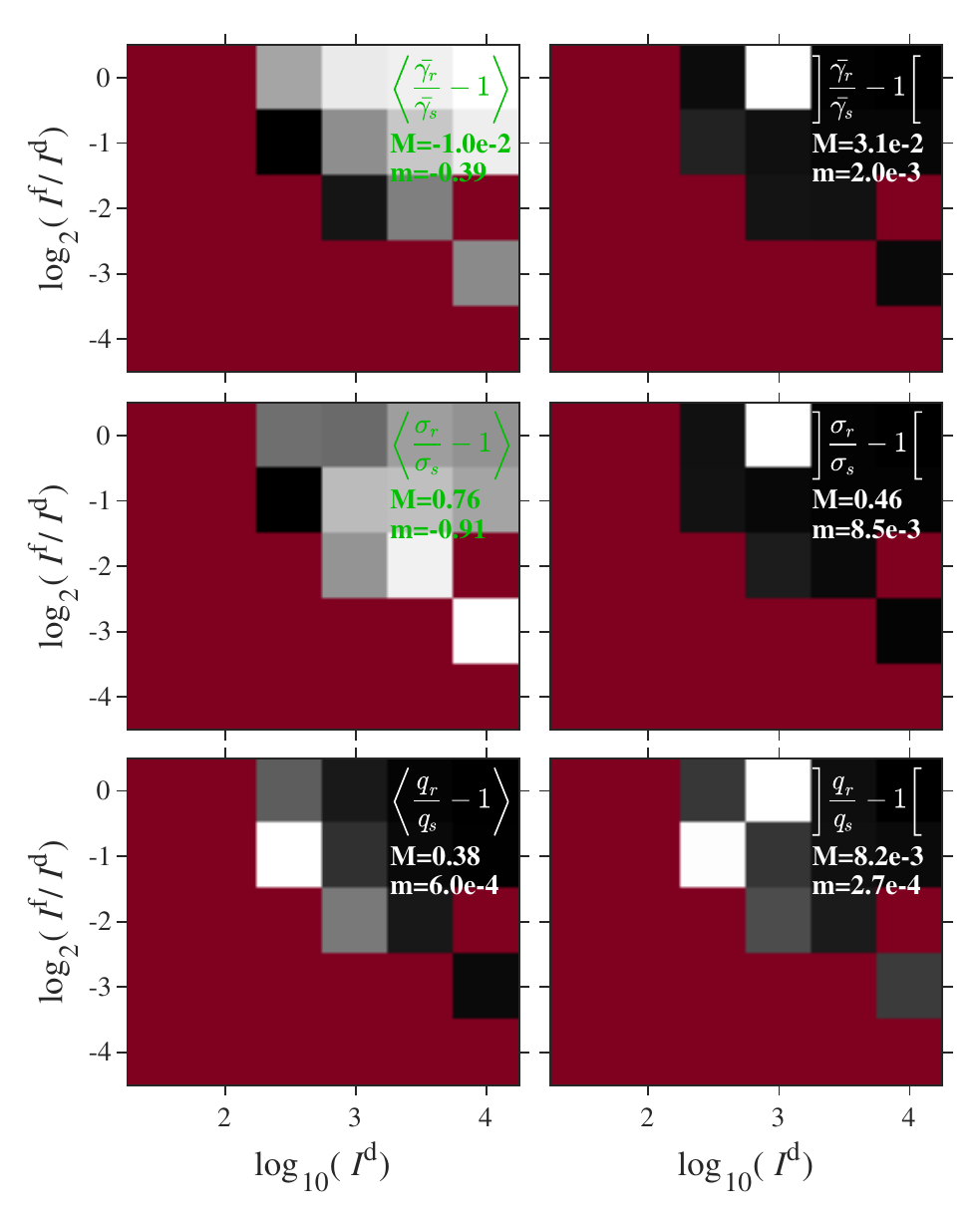}
	\caption{Same as \Fig{FigSM_FRET_ParamError_g=1_k=1_Ib=0} for $\gbs=0.1/$\,ns, $\Ib=20$ and $\kappa=0$: \add{Relative error of retrieved FRET parameters, versus \Id\ and \Idap, for the analysis of synthetic data generated with $\gbs=0.1$/ns, $\sgs=0.5$, $\qs=1$, $\Ib=20$ and $\kappa=0$. The mean values over data realisations are shown on the left ($\langle\gbf/\gbs-1\rangle$, $\langle\sgf/\sgs-1\rangle$ and $\langle\qf/\qs-1\rangle$ from top to bottom), while the standard deviations are shown on the right.}}\label{FigSM_FRET_ParamError_g=1_k=0_Ib=20}
\end{figure}

\begin{figure}
	\includegraphics[width=0.9\textwidth]{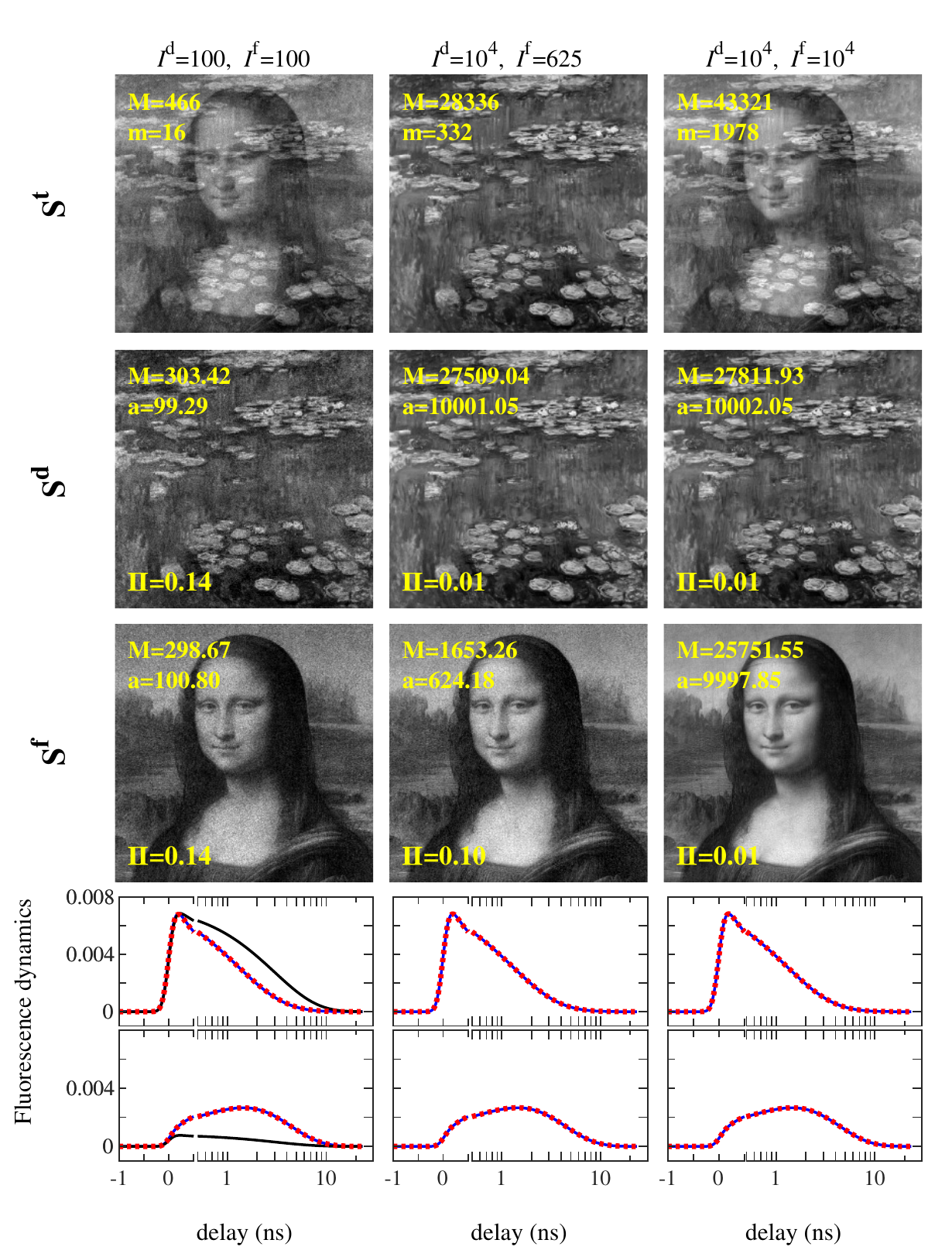}
\end{figure}
\begin{figure}
	\caption{Same as \Fig{Fig_FRET_g=5_k=1_Ib=2} but for $\gbs=0.5/$\,ns, $\Ib=0$ and $\kappa=0$: \add{Results of uFLIFRET for synthetic data generated using $\gbs=0.5$/ns, $\sgs=0.5$, $\qs=1$, $\Ib=0$ and $\kappa=0$. The three columns refer to different intensities as indicated. Top row: the time summed data $\St=\Nt\Sav$, on a linear grey scale from a minimum $m$ (black) to a maximum $M$ (white) as indicated. The second to fourth rows show the retrieved spatial distributions of the donor \Sd, acceptor \Sa, and FRET \Sf. Here $m=0$ and $a$ is the average pixel value over the image. The bottom panels show the synthetic original dynamics of donor \bTd (black), acceptor \bTa (green), and DAPs undergoing FRET \bbT (blue), with the retrieved FRET dynamics given as red dashed lines. The signal acquired at the donor (acceptor) detector are given in the top (bottom) panel, respectively. The dynamics are normalized to have a sum of unity over the 2000 temporal points of both detectors.}\label{FigSM_FRET_g=5_k=0_Ib=0}}
\end{figure}
\begin{figure}
	\includegraphics[width=0.6\textwidth]{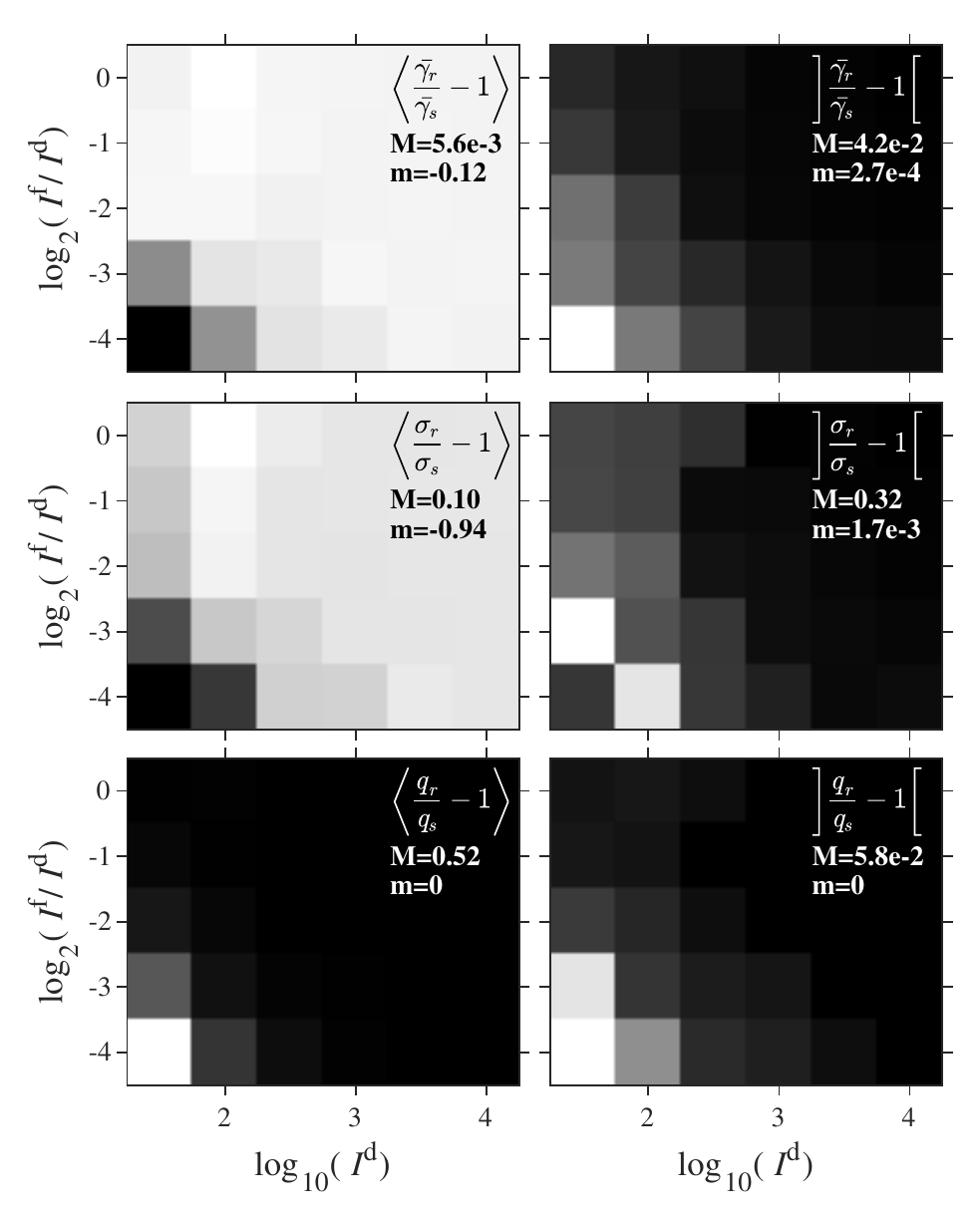}
	\caption{Same as \Fig{FigSM_FRET_ParamError_g=1_k=1_Ib=0} but for $\gbs=0.5/$\,ns, $\Ib=0$ and $\kappa=0$: \add{Relative error of retrieved FRET parameters, versus \Id\ and \Idap, for the analysis of synthetic data generated with $\gbs=0.5$/ns, $\sgs=0.5$, $\qs=1$, $\Ib=0$ and $\kappa=0$. The mean values over data realisations are shown on the left ($\langle\gbf/\gbs-1\rangle$, $\langle\sgf/\sgs-1\rangle$ and $\langle\qf/\qs-1\rangle$ from top to bottom), while the standard deviations are shown on the right.}}\label{FigSM_FRET_ParamError_g=5_k=0_Ib=0}
\end{figure}

\begin{figure}
	\includegraphics[width=0.9\textwidth]{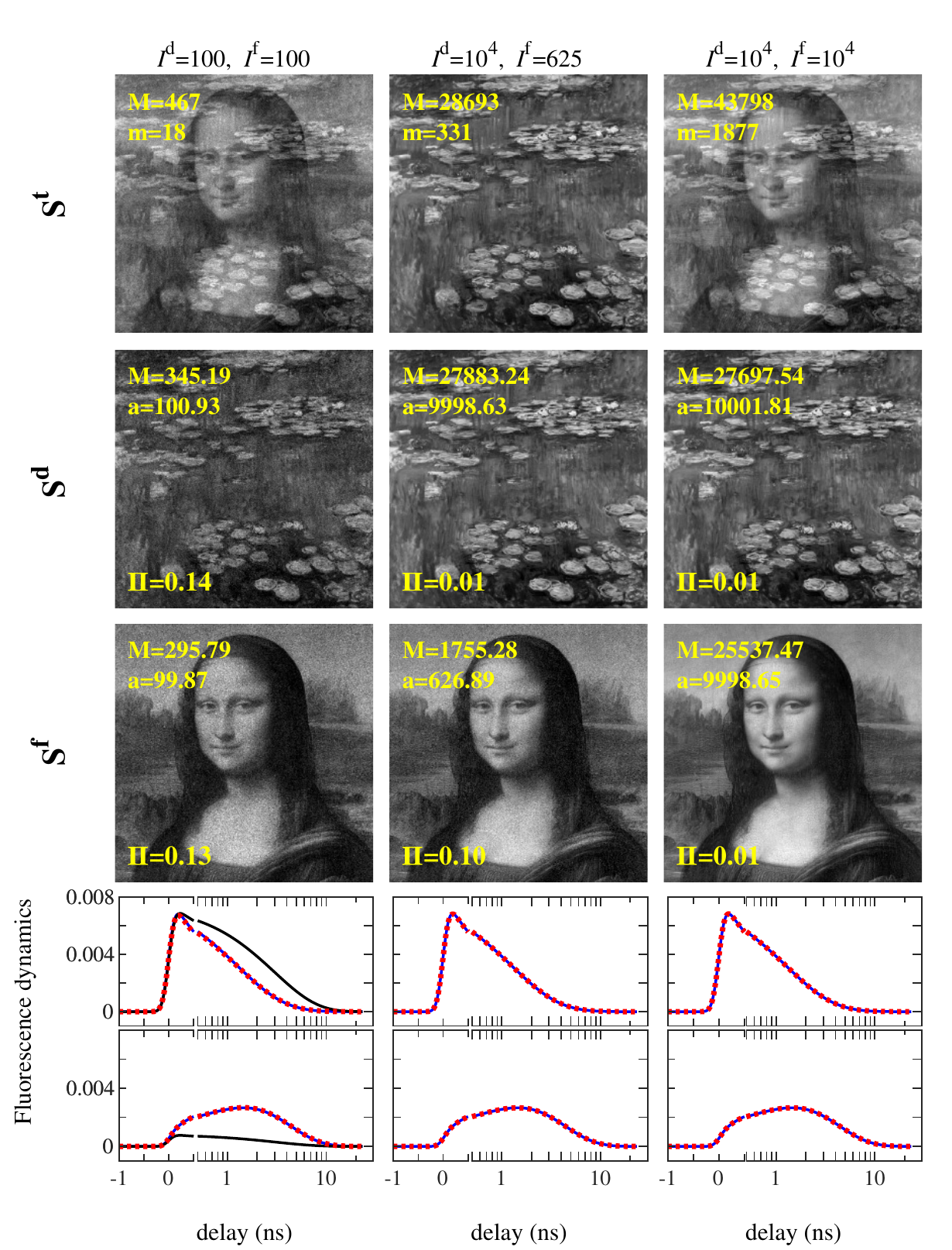}
\end{figure}
\begin{figure}
	\caption{Same as \Fig{Fig_FRET_g=5_k=1_Ib=2} but for $\gbs=0.5/$\,ns, $\Ib=2$ and $\kappa=0$: \add{Results of uFLIFRET for synthetic data generated using $\gbs=0.5$/ns, $\sgs=0.5$, $\qs=1$, $\Ib=2$ and $\kappa=0$. The three columns refer to different intensities as indicated. Top row: the time summed data $\St=\Nt\Sav$, on a linear grey scale from a minimum $m$ (black) to a maximum $M$ (white) as indicated. The second to fourth rows show the retrieved spatial distributions of the donor \Sd, acceptor \Sa, and FRET \Sf. Here $m=0$ and $a$ is the average pixel value over the image. The bottom panels show the synthetic original dynamics of donor \bTd (black), acceptor \bTa (green), and DAPs undergoing FRET \bbT (blue), with the retrieved FRET dynamics given as red dashed lines. The signal acquired at the donor (acceptor) detector are given in the top (bottom) panel, respectively. The dynamics are normalized to have a sum of unity over the 2000 temporal points of both detectors.}\label{FigSM_FRET_g=5_k=0_Ib=2}}
\end{figure}
\begin{figure}
	\includegraphics[width=0.6\textwidth]{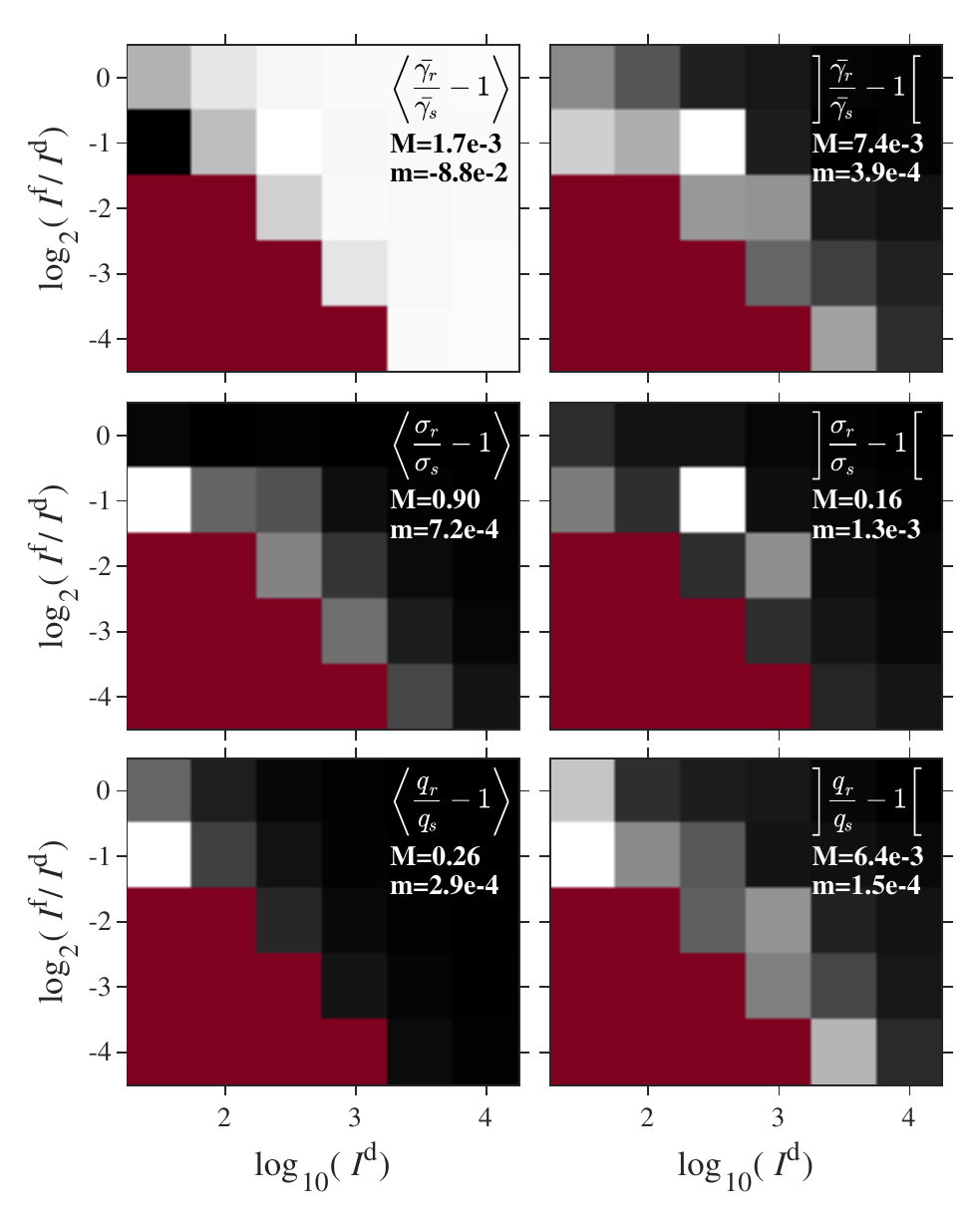}
	\caption{Same as \Fig{FigSM_FRET_ParamError_g=1_k=1_Ib=0} but for $\gbs=0.5/$\,ns, $\Ib=2$ and $\kappa=0$: \add{Relative error of retrieved FRET parameters, versus \Id\ and \Idap, for the analysis of synthetic data generated with $\gbs=0.5$/ns, $\sgs=0.5$, $\qs=1$, $\Ib=2$ and $\kappa=0$. The mean values over data realisations are shown on the left ($\langle\gbf/\gbs-1\rangle$, $\langle\sgf/\sgs-1\rangle$ and $\langle\qf/\qs-1\rangle$ from top to bottom), while the standard deviations are shown on the right.}}\label{FigSM_FRET_ParamError_g=5_k=0_Ib=2}
\end{figure}

\begin{figure}
	\includegraphics[width=0.7\textwidth]{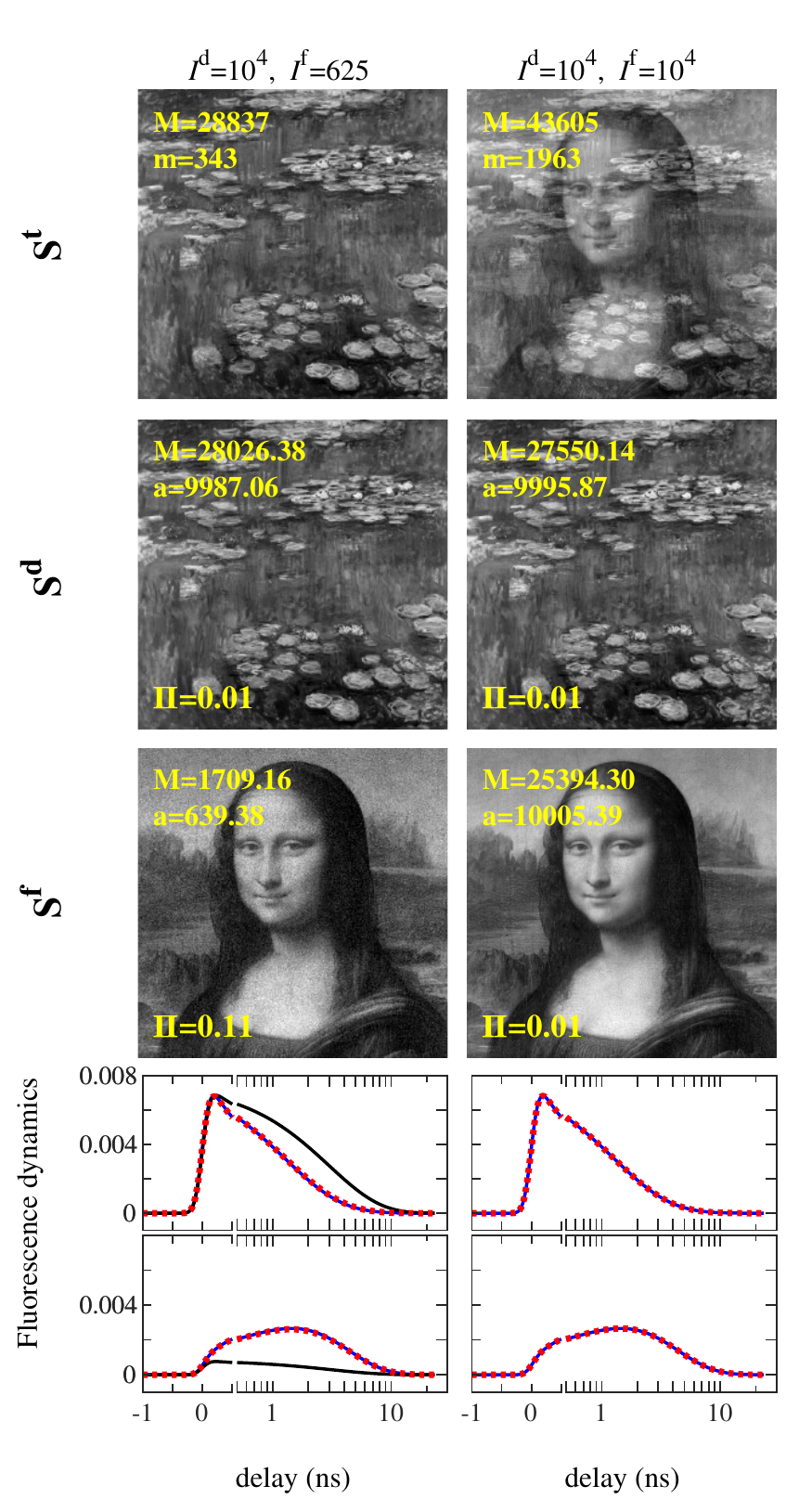}
\end{figure}
\begin{figure}
	\caption{Same as \Fig{Fig_FRET_g=5_k=1_Ib=2} but for $\gbs=0.5/$\,ns, $\Ib=20$ and $\kappa=0$: \add{Results of uFLIFRET for synthetic data generated using $\gbs=0.5$/ns, $\sgs=0.5$, $\qs=1$, $\Ib=20$ and $\kappa=0$. The two columns refer to different intensities as indicated. Top row: the time summed data $\St=\Nt\Sav$, on a linear grey scale from a minimum $m$ (black) to a maximum $M$ (white) as indicated. The second to fourth rows show the retrieved spatial distributions of the donor \Sd, acceptor \Sa, and FRET \Sf. Here $m=0$ and $a$ is the average pixel value over the image. The bottom panels show the synthetic original dynamics of donor \bTd (black), acceptor \bTa (green), and DAPs undergoing FRET \bbT (blue), with the retrieved FRET dynamics given as red dashed lines. The signal acquired at the donor (acceptor) detector are given in the top (bottom) panel, respectively. The dynamics are normalized to have a sum of unity over the 2000 temporal points of both detectors.}\label{FigSM_FRET_g=5_k=0_Ib=20}}
\end{figure}
\begin{figure}
	\includegraphics[width=0.6\textwidth]{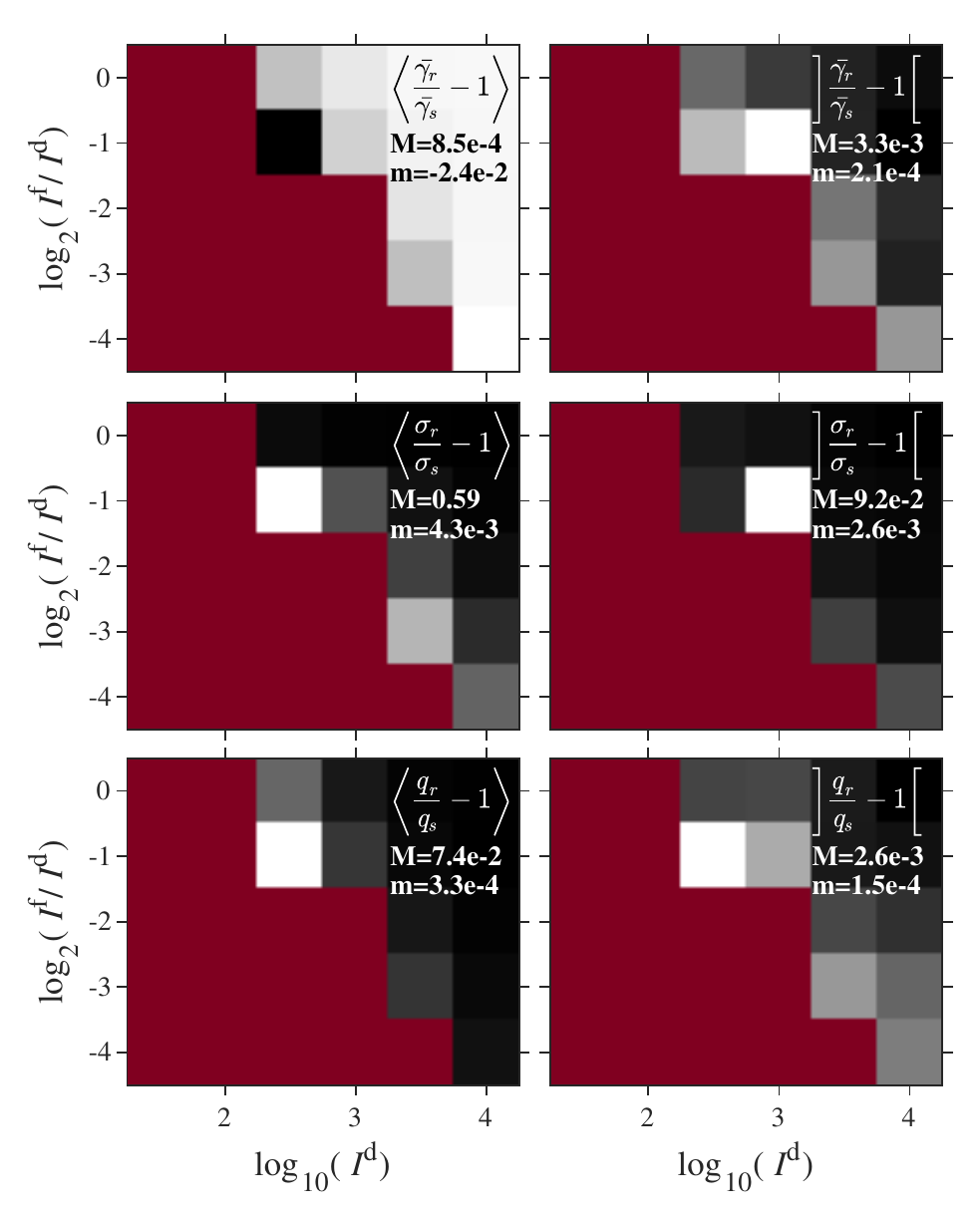}
	\caption{Same as \Fig{FigSM_FRET_ParamError_g=1_k=1_Ib=0} but for $\gbs=0.5/$\,ns, $\Ib=20$ and $\kappa=0$: \add{Relative error of retrieved FRET parameters, versus \Id\ and \Idap, for the analysis of synthetic data generated with $\gbs=0.5$/ns, $\sgs=0.5$, $\qs=1$, $\Ib=20$ and $\kappa=0$. The mean values over data realisations are shown on the left ($\langle\gbf/\gbs-1\rangle$, $\langle\sgf/\sgs-1\rangle$ and $\langle\qf/\qs-1\rangle$ from top to bottom), while the standard deviations are shown on the right.}}\label{FigSM_FRET_ParamError_g=5_k=0_Ib=20}
\end{figure}

\begin{figure}
	\includegraphics[width=0.9\textwidth]{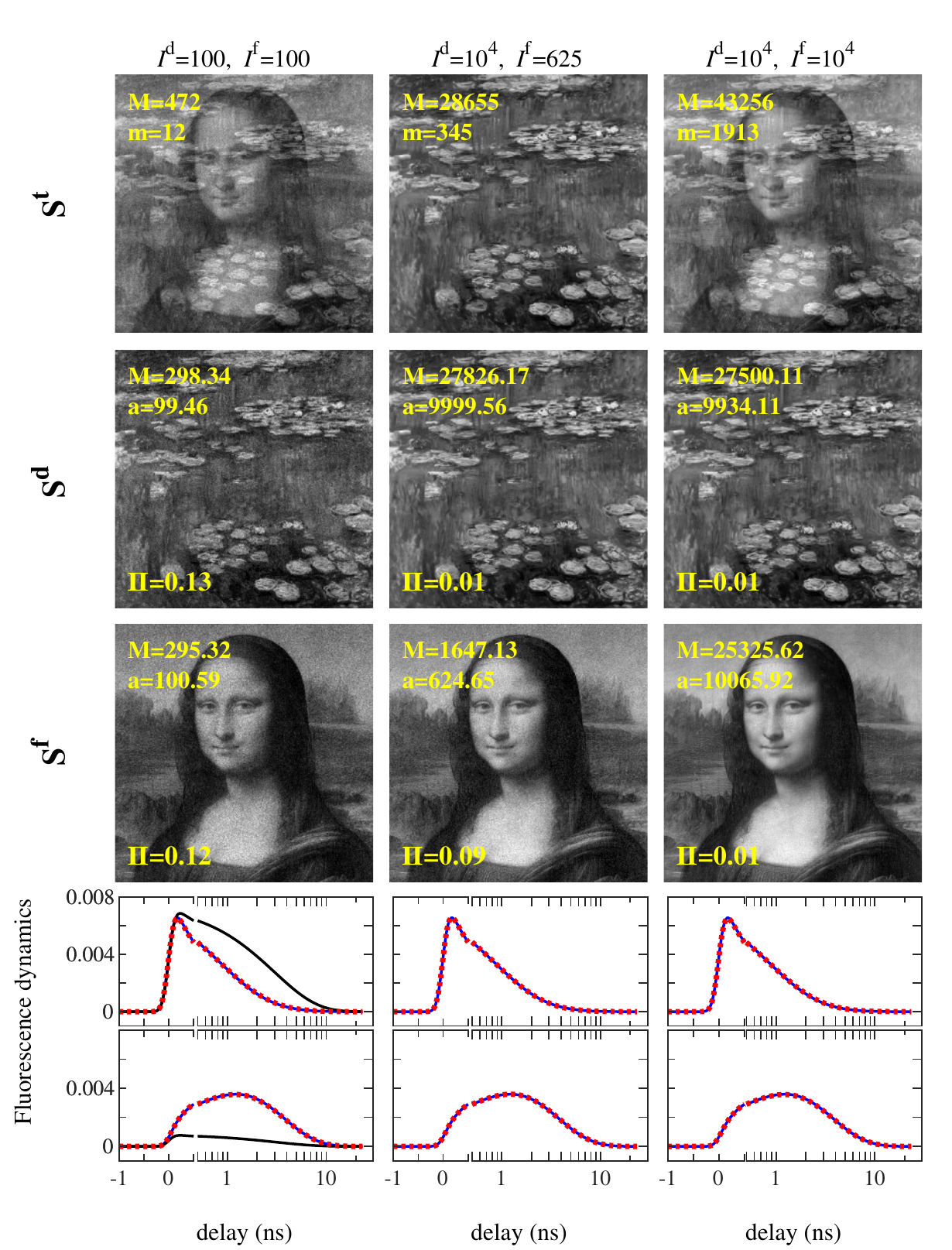}
\end{figure}
\begin{figure}
	\caption{Same as \Fig{Fig_FRET_g=5_k=1_Ib=2} but for $\gbs=0.9/$\,ns, $\Ib=0$ and $\kappa=0$: \add{Results of uFLIFRET for synthetic data generated using $\gbs=0.9$/ns, $\sgs=0.5$, $\qs=1$, $\Ib=0$ and $\kappa=0$. The three columns refer to different intensities as indicated. Top row: the time summed data $\St=\Nt\Sav$, on a linear grey scale from a minimum $m$ (black) to a maximum $M$ (white) as indicated. The second to fourth rows show the retrieved spatial distributions of the donor \Sd, acceptor \Sa, and FRET \Sf. Here $m=0$ and $a$ is the average pixel value over the image. The bottom panels show the synthetic original dynamics of donor \bTd (black), acceptor \bTa (green), and DAPs undergoing FRET \bbT (blue), with the retrieved FRET dynamics given as red dashed lines. The signal acquired at the donor (acceptor) detector are given in the top (bottom) panel, respectively. The dynamics are normalized to have a sum of unity over the 2000 temporal points of both detectors.}\label{FigSM_FRET_g=9_k=0_Ib=0}}
\end{figure}
\begin{figure}
	\includegraphics[width=0.6\textwidth]{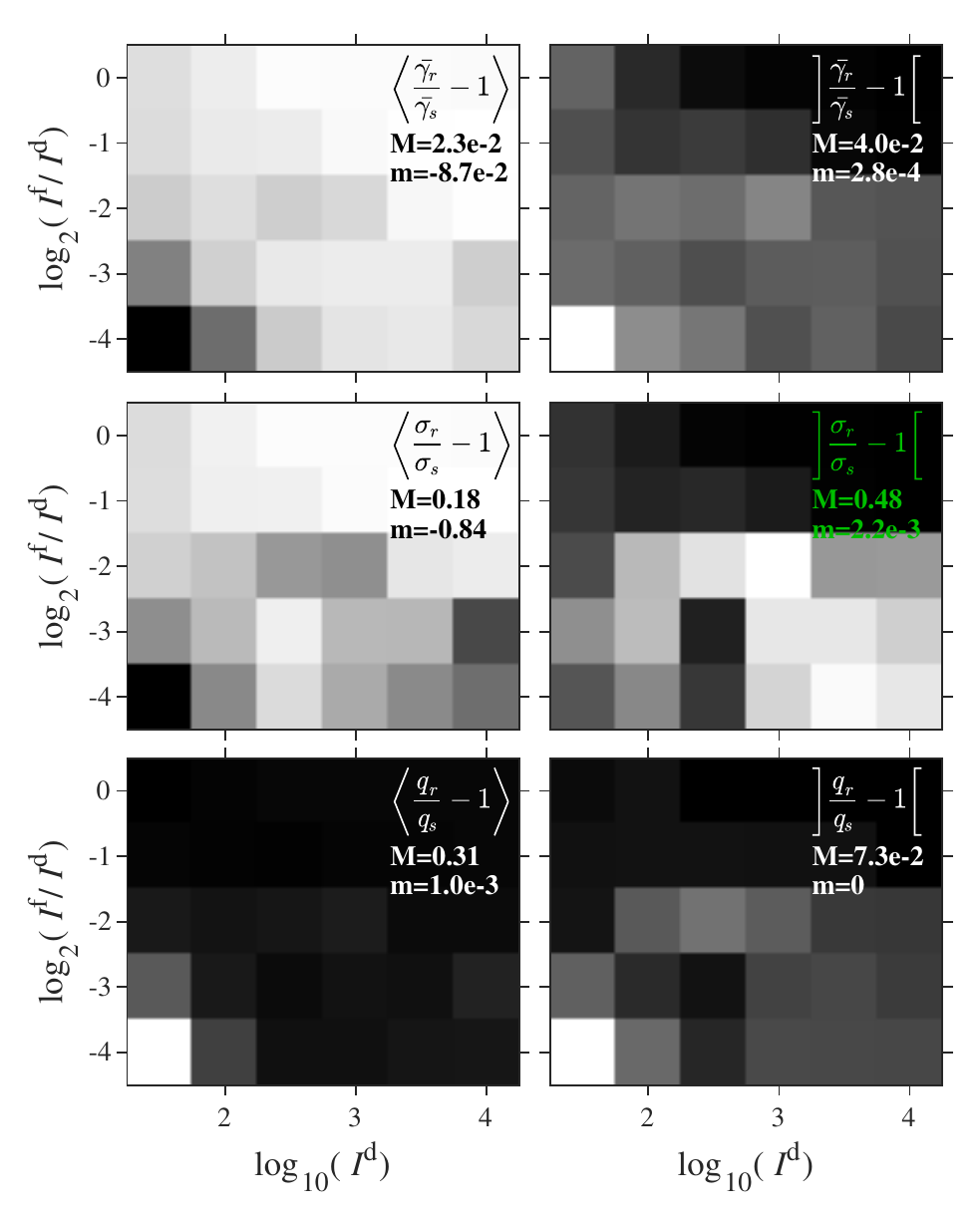}
	\caption{Same as \Fig{FigSM_FRET_ParamError_g=1_k=1_Ib=0} but for $\gbs=0.9/$\,ns, $\Ib=0$ and $\kappa=0$: \add{Relative error of retrieved FRET parameters, versus \Id\ and \Idap, for the analysis of synthetic data generated with $\gbs=0.9$/ns, $\sgs=0.5$, $\qs=1$, $\Ib=0$ and $\kappa=0$. The mean values over data realisations are shown on the left ($\langle\gbf/\gbs-1\rangle$, $\langle\sgf/\sgs-1\rangle$ and $\langle\qf/\qs-1\rangle$ from top to bottom), while the standard deviations are shown on the right.}}\label{FigSM_FRET_ParamError_g=9_k=0_Ib=0}
\end{figure}
\clearpage
\begin{figure}
	\includegraphics[width=0.9\textwidth]{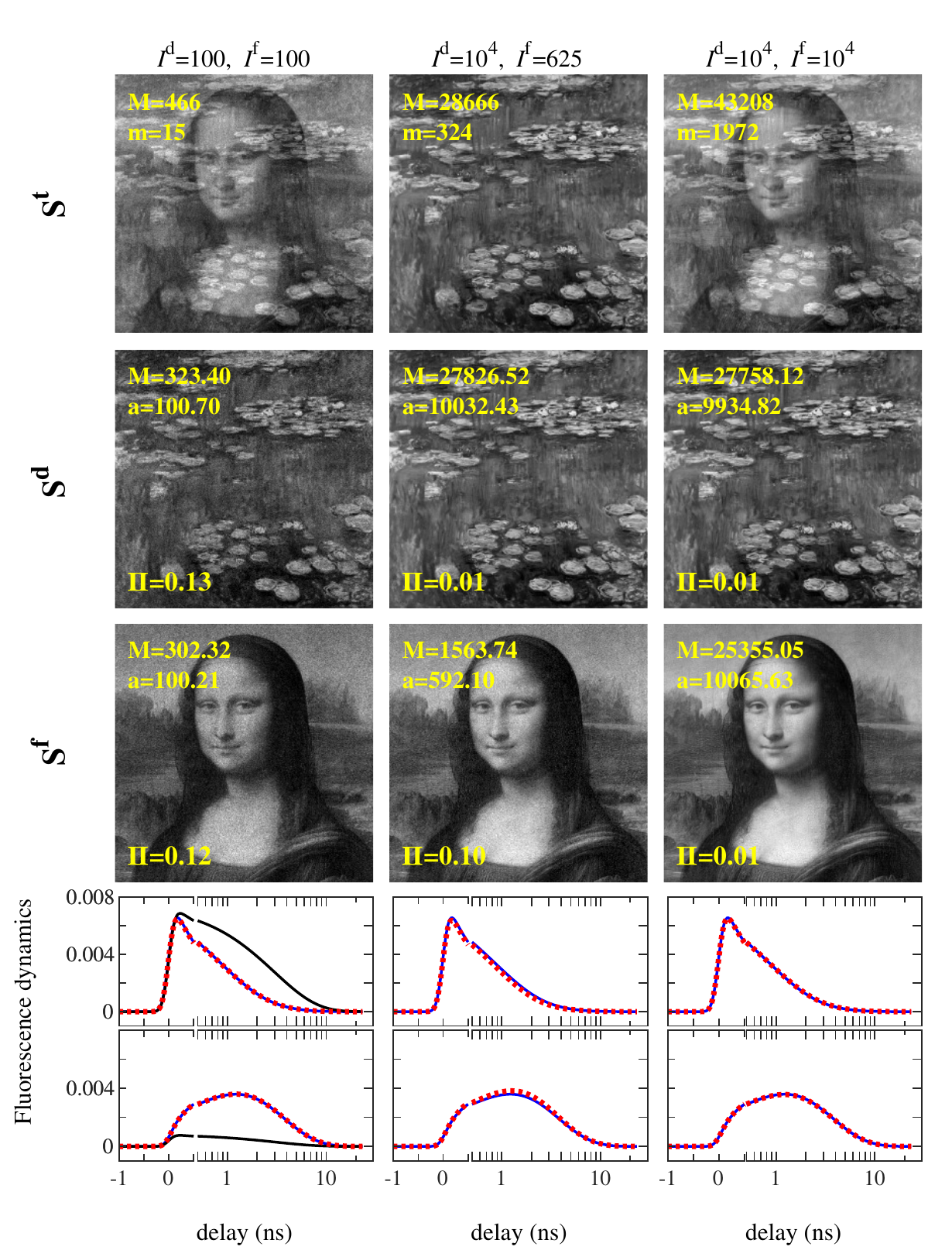}
\end{figure}
\begin{figure}
	\caption{Same as \Fig{Fig_FRET_g=5_k=1_Ib=2} but for $\gbs=0.9/$\,ns, $\Ib=2$ and $\kappa=0$: \add{Results of uFLIFRET for synthetic data generated using $\gbs=0.9$/ns, $\sgs=0.5$, $\qs=1$, $\Ib=2$ and $\kappa=0$. The three columns refer to different intensities as indicated. Top row: the time summed data $\St=\Nt\Sav$, on a linear grey scale from a minimum $m$ (black) to a maximum $M$ (white) as indicated. The second to fourth rows show the retrieved spatial distributions of the donor \Sd, acceptor \Sa, and FRET \Sf. Here $m=0$ and $a$ is the average pixel value over the image. The bottom panels show the synthetic original dynamics of donor \bTd (black), acceptor \bTa (green), and DAPs undergoing FRET \bbT (blue), with the retrieved FRET dynamics given as red dashed lines. The signal acquired at the donor (acceptor) detector are given in the top (bottom) panel, respectively. The dynamics are normalized to have a sum of unity over the 2000 temporal points of both detectors.}\label{FigSM_FRET_g=9_k=0_Ib=2}}
\end{figure}
\clearpage
\begin{figure}
	\includegraphics[width=0.6\textwidth]{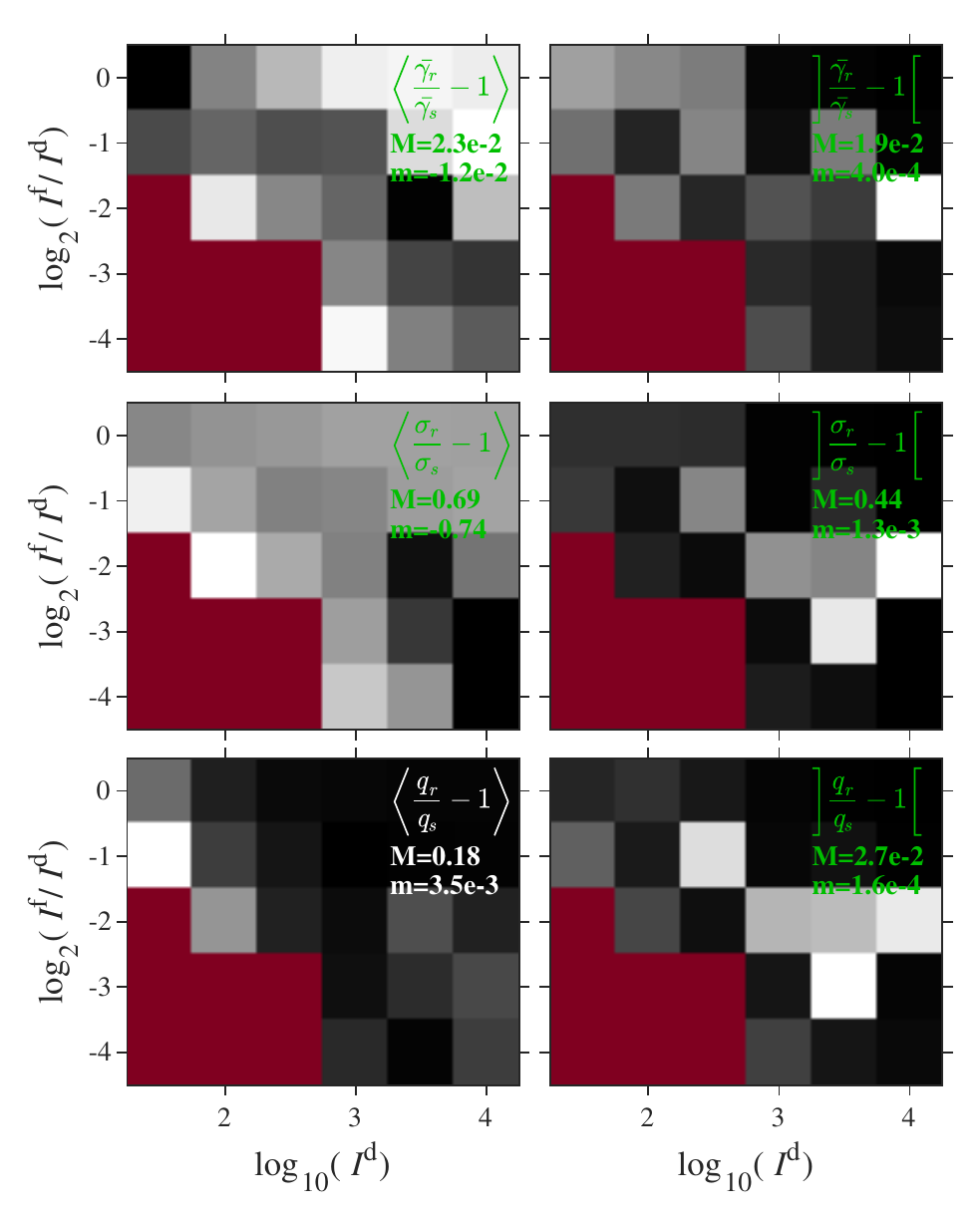}
	\caption{Same as \Fig{FigSM_FRET_ParamError_g=1_k=1_Ib=0} but for $\gbs=0.9/$\,ns, $\Ib=2$ and $\kappa=0$: \add{Relative error of retrieved FRET parameters, versus \Id\ and \Idap, for the analysis of synthetic data generated with $\gbs=0.9$/ns, $\sgs=0.5$, $\qs=1$, $\Ib=2$ and $\kappa=0$. The mean values over data realisations are shown on the left ($\langle\gbf/\gbs-1\rangle$, $\langle\sgf/\sgs-1\rangle$ and $\langle\qf/\qs-1\rangle$ from top to bottom), while the standard deviations are shown on the right.}}\label{FigSM_FRET_ParamError_g=9_k=0_Ib=2}
\end{figure}
\clearpage
\begin{figure}
	\includegraphics[width=0.9\textwidth]{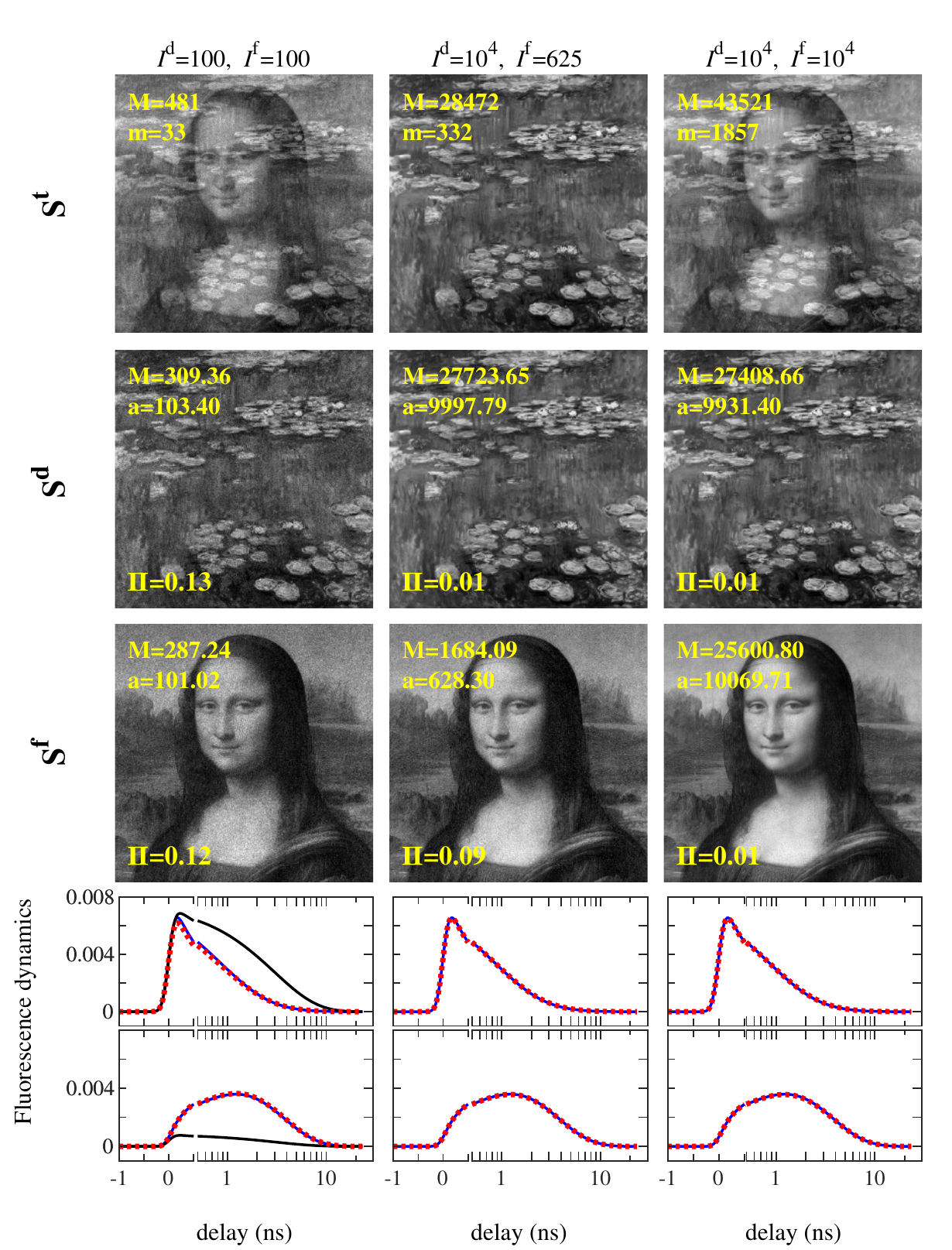}
\end{figure}
\clearpage
\begin{figure}
	\caption{Same as \Fig{Fig_FRET_g=5_k=1_Ib=2} but for $\gbs=0.9/$\,ns, $\Ib=20$ and $\kappa=0$: \add{Results of uFLIFRET for synthetic data generated using $\gbs=0.9$/ns, $\sgs=0.5$, $\qs=1$, $\Ib=20$ and $\kappa=0$. The three columns refer to different intensities as indicated. Top row: the time summed data $\St=\Nt\Sav$, on a linear grey scale from a minimum $m$ (black) to a maximum $M$ (white) as indicated. The second to fourth rows show the retrieved spatial distributions of the donor \Sd, acceptor \Sa, and FRET \Sf. Here, $m=0$ and $a$ is the average pixel value over the image. The bottom panels show the synthetic original dynamics of donor \bTd (black), acceptor \bTa (green), and DAPs undergoing FRET \bbT (blue), with the retrieved FRET dynamics given as red dashed lines. The signal acquired at the donor (acceptor) detector are given in the top (bottom) panel, respectively. The dynamics are normalized to have a sum of unity over the 2000 temporal points of both detectors.}\label{FigSM_FRET_g=9_k=0_Ib=20}}
\end{figure}
\clearpage
\begin{figure}
	\includegraphics[width=0.6\textwidth]{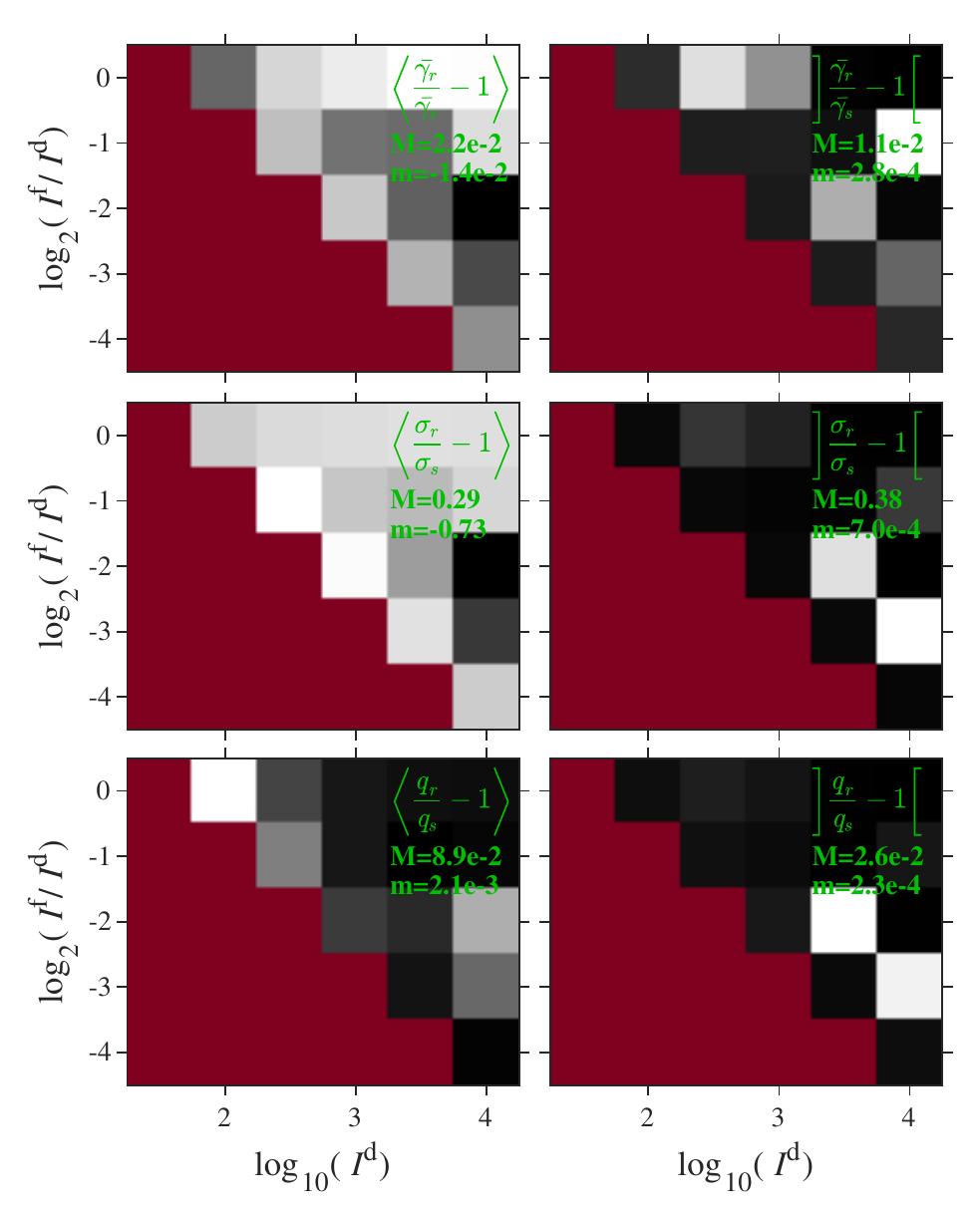}
	\caption{Same as \Fig{FigSM_FRET_ParamError_g=1_k=1_Ib=0} but for $\gbs=0.9/$\,ns, $\Ib=20$ and $\kappa=0$: \add{Relative error of retrieved FRET parameters, versus \Id\ and \Idap, for the analysis of synthetic data generated with $\gbs=0.9$/ns, $\sgs=0.5$, $\qs=1$, $\Ib=20$ and $\kappa=0$. The mean values over data realisations are shown on the left ($\langle\gbf/\gbs-1\rangle$, $\langle\sgf/\sgs-1\rangle$ and $\langle\qf/\qs-1\rangle$ from top to bottom), while the standard deviations are shown on the right.}}
\label{FigSM_FRET_ParamError_g=9_k=0_Ib=20}
\end{figure}
\clearpage
\section{Dependence of quadratic coefficients on signal intensities}\label{sec:fretquad}
\begin{figure}[b]
	\includegraphics[width=0.6\textwidth]{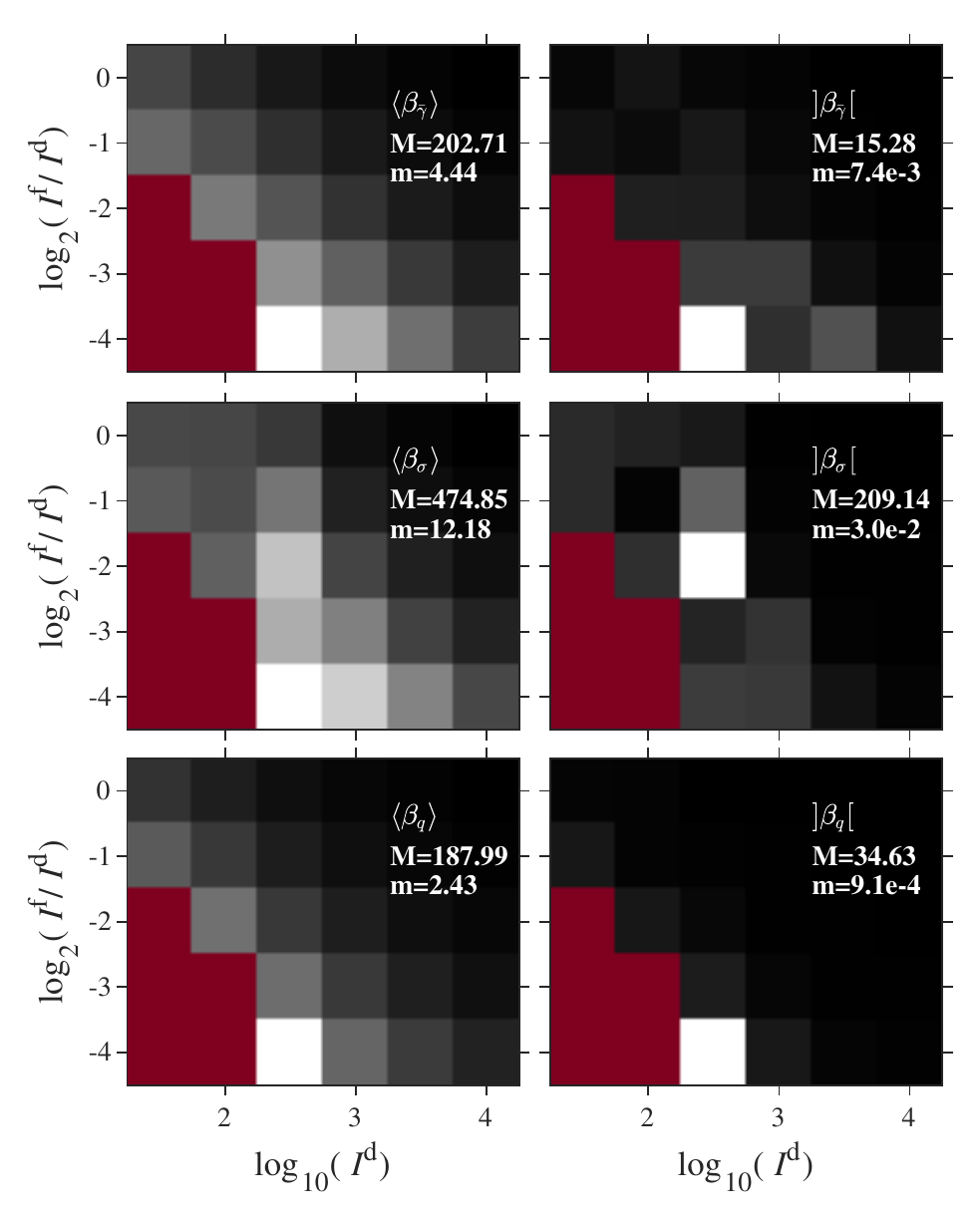}
	\caption{Average (left column) and standard deviation (right column) of the quadratic coefficients of the final fit to the residual distribution \add{for the analysis of synthetic data generated with $\gbs=0.9$/ns, $\sgs=0.5$, $\qs=1$, $\Ib=2$ and $\kappa=1$ as in \Fig{Fig_FRET_g=5_k=1_Ib=2}.}}
	\label{FigSM_FRET_QuadCoeff_g=5_k=1_Ib=2}
\end{figure}

To understand the difference in the accuracy of the retrieval of the different FRET parameters, we show in \Fig{FigSM_FRET_QuadCoeff_g=5_k=1_Ib=2} the average and standard deviation of the coefficients of the quadratic terms of the polynomial used to fit the reconstruction error, i.e. 
\be
E = E_0     \left(1+\left(\frac{\ln(\gb)-\ln(\gbf)}{\beta_{\gb}}\right)^2+\left(\frac{\sg-\sgf}{\beta_{\sg}}\right)^2+
\left(\frac{q-\qf}{\beta_q}\right)^2+{\rm mixed\ terms}\right)
\ee
over the data realisations. As anticipated, $\beta_{\sg}$ is larger than the other two, resulting in a lower precision of the width \sgf\ of the retrieved FRET rate distribution.

\clearpage
\section{Effect of autofluoresence on FRET retrieval}\label{sec:resultsFRETAF} 
We have tested the ability of uFLIFRET to retrieve the FRET parameters and the DAP spatial distribution in presence of an additional component, such as autofluorescence. We used $\Id=10^4$, $\Ia=0.8\Id$, $\Idap=500$, $\gbs=0.5$/ns, $\sgs=0.5$, $\qs=1$, and $\Ib=2$. The autofluorescence component uses a spatial distribution $\SAF$ given by Bronzino's {\it Portrait of Nano Morgante}\,\cite{WikipediaS} with a spatially-averaged time-integrated photon count $\IAF$, and a dynamics $\bTAF$ given by \Eq{eq:decay} with a large decay rate of 10/ns close to the temporal resolution of the system, typical for autofluorescence, and $R^{\rm u}=0.5$ corresponding to a broad spectral emission. The strength of the component, expressed by its fraction $\eta=\IAF/(\IAF + \Id + \Ia + \Idap)$ of the total intensity, was varied between 0.1 and 0.9. The available simulations are summarized in \Tab{tab:simfigsauto} with the retrieved FRET parameters. 

We first assume that $\bTAF$ is known, which is realistic considering that the autofluorescence dynamics can be measured in a sample with no exogenous molecules. Results of the factorization when acceptor molecules are excited as strong as the donor, $\kappa=1$, are given in \Fig{FigSM_FRET_AF_k=1}. The spatial distribution and dynamics of the FRET component are correctly retrieved even for dominant \IAF. This result is encouraging for the analysis of data containing strong autofluorescence, as it is often the case in plants. \Fig{FigSM_FRET_AF_k=0} shows the results when the acceptor molecules are not excited ($\kappa=0$), leading to an improved retrieval, specifically of the width of the FRET distribution \sgf.

\begin{table}
		\renewcommand*{\arraystretch}{1.1}
	\begin{tabular*}{\textwidth}{@{\extracolsep{\fill}}cccccccc}
		\hline
		Image & Free AF & $\kappa$ & $\eta$ & $\gbf$ ($\mu$s$^{-1}$) & $\sgf$ & $\qf$\\
		\hline
		Ground truth & & & & $500$ & $0.5$ & $1.0$ \\
		\hline
		\Fig{FigSM_FRET_AF_k=1} & N & $1$ & $0.1$ & $484.48\pm 0.13$ & $0.401 \pm 0.0007$ & $1.00\pm 0.001$ \\
		\Fig{FigSM_FRET_AF_k=1} & N & $1$ & $0.2$ & $475.68\pm 0.08$ & $0.3425 \pm 0.0006$ & $1.00\pm 0.001$ \\
		\Fig{FigSM_FRET_AF_k=1} & N & $1$ & $0.5$ & $412.11\pm 0.23$ & $0.002 \pm 0.05$ & $1.05\pm 0.01$ \\
		\Fig{FigSM_FRET_AF_k=1} & N & $1$ & $0.9$ & $389.25\pm 0.07$ & $0.010 \pm 0.003$ & $0.98\pm 0.001$ \\
		\hline
		\Fig{FigSM_FRET_AF_k=0} & N & $0$ & $0.1$ & $499.71\pm 0.08$ & $0.5265 \pm 0.0005$ & $1.01\pm 0.001$ \\
		\Fig{FigSM_FRET_AF_k=0} & N & $0$ & $0.2$ & $499.3\pm 0.1$ & $0.5171 \pm 0.0006$ & $1.00\pm 0.001$ \\	
		\Fig{FigSM_FRET_AF_k=0} & N & $0$ & $0.5$ & $500.63\pm 0.11$ & $0.5104 \pm 0.0005$ & $1.01\pm 0.001$ \\	
		\Fig{FigSM_FRET_AF_k=0} & N & $0$ & $0.9$ & $496.66\pm 0.27$ & $0.5313 \pm 0.0012$ & $1.03\pm 0.001$ \\
		\hline		
		\Fig{FigSM_FRET_UnkAF_k=1} & Y & $1$ & $0.1$ & $477.12\pm 0.92$ & $0.3028\pm 0.0086$ & $0.9866\pm 0.0003$ \\
		\Fig{FigSM_FRET_UnkAF_k=1} & Y & $1$ & $0.2$ & $477.15\pm 0.16$ & $0.306 \pm 0.022$ & $0.9813\pm 0.0003$ \\
		\Fig{FigSM_FRET_UnkAF_k=1} & Y & $1$ & $0.5$ & $446.28\pm 6.6$ & $0.237 \pm 0.278$ & $0.98\pm 0.28$ \\
		\Fig{FigSM_FRET_UnkAF_k=1} & Y & $1$ & $0.9$ & $333.81\pm 31.3$ & $0.0683 \pm 0.0073$ & $0.9021\pm 0.0026$ \\
		\hline		
		\Fig{FigSM_FRET_UnkAF_k=0} & Y & $0$ & $0.1$ & $499.68\pm 0.11$ & $0.5040 \pm 0.0008$ & $1.0026\pm 0.00015$ \\
		\Fig{FigSM_FRET_UnkAF_k=0} & Y & $0$ & $0.2$ & $500.74 \pm 0.16$ & $0.4942 \pm 0.0010$ & $1.0038\pm 0.0002$ \\
		\Fig{FigSM_FRET_UnkAF_k=0} & Y & $0$ & $0.5$ & $500.63\pm 0.53$ & $0.5096 \pm 0.0033$ & $1.007 \pm 0.0007$ \\
		\Fig{FigSM_FRET_UnkAF_k=0} & Y & $0$ & $0.9$ & $502.20 \pm 13.5$ & $0.4519 \pm 0.0068$ & $0.98795\pm 0.00098$ \\
		\hline
	\end{tabular*}
	\caption{Overview over uFLIFRET results including autofluorescence.}
	\label{tab:simfigsauto}
\end{table}

So far, we have shown uFLIFRET on data using components with known dynamics, specifically a component describing the unperturbed donor, a component for the acceptor, a component for donor-acceptor pairs undergoing FRET, and optionally a component to consider autofluorescence. However, we can also use the method to retrieve additional unknown components, in the same way as in uFLIM. We demonstrate this here, using the data which include autofluorescence. 
To determine the dynamics of the (for the algorithm) unknown autofluorescence component, we use the same approach as in uFLIM. The spatial distributions of all the components and the dynamics of the unknown component are initialized as random numbers uniformly distributed by 0 and 1. First the spatial distribution are estimated to minimise the reconstruction error by keeping fixed the dynamics. Then we estimate the dynamics which minimise the reconstruction error. The obtained dynamics of the known components are replaced with the known ones.  We repeat these steps for 30 iterations to ensure convergence (typically these correspond to a relative change of the reconstruction error between iterations smaller than 1e-4). 
The results are shown in \Fig{FigSM_FRET_UnkAF_k=1} (\Fig{FigSM_FRET_UnkAF_k=0}) for different relative strength of the autofluorescence component $\eta$ and for $\kappa=1$ ($\kappa=0$), respectively.

\begin{figure}
	\includegraphics[width=0.95\textwidth]{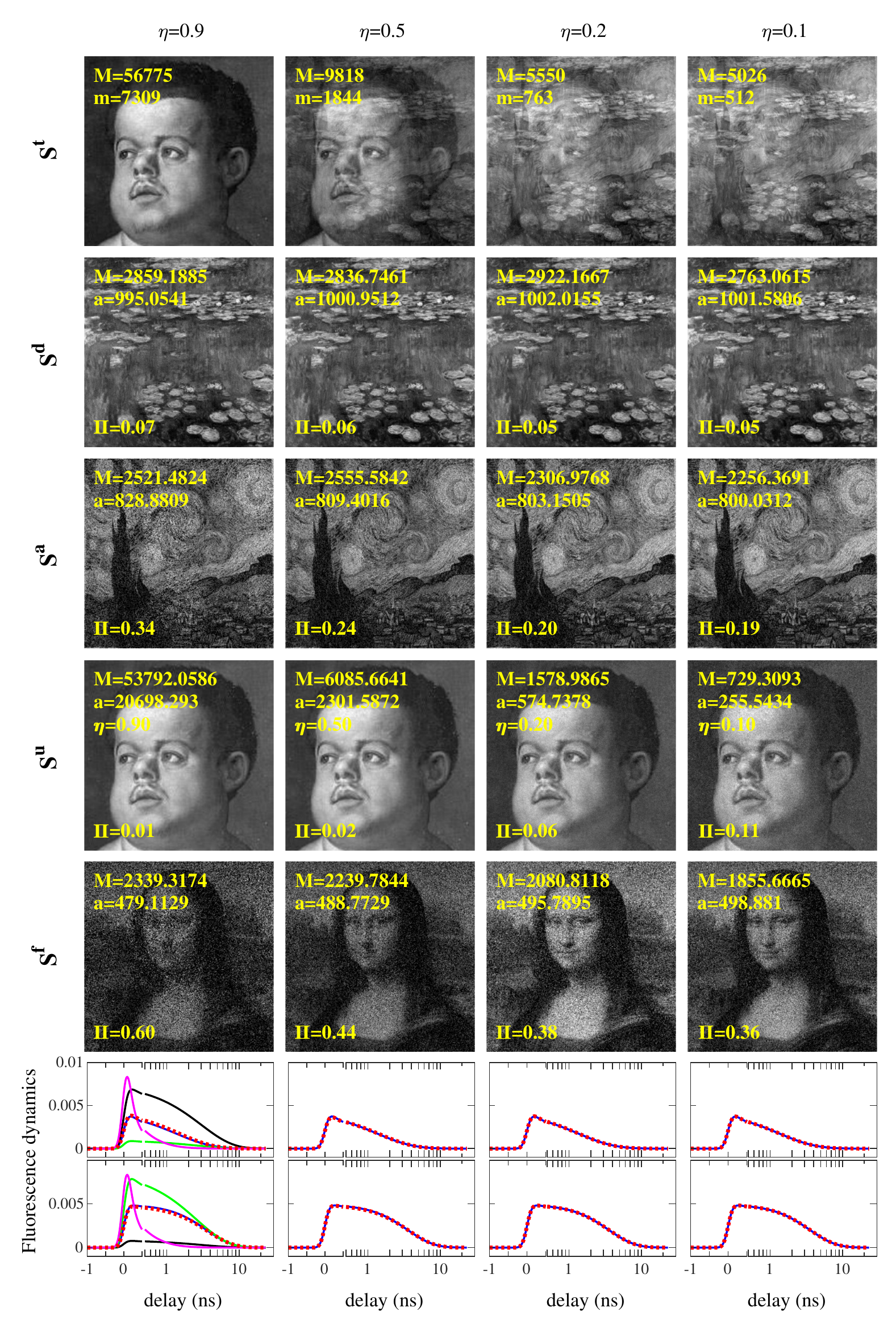}	
\end{figure}
\begin{figure}
	\caption{Same as \Fig{Fig_FRET_g=5_k=1_Ib=2}, but including an additional autofluorescence component, for $\Id=10^4$ and $\Idap=500$: \add{Results of uFLIFRET for synthetic data generated using $\gbs=0.5$/ns, $\sgs=0.5$, $\qs=1$, $\Ib=2$ and $\kappa=1$ and including an additional autofluorescence component. We have used $\Id=10^4$ and $\Idap=500$. The four columns refer to different fractions $\eta$ of the autofluorescence intensity \IAF\ in the total intensity, as labelled. The fourth rows shows the retrieved spatial distribution of the autofluorescence \SAF, and the retrieved $\eta$. In the bottom row the dynamics of the autofluorescence \bTAF\ (divided by three, magenta) has been added. The signal acquired at the donor (acceptor) detector are given in the top (bottom) panel, respectively. The dynamics are normalized to have a sum of unity over the 2000 temporal points of both detectors.}\label{FigSM_FRET_AF_k=1}}
\end{figure}

\begin{figure}
	\includegraphics[width=\textwidth]{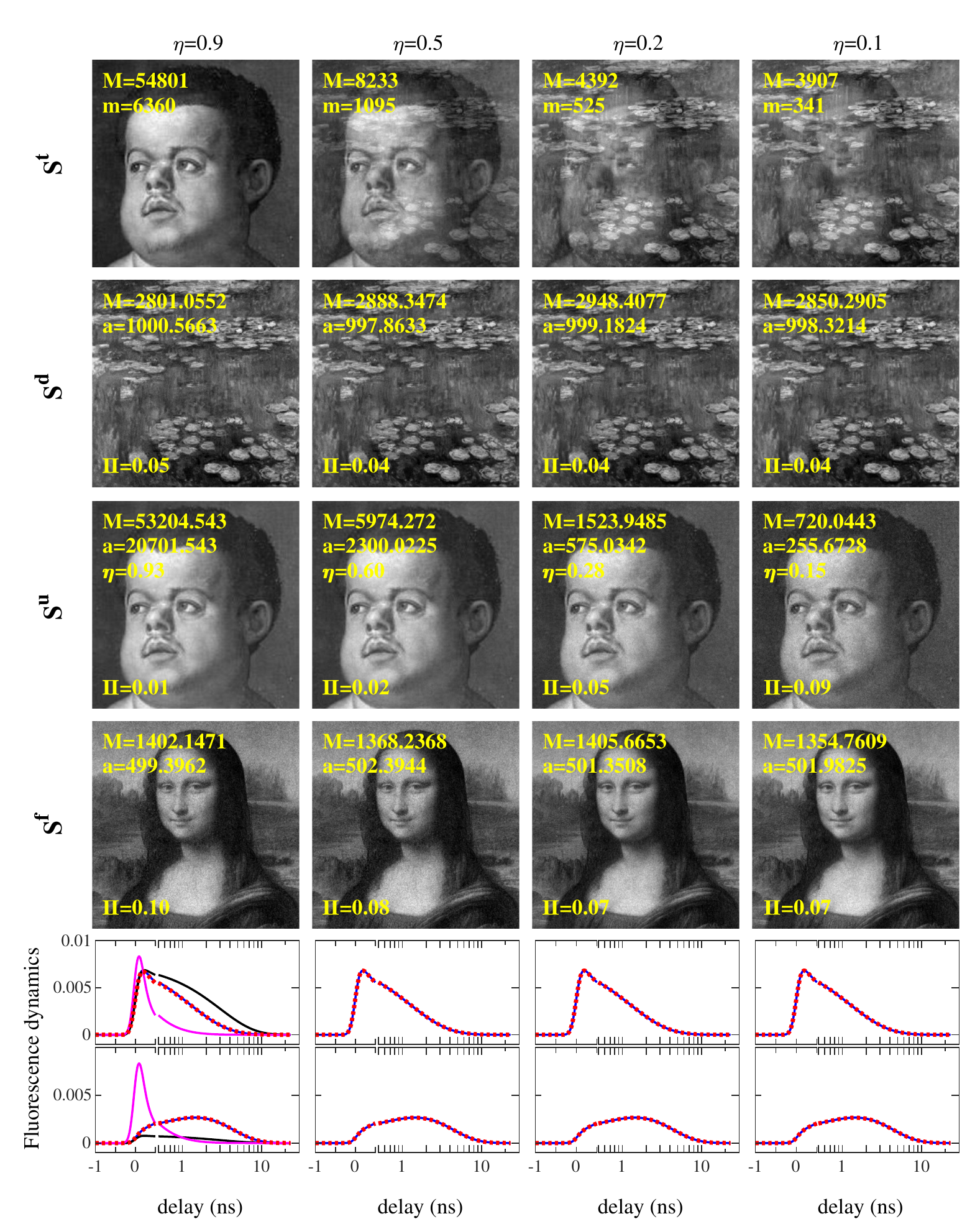}
\end{figure}
\begin{figure}
	\caption{Same as \Fig{FigSM_FRET_AF_k=1}, but for $\kappa=0$: \add{Results of uFLIFRET for synthetic data generated using $\gbs=0.5$/ns, $\sgs=0.5$, $\qs=1$, $\Ib=2$ and $\kappa=0$ and including an additional autofluorescence component. We have used $\Id=10^4$ and $\Idap=500$. The three columns refer to different fractions $\eta$ of the autofluorescence intensity \IAF\ in the total intensity, as labelled. The fourth row shows the retrieved spatial distribution of the autofluorescence \SAF, and the retrieved $\eta$. In the bottom row the dynamics of the autofluorescence \bTAF\ (divided by three, magenta) has been added. The signal acquired at the donor (acceptor) detector are given in the top (bottom) panel, respectively. The dynamics are normalized to have a sum of unity over the 2000 temporal points of both detectors.}\label{FigSM_FRET_AF_k=0}}
\end{figure}

\begin{figure}
	\includegraphics[width=0.98\textwidth]{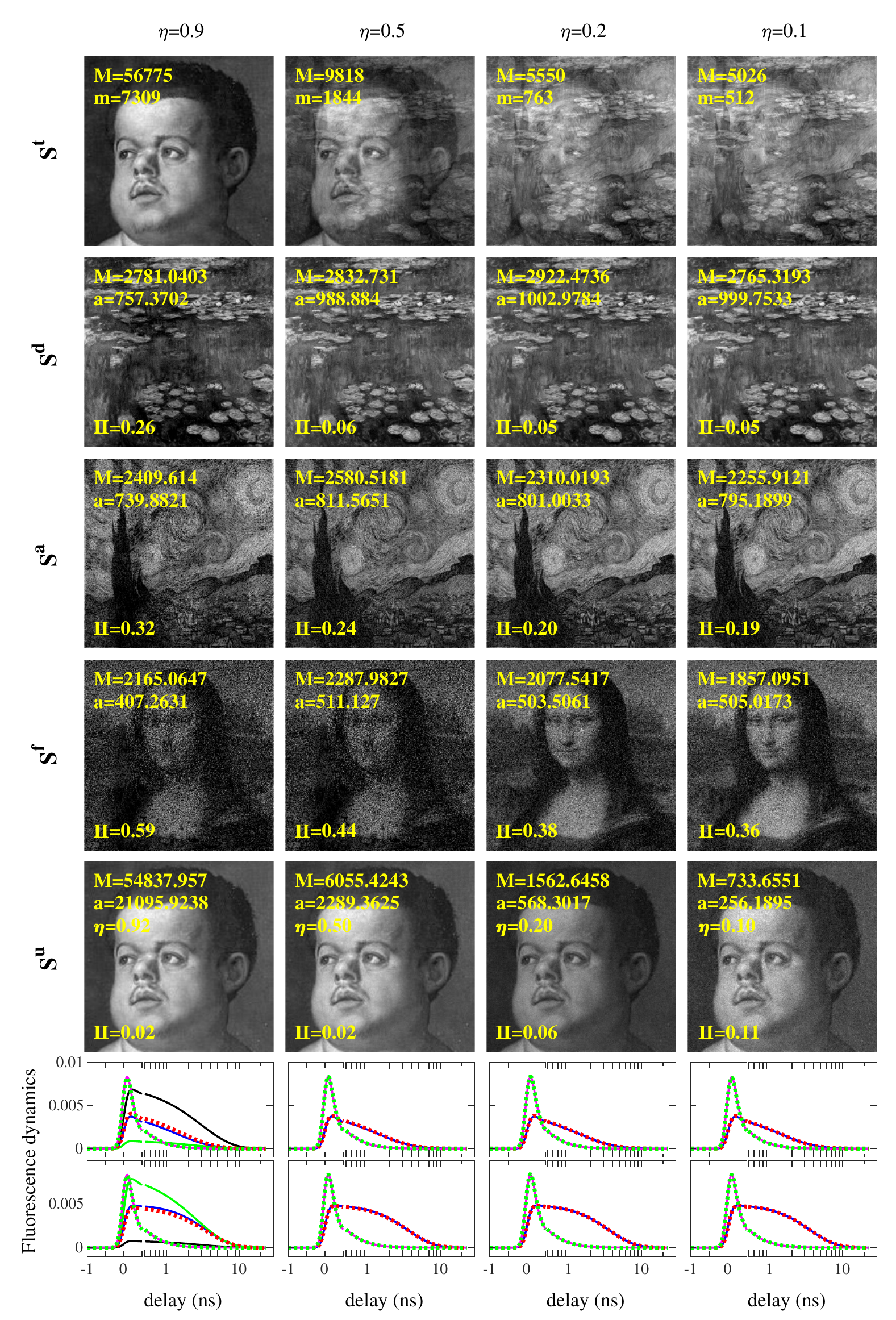}
\end{figure}
\begin{figure}
	\caption{Same as \Fig{FigSM_FRET_AF_k=1}, but for an unknown autofluorescence component and  $\kappa=1$: \add{Results of uFLIFRET for synthetic data generated using $\gbs=0.5$/ns, $\sgs=0.5$, $\qs=1$, $\Ib=2$ and $\kappa=1$ and including an additional autofluorescence component. We have used $\Id=10^4$ and $\Idap=500$. In this analysis, the autofluorescence dynamics is unknwown and retrieved from the algorithm.  The four columns refer to different fractions $\eta$ of the autofluorescence intensity \IAF\ in the total intensity, as labelled. The fifth row shows the retrieved spatial distribution of the autofluorescence \SAF, and the retrieved $\eta$. In the bottom row the nominal dynamics of the autofluorescence \bTAF\ (divided by three, magenta) has been added together with the retrieved dynamics (dashed green). The signal acquired at the donor (acceptor) detector are given in the top (bottom) panel, respectively. The dynamics are normalized to have a sum of unity over the 2000 temporal points of both detectors.}\label{FigSM_FRET_UnkAF_k=1}}
\end{figure}

As a result of the additional degrees of freedom, given by the 2\Nt\ points of the dynamics, the error of the retrieved autofluorescence component is higher, but notably the retrieved FRET parameters are still close to the nominal values. As expected, the unknown component dynamics and distribution is better retrieved when it is more prominent in the data, that is for larger $\eta$. At the same time, the error of the calculated integrated intensities of  the known components and of the retrieved FRET parameters is increasing -- for $\eta=0.9$ we find $\gbf=0.334$/ns, $\sgf=0.068$ and $\qf=0.902$, while for $\eta=0.1$ we find $\gbf=0.477$/ns, $\sgf=0.303$ and $\qf=0.987$, noting that the simulation values are $\gbs=0.5$/ns, $\sgf=0.5$ and $\qs=1$. When the pure acceptor component is not present ($\kappa=0$), the retrieval of the FRET parameters is improved -- for $\eta=0.9$ we find $\gbf=0.502$/ns, $\sgf=0.452$ and $\qf=0.988$, and $\gbf=0.50$/ns, $\sgf=0.504$ and $\qf=1.003$ for $\eta=0.1$.

\begin{figure}
	\includegraphics[width=\textwidth]{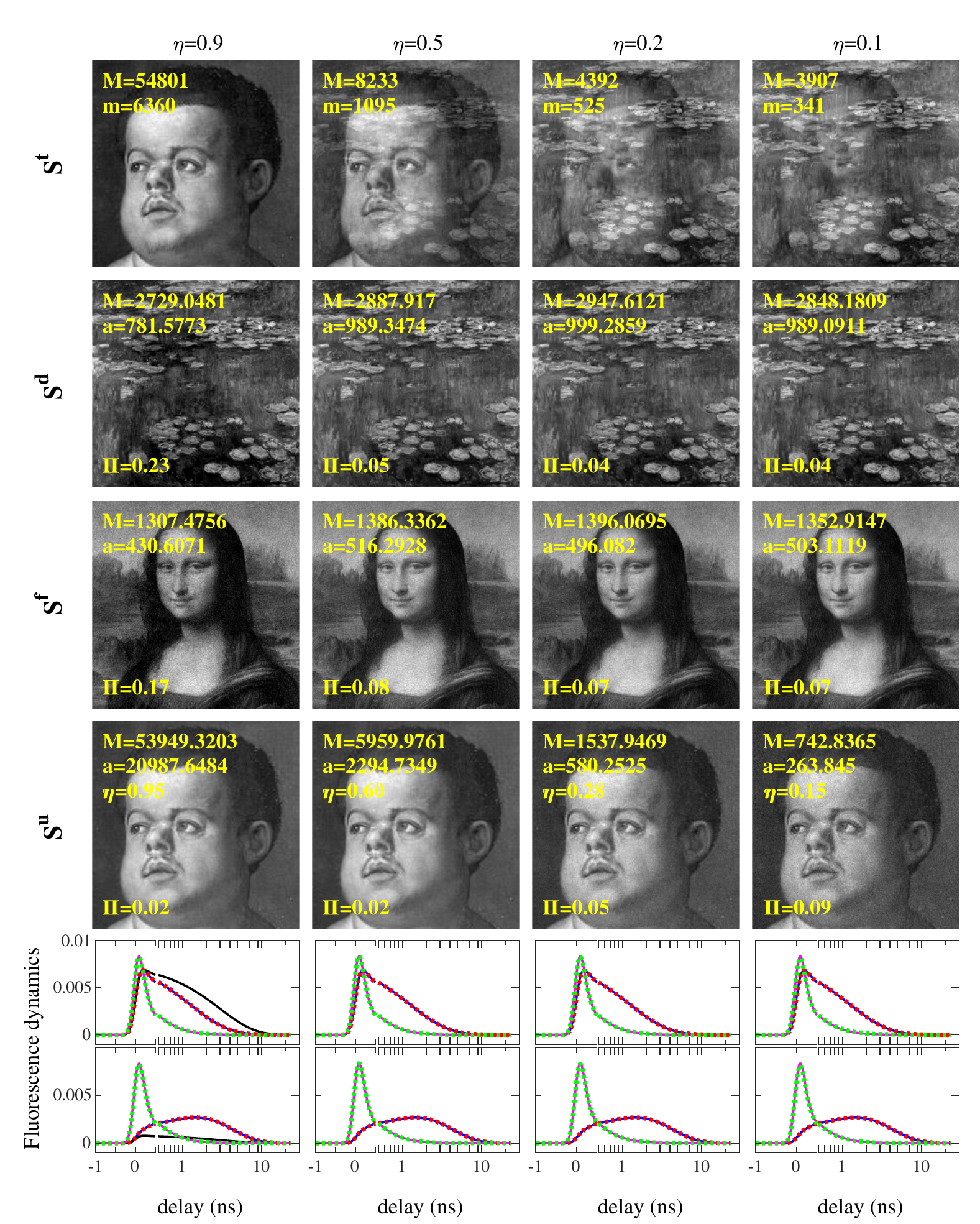}
\end{figure}
\begin{figure}
	\caption{Same as \Fig{FigSM_FRET_UnkAF_k=1}, but for $\kappa=0$: \add{Results of uFLIFRET for synthetic data generated using $\gbs=0.5$/ns, $\sgs=0.5$, $\qs=1$, $\Ib=2$ and $\kappa=0$ and including an additional autofluorescence component. We have used $\Id=10^4$ and $\Idap=500$. In this analysis, the autofluorescence dynamics is unknwown and retrieved from the algorithm.  The four columns refer to different fractions $\eta$ of the autofluorescence intensity \IAF\ in the total intensity, as labelled. The fifth row shows the retrieved spatial distribution of the autofluorescence \SAF, and the retrieved $\eta$. In the bottom row the nominal dynamics of the autofluorescence \bTAF\ (divided by three, magenta) has been added together with the retrieved dynamics (dashed green). The signal acquired at the donor (acceptor) detector are given in the top (bottom) panel, respectively. The dynamics are normalized to have a sum of unity over the 2000 temporal points of both detectors.}\label{FigSM_FRET_UnkAF_k=0}}
\end{figure}
\clearpage

\add{\section{Retrieval of FRET parameters in presence of multiple autofluorescent species.}\label{sec:resultsFRETNADHFAD} 
It is not uncommon in FLIM that multiple endogenous molecules contribute to the autofluorescence signal. For example, electron carriers, such as nicotinamide adenine dinucleotide (NADH) and flavin adenine dinucleotide (FAD), which are involved in the mitochondrial electron chain, can be used as endogenous reporter of mitochondrial functionality. The presence of multiple autofluorescence components is a challenging scenario for all methods presently used in the analysis of FLIFRET data, specifically considering that the binding of NADH or FAD to proteins alters their fluorescence lifetime, introducing an additional complexity. However, on can exploit the differences in the spectro-temporal properties of bound and unbound NADH and FAD fluorescence and the fluorescent proteins used as donors and acceptors in FLIFRET experiments to disentangle the various contributions, as we show here in silico. We employ three excitation wavelengths and three detection channels to provide sufficiently orthogonal information for uFLIFRET to analyze data which includes NADH and FAD, both bound and unbound, unpaired donors and acceptors, and DAPs. Synthetic data have been generated using selected paintings as the spatial distributions as follows: {\it Water Lilies} (unpaired donor), {\it Starry Night} (unpaired acceptor), {\it La Gioconda} (DAP), {\it Venus} (unbound NADH), {\it Nano Morgante} (bound NADH), {\it The Ambassadors} (unbound FAD), and {\it The Great Wave off Kanagawa} (bound FAD). \Tab{tab:exc} summarises the normalised absorbance of the different species as a function of the excitation wavelength $\lambda_e$.\,\cite{BeckerProcSPIE19S} We have assumed no change  of the spectral properties for the NADH and FAD upon binding.}

\begin{table}[]
	\begin{center}		
		\begin{tabular}{l|c|c|c} 
			\hline
			Fluorescent species & $\lambda_e=350$\,nm & $\lambda_e=440$\,nm & $\lambda_e=510$\,nm\\
			\hline\hline
			Donor (mNeonGreen) & 0.01 & 0.11 & 1\\
			\hline
			Acceptor (mRuby) & 0.125 & 0.0325 & 1\\
			\hline
			NADH & 1 & 0 & 0\\
			\hline
			FAD & 0.59 & 1 & 0.1\\
\hline		
\end{tabular}
		\caption{Relative absorbance of the different species at the excitation wavelengths.\label{tab:exc}}		
	\end{center}
\end{table}

\begin{table}[]
	\begin{center}		
		\begin{tabular}{l|c|c|c} 
			\hline
			Fluorescent species & $\ld=450-475$\,nm & $\ld=525-550$\,nm & $\ld=590-615$\,nm\\
			\hline\hline
			Donor (mNeonGreen) & 0 & 0.94 & 0.06\\
			\hline
			Acceptor (mRuby) & 0 & 0 & 1\\
			\hline
			NADH & 0.74 & 0.26 & 0\\
			\hline
			FAD & 0.01 & 0.77 & 0.22\\
			\hline		
		\end{tabular}
		\caption{Fraction of photons emitted by the different species into the detecion channels.\label{tab:det}}		
	\end{center}
\end{table}

\begin{table}
	\begin{center}		
		\begin{tabular}{l|c|c} 
			\hline
			Fluorescent species & $\tau$\,(ns) & $I$\\
			\hline\hline
			Donor (mNeonGreen) & 3 & 1000\\
			\hline
			Acceptor (mRuby) & 2.6 & 1000\\
			\hline
			DAP & 0.13 & 1000\\
			\hline
			unbound NADH & 0.4 & 200\\
			\hline
			bound NADH & 1 & 200\\
			\hline
			unbound FAD & 2.33 & 200\\
			\hline
		bound FAD & 0.13 & 200\\
			\hline		
		\end{tabular}
		\caption{Lifetimes $\tau$ and spatially-averaged spectro-temporally integrated photons $I$ of the different species used in the generation of synthetic data.\label{tab:lifet}}		
	\end{center}
\end{table}

\begin{figure}[]
	\includegraphics[width=\textwidth]{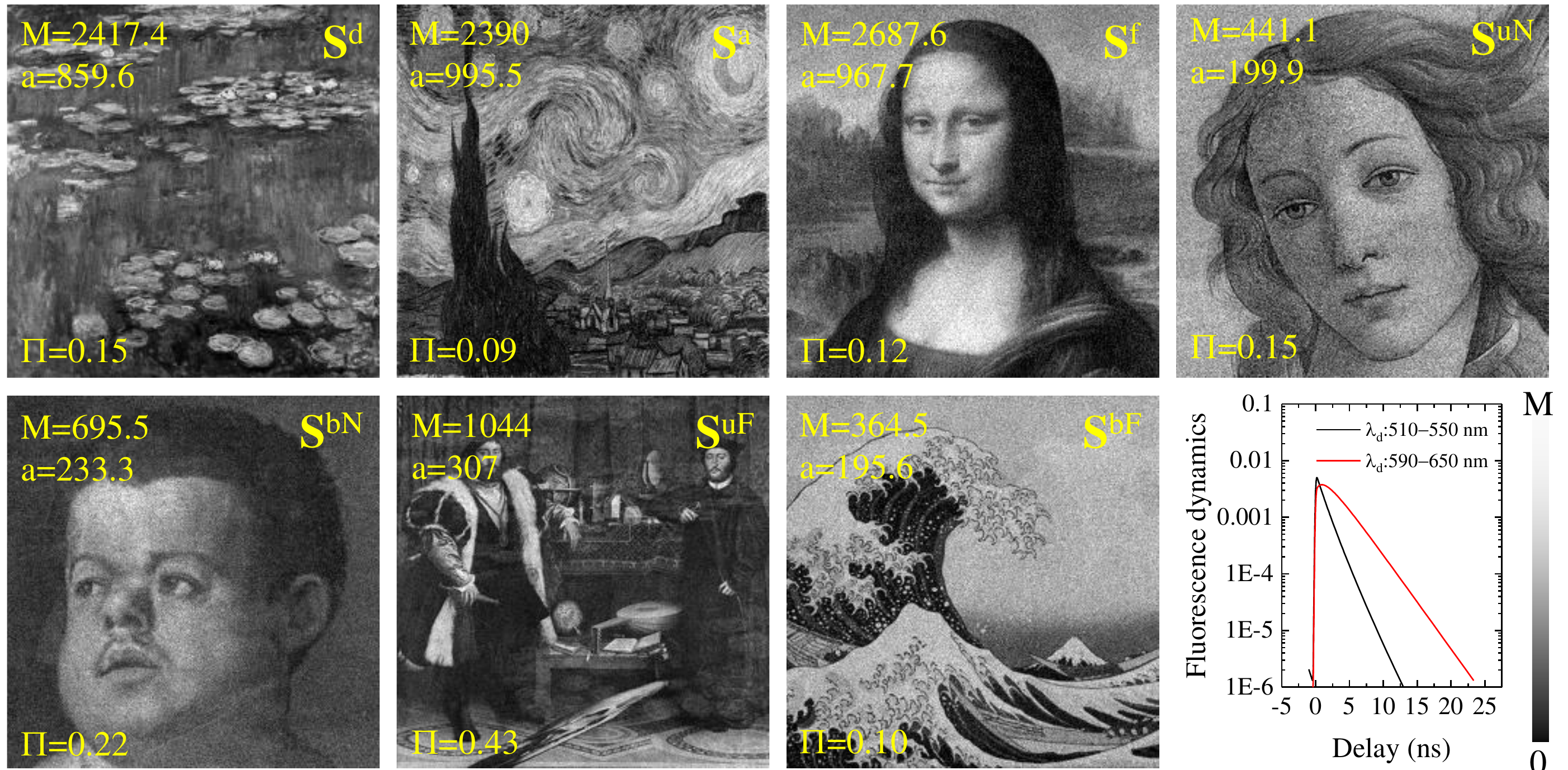}
	\caption{\add{Results of uFLIFRET of synthetic FLIT-FRET data including four different autofluorescence species. The grayscale images show the retrieved spatial distribution of the seven components (donor: D, acceptor: A, DAP: F, unbound NADH: uN, bound NADH: bN, unbound FAD: uF and bound FAD: bF). M indicates the maximum, while a the spatially averaged spectro-temporal integrated photons, $\Pi$ is the error with respect to the ground truth as discussed in the main manuscript. The plot shows the retrieved DAP dynamics as detected in the range $\ld=525-550$\,nm (black line) and $\ld=590-615$\,nm (red line). As expected, no signal observed over the $\ld=450-475$\,nm range.}}\label{FigSM_NADHFAD}
\end{figure}

\add{We have considered a relative peak absorbance between the different species to the donor of 0.32 (Acceptor, $\kappa$) and 1.0 (NADH and FAD), respectively.
	As detection channels, we have chosen three wavelength ranges ($\ld$), probing the difference in the emission spectra of the various species, and excluding the excitation wavelengths. The fraction of photons emitted into these three ranges by the various species are summaries in \Tab{tab:det}.}
\add{The time constants \cite{LakowiczPNAS92S,KolencARS19S,NakashimaJBC80S} used to generate the dynamics are shown in \Tab{tab:lifet}.}
\add{The DAP dynamics is calculated using the model described in the main manuscript, with parameters $\gbs=0.5$/ns, $\sgs=0.5$ and $q_s$=1.0. The spatially-averaged spectro-temporal-integrated detected photons $I$ for the different components are given in \Tab{tab:lifet}. Poisson noise is included in the generation of the hyperspectral image which undergo time binning (using \tb=0.025\,ns and \rb=0.05) before being analysed using uFLIFRET. The spectro-temporal properties of unpaired donor and acceptor as well as bound and unbound NADH and FAD were used as prior knowledge, which can be determined by uFLIM is suited reference samples. uFLIFRET then retrieves all spatial distributions and the FRET parameters. The resulting spatial distributions of the different components are shown in \Fig{FigSM_NADHFAD}, together with the retrieved DAP dynamics. All spatial distributions and the retrieved FRET parameters (\gbf$=0.52953\pm0.00012$\,ns, $\sgf=0.55289\pm0.00063$, $q=1.02310\pm0.00002$) are in good agreement with the ground truth.}

\clearpage
\section{Disentangling FRET from environment-induced lifetime changes}\label{sec:resultsFRETpH}
\add{The lifetimes of fluorescent molecules is affected by their local environment, which can be used as a sensing mechanism. For example, it has been shown that a variation in pH changes the lifetime of red fluorescing proteins\,\cite{TantamaJACS11S}. This effect however can confuse the analysis of FRET which also changes the dynamics. Disentangling the environmental effect from FRET is challenging for any analysis method. Here, we deal with this challenge in uFLIFRET by using two end-point components for each species, which are reflecting the properties at the minimum and maximum environmental parameter value considered. We assume that the effect of the environmental parameter can be approximated as linear mixing between the endpoints. While this is suited for small variations, larger variations could be treated extending the scheme to include not only endpoints but additional points within the dependence.}

\add{We treat here a single varying environmental parameter, and accordingly lock the relative admixtures between the endpoints for all species. We also assume here that the FRET rate is not depending on the environmental parameter, but note that this could be dropped by again defining different FRET at the two endpoints and locking also this. 
The scheme could also be extended to multiple environmental parameters, e.g. pH and temperature, in a straightforward way.}

\add{We consider the case a single donor and acceptor contributing to the emission. This provides two endpoints of unpaired donor, and two endpoints of unpaired acceptor, with their dynamics defined by the two extreme environmental conditions, e.g. the lowest and highest pH present in the sample. Using donor and acceptor endpoints, the two DAP endpoints for given FRET parameters are calculated. More complex cases with autofluorescence could be treated by adding such components as shown before.}

\add{We assume prior knowledge of the unpaired donor and acceptor dynamics, which can be obtained by unsupervised uFLIM of control samples. In uFLIFRET, the spatial distributions $\bS^{j}_{1,2}$ are calculated for a specific set of FRET parameters, where the index 1,2 refers to the two endpoints and the superscript $j\in\{{\rm d,a,f}\}$ refers to the species (donor, acceptor or DAP). Using  these, the spatial distribution of the enviromental parameter $\Se$ changing from $1$ to $-1$ between the endpoints is calculated point-wise by}
\be
\Se=\frac{\left(\Sd_1+\Sa_1+2\Sf_1\right)-\left(\Sd_2+\Sa_2+2\Sf_2\right)}{\left(\Sd_1+\Sa_1+2\Sf_1\right)+\left(\Sd_2+\Sa_2+2\Sf_2\right)}\,.
\ee
\add{Note that this expression uses an average between species, recognizing that the retrieved parameter for each individual species is varying due to noise and systematics. Using \Se, the end-point spatial distributions of the species are constraint to}
\be
\bS^{j,c}_1=(1+\Se)\frac{\bS^j_1+\bS^j_2}{2}\,, \quad
\bS^{j,c}_2=(1-\Se)\frac{\bS^j_1+\bS^j_2}{2}\,,
\ee
 \add{which enforces the condition of a common environmental distribution \Se. The factorisation error is calculated for the constraint distributions and the FRET parameters minimising this error are found.}
 
 \add{To validate this method in silico, we have generated synthetic data using monoexponential dynamics for the unpaired donor (acceptor) with a linear gradient of lifetime varying from 3 to 2.5\,ns (2.6 to 2.16\,ns), to simulate the effect of the environment on the dynamics. This choice is reflecting the typical effect of the local pH changing between 7 and 6\,\cite{TantamaJACS11S}. We have used the following FRET parameters: $\gbs=0.9$/ns, $\sgs=0.5$ and \qs=1, and allowed for direct excitation of the acceptor ($\kappa=1$). The temporal integrated photons for the different species were $\Id=\Idap=10^4$ and $\Ia=8\times10^3$, respectively. Two detection channels, centred at the emission of the donor and  acceptor, respectively, are considered, with a 10\% bleed-through ($R^d=R^a=0.9$). A time binning with \tb=0.025\,ns and \rb=0.05 and a partial whitening with $xi=0$ is used in the uFLIFRET analysis. The spatial distribution of donor, acceptor, and DAP are given by the same paintings as in the main text, and the spatial distribution of the environmental parameter was chosen as a linear gradient from 1 to -1 from left to right.}
 
\begin{figure}
	\includegraphics[width=\textwidth]{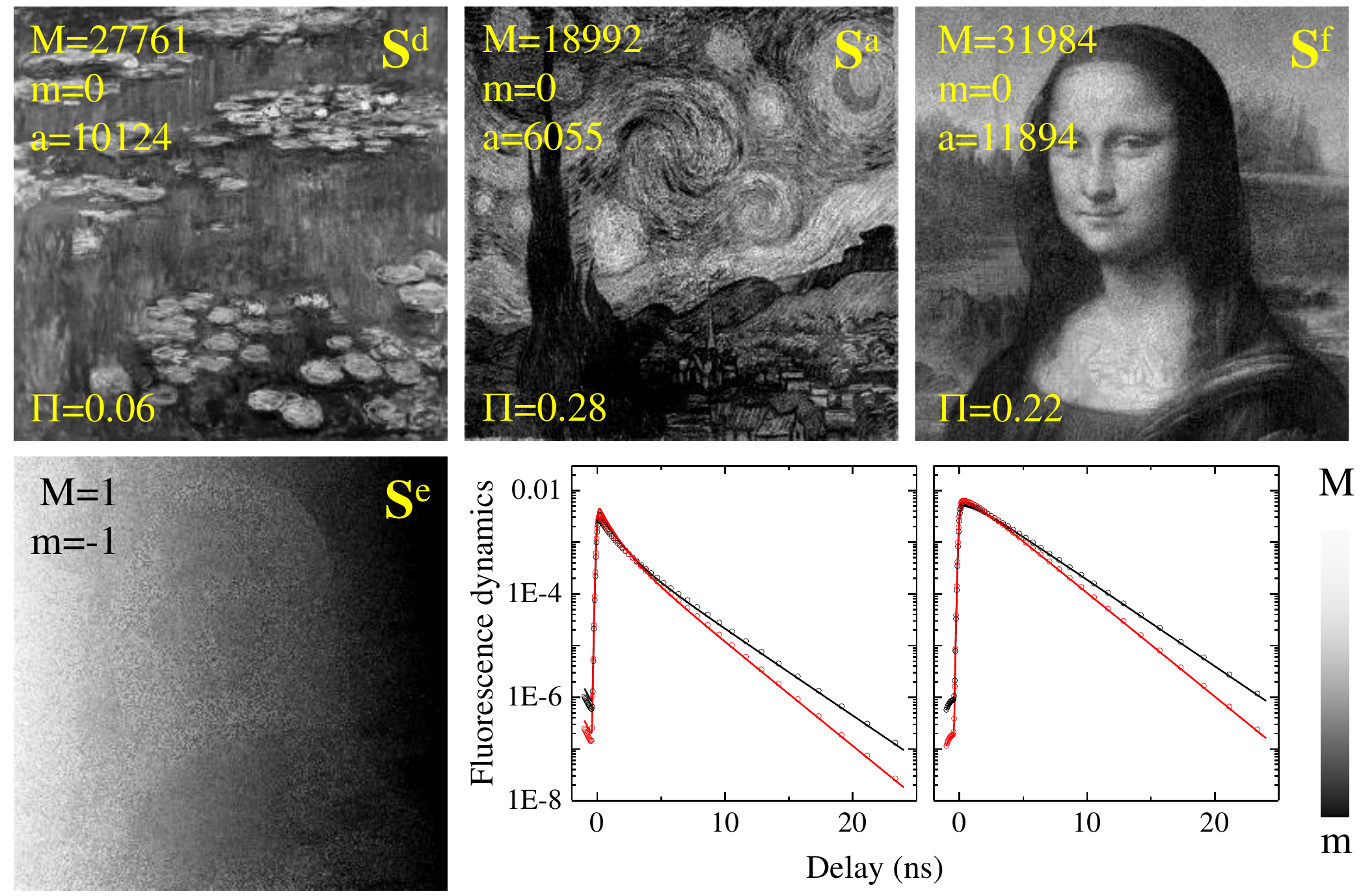}
	\caption{\add{Results of uFLIFRET on synthetic data including environment-induced changes of donor and acceptor lifetimes. The grayscale images show the spatial distribution of the retrieved components, from m to M as indicated. The spatially averaged spectro-temporal integrated number of photons $a$, $\Pi$ is the error with respect to the ground truth as discussed in the main manuscript. \Se\ shows the spatial distribution of the environment parameter. The DAP dynamics is shown in the graph with lines giving the ground truth and symbols the giving the uFLIFRET result. The dynamics for both endpoints are shown, corresponding to the longest (black) and  shortest (red) lifetime. The left (right) panel refers to the detector centred at the donor (acceptor) emission.}}\label{FigSM_Environment}
\end{figure}

\add{\Fig{FigSM_Environment} shows the results of the analysis, with the retrieved total unpaired donor ($\mathbf{S}^{\rm d,c}=\mathbf{S}^{\rm d,c}_1+\mathbf{S}^{\rm d,c}_2$), acceptor ($\mathbf{S}^{\rm a,c}=\mathbf{S}^{\rm a,c}_1+\mathbf{S}^{\rm a,c}_2$) and DAP ($\mathbf{S}^{\rm f,c}=\mathbf{S}^{\rm f,c}_1+\mathbf{S}^{\rm f,c}_2$) and the  environment spatial distributions \Se\ in good agreement with the ground truth. The FRET parameters minimising the factorisation error are $\gbf=1.06\pm0.01$/ns, $\sgf=0.86\pm0.01$ and $\qf=1.443\pm0.005$, in resonable agreement with the ground truth.}

\add{These in-silico results show that it is in principle possible to disentangle environmental effects on lifetimes from FRET using uFLIFRET. It will be interesting to see the method applied to experimental data in the near future.}

\clearpage

\section{Lifetime map from uFLIFRET analysis of Arabidopsis root experiment}\label{sec:HeatMap}
\begin{figure}[b]
	\includegraphics[width=0.5\textwidth]{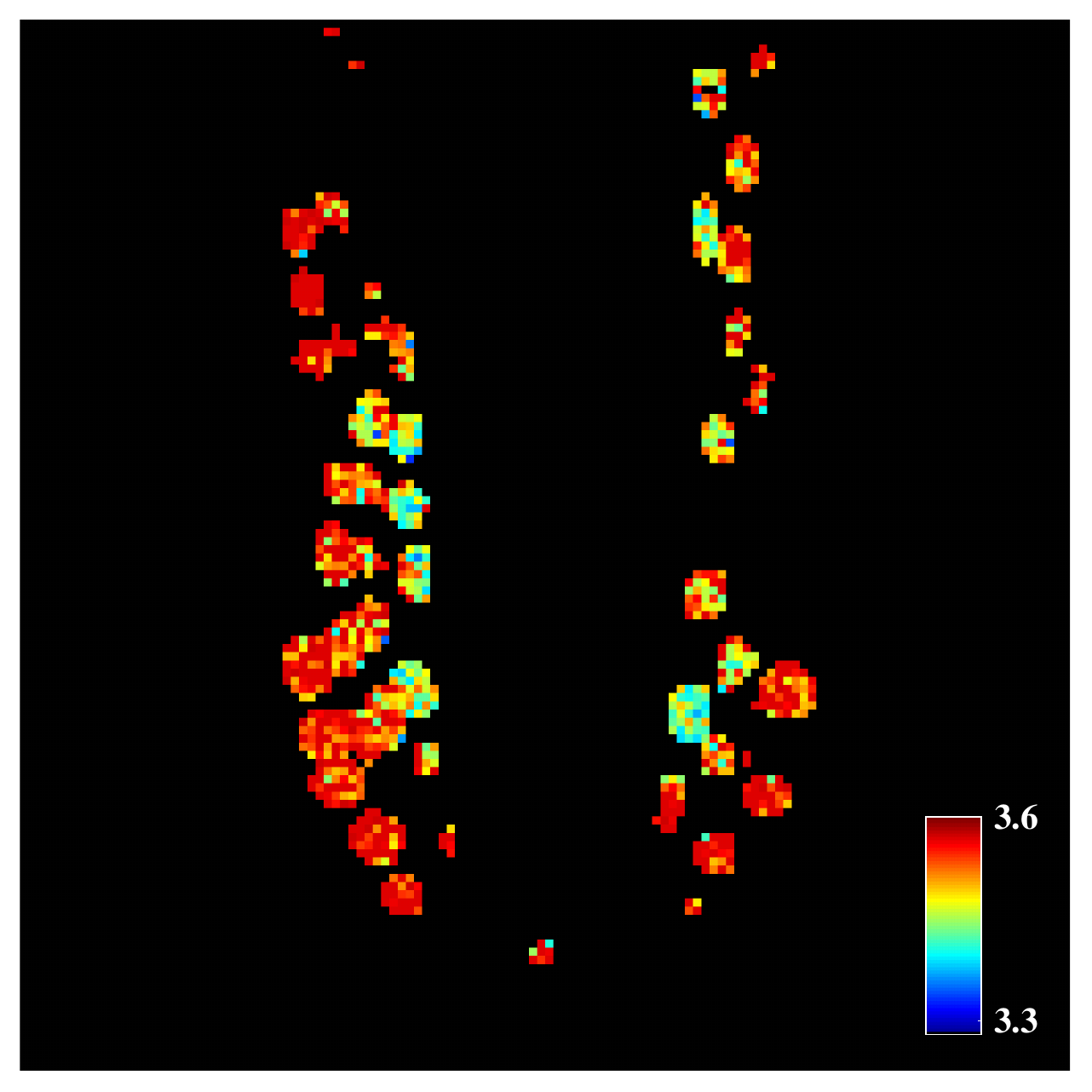}
	\caption{Map of the average lifetime $\langle\tau\rangle$ (in ns) as defined in the text. We report values only for pixels with $\Sd>1700$.}
	\label{FigSM_Arabidopsis_HeatMap}
\end{figure}

To compare the results of the uFLIFRET analysis on the experiments on Arabidopsis root with the single pixel lifetime fitting method (see \Onlinecite{LongN17}), we have calculated the average lifetime using the obtained distributions as weights, i.e.
\be
\langle\tau\rangle=\frac{\Sd\tau_{\rm d}+\Sf\tau_{\rm f}}{\Sd+\Sf}.
\ee
The resulting values are shown in \Fig{FigSM_Arabidopsis_HeatMap}

\clearpage

\section{Analysis of FRET-FLIM data using gradient descent method}
\Fig{FigSM_FRET_g=5_k=1_Ib=2_GD} shows the results of the factorisation of the data presented in \Fig{Fig_FRET_g=5_k=1_Ib=2} using the gradient descent method. The spatial distribution and FRET parameters obtained are very similar to the case of fast NMF. At low signal to noise ratio the retrieved value of \sgf\ is similar to what observed with the fast NMF, suggesting that the systematic error observed can not be attributed to the approximate treatment of the noise by the fast NMF. However, we found that the gradient descent can be superior to the fast NMF for data with significant dark counts. For example, for the data with $\Id=100, \Idap=50, \Ib=20$ and $\gbs=0.5$/ns, $\sgf=0.5$ and $\qs=1$ where the fast NMF could not find a solution within the allowed parameter range (see \Fig{FigSM_FRET_ParamError_g=5_k=1_Ib=20}), the gradient descent retrieved \gbf\ and \qf\ with a relative error of $\sim10\%$. The retrieved distribution width \sgf\ however was close to zero, similar to the fast NMF results with low signal to noise ratio. The gradient descent in this case was approximately three orders of magnitude more computationally expensive than the fast NMF.

\begin{figure}
	\includegraphics[width=0.7\textwidth]{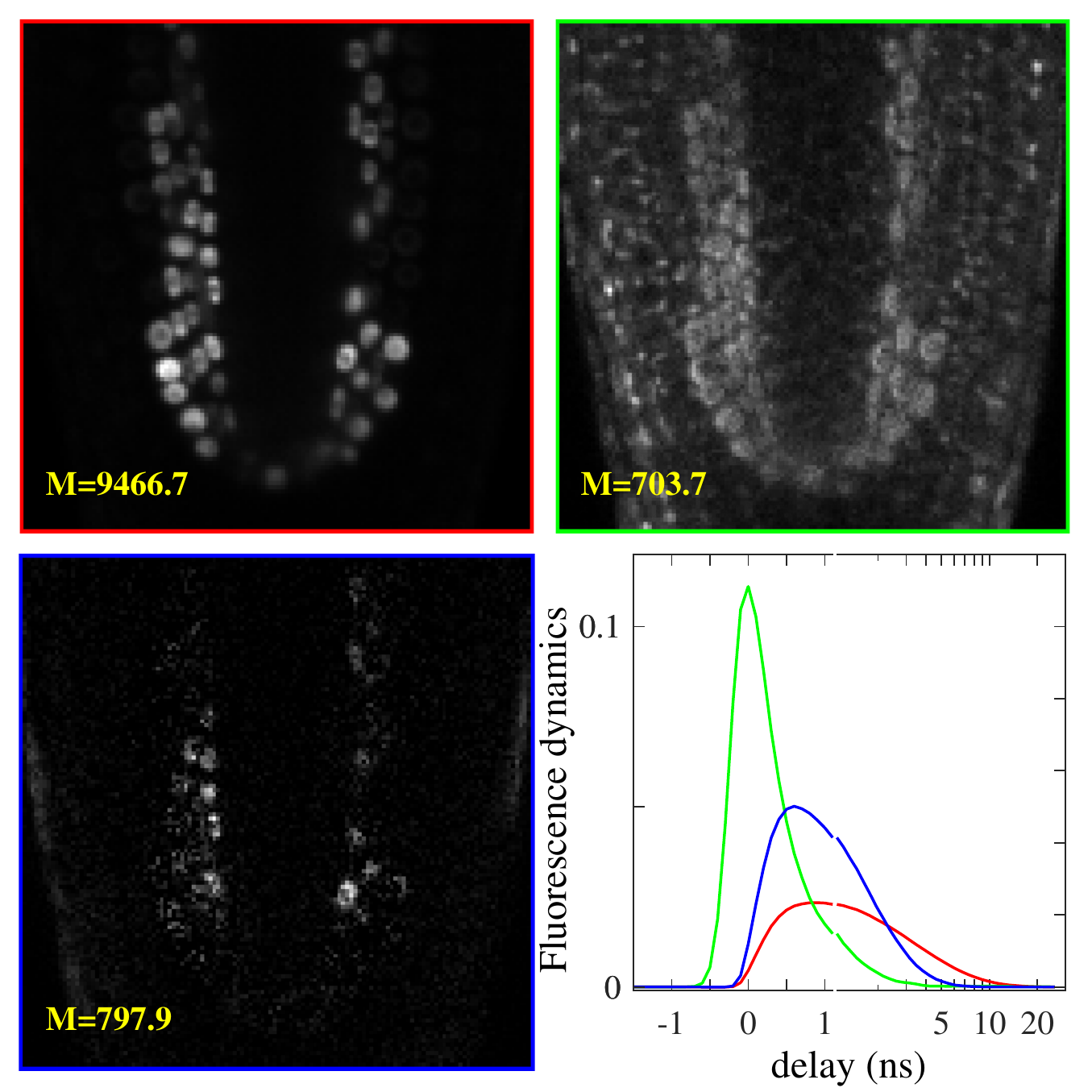}
	\caption{\add{Same as \Fig{Fig_Arabidopsis} but using the KLD and gradient descent: Results of the analysis of the \add{{\it Arabidopsis} root} data using the the KLD and gradient descent. } The retrieved FRET rate distribution parameters are $\gbf=0.46/$ns and $\sgf \approx 0$.}\label{FigSM_Arabidopsis_GD}
\end{figure}

\begin{figure}
	\includegraphics[width=0.5\textwidth]{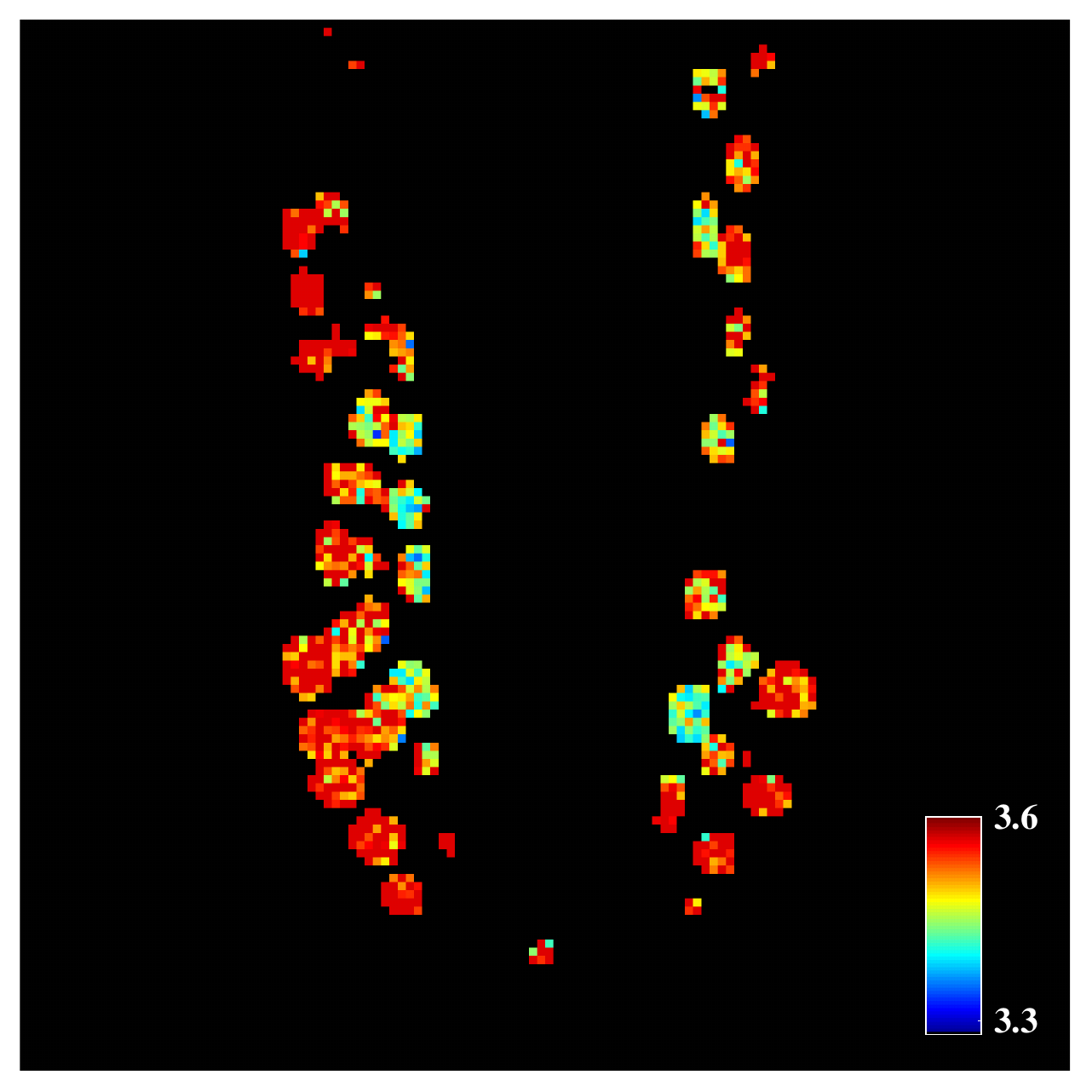}
	\caption{\edit{Same as \Fig{FigSM_Arabidopsis_HeatMap}}{Map of the average lifetime $\langle\tau\rangle$ (in ns)} obtained from the factorisation results using the gradient descent.}
	\label{FigSM_ArabidopsisGD_HeatMap}
\end{figure}

The results on experimental data show a good agreement between the fast NMF and the gradient descent (see \Fig{FigSM_Arabidopsis_GD}). The average lifetime map obtained with the gradient descent results is shown in \Fig{FigSM_ArabidopsisGD_HeatMap}

\begin{figure}
	\includegraphics[width=0.9\textwidth]{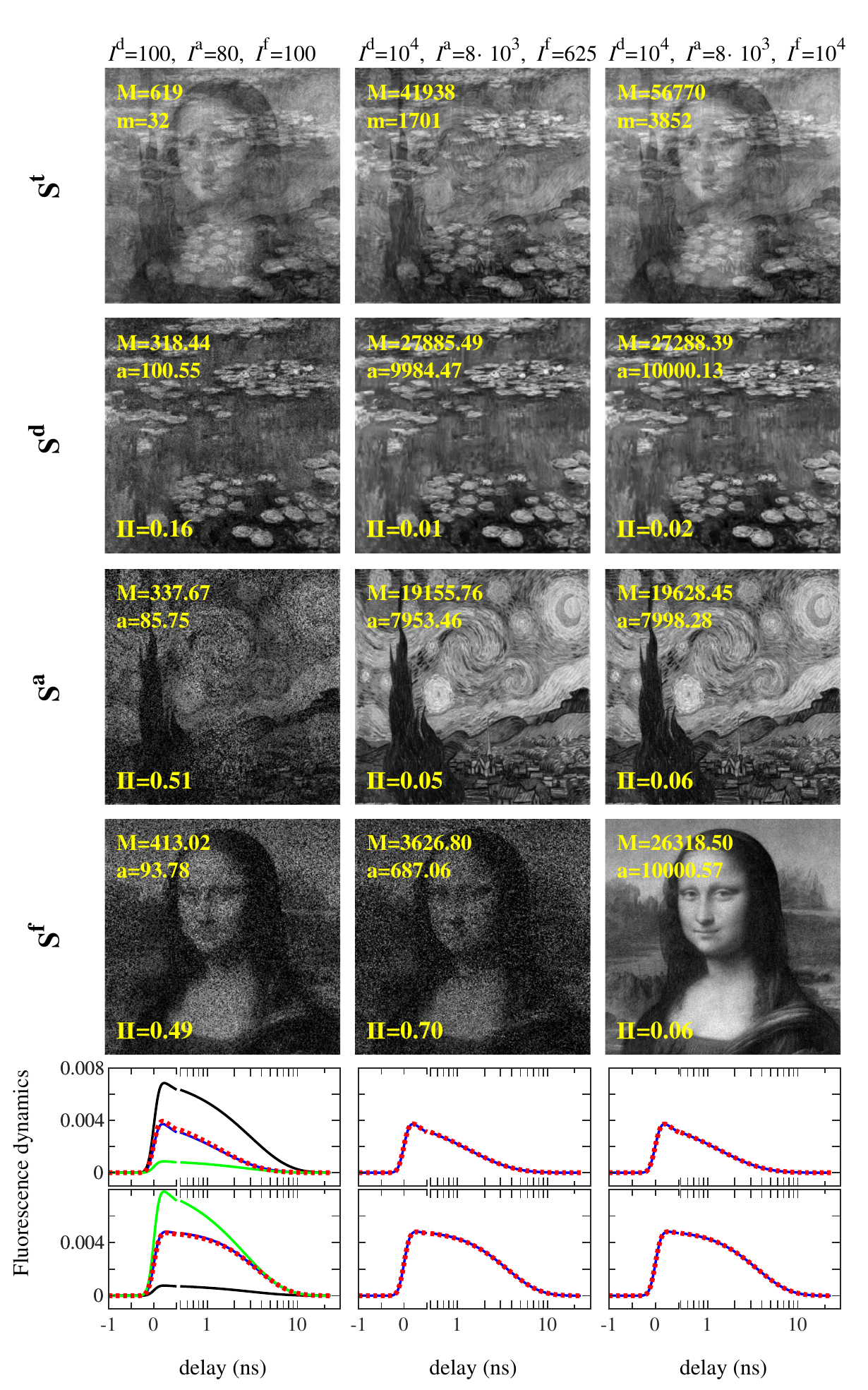}
\end{figure}
\begin{figure}
	\caption{\add{Same as \Fig{Fig_FRET_g=5_k=1_Ib=2} but using the KLD and gradient descent: Results of uFLIFRET analysis. Synthetic data generated with $\gbs=0.5$/ns, $\sgs=0.5$, $\qs=1$, $\Ib=2$ and $\kappa=1$}.  The retrieved FRET rate distribution parameters are $\gbf=(462.86\pm 0.21)/\mu$s, $\sgf=0 \pm 0.0015$ and $\qf=0.92506\pm 0.00018$ for the first column, $\gbf=(485.92\pm 0.10)/\mu$s, $\sgf=0.47312 \pm 0.00006 $ and $\qf=0.99829  \pm 0.00003 $ for the second column, and $\gbf=(499.90\pm 0.097)/\mu$s, $\sgf=0.4998 \pm 0.00065$, and $\qf=1.00035 \pm 0.00002$ for the third column.}\label{FigSM_FRET_g=5_k=1_Ib=2_GD}
\end{figure}

\clearpage

\end{document}